\documentclass[
fontsize=12pt,          
numbers=noenddot,      
listof=totoc,        	
bibliography=totoc,  	
headsepline=true,       
footsepline=false, 		
DIV=12                	
]{scrartcl}
\addtokomafont{disposition}{\boldmath}

\usepackage[utf8]{inputenc}
\usepackage[T1]{fontenc}
\usepackage[english]{babel}

\usepackage{enumitem}

\usepackage{graphicx}
\usepackage[table,x11names]{xcolor}

\usepackage{amsmath, amsfonts, amssymb, bbm, bm, mathabx}
\usepackage{csquotes}
\usepackage[citestyle = authoryear,
bibstyle = authoryear, dashed = false, giveninits, 
backend=bibtex, maxnames = 2, maxbibnames = 10, uniquename = false, uniquelist = false]{biblatex} 
\renewbibmacro*{volume+number+eid}{%
	\printfield{volume}%
	\setunit*{\addnbthinspace}%
	\printfield{number}%
	\setunit{\addcomma\space}%
	\printfield{eid}}
\DeclareFieldFormat[article]{number}{\mkbibparens{#1}}

\usepackage[a4paper, left = 2cm, right = 2cm, top = 2cm, bottom = 2cm]{geometry}

\usepackage{booktabs} 	
\usepackage{longtable} 
\usepackage{tabularx}
\usepackage{xltabular}
\usepackage{lscape}
\usepackage{hyperref} 
\usepackage{cleveref}

\usepackage{algorithm}
\usepackage{algpseudocode}

\usepackage{todonotes}
\usepackage{lscape}
\usepackage{nowidow}
\usepackage{placeins}
\usepackage[bf, format = plain]{caption}
\usepackage{subcaption}

\usepackage{fontawesome5}
\usepackage{tikz}
\usetikzlibrary {graphs} 

\usepackage{orcidlink}

\DeclareNameAlias{sortname}{family-given}

\renewbibmacro*{publisher+location+date}{%
	\iflistundef{publisher}
	{\setunit*{\addcomma\space}}
	{\setunit*{\addcolon\space}}%
	\printlist{publisher}%
	\setunit*{\addcomma\space}%
	\printlist{location}%
	\usebibmacro{date}%
	\newunit}

\renewbibmacro*{volume+number+eid}{%
	\printfield{volume}%
	\setunit*{\addnbspace}
	\printfield{number}%
	\setunit{\addcomma\space}%
	\printfield{eid}}
\DeclareFieldFormat[article]{number}{\mkbibparens{#1}}

\AtEveryBibitem{\clearlist{language}}
\AtEveryBibitem{\clearfield{month}}
\AtEveryCitekey{\clearfield{month}}

\AtEveryBibitem{%
	\ifentrytype{book}{
		\clearfield{url}%
		\clearfield{urlyear}%
	}{}
	\ifentrytype{article}{
		\clearfield{url}%
		\clearfield{urlyear}%
	}{}
	\ifentrytype{collection}{
		\clearfield{url}%
		\clearfield{urlyear}%
	}{}
	\ifentrytype{incollection}{
		\clearfield{url}%
		\clearfield{urlyear}%
	}{}
}

\DeclareMathOperator{\Var}{\mathbb{V}\text{ar}}
\DeclareMathOperator{\E}{\mathbb{E}}
\DeclareMathOperator{\Prob}{\mathbb{P}}
\DeclareMathOperator{\Cov}{\mathbb{C}\text{ov}}

\newcommand{\dif}{\mathop{}\!\mathrm{d}}

\title{\Large An Empirical Comparison of Methods for Quantifying the Similarity of Categorical Datasets}

\author{\normalsize Marieke Stolte$^1$\thanks{Corresponding author, e-mail: \texttt{stolte@statistik.tu-dortmund.de}}\orcidlink{https://orcid.org/0009-0002-0711-6789} \and\normalsize Jörg Rahnenführer$^1$\orcidlink{https://orcid.org/0000-0002-8947-440X} \and\normalsize Andrea Bommert$^1$\orcidlink{https://orcid.org/0000-0002-1005-9351}}
\date{\small$^1$Department of Statistics, TU Dortmund University}

\begin{document}
	\maketitle
	\section*{Abstract}
	Quantifying the similarity of two or more datasets has widespread applications in statistics and machine learning. 
	The method choice is, however, difficult due to the abundance of proposed methods and the lack of neutral comparison studies, especially for categorical data. 
	Here, the most promising methods are compared concerning their ability to detect certain differences between datasets and their resource consumption.  
	The results show that the edge count tests perform well when comparing two datasets (i.e., the two-sample case). 
	For certain scenarios, the constrained minimum (CM) distance performs even better.
	For categorical data consisting of variables with five categories each, the best method depends on the type of difference between the distributions, with either the CM distance and certain graph-based tests performing best, or the classifier-based tests (C2ST). 
	This tendency is even clearer for multiple datasets.
	Overall, the Friedman-Rafsky test can be recommended for two samples as a compromise of high performance, acceptable resource consumption, and computational error occurrences. 
	For the multi-sample case, the Multi-Sample Mahalanobis Cross-Match (MMCM) test can be recommended due to its comparably good performance and low resource consumption.
	\paragraph{Keywords:} dataset similarity; two-sample test; multi-sample test; neutral comparison

	\section{Introduction}
	Methods for quantifying the similarity of two or more datasets are relevant in many applications in statistics and machine learning. 
	One typical application is two- and $k$-sample testing, where the hypothesis of equal distributions is checked. 
	These tests can, for example, be used to compare groups, e.g.\ treated vs.\ untreated which is a common task in many applications in the biomedical or social sciences.
	There are, however, other applications like meta- or transfer learning or the comparison of simulated and real-world datasets that do not require a hypothesis test but are gaining relevance in practice. 
	With the rise of machine learning in all fields of data analysis, the former task becomes more relevant in many fields. 
	The latter task of comparing simulated and real-world datasets is becoming more important with the development of synthetic datasets in fields where privacy is of concern, e.g.\ again in medical applications where patient data cannot be published. 
	
	\textcite{stolte_methods_2024} performed an extensive review of methods for quantifying the similarity of multivariate datasets and presented a taxonomy of such methods as well as a theoretical comparison that rated the applicability, interpretability, and theoretical properties of each method. 
	This comparison does, however, not cover the performance of the methods in practice. 
	There are some simulations in the literature that evaluate the performance of newly presented methods or compare new methods with parametric or univariate alternatives \textcite{friedman_multivariate_1979,  schilling_multivariate_1986, baringhaus_new_2004, rosenbaum_exact_2005, yu_two-sample_2007, baringhaus_rigid_2010, zhang_graph-based_2022}.
	There are also some limited comparisons of dataset similarity methods \parencite{szekely_testing_2004, gretton_kernel_2012, biswas_distribution-free_2014, biswas_nonparametric_2014, petrie_graph-theoretic_2016, chen_new_2017, lopez-paz_revisiting_2017, chen_weighted_2018, pan_ball_2018, mukhopadhyay_nonparametric_2020, hediger_use_2021, li_measuring_2022, mukherjee_distribution-free_2022, song_gtestsmulti_2022, zaremba_b_2022, huang_kernel_2022, song_generalized_2021}. 
	None of these is a neutral comparison study in the sense of \textcite{boulesteix_plea_2013}, though, since all are conducted in the context of presenting new methods. 
	Neutral comparison studies are, however, very important for making informed method choices \textcite{boulesteix_plea_2013}. 
	Moreover, previous studies almost exclusively compare the power of asymptotic or permutation / Bootstrap two- or $k$-sample tests for numeric data. 
	
	There is a lack of studies for categorical data and for methods that do not define a two- or $k$-sample test. 
	However, in practice, many datasets consist of categorical variables, e.g.\ nominal variables like gender, or ordinal Likert scale data, especially in biomedical or social sciences as relevant application areas \textcite{preisser_categorical_1997, agresti_categorical_2013, azen_categorical_2021, chen_categorical_2023}. 
	For example, many datasets include demographic data like gender, highest level of education, or marital status that are typically categorical. 
	Medical datasets commonly include nominal variables like smoking status, blood type, and pre-existing conditions, as well as ordinal variables like pain levels or tumor stages.
	In social sciences, questionnaire data is often recorded on Likert scales. 
	Moreover, it is common practice in many fields to categorize variables like age or monthly salary for privacy reasons.  
	
	To close the gap of missing guidance on choosing a method for comparing two or more categorical datasets, an extensive comparison of methods for quantifying the similarity of categorical datasets is performed. 
	This comparison study is neutral in the sense that its focus is the comparison itself, and none of the authors of the current study were involved in the development of any of the compared methods. 
	The most promising methods that are applicable to categorical data are selected from the theoretical comparison. 
	
	The study aims are to compare how good the methods are at detecting certain differences between datasets. It is not expected to find a single method that compares best in all scenarios \textcite{strobl_against_2024}. Therefore, the goal is rather to identify groups of methods that act similarly across different analysis scenarios and determine which deviations between datasets these groups of methods can detect well.
	An additional goal is to find out which methods are computationally feasible and numerically stable. 
	Note that since the comparison is not limited to two- and $k$-sample tests, this study does not conduct power comparisons. 
	A similar quantity is used instead, which compares the statistic values simulated for datasets drawn from different distributions to those values simulated for datasets drawn from the same distribution. 
	
	The remaining manuscript is structured as follows. 
	In Section~\ref{sec:setup}, the simulation setup is presented according to the ADEMP structure \textcite{morris_using_2019}. 
	Next, in Section~\ref{sec:sensitivity}, the results of the method comparison with respect to the ability to detect certain differences in datasets are presented. 
	Afterward, in Section~\ref{sec:applicability}, the runtime, memory consumption, and occurring numerical errors of the methods are compared. 
	Section~\ref{sec:best.meth} summarizes the findings in the preceding sections into recommendations for choosing an appropriate method.
	Last, in Section~\ref{sec:summary}, the results are discussed, and an outlook on open research questions is given.

	\section{Simulation Setup}\label{sec:setup}
	The following describes the simulation setup according to the \textit{ADEMP} \textcite{morris_using_2019} structure (aims, data-generating mechanisms, estimands / other targets, methods, and performance measures).
	
	\subsection{Aims}
	The aims of the simulation study are to: 
	\begin{enumerate}
		\item Compare dataset similarity measures with respect to their performance in detecting differences of datasets drawn from distributions that differ in certain aspects, and to identify groups of dataset similarity measures that act similarly across different alternatives.
		\item Compare dataset similarity measures with respect to their consumption of computational resources.
	\end{enumerate}
	
	\subsection{Data-Generating Mechanisms}
	In the following, the data-generating mechanisms of the simulation study are explained. 
	The data-generating mechanisms can be divided into the following cases: 
	\begin{enumerate}
		\item Comparison of two datasets
		\begin{enumerate}
			\item without a target variable,
			\item with a target variable.
		\end{enumerate}
		\item Comparison of four datasets (without a target variable).
	\end{enumerate}
	In each case, multiple true data-generating mechanisms for the datasets are considered. 
	Two or more datasets are generated from the same underlying distributions or different underlying distributions. 
	For the two-sample case, datasets with an additional target variable are created, as well as datasets without such a target variable. 
	For the $k$-sample case, none of the methods for multiple samples can appropriately consider a target variable in the data. 
	The $k$-sample case has different options for how many and which distributions can differ. 
	The number of possible settings increases with increasing $k$. 
	Here, only $k = 4$ is considered for the $k$-sample case as a compromise between comparing multiple samples but still having a reasonably low number of possible settings.
	For $k = 4$, there are four possible settings for how many distributions differ from each other: 
	\begin{enumerate}[label=\alph*)]
		\item $3 + 1$:
		One distribution differs from the others, which are equal, e.g.\ $F_1 = F_2 = F_3 \ne F_4$.
		\item $2 + 2$:
		Two groups of two distributions each, where the distributions within the groups are equal but the distributions between the groups differ, e.g.\ $F_1 = F_2 \ne F_3 = F_4$.
		\item $2 + 1 + 1$:
		Two distributions are equal, and the other two distributions are different from these and each other, e.g.\ $F_1 = F_2 \ne F_3 \ne F_4, F_1\ne F_4$. 
		\item $1 + 1 + 1 + 1$:
		All distributions differ, $F_i \ne F_j, i\ne j\in\{1,\dots,4\}$.
	\end{enumerate}
	Settings a)--c) have mostly been neglected previously \textcite[e.g.][]{mukherjee_distribution-free_2022, song_new_2022}. 
	
	\subsubsection{Numbers of Observations and Variables}
	For each setting of the underlying distributions, different numbers of variables and observations are considered. 
	Additionally, the imbalance of the number of observations of different datasets is also varied, as this might impact the method performance \textcite{chen_weighted_2018}.
	The number of variables $p$ is varied over $p \in \{2, 10, 50\}$, which represents low- to middle-dimensional data.
	The lower-dimensional data is chosen since not all methods are intended for high-dimensional data, and many previous studies only considered lower numbers of variables as well \textcite{friedman_multivariate_1979, schilling_multivariate_1986, baringhaus_new_2004, szekely_testing_2004, rosenbaum_exact_2005, baringhaus_rigid_2010, lopez-paz_revisiting_2017, pan_ball_2018,  li_measuring_2022}.
	Moreover, the runtime of many methods increases both in $p$ and in the sample size $N$ such that high values are infeasible in the scope of a simulation study.
	For $k = 2$, the overall sample size is varied as $N \in \{50, 100, 200, 500, 1000\}$, and the individual sample sizes are set to $n_1 = \pi\cdot N$ and $n_2 = (1-\pi)\cdot N$ with $\pi\in\{0.2, 0.5\}$ to cover typical sample sizes and one balanced and unbalanced sample size setting, respectively.
	For $k = 4$, the sample size is varied as $N \in \{100, 200, 400\}$ and the individual sample sizes are set to $n_1 = n_2 = n_3 = n_4 = 0.25\cdot N$ or $n_1 = 0.1\cdot N, n_2 = 0.2\cdot N, n_3 = 0.3\cdot N, n_4 = 0.4\cdot N$.
	A full factorial design is used, i.e.\ all combinations of $p$, $N$, and the settings for the individual sample sizes are used in each scenario, and in the four-sample case, also for each of the settings a)--d).

	\subsubsection{Generation of Categorical Data}
	For categorical data, there are not many simulation studies comparing dataset similarity methods. 
	\textcite{chen_ensemble_2013} compare their tests to the $\chi^2$ test for $p = 1$ and use normal and uniform distributions discretized into twelve classes. 
	For the alternatives, the second distribution is shifted, which changes the class distribution.
	This approach is inflexible with respect to the resulting deviations of the class distributions.
	Here, data are generated in a more flexible way using Bernoulli distributions and multinomial distributions with five classes, motivated by the common use of binary and 5-point Likert scale-like data, e.g.\ in questionnaire data in social sciences or for rating the severity of disease in medical applications. 
	For most methods, a potential ordering of the categories is not taken into account. 
	Some methods use the ordering indirectly by using the Euclidean distances of the observations.
	Only independent variables are considered in this study. 
	
	\begin{figure}[!b]
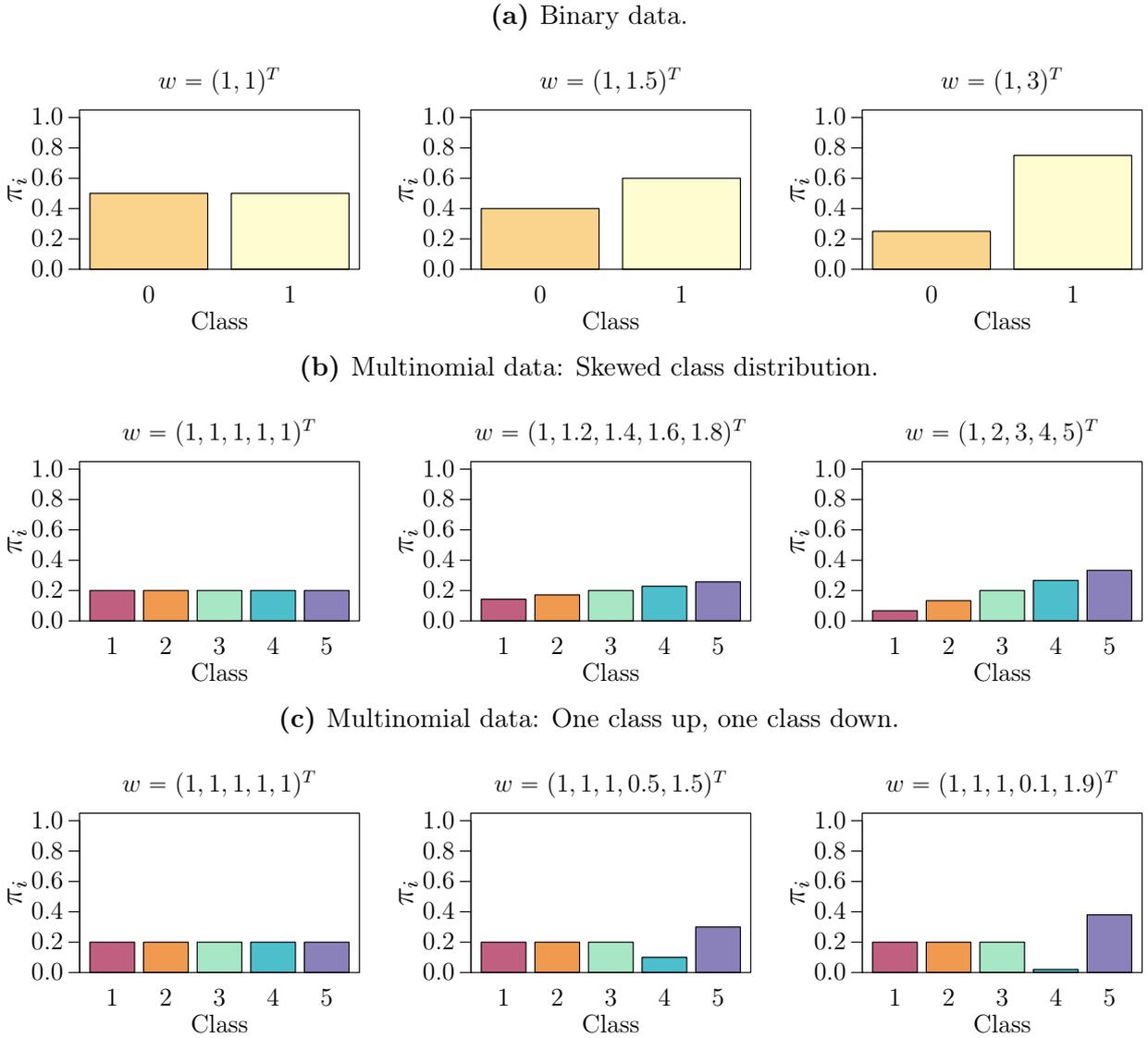

		\centering
		\begin{subfigure}{\linewidth}
			\caption{Binary data.}\label{fig:prob.bin}\smallskip
			\centering
			\resizebox{\linewidth}{!}{%
\begin{tikzpicture}[x=1pt,y=1pt]
\definecolor{fillColor}{RGB}{255,255,255}
\path[use as bounding box,fill=fillColor,fill opacity=0.00] (0,0) rectangle (433.62,108.41);
\begin{scope}
\path[clip] (  0.00,  0.00) rectangle (144.54,108.41);
\definecolor{drawColor}{RGB}{0,0,0}
\definecolor{fillColor}{RGB}{249,194,92}

\path[draw=drawColor,line width= 0.4pt,line join=round,line cap=round,fill=fillColor,fill opacity=0.70] ( 36.30, 24.55) rectangle ( 79.80, 52.79);
\definecolor{fillColor}{RGB}{254,253,190}

\path[draw=drawColor,line width= 0.4pt,line join=round,line cap=round,fill=fillColor,fill opacity=0.70] ( 88.50, 24.55) rectangle (132.00, 52.79);
\end{scope}
\begin{scope}
\path[clip] (  0.00,  0.00) rectangle (433.62,108.41);
\definecolor{drawColor}{RGB}{0,0,0}

\node[text=drawColor,anchor=base,inner sep=0pt, outer sep=0pt, scale=  0.79] at ( 58.05, 12.28) {0};

\node[text=drawColor,anchor=base,inner sep=0pt, outer sep=0pt, scale=  0.79] at (110.25, 12.28) {1};
\end{scope}
\begin{scope}
\path[clip] (  0.00,  0.00) rectangle (144.54,108.41);
\definecolor{drawColor}{RGB}{0,0,0}

\node[text=drawColor,rotate= 90.00,anchor=base,inner sep=0pt, outer sep=0pt, scale=  0.99] at ( 11.09, 54.20) {$\pi_i$};
\end{scope}
\begin{scope}
\path[clip] (  0.00,  0.00) rectangle (433.62,108.41);
\definecolor{drawColor}{RGB}{0,0,0}

\path[draw=drawColor,line width= 0.4pt,line join=round,line cap=round] ( 32.47, 24.55) -- ( 32.47, 81.03);

\path[draw=drawColor,line width= 0.4pt,line join=round,line cap=round] ( 32.47, 24.55) -- ( 28.51, 24.55);

\path[draw=drawColor,line width= 0.4pt,line join=round,line cap=round] ( 32.47, 35.85) -- ( 28.51, 35.85);

\path[draw=drawColor,line width= 0.4pt,line join=round,line cap=round] ( 32.47, 47.14) -- ( 28.51, 47.14);

\path[draw=drawColor,line width= 0.4pt,line join=round,line cap=round] ( 32.47, 58.44) -- ( 28.51, 58.44);

\path[draw=drawColor,line width= 0.4pt,line join=round,line cap=round] ( 32.47, 69.73) -- ( 28.51, 69.73);

\path[draw=drawColor,line width= 0.4pt,line join=round,line cap=round] ( 32.47, 81.03) -- ( 28.51, 81.03);

\node[text=drawColor,anchor=base east,inner sep=0pt, outer sep=0pt, scale=  0.79] at ( 26.53, 21.82) {0.0};

\node[text=drawColor,anchor=base east,inner sep=0pt, outer sep=0pt, scale=  0.79] at ( 26.53, 33.12) {0.2};

\node[text=drawColor,anchor=base east,inner sep=0pt, outer sep=0pt, scale=  0.79] at ( 26.53, 44.42) {0.4};

\node[text=drawColor,anchor=base east,inner sep=0pt, outer sep=0pt, scale=  0.79] at ( 26.53, 55.71) {0.6};

\node[text=drawColor,anchor=base east,inner sep=0pt, outer sep=0pt, scale=  0.79] at ( 26.53, 67.01) {0.8};

\node[text=drawColor,anchor=base east,inner sep=0pt, outer sep=0pt, scale=  0.79] at ( 26.53, 78.30) {1.0};
\end{scope}
\begin{scope}
\path[clip] (  0.00,  0.00) rectangle (144.54,108.41);
\definecolor{drawColor}{RGB}{0,0,0}

\node[text=drawColor,anchor=base,inner sep=0pt, outer sep=0pt, scale=  0.79] at ( 84.15,  2.38) {Class};

\node[text=drawColor,anchor=base,inner sep=0pt, outer sep=0pt, scale=  0.79] at ( 84.15, 91.77) {\bfseries $w = (1, 1)^T$};
\end{scope}
\begin{scope}
\path[clip] (  0.00,  0.00) rectangle (433.62,108.41);
\definecolor{drawColor}{RGB}{0,0,0}

\path[draw=drawColor,line width= 0.4pt,line join=round,line cap=round] ( 32.47, 24.55) --
	(135.83, 24.55) --
	(135.83, 83.85) --
	( 32.47, 83.85) --
	cycle;
\end{scope}
\begin{scope}
\path[clip] (144.54,  0.00) rectangle (289.08,108.41);
\definecolor{drawColor}{RGB}{0,0,0}
\definecolor{fillColor}{RGB}{249,194,92}

\path[draw=drawColor,line width= 0.4pt,line join=round,line cap=round,fill=fillColor,fill opacity=0.70] (180.84, 24.55) rectangle (224.34, 47.14);
\definecolor{fillColor}{RGB}{254,253,190}

\path[draw=drawColor,line width= 0.4pt,line join=round,line cap=round,fill=fillColor,fill opacity=0.70] (233.04, 24.55) rectangle (276.54, 58.44);
\end{scope}
\begin{scope}
\path[clip] (  0.00,  0.00) rectangle (433.62,108.41);
\definecolor{drawColor}{RGB}{0,0,0}

\node[text=drawColor,anchor=base,inner sep=0pt, outer sep=0pt, scale=  0.79] at (202.59, 12.28) {0};

\node[text=drawColor,anchor=base,inner sep=0pt, outer sep=0pt, scale=  0.79] at (254.79, 12.28) {1};
\end{scope}
\begin{scope}
\path[clip] (144.54,  0.00) rectangle (289.08,108.41);
\definecolor{drawColor}{RGB}{0,0,0}

\node[text=drawColor,rotate= 90.00,anchor=base,inner sep=0pt, outer sep=0pt, scale=  0.99] at (155.63, 54.20) {$\pi_i$};
\end{scope}
\begin{scope}
\path[clip] (  0.00,  0.00) rectangle (433.62,108.41);
\definecolor{drawColor}{RGB}{0,0,0}

\path[draw=drawColor,line width= 0.4pt,line join=round,line cap=round] (177.01, 24.55) -- (177.01, 81.03);

\path[draw=drawColor,line width= 0.4pt,line join=round,line cap=round] (177.01, 24.55) -- (173.05, 24.55);

\path[draw=drawColor,line width= 0.4pt,line join=round,line cap=round] (177.01, 35.85) -- (173.05, 35.85);

\path[draw=drawColor,line width= 0.4pt,line join=round,line cap=round] (177.01, 47.14) -- (173.05, 47.14);

\path[draw=drawColor,line width= 0.4pt,line join=round,line cap=round] (177.01, 58.44) -- (173.05, 58.44);

\path[draw=drawColor,line width= 0.4pt,line join=round,line cap=round] (177.01, 69.73) -- (173.05, 69.73);

\path[draw=drawColor,line width= 0.4pt,line join=round,line cap=round] (177.01, 81.03) -- (173.05, 81.03);

\node[text=drawColor,anchor=base east,inner sep=0pt, outer sep=0pt, scale=  0.79] at (171.07, 21.82) {0.0};

\node[text=drawColor,anchor=base east,inner sep=0pt, outer sep=0pt, scale=  0.79] at (171.07, 33.12) {0.2};

\node[text=drawColor,anchor=base east,inner sep=0pt, outer sep=0pt, scale=  0.79] at (171.07, 44.42) {0.4};

\node[text=drawColor,anchor=base east,inner sep=0pt, outer sep=0pt, scale=  0.79] at (171.07, 55.71) {0.6};

\node[text=drawColor,anchor=base east,inner sep=0pt, outer sep=0pt, scale=  0.79] at (171.07, 67.01) {0.8};

\node[text=drawColor,anchor=base east,inner sep=0pt, outer sep=0pt, scale=  0.79] at (171.07, 78.30) {1.0};
\end{scope}
\begin{scope}
\path[clip] (144.54,  0.00) rectangle (289.08,108.41);
\definecolor{drawColor}{RGB}{0,0,0}

\node[text=drawColor,anchor=base,inner sep=0pt, outer sep=0pt, scale=  0.79] at (228.69,  2.38) {Class};

\node[text=drawColor,anchor=base,inner sep=0pt, outer sep=0pt, scale=  0.79] at (228.69, 91.77) {\bfseries $w = (1, 1.5)^T$};
\end{scope}
\begin{scope}
\path[clip] (  0.00,  0.00) rectangle (433.62,108.41);
\definecolor{drawColor}{RGB}{0,0,0}

\path[draw=drawColor,line width= 0.4pt,line join=round,line cap=round] (177.01, 24.55) --
	(280.37, 24.55) --
	(280.37, 83.85) --
	(177.01, 83.85) --
	cycle;
\end{scope}
\begin{scope}
\path[clip] (289.08,  0.00) rectangle (433.62,108.41);
\definecolor{drawColor}{RGB}{0,0,0}
\definecolor{fillColor}{RGB}{249,194,92}

\path[draw=drawColor,line width= 0.4pt,line join=round,line cap=round,fill=fillColor,fill opacity=0.70] (325.38, 24.55) rectangle (368.88, 38.67);
\definecolor{fillColor}{RGB}{254,253,190}

\path[draw=drawColor,line width= 0.4pt,line join=round,line cap=round,fill=fillColor,fill opacity=0.70] (377.58, 24.55) rectangle (421.08, 66.91);
\end{scope}
\begin{scope}
\path[clip] (  0.00,  0.00) rectangle (433.62,108.41);
\definecolor{drawColor}{RGB}{0,0,0}

\node[text=drawColor,anchor=base,inner sep=0pt, outer sep=0pt, scale=  0.79] at (347.13, 12.28) {0};

\node[text=drawColor,anchor=base,inner sep=0pt, outer sep=0pt, scale=  0.79] at (399.33, 12.28) {1};
\end{scope}
\begin{scope}
\path[clip] (289.08,  0.00) rectangle (433.62,108.41);
\definecolor{drawColor}{RGB}{0,0,0}

\node[text=drawColor,rotate= 90.00,anchor=base,inner sep=0pt, outer sep=0pt, scale=  0.99] at (300.17, 54.20) {$\pi_i$};
\end{scope}
\begin{scope}
\path[clip] (  0.00,  0.00) rectangle (433.62,108.41);
\definecolor{drawColor}{RGB}{0,0,0}

\path[draw=drawColor,line width= 0.4pt,line join=round,line cap=round] (321.55, 24.55) -- (321.55, 81.03);

\path[draw=drawColor,line width= 0.4pt,line join=round,line cap=round] (321.55, 24.55) -- (317.59, 24.55);

\path[draw=drawColor,line width= 0.4pt,line join=round,line cap=round] (321.55, 35.85) -- (317.59, 35.85);

\path[draw=drawColor,line width= 0.4pt,line join=round,line cap=round] (321.55, 47.14) -- (317.59, 47.14);

\path[draw=drawColor,line width= 0.4pt,line join=round,line cap=round] (321.55, 58.44) -- (317.59, 58.44);

\path[draw=drawColor,line width= 0.4pt,line join=round,line cap=round] (321.55, 69.73) -- (317.59, 69.73);

\path[draw=drawColor,line width= 0.4pt,line join=round,line cap=round] (321.55, 81.03) -- (317.59, 81.03);

\node[text=drawColor,anchor=base east,inner sep=0pt, outer sep=0pt, scale=  0.79] at (315.61, 21.82) {0.0};

\node[text=drawColor,anchor=base east,inner sep=0pt, outer sep=0pt, scale=  0.79] at (315.61, 33.12) {0.2};

\node[text=drawColor,anchor=base east,inner sep=0pt, outer sep=0pt, scale=  0.79] at (315.61, 44.42) {0.4};

\node[text=drawColor,anchor=base east,inner sep=0pt, outer sep=0pt, scale=  0.79] at (315.61, 55.71) {0.6};

\node[text=drawColor,anchor=base east,inner sep=0pt, outer sep=0pt, scale=  0.79] at (315.61, 67.01) {0.8};

\node[text=drawColor,anchor=base east,inner sep=0pt, outer sep=0pt, scale=  0.79] at (315.61, 78.30) {1.0};
\end{scope}
\begin{scope}
\path[clip] (289.08,  0.00) rectangle (433.62,108.41);
\definecolor{drawColor}{RGB}{0,0,0}

\node[text=drawColor,anchor=base,inner sep=0pt, outer sep=0pt, scale=  0.79] at (373.23,  2.38) {Class};

\node[text=drawColor,anchor=base,inner sep=0pt, outer sep=0pt, scale=  0.79] at (373.23, 91.77) {\bfseries $w = (1, 3)^T$};
\end{scope}
\begin{scope}
\path[clip] (  0.00,  0.00) rectangle (433.62,108.41);
\definecolor{drawColor}{RGB}{0,0,0}

\path[draw=drawColor,line width= 0.4pt,line join=round,line cap=round] (321.55, 24.55) --
	(424.91, 24.55) --
	(424.91, 83.85) --
	(321.55, 83.85) --
	cycle;
\end{scope}
\end{tikzpicture}
			}
		\end{subfigure}
		\begin{subfigure}{\linewidth}
			\caption{Multinomial data: Skewed class distribution.}\label{fig:prob.mult.skew}\smallskip
			\centering
			\resizebox{\linewidth}{!}{%
			\input{barplot_probs_mult_skew.tex}
			}
		\end{subfigure}
		\begin{subfigure}{\linewidth}
			\caption{Multinomial data: One class up, one class down.}\label{fig:prob.mult.1u1d}\smallskip
			\centering
			\resizebox{\linewidth}{!}{%
			\input{barplot_probs_mult_1up_1down.tex}
			}
		\end{subfigure}
		\smallskip
		\caption{Class probability distribution for selected scenarios given by the weight vectors $w$. It holds $\pi_i = w_i / \sum_j w_j$.}
		\label{fig:probs}
	\end{figure}
	
	The probability distributions of each variable in the datasets are varied.
	Some selected probability distributions are visualized in Figure~\ref{fig:probs}.
	For the null situation, all classes are equally likely. 
	For binary data, as an alternative, the class probabilities are varied to gradually become more unbalanced.
	This is shown for two example scenarios in Figure~\ref{fig:prob.bin}.
	For the multinomial data, two types of alternatives are considered. 
	First, the class probability distribution is varied in such a way that more probability mass is given to the higher classes. 
	This is referred to as ``skewed'' for convenience, even though no strict ordering of the classes is assumed.
	Second, the class probability of one class is increased while the other is decreased by the same amount.
	The two cases are visualized for two example scenarios each in Figure~\ref{fig:prob.mult.skew} and Figure~\ref{fig:prob.mult.1u1d}, respectively. 
	For the concrete class probabilities for all cases, see Tables~\ref{tab:scen.cat.no.y} to~\ref{tab:scen.cat.multi} in Appendix~\ref{app:scen.tabs}.
	
	For the case of categorical datasets with a target variable, there is one true outcome-generating model, and then certain deviations from this. 
	The true outcome-generating model is chosen as a logistic regression model since all methods need a categorical outcome variable. 
	Moreover, the number of classes should be low, as the smallest generated datasets consist of only 25 data points and it has to be ensured that there are observations from all classes in all datasets. 
	Therefore, the logistic model was chosen such that for all distributions of the covariates, a reasonable number of ones and zeroes is expected to be generated.
	
	As deviations of the outcome-generating model (OGM), once the signs of all coefficients except for the intercept are switched, and once each coefficient is divided by two. 
	In practice, this would present an extreme change in the interpretation of the relationship between the covariates and the target variable if the effect of each covariate is inverted or halved. 
	Since the division of all coefficients by two leads to the same decision boundary with more uncertainty in that decision (see Figure~\ref{fig:dec.bound} in Appendix~\ref{app:scen.tabs}) and for changing all signs except for the intercept the roles for ones and zeroes in the target variable are inverted except for some shift in the decision boundary (see Figure~\ref{fig:dec.bound} in Appendix~\ref{app:scen.tabs}), a third wrong OGM was added later in the simulation process. 
	The coefficients in this third OGM are completely different from those in the true OGM and lead to a clearly different, axis-parallel decision boundary (see Figure~\ref{fig:dec.bound} in Appendix~\ref{app:scen.tabs}). 
	The intention was that this axis-parallel decision boundary should be picked up easily by the decision trees that are used for one of the methods, such that it would be expected that the method can identify this difference in the OGM. 
	
	The categorical datasets are dummy-coded for the generation of the outcome variables according to the logistic model. 
	For variables with five categories, all four categories that are not the reference have the same coefficient values in the OGMs to make the setting as comparable to the binary case as possible. 
	Since this might be an unrealistic assumption in practice, an OGM where the coefficients for increasing categories have increasing absolute values was added. 
	This corresponds to the case where a higher class of each variable leads to a higher (or lower) odds ratio. 
	The results for this OGM did not differ from those for the other OGM, but this specification led to more numerical problems.
	Therefore, the results are not presented here.
	For the concrete model specifications of the models analyzed in the simulation study, see Table~\ref{tab:scen.cat.y} in Appendix~\ref{app:scen.tabs}.
	
	\subsection{Estimands}
	The population quantity of interest is the similarity or, equivalently, distance of the underlying distributions of the datasets. 
	This is estimated by each method.

	\subsection{Methods}
	The most promising methods from the previous review and theoretical comparison \textcite{stolte_methods_2024} are included in this empirical comparison. 
	Methods are selected from the review if any of the following criteria are fulfilled:
	\begin{enumerate}
		\item The method is implemented in R.
		\item The method fulfills at least 11 (i.e.\ more than half of the) criteria in the theoretical comparison, excluding the consistency criteria. 
		\item The method is the best in its subclass in the theoretical comparison, and no other method from this subclass was chosen with the first two criteria.
	\end{enumerate}
	An overview of all methods that fulfill these criteria is provided in Table~\ref{tab:methods} in Appendix~\ref{app:meth.tabs}.
	The analysis here is restricted to the subset of those methods that are applicable to categorical data.
	These 12 methods are explained in the following.
	Most of these methods are only applicable to two samples. 
	For the methods that are applicable to multiple samples, this is explicitly stated. 
	All methods are used with default parameters based on recommendations from the literature, if available. 
	If no sensible default is available, different options are compared.
	The choices of these are explained in Appendix~\ref{app:meth.pars}. 
	The methods are applied using parameter choices that a practitioner with good knowledge of the underlying literature, but without expert knowledge of the methods, could use.
	\begin{itemize}
		\item Classifier two-sample test (C2ST, \cite{lopez-paz_revisiting_2017}): The pooled dataset is split into a training and test set, and a classifier is trained on the training set to distinguish between the datasets. 
		The classifier's accuracy on the test set is used as the statistic. 
		For similar datasets, an accuracy close to the accuracy of the naive prediction of the larger dataset is expected, while for different datasets, higher accuracies are expected. 
		The procedure can also be used for multiple samples.
		\item Random forest-based test by \textcite{hediger_use_2021} (HMN): A random forest is trained on the entire pooled dataset to distinguish between the individual datasets. 
		The out-of-bag prediction error is used as a test statistic. 
		If the datasets are similar, the error should be close to that expected for always predicting the larger dataset. 
		\item Tree-based test by \textcite{yu_two-sample_2007} (YMRZL): A classification tree is trained to distinguish between the datasets using a training dataset that is a subset of the pooled sample. 
		Its classification error on the left-out test set is used as the statistic. 
		If the datasets are similar, the error should be close to that expected for always predicting the larger sample.
		\item Original edge count test by \textcite{friedman_multivariate_1979} (FR): A graph (originally the minimum spanning tree, MST) is constructed on the pooled sample using an appropriate distance measure. 
		Here, the Hamming distance is used.
		For categorical data, the optimal graph might not be unique due to ties in the inter-point distances. 
		Therefore, either the arithmetic mean of the test statistics on all optimal graphs (``a'') or the test statistic on the union of all optimal graphs (``u''), i.e.\ the graph that includes all edges of all optimal graphs, is calculated \textcite{chen_ensemble_2013}.
		For calculating the test statistic for a given similarity graph, the number of edges connecting points from different samples is counted.
		The expectation and variance of this edge-count statistic under the null hypothesis of equal distributions are known and can be calculated analytically.
		The standardized edge count using this null expectation and standard deviation is used as the test statistic. 
		For similar datasets, higher numbers of edges connecting points from different samples are expected. 
		\item Generalized edge count test by \textcite{chen_new_2017} (CF): The Friedman-Rafsky test is generalized to improve the power for detecting both location and scale alternatives. 
		The number of edges connecting points within each of the two samples, $R_1, R_2,$ respectively, is counted in a similarity graph on the pooled sample.
		The Mahalanobis distance 
		\[
		(R_1 - \E_{H_0}(R_1), R_2 - \E_{H_0}(R_2)) \Cov_{H_0}^{-1}(R) \begin{pmatrix}
			R_1 - \E_{H_0}(R_1)\\ R_2 - \E_{H_0}(R_2)
		\end{pmatrix}
		\] 
		of the vector $R = (R_1, R_2)^T$ is used as the test statistic. 
		Small values of the statistic indicate the similarity of the datasets.
		Again, for categorical data, averaging (``a'') or the union (``u'') can be used \textcite{zhang_graph-based_2022}. 
		\item Weighted edge count test by \textcite{chen_weighted_2018} (CCS): The Friedman-Rafsky test is generalized to improve the power in settings with unequal sample sizes. 
		The weighted statistic is defined as 
		\[
		R_w = \frac{n_1}{N} R_1 + \frac{n_2}{N} R_2,
		\]
		where $R_1, R_2$ are defined as above and $n_i$ denotes the sample size of the $i$-th sample, $i = 1, 2$, $N = n_1 + n_2$. 
		Again, the expectation and standard deviation of $R_w$ can be calculated analytically and are used to define a standardized test statistic. 
		Small numbers of edges connecting points within the same sample indicate similar datasets.
		Therefore, small values of $R_w$ or its standardized version indicate similarity. 
		Again, for categorical data, averaging (``a'') or the union (``u'') can be used \textcite{zhang_graph-based_2022}. 
		\item Max-type edge count test by \textcite{zhang_graph-based_2022} (ZC): The test is yet another generalization of the Friedman-Rafsky test. 
		The test statistic is given by 
		\[
		R_m = \max\{\kappa R_w, |R_1 - R_2|\},
		\]
		where $\kappa$ is a parameter that has to be chosen prior to testing. 
		Again, a standardized version is given by standardizing $R_w$ and $R_d = |R_1 - R_2|$ with their expectations and standard deviations under the null. 
		Small values of the statistic indicate similarity. 
		Again, for categorical data, averaging (``a'') or the union (``u'') can be used \textcite{zhang_graph-based_2022}. 
		\item Multi-sample Cross-Match statistic by \textcite{petrie_graph-theoretic_2016}: The optimal non-bipartite matching is calculated on the pooled sample, and the overall number of edges connecting points from different samples is calculated. 
		It is standardized by the analytical expectation and standard deviation under the null hypothesis. 
		High values of the cross-match statistic indicate similarity between the datasets.  
		\item Multi-sample Mahalanobis Cross-Match (MMCM) statistic \textcite{mukherjee_distribution-free_2022}: The optimal non-bipartite matching is calculated on the pooled sample.
		The numbers of edges $a_{ij}$ connecting points from sample $i$ and sample $j$, $i\ne j \in \{1,\dots,k\}$, is calculated. 
		The Mahalanobis distance of the cross-match vector $A = a_{12}$ in the two-sample case and $A = (a_{12}, a_{13}, a_{23}, a_{24})^T$ in the four-sample case, respectively, is used as the test statistic 
		\[
		\text{MMCM} = (A - \E_{H_0}(A))^T \Cov_{H_0}^{-1}(A) (A - \E_{H_0}(A)),
		\]
		where again expectations and covariances under the null can be calculated analytically. 
		For similar datasets, low MMCM values are expected. 
		For two samples, the MMCM test is analytically equivalent to Petrie's test.
		\item Constrained Minimum (CM) Distance \textcite{tatti_distances_2007}: The CM distance is based on a \emph{feature function} $S:\mathcal{X}\to\mathbb{R}^m$ that maps points from the sample space $\mathcal{X}$ to a real vector. 
		The \emph{frequency} $\theta\in\mathbb{R}^m$ of $S$ with respect to dataset $X^{(j)}$ is the average of the values of $S$
		\[
		\theta_j= \frac{1}{N}\sum_{i = 1}^{n_j} S(X_{i}^{(j)}), j = 1, 2.
		\]
		The CM distance is then defined as
		\[
		D_{\text{CM}}(X^{(1)}, X^{(2)}|S)^2 = (\theta_1 - \theta_2)^{T}\Cov^{-1}(S)(\theta_1 - \theta_2),
		\]
		with
		\[
		\Cov(S) = \frac{1}{|\mathcal{X}|}\sum_{\omega\in\mathcal{X}} S(\omega)S(\omega)^{T} - \left(\frac{1}{|\mathcal{X}|}\sum_{\omega\in\mathcal{X}}S(\omega)\right)\left(\frac{1}{|\mathcal{X}|}\sum_{\omega\in\mathcal{X}}S(\omega)\right)^{T}.
		\] 
		The recommended feature function $S$ is used here, i.e.\ the independent means of the variables are considered (see Appendix~\ref{app:meth.pars}).
		\item Decision tree-based dataset distance by \textcite{ganti_framework_1999} (GGRL): The GGRL requires the datasets to include a target variable. 
		A decision tree is fit to each dataset. 
		The partitions of the sample space induced by these trees are intersected and the proportions of data points falling into each segment of this so-called greatest common refinement (GCR) are determined for each dataset. 
		The resulting probability vectors $p$ and $q$ are then compared using a difference function $f$, and the results are aggregated using an aggregate function $g$: 
		\[
		\text{GGRL} = g(f(p, q)).
		\]
		Proposed choices for the difference function are the absolute component-wise differences ($f_a$) or the absolute component-wise differences scaled by their means ($f_s$). 
		For the aggregate function, the sum or the maximum is proposed.
		\item Optimal Transport Dataset Distance \textcite{alvarez-melis_geometric_2020} (OTDD): 
		The OTDD is a distance between datasets that takes into account a target variable $y$ included in the datasets. It is defined as 
		\[
		d_{\text{OT}}(X^{(1)}, X^{(2)}) = \min_{\pi\in\Pi(F_1, F_2)} \int_{\mathcal{Z}\times\mathcal{Z}} d_{\mathcal{Z}}(z, z^\prime)^q \dif \pi(z, z^\prime),\text{ where}
		\]
		 \[
		\Pi(F_1, F_2) := \{\pi_{1,2}\in\mathcal{P}(\mathcal{Z}\times\mathcal{Z})\arrowvert \pi_1 = F_1, \pi_2 = F_2\}
		\]
		is the set of joint distributions over the product space $\mathcal{Z}\times\mathcal{Z}$ over the sample space of the pooled sample with marginal distributions $F_1$ and $F_2$, and
		\[
		d_{\mathcal{Z}}(z, z^\prime) := (d_{\mathcal{X}}(x, x^{\prime})^{q^\prime} + W_{q^\prime}(\alpha_y, \alpha_{y^{\prime}})^{q^\prime})^{1/{q^\prime}}.
		\]
		defines a distance of two points $z^{\top} = (x^{\top}, y)$, and ${z^\prime}^{\top} = ({x^\prime}^{\top}, y^\prime)$ in the pooled sample.
		$d_{\mathcal{X}}$ defines a distance on the covariate space, e.g., the Hamilton distance, and  $W_{q^\prime}(\alpha_y, \alpha_{y^{\prime}})$ is the $q^\prime$-Wasserstein distance of the distribution of the subset of covariate data $x$ with corresponding response value $y$ and the distribution of the subset of covariate data $x^\prime$ with corresponding response value $y^\prime$. 
	\end{itemize}
	All of these methods are applied in the two-sample case. 
	The C2ST, MMCM, and the method of \textcite{petrie_graph-theoretic_2016} are also applied in the multi-sample case. 
	For the other methods, no generalizations to the multi-sample case are available in the literature.
	For all methods except for the CM distance and OTDD, a permutation test is proposed in addition.
	However, no permutation test is performed here due to the high runtime. 
	More details on the methods can be found in \textcite{stolte_methods_2024} and the references therein.  
	
	\subsection{Performance Measures}
\begin{figure}[!b]
		\centering
		\includegraphics[width=1\linewidth]{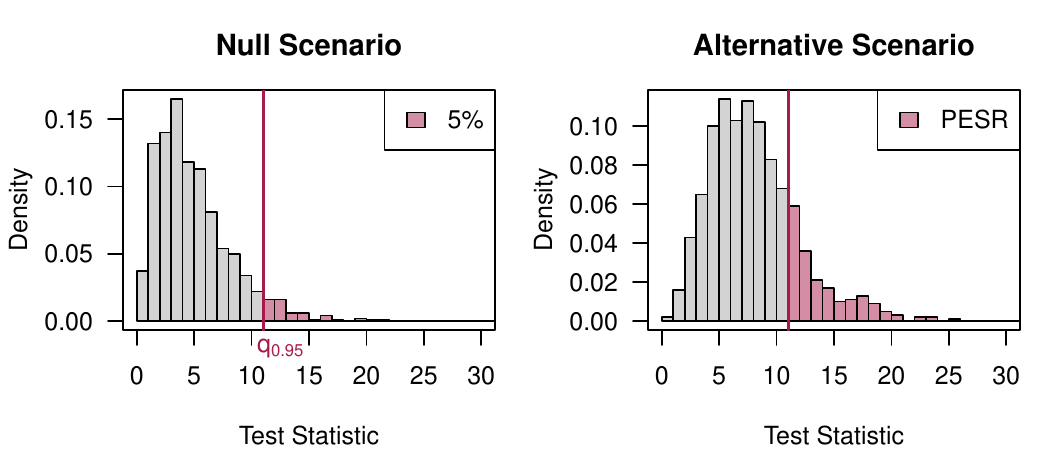}
		\caption{Illustration of PESR calculation for the case where low test statistic values correspond to similarity between the datasets. The 95\,\% quantile $q_{0.95}$ of the test statistic values simulated under a null scenario is used as the threshold under the alternative scenario. The proportion of values that are more extreme than this threshold is evaluated and denoted as the PESR. Test statistic values for this figure were artificially generated from a $\chi^2_5$ and $\chi^2_8$ distribution, respectively, for demonstration.}
		\label{fig:example-pesr}
	\end{figure}

	Two aspects are evaluated here. 	
	First, it is evaluated how well the methods can detect the differences between the distributions that were described in the previous subsections. 
	Second, the computational costs of the methods are compared.
	For the first aim, no classical power comparison can be conducted since not all methods define a test.
	Moreover, such a comparison would not be possible due to the very high runtimes of the many permutation tests. 
	Instead, the methods are compared as follows. 
	As the ranges of the method values vary heavily from method to method, the values cannot be compared directly. 
	Therefore, the performance of the methods has to be made comparable. 
	The approach here is based on the observation that a typical power comparison evaluates the proportion of simulation repetitions in which the observed test statistic is more extreme than some quantile of the (permutation) null distribution. 
	The approach is illustrated in Figure~\ref{fig:example-pesr}. 
	For each setting in which the distributions do not differ, a quantile of the simulated statistic values for a certain method is calculated. 
	The statistic values of that method for deviations of one (or more) distribution(s) from the first are then compared to this quantile. 
	The proportion of simulation repetitions in which a more extreme statistic value than this threshold is observed is used to quantify the performance of the method. 
	For methods for which high values correspond to the similarity of the distributions, the proportion of simulation repetitions is used for which the resulting statistic value is smaller than the 5\,\% quantile of the corresponding statistic values simulated under equal distributions. 
	For methods for which low values correspond to similarity, the proportion of repetitions with values higher than the 95\,\% quantile is used.
	This \textit{Proportion of Extreme Simulation Repetitions} w.r.t.\ the null threshold is abbreviated as \textit{PESR} in the following.
	It can be used to evaluate how well the methods detect the difference in the distributions, and it can be compared between different methods. 
	Note that the determined PESR does not directly equal the testing power since only one specific null situation is considered. 
	Moreover, the PESR might still show high values in cases where an asymptotic test does not find any differences or in cases where the asymptotic test does not hold its $\alpha$ level. 
	This can happen when the asymptotic null distribution does not match the empirical distribution.
	For methods for which an asymptotic test exists, the asymptotic testing power can be simulated without relevant increases in runtime.
	See Appendix~\ref{app:comp.pesr.power} for a comparison and discussion of simulated asymptotic power and the PESR for selected methods. 	
	
	The number of simulation repetitions is set to $500$ repetitions per scenario. 
	The Monte Carlo standard error (MCSE) for proportions like the PESR is highest for a proportion of $0.5$. 
	For $500$ iterations, the MCSE is then $\sqrt{0.5(1-0.5) / 500}\approx 0.022$. 
	For a proportion of $0.05$, which corresponds to the PESR in null situations, the MCSE is $\sqrt{0.05(1-0.05) / 500}\approx 0.01$. 
	This is considered sufficiently small here. 
	Using $1000$ iterations would only bring down the MCSEs to $\approx0.015$ and $\approx 0.006$, respectively, but this would double the runtime.
	
	In case of computational errors resulting in missing or infinite values of the statistics, the affected repetitions are excluded from the PESR calculation. 
	If there are missing values in more than $100$ of the $500$ iterations, the corresponding PESR value is set to missing. 
	This ensures that the calculated PESR values are based on a reasonably high number of repetitions.
	
	In addition to the PESR, the applicability of the methods in practice is considered.
	To do this, runtime, memory consumption, and any numerical problems are taken into account. 
	These are measured for selected scenarios only. 
	Here, the null situations for two balanced classes are chosen. 
	All combinations of $N$ and $p$ as discussed before are used except for $N = 1000$ for the two-sample case, since it is infeasible with the RAM configuration of the used computer.
	Datasets are generated once for each combination of $N$ and $p$ for that scenario. 
	On each dataset, each similarity method is then applied once to measure the memory consumption and afterward at least $10$~times to measure the runtime per method call. 
	For methods with low runtimes, the number of repetitions is increased such that the method is run for at least $1$~second to get stable estimates of the runtime per method call. 
	Each method is called once before starting the benchmark to ensure that all required packages and objects are already loaded at the start of the benchmark, and the results are not distorted by lazy loading.
	The benchmarks are performed on a Lenovo ThinkPad laptop with an AMD Ryzen~5 PRO 4650U processor with six cores and 16\,GB of RAM under Windows~11. 
	Benchmarks are run during the nighttime when the laptop is not used for any other work to ensure that the results are not disturbed by other computations. 
	
	\subsection{Software}
	All simulations are performed using \texttt{R} version 4.4.0 \textcite{R} on the Linux-HPC-Cluster (LiDO3) at TU Dortmund University.
	Further analyses and benchmarking are performed using \texttt{R} version 4.5.1 \textcite{R} on a personal computer.
	The implementation of all methods can be found in the \texttt{DataSimilarity}
	package \textcite{DataSimilarity}. 
	The \texttt{bench} package \textcite{bench} is used for measuring runtime and memory consumption. 
	The \texttt{pheatmap} \textcite{pheatmap} and the \texttt{cba} \textcite{cba} packages are used for visualizing and clustering the PESR values of the methods across scenarios. 
	The \texttt{rpart.plot} \textcite{rpart.plot} package is used for visualizing the decision rules for finding the best-performing methods for a specific scenario.
	The full \texttt{R} code of the study can be found on Zenodo \textcite{code_zenodo}.
	\FloatBarrier

	\section{Sensitivity in Detecting Differences Between Datasets}\label{sec:sensitivity}
	In the following, the proportions of extreme simulation repetitions (PESR) are compared between the methods, first for the two-sample and then for the multi-sample setting. 
	
	\subsection{Two-sample Setting}
	In the following, the results for the two-sample setting ($k = 2$) are discussed.
	For a detailed discussion of the occurring errors, see Appendix~\ref{sec:err}.
	In short, for the case without a target variable, mostly no errors occur or a method is not applicable at all for certain dataset dimensions. 
	For the case with a target variable, more errors occured.
	Often these occured due to too high imbalance in the generated binary target variable.
	Moreover, for some scenarios with a target variable, not all repetitions could be finished in time, reducing the number of repetitions to 450 for certain scenarios, mostly for unbalanced datasets with low numbers of variables, unequal and low sample sizes, and five classes.
	First, a pre-selection of methods is performed to exclude variants that are inferior in all scenarios from the following comparisons for clarity.
	See Appendix~\ref{app:presel} for the details of this pre-selection.  
	The number of variants is reduced from 62 variants overall to 19 selected variants for the case of unbalanced sample sizes and 16 variants for balanced sample sizes. 
	Next, the results of the selected methods are discussed for datasets consisting of binary data. 
	Afterward, the results for datasets consisting of categorical data with five categories are discussed. 
	In the former case, the success probability in the second dataset is varied. 
	In the latter case, two alternatives for varying the probabilities of the five classes are considered: a skewed probability distribution and increasing the probability of one class while decreasing the probability of another class (see Figure~\ref{fig:probs} for an illustration).

	\subsubsection{Binary Data}
\begin{figure}[!t]
		\centering
		\includegraphics[width=\linewidth]{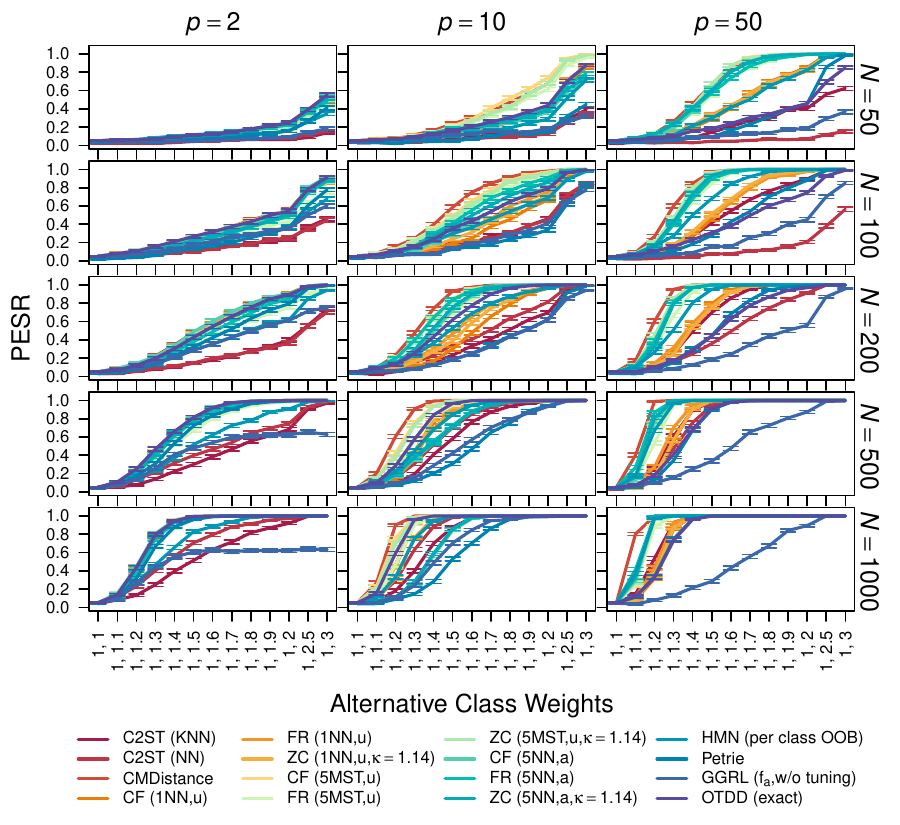}
		\caption{PESR (proportion of extreme simulation repetitions) for $k = 2$ binary datasets of equal sample sizes. The class weights give the unnormalized probabilities $(1, 1+\delta)$ for the values 0 and 1 for each variable in the second dataset. This means the weights in the first dataset are set to $(1, 1)$, and in the second dataset to $(1, 1+\delta)$. Error bars indicate Monte Carlo standard errors.}
		\label{fig:pow.cat.no.y.bin.bal}
	\end{figure}

	Figure~\ref{fig:pow.cat.no.y.bin.bal} shows the proportions of extreme simulation repetitions (PESR) for two binary datasets of equal sample sizes for each of the pre-selected methods. 
	The Monte Carlo standard errors (MCSEs) are displayed by error bars. 
	The MCSEs are small here and in all following analyses, indicating low sampling variability of the simulation results. 
	For $p = 2$, many methods perform similarly well.
	In particular, the CM distance, the edge count tests for 1NN, ``u'', and the OTDD perform comparably well. 
	The HMN, C2ST, Petrie's method, and GGRL perform considerably worse than the rest, with GGRL plateauing at a PESR of around 0.6. 
	For $p>2$, the CM distance performs best. 
	The edge count tests for the $K = 5$ graphs also perform well. 
	For $p = 10$, the ZC~(5MST, u) is the second best method, for $p = 50$, it is the CCS (5NN, a).
	The classifier-based tests HMN and C2ST, as well as Petrie's method and GGRL, perform worse.
	The OTDD is in the middle-field for $p = 10$ and among the worst methods for $p = 50$.
	As expected, for all methods, the PESR increases with an increasing number of observations. 
	
\begin{figure}[!t]
		\centering
		\includegraphics[width=\linewidth]{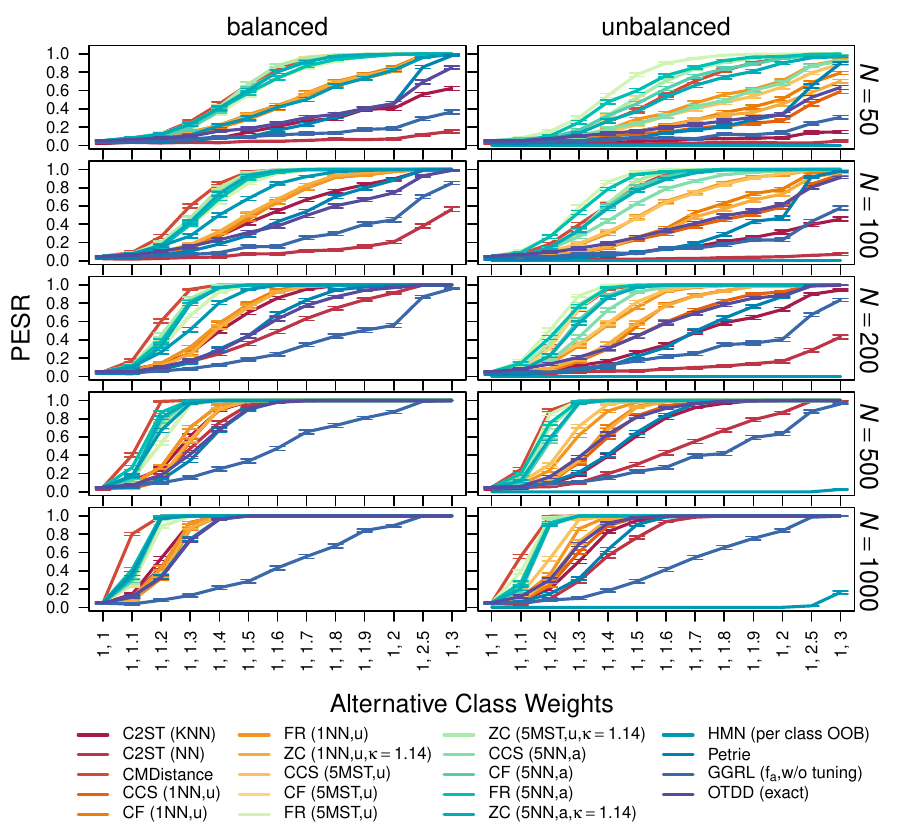}
		\caption{PESR (proportion of extreme simulation repetitions) for $k = 2$ binary datasets of equal (``balanced'') and unequal (``unbalanced'') sample sizes with $p = 50$ variables. The class weights give the unnormalized probabilities $(1, 1+\delta)$ for the values 0 and 1 for each variable in the second dataset. This means the weights in the first dataset are set to $(1, 1)$, and in the second dataset to $(1, 1+\delta)$. Error bars indicate Monte Carlo standard errors.}
		\label{fig:pow.cat.no.y.bin.p.50}
	\end{figure}

	For unbalanced sample sizes, the PESR of all methods decreases. 
	This is, for example, illustrated in Figure~\ref{fig:pow.cat.no.y.bin.p.50} for $p = 50$.
	The full results for the unbalanced case are shown in Figure~\ref{fig:pow.cat.no.y.bin.unbal} in Appendix~\ref{app:add.figs.cat.no.y.unbal}. 
	The performance of the CM distance and the classifier-based tests is more heavily impacted than that of the graph-based tests. 
	Therefore, the CM distance is no longer the best-performing method, but typically, CF (5MST, u) becomes the best method.
	The method CF, which was specifically designed as an alternative to FR for unbalanced sample sizes, is less affected than that of FR; see e.g.\ Figure~\ref{fig:pow.cat.no.y.bin.p.50}. 
	The HMN method breaks down completely for unbalanced sample sizes. 
	
\begin{figure}[!t]
		\centering
		\includegraphics[width=\linewidth]{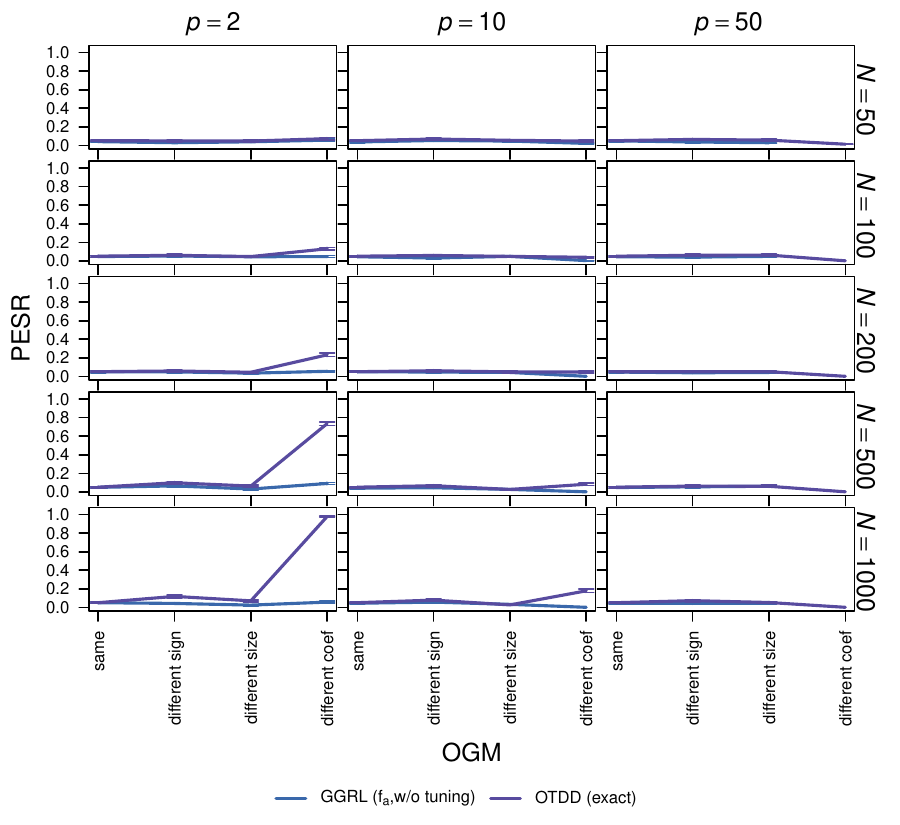}
		\caption{PESR (proportion of extreme simulation repetitions) for $k = 2$ binary datasets of equal sample sizes. A target variable is generated in both datasets, either using the same outcome-generating model (OGM) or the sign or size of the coefficients in the logistic model in the second dataset are changed. The OGM in the first dataset is a logistic model with the first half of the coefficients equal to $0.5$ and the second half and the intercept equal to $-0.5$. For ``different sign'' the signs of all coefficients except for the intercept are changed in the second dataset. For ``different size'' all coefficients except for the intercept are halved in the second dataset. For ``different coef'' completely different coefficients are chosen. Error bars indicate Monte Carlo standard errors.}
		\label{fig:pow.cat.y.bin.bal}
	\end{figure}

	For datasets that include a target variable, the outcome-generating model (OGM) that defines the relationship of the other variables with that target variable can also differ between the datasets. 
	Figure~\ref{fig:pow.cat.y.bin.bal} shows the PESR values for the cases where the coefficients of the logistic OGM differ between the datasets by their sign, size, or completely. 
	Only GGRL and OTDD are compared since the other methods cannot handle the target variable. 
	Both methods show very poor performance for detecting these deviations. 
	The PESR values mostly fluctuate around $5\,\%$. 
	Only in the case of completely different coefficients, low $p = 2$ and the highest $N = 500, 1000$, the OTDD can detect a difference with a high PESR.
	Note that for $p=50$ and the ``different coef'' OGM, there are very few ones generated for the target variable, which leads to computational errors in the calculation of the GGRL. 
	As this happened in many cases, no PESR values were computed (for details, see Section~\ref{sec:err}). 
	
	In addition to the PESR curve comparisons, the PESR values of the methods are clustered per $N$ and $p$ across the different dataset deviations. 
	For details, refer to Section~\ref{sec:clustering} in Appendix~\ref{app:add.figs}.
	In short, the clustering reveals no cluster structure in this case. 
	The methods are ordered by their performance, which can be summarized as the CM distance and edge count tests performing better than the MMCM, Petrie, C2ST, and YMRZL. 
	HMN is competitive for balanced sample sizes but breaks down for unbalanced sample sizes. 
	Out of the methods that take a target variable into account, OTDD performs considerably better than the GGRL.
	
	\subsubsection{Multinomial Data}
	\begin{figure}[!t]
		\centering
		\includegraphics[width=\linewidth]{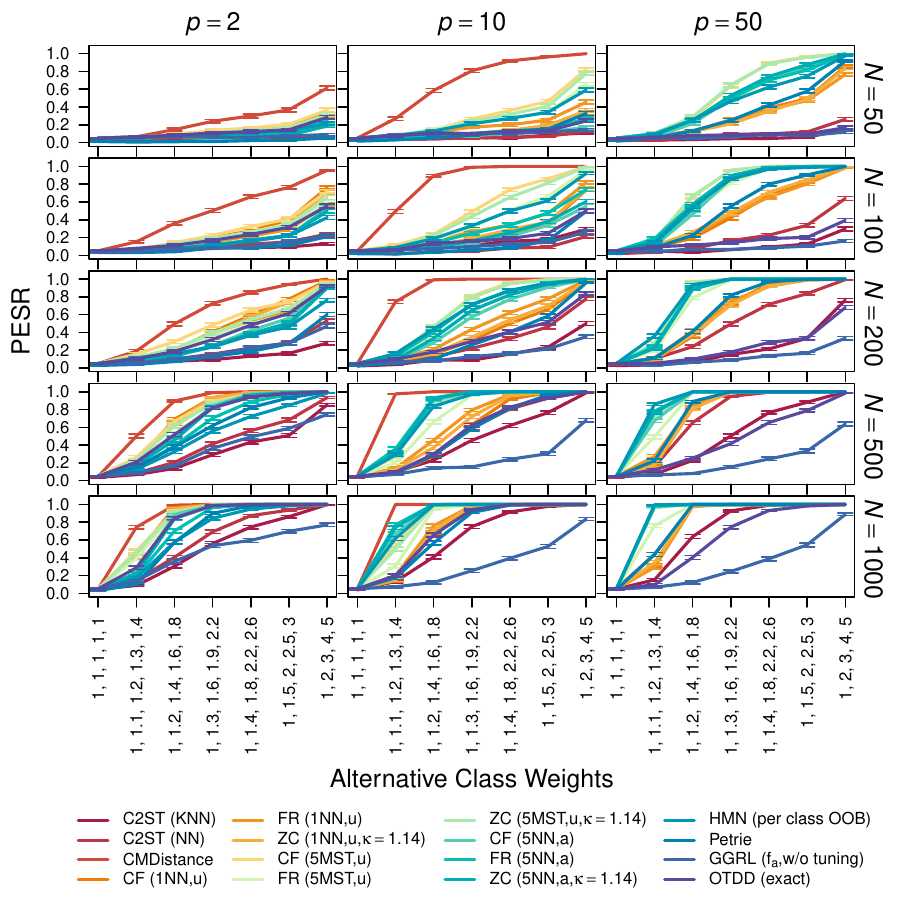}
		\caption{PESR (proportion of extreme simulation repetitions) for $k = 2$ datasets of equal sample sizes with categorical variables. The class weights give the unnormalized probabilities $(1, 1+\delta, 1+2\delta, 1+3\delta, 1+4\delta)$ for the values $1$ to $5$ for each variable in the second dataset. The weights in the first dataset are always set to $(1, 1, 1, 1, 1)$. Error bars indicate Monte Carlo standard errors.}
		\label{fig:pow.cat.no.y.multinom.skewed.bal}
	\end{figure}
	Figure~\ref{fig:pow.cat.no.y.multinom.skewed.bal} shows the proportions of extreme simulation repetitions (PESR) for two datasets of equal sample sizes with categorical variables. 
	The probability distribution for the five categories in the second dataset gets more and more skewed (from left to right in each panel of the figure), while the probability distribution for the five classes in the first dataset is a uniform distribution. 
	For $p = 2$ and $p = 10$, the CM distance is again the best-performing method. 
	However, for $p = 50$, it cannot be computed anymore as it requires the enumeration of the whole sample space, which becomes numerically infeasible for the $5^{50} \approx 8.9\cdot 10^{34}$ possible values due to memory restrictions.
	For $p = 2$ and $10$, the edge count tests FR, CCS, and ZC~(1NN, u), and the OTDD perform quite well. 
	The classifier-based tests HMN, C2ST variants, GGRL, and Petrie's method perform poorly in the comparison. 
	For $p = 50$, the graph-based tests with $K = 5$ and the HMN work best, followed by the $K = 1$ versions and Petrie's method. 
	The C2ST versions, especially the C2ST~(KNN), perform worse, and the OTDD and GGRL  show the lowest PESR values. 
	
	For unbalanced sample sizes, the PESR decreases again for all methods (see Figure~\ref{fig:pow.cat.no.y.multinom.skewed.unbal} in Appendix~\ref{app:add.figs.cat.no.y.unbal}). 
	As for binary data, the C2ST and the CM distance are more affected than the edge count tests. 
	The HMN fails again in the case of unbalanced sample sizes. 
	The CM Distance is still the best-performing method for $p = 10$.
	In the other cases, the FR performs best (1NN, ``u'' for $p = 2$ and 5MST, ``u'' for $p = 50$). 
	Here, the CCS has no improvement for the FR performance. 
	
	\begin{figure}[!t]
		\centering
		\includegraphics[width=\linewidth]{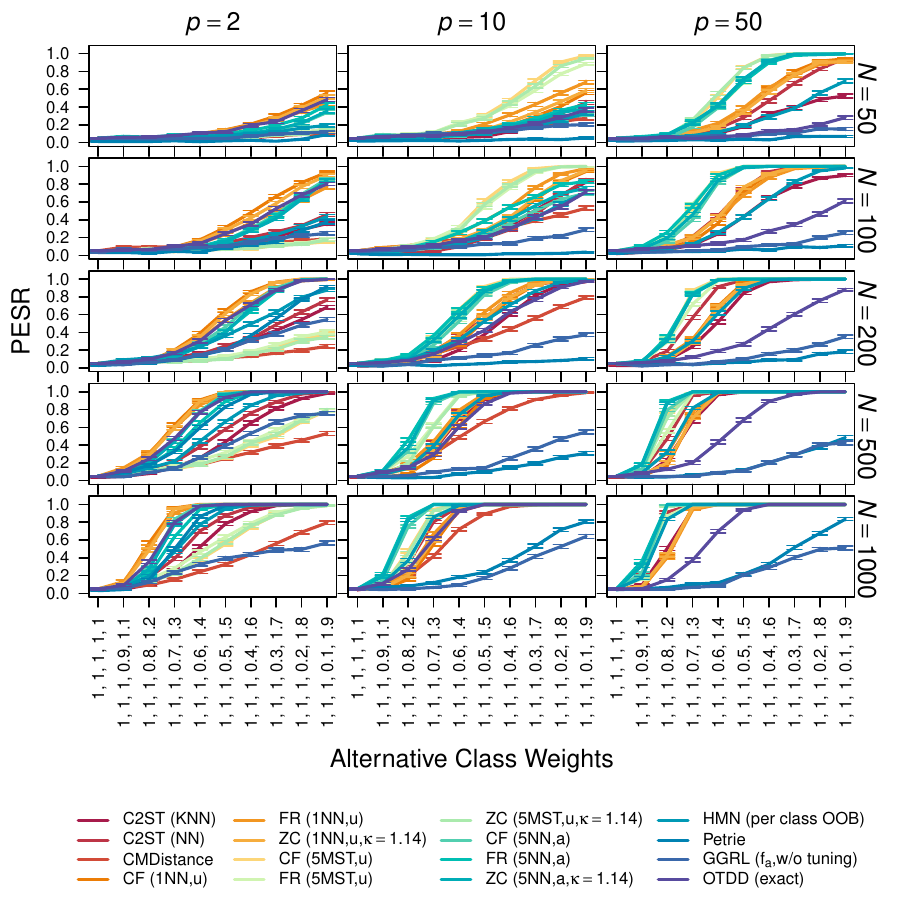}
		\caption{PESR (proportion of extreme simulation repetitions) for $k = 2$ datasets of equal sample sizes with categorical variables. The class weights give the unnormalized probabilities $(1, 1, 1, 1+\delta, 1-\delta)$ for the values $1$ to $5$ for each variable in the second dataset. The weights in the first dataset are always set to $(1, 1, 1, 1, 1)$. Error bars indicate Monte Carlo standard errors.}
		\label{fig:pow.cat.no.y.multinom.1u1d.bal}
	\end{figure}
	
	Figure~\ref{fig:pow.cat.no.y.multinom.1u1d.bal} shows proportions of extreme simulation repetitions (PESR) for two datasets of equal sample sizes with categorical variables when the probability of one class increases while the probability of another class decreases in the second dataset.
	Contrary to the results for the other settings, here, the CM distance is not the best-performing method but almost the worst. 
	For $p = 50$, it is again infeasible to compute. 
	For $p = 2$, the edge count tests using 1NN, ``u'' are best, with the ZC slightly outperforming the others.
	The CM distance achieves the lowest PESR values.
	For $p = 10, 50$, and small $N$, the edge count tests using 5MST, ``u'' perform best.
	For larger $N$, their performances are exceeded by that for the 5NN, ``a''. 
	The CM distance (for $p = 10$), the method by \textcite{petrie_graph-theoretic_2016}, and the GGRL perform the worst. 
	For $p = 50$, the OTDD is also among the worst methods.
	The C2ST methods perform comparatively better than for the skewed probability distribution alternative. 
	
	For unbalanced sample sizes, the performances of all methods decrease again (see Figure~\ref{fig:pow.cat.no.y.multinom.1u1d.unbal} in Appendix~\ref{app:add.figs.cat.no.y.unbal}). 
	The classifier-based methods especially have very low performance in that case. 
	The FR (1NN, u for $p = 2$, 5MST, u for $p > 2$) performs best, followed by the CF and ZC. 
	
	For changes in the OGM, the OTDD and GGRL show very poor performance again (see Figures~\ref{fig:pow.y.multi.bal} and \ref{fig:pow.cat.y.bin.unbal} in Appendix~\ref{app:add.figs.cat.y}).
	The PESR values for detecting changes in the signs or magnitude of the coefficients are around 5\,\%. 
	
	The PESR values for the methods on multinomial data are again clustered to identify groups of methods with similar performance.
	The resulting clustering reveals some differences between the types of deviations, at least for higher $N$ and $p$.
	The edge count tests perform best across all deviations. 
	HMN is competitive for the balanced sample size settings but not for the unbalanced cases. 
	The CM distance performs very well for detecting the ``skewed'' deviations only. 
	The C2ST variants, on the other hand, are only competitive for the ``1 up, 1 down'' alternatives. 
	The OTDD and GGRL perform rather poorly for all considered scenarios with five categories in the covariates, with the OTDD performing notably better than the GGRL.
	For a detailed analysis of the clustering results, refer to Section~\ref{sec:clustering} in Appendix~\ref{app:add.figs}.

	\subsection{Multi-sample Setting}
	In the following, the results for the multi-sample setting ($k = 4$) are discussed. 
	No errors occurred during the simulations. 
	First, the results for datasets consisting of binary data are discussed. 
	Afterward, the results for datasets consisting of categorical data with five categories are discussed. 
	
	\subsubsection{Binary Data}
	\begin{figure}[!t]
		\centering
		\includegraphics[width=\linewidth]{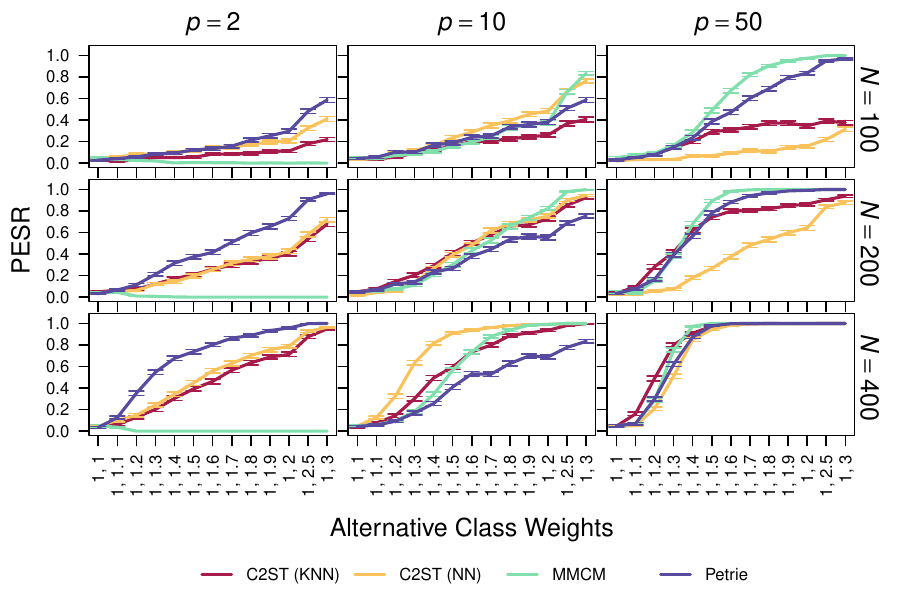}
		\caption{PESR (proportion of extreme simulation repetitions) for $k = 4$ binary datasets of equal sample sizes. The class weights give the unnormalized probabilities $(1, 1+\delta)$ for the values 0 and 1 for each variable in the second dataset. This means the weights in the first dataset are set to $(1, 1)$, in the second dataset to $(1, 1+\delta)$, in the third to $(1, 1+2\delta)$, and in the fourth to $(1, 1+3\delta)$. Error bars indicate Monte Carlo standard errors.}
		\label{fig:pow.cat.multi.bin.bal.1+1+1+1}
	\end{figure}
	Figure \ref{fig:pow.cat.multi.bin.bal.1+1+1+1} shows the PESR for binary data and the ``1+1+1+1'' case with balanced sample sizes. 
	Similar results can be observed for other groupings except where explicitly mentioned below (see Figures~\ref{fig:pow.cat.multi.bin.bal.3+1} to~\ref{fig:pow.cat.multi.bin.bal.2+1+1} in Appendix~\ref{app:add.figs.cat.multi.bin.bal}).
	In general, the proportions increase with increasing differences in the class weights of the datasets. 
	Moreover, the proportions increase with the overall sample size $N$. 
	The proportion also increases with the number of variables $p$, except for the method C2ST~(NN), for which the proportions are highest for $p = 10$. 
	The alternatives here are chosen to affect all variables in the datasets such that the true differences between the datasets increase with increasing dimension, which might explain the generally increased proportions for higher-dimensional data. 
	For C2ST~(NN) using the multilayer perceptron as the classifier, this effect of the larger true differences might be canceled out by the decreasing performance of the classifier for higher-dimensional data.
	
\begin{figure}[!t]
		\centering
		\includegraphics[width=\linewidth]{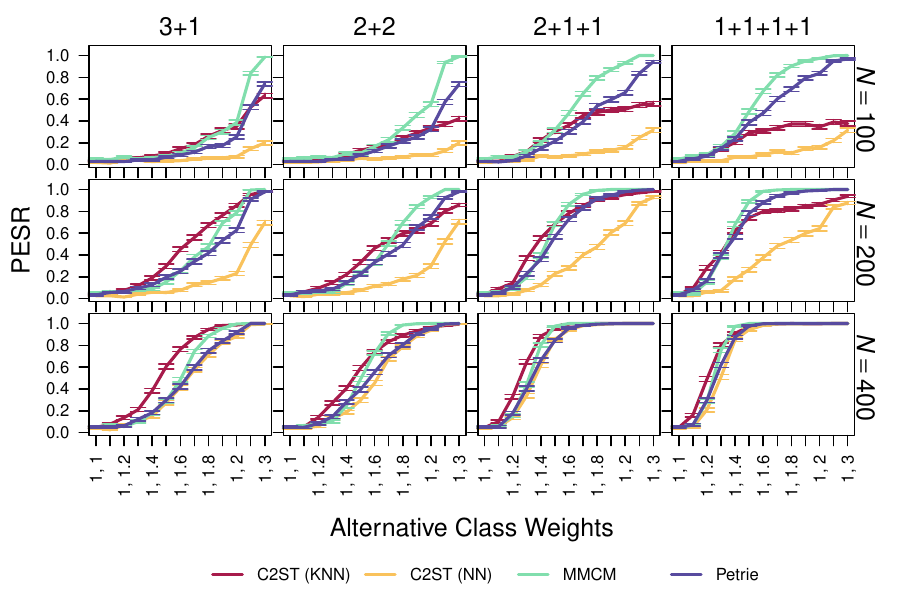}
		\caption{PESR (proportion of extreme simulation repetitions) for $k = 4$ binary datasets of equal sample sizes with $p = 50$ variables. The class weights give the unnormalized probabilities $(1, 1+\delta)$. The weights in the first dataset are always set to $(1, 1)$. For ``3+1'', the weights in the second and third datasets are $(1, 1)$, and in the fourth dataset $(1, 1+\delta)$. For ``2+2'', the weights in the third dataset are also $(1, 1+\delta)$. For ``2+1+1'', the weights in the third dataset are $(1, 1+\delta)$, and in the fourth $(1, 1+2\delta)$. For ``1+1+1+1'', the weights are $(1, 1+\delta)$ in the second dataset, $(1, 1+2\delta)$ in the third, and $(1, 1+3\delta)$ in the fourth. Error bars indicate Monte Carlo standard errors.}
		\label{fig:pow.cat.multi.bin.bal.p.50}
	\end{figure}

	For $p = 2$, MMCM fails, the method of \textcite{petrie_graph-theoretic_2016} outperforms the other methods, and C2ST~(NN) either outperforms C2ST~(KNN) or these two are comparable. 
	This also holds for the other groupings, except for the ``3+1'' case where none of the methods perform well. 
	For $p = 10$ and $N < 400$, the methods perform similarly for small to medium deviations. 
	For large deviations, clearer differences are visible, and C2ST~(NN) or MMCM outperforms the other methods.
	For $p = 10$ and $N = 400$, C2ST~(NN) performs better than other methods, the method of \textcite{petrie_graph-theoretic_2016} often performs worst, and C2ST~(KNN) performs similarly to MMCM, with the latter performing slightly worse for small deviations and slightly better for large deviations.
	For $p = 50$, MMCM outperforms the other methods for the small value $N=100$. 
	For larger $N$, C2ST~(KNN) performs better, and the method of \textcite{petrie_graph-theoretic_2016} is equal to MMCM for small deviations. 
	For $N = 100$, C2ST~(NN) and C2ST~(KNN) perform poorly compared to the other methods. 
	C2ST~(NN) still performs poorly in the comparison for $N = 200$ while C2ST~(KNN) is competitive there for small and medium deviations.
	For other groupings with fewer numbers of differing datasets, MMCM (and Petrie's method) tends to perform worse than the C2ST~(KNN), especially for small deviations and large $N$.
	This can be seen in Figure~\ref{fig:pow.cat.multi.bin.bal.p.50}, where the PESR curves are compared for the different groupings for $p = 50$. 
	
	Overall, the PESRs are increasing with an increasing number of differing datasets, so the performances of the methods for fixed $N$, $p$, and alternative class weights are increasing from ``3+1'' to ``2+2'' to ``2+1+1'' to ``1+1+1+1'' as can be seen in Figure~\ref{fig:pow.cat.multi.bin.bal.p.50} for the special case of $p = 50$ (see Figures~\ref{fig:pow.cat.multi.bin.bal.3+1} to~\ref{fig:pow.cat.multi.bin.bal.2+1+1} in Appendix~\ref{app:add.figs.cat.multi.bin.bal} for full results). 
	However, the largest difference between the two datasets seems to be crucial, and this grows with the number of differing datasets.
	Often, the proportions for a certain combination of $N$ and $p$ and a certain weight vector for ``3+1'' (see e.g.\ Figure~\ref{fig:pow.cat.multi.bin.bal.3+1} in Appendix~\ref{app:add.figs.cat.multi.bin.bal}) are comparable to the proportions for the same combination of $N$ and $p$ for ``1+1+1+1'' and the second weight in the weight vector minus $0.2$. 
	In that case, the weights for the fourth dataset of the ``1+1+1+1'' case coincide with the ones in the ``3+1'' case.
	
\begin{figure}[!t]
		\centering
		\includegraphics[width=\linewidth]{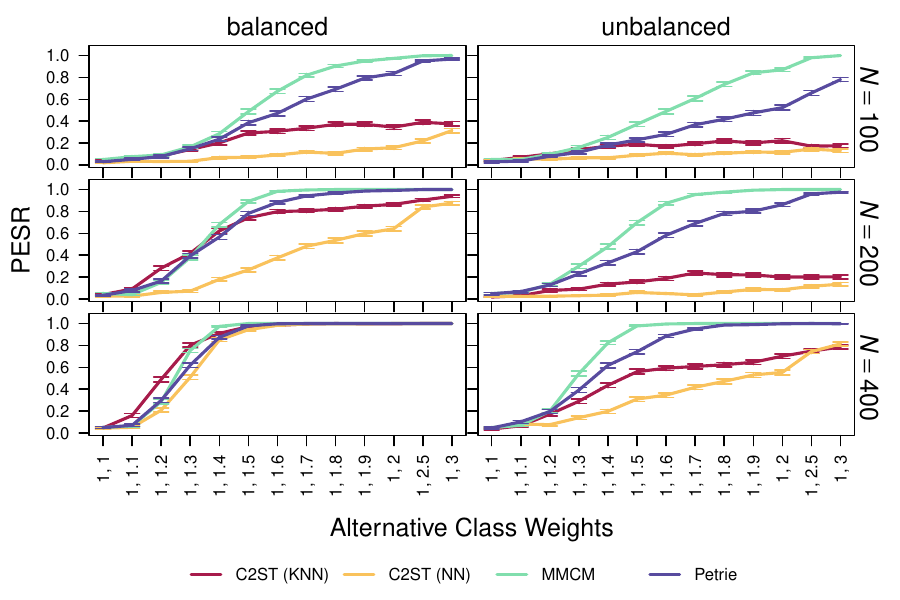}
		\caption{PESR (proportion of extreme simulation repetitions) for $k = 4$ binary datasets of equal (``balanced'') or unequal (``unbalanced'') sample sizes with $p = 50$ variables. The class weights give the unnormalized probabilities $(1, 1+\delta)$ for the values 0 and 1 for each variable in the second dataset. This means the weights in the first dataset are set to $(1, 1)$, in the second dataset to $(1, 1+\delta)$, in the third to $(1, 1+2\delta)$, and in the fourth to $(1, 1+3\delta)$. Error bars indicate Monte Carlo standard errors.}
		\label{fig:pow.cat.multi.bin.p.50.1+1+1+1}
	\end{figure}

	Regarding the balance of the sample sizes, typically, the performance of each method for a certain combination of $N$, $p$, and the alternative class weights is higher for equal sample sizes than for unbalanced sample sizes.
	This is illustrated in Figure~\ref{fig:pow.cat.multi.bin.p.50.1+1+1+1} for $p = 50$ and the ``1+1+1+1'' grouping.
	See Figures~\ref{fig:pow.cat.multi.bin.unbal.3+1} to~\ref{fig:pow.cat.multi.bin.unbal.1+1+1+1} in Appendix~\ref{app:add.figs.cat.multi.bin.unbal} for the other cases.
	Especially, the C2ST's performance suffers severely from unbalanced sample sizes.
	The performance of MMCM and the method of \textcite{petrie_graph-theoretic_2016} decreases only slightly, except for ``2+1+1'', ``1+1+1+1'' and $p = 2$, where the performance of MMCM breaks down, and for ``1+1+1+1'' and $p = 10$, where the proportion decreases for large deviations for the method of \textcite{petrie_graph-theoretic_2016}. 
	
	Like in the two-sample case, the clustering of the results (see Section~\ref{sec:clustering} in Appendix~\ref{app:add.figs}) reveals no cluster structure but rather an ordering of the methods according to their performance.
	
	\subsubsection{Multinomial Data}
\begin{figure}[!t]
		\centering
		\includegraphics[width=\linewidth]{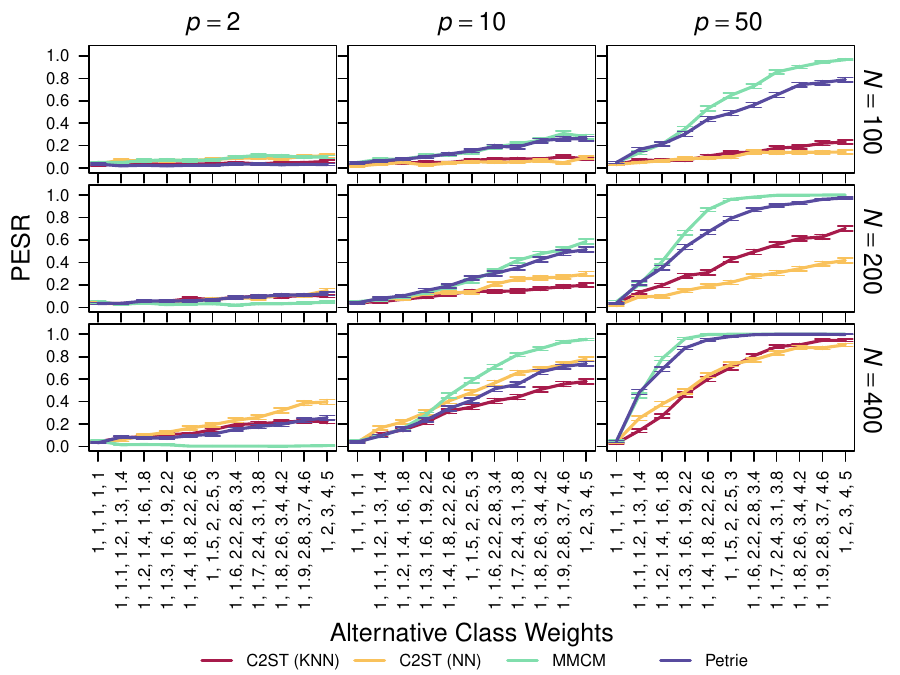}
		\caption{PESR (proportion of extreme simulation repetitions) for $k = 4$ multinomial datasets of equal sample sizes. The class weights, i.e.\ the unnormalized probabilities for the values $1$ to $5$, in the first dataset are always set to $(1, 1, 1, 1, 1)$. The class weights on the $x$-axis give the unnormalized probabilities $(1, 1+\delta, 1+2\delta, 1+3\delta, 1+4\delta)$ for each variable in the second dataset. The weights in the third dataset are given by $(1, 1+(\delta+0.1), 1+2(\delta+0.1), 1+3(\delta+0.1), 1+4(\delta+0.1))$, and the weights for the fourth dataset are given by $(1, 1+(\delta+0.2), 1+2(\delta+0.2), 1+3(\delta+0.2), 1+4(\delta+0.2))$. Error bars indicate Monte Carlo standard errors.}
		\label{fig:pow.cat.multi.multi.skew.bal.1+1+1+1}
	\end{figure}

	Figure \ref{fig:pow.cat.multi.multi.skew.bal.1+1+1+1} shows the method performances for the ``1+1+1+1'' case and balanced sample sizes for increasing skewness of the class probability distribution. 
	Again, the performance increases with the number of differing datasets (see Figure~\ref{fig:pow.cat.multi.multi.skew.bal.3+1} to~\ref{fig:pow.cat.multi.multi.skew.bal.2+1+1} in Appendix~\ref{app:add.figs.cat.multi.skew.bal}).
	Moreover, the method performance generally increases with increasing $N$ and $p$ as before.
	The MMCM is not working as intended for $p = 2$ as the proportions are close to zero regardless of the alternative class weights. 
	The other methods are also performing poorly for $p = 2$ as the proportions are quite low. 
	Only for the highest sample size $N = 400$, the C2ST variants and Petrie's method show some increase in the PESR for increasing class weights.
	For $p = 10$ and $p = 50$, MMCM typically performs best, followed by Petrie's method, then C2ST~(NN), and C2ST~(KNN) performs worst. 
	For small deviations, the MMCM and Petrie's method often perform similarly, but for larger deviations, MMCM is mostly clearly superior.
	For $p = 10$ and $N = 100$, C2ST~(NN) performs best for small deviations. 
	
	As in the binary case, the methods suffer from imbalance, except for the method of \textcite{petrie_graph-theoretic_2016} in the ``3+1'' case, which gets better and outperforms other methods in that case. 
	The performances of the MMCM and Petrie's method are not as severely impacted as those of the C2ST methods.
	
	\begin{figure}[!t]
		\centering
		\includegraphics[width=\linewidth]{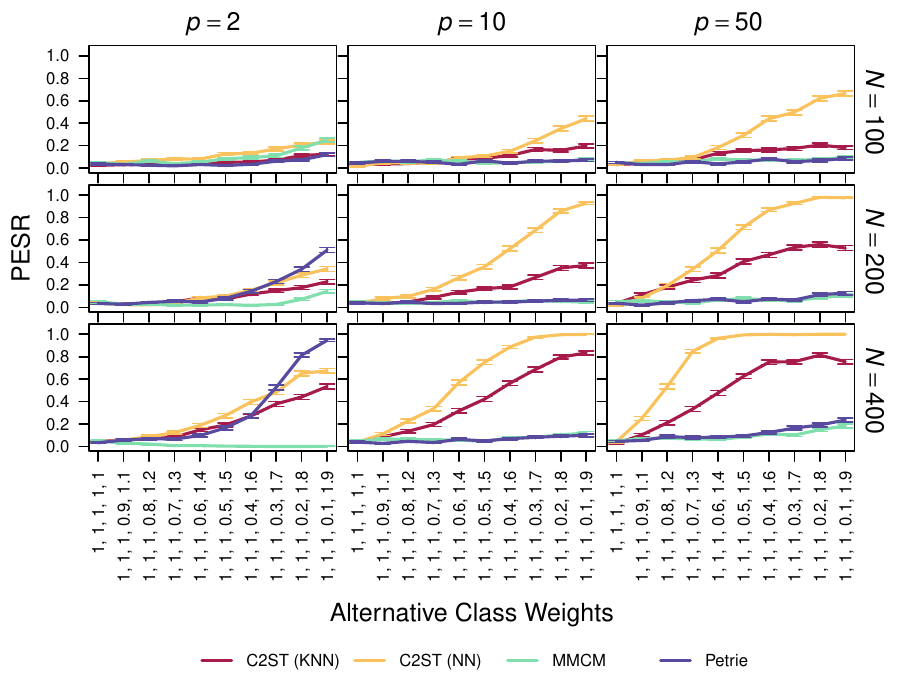}
		\caption{PESR (proportion of extreme simulation repetitions) for $k = 4$ binary datasets of equal sample sizes.  The class weights, i.e.\ the unnormalized probabilities for the values $1$ to $5$, in the first dataset are always set to $(1, 1, 1, 1, 1)$.
			The class weights on the $x$-axis give the unnormalized probabilities $(1, 1, 1, 1+\delta, 1-\delta)$ in the second dataset. The weights in the third dataset are given by $(1, 1, 1, 1+\delta+0.1, 1-\delta-0.1)$, and the weights for the fourth dataset are given by $(1, 1, 1, 1+\delta+0.2, 1-\delta-0.2)$. Error bars indicate Monte Carlo standard errors.}
		\label{fig:pow.cat.multi.multi.1up1down.bal}
	\end{figure}
	
	Figure \ref{fig:pow.cat.multi.multi.1up1down.bal} shows the method performances for the ``1+1+1+1'' case and balanced sample sizes for increasing the class probability of one class and decreasing the class probability of another class. 
	The performance for the other groupings is again increasing in the number of differing datasets (see Figure~\ref{fig:pow.cat.multi.multi.1u1d.bal.3+1} to~\ref{fig:pow.cat.multi.multi.1u1d.bal.2+1+1} in Appendix~\ref{app:add.figs.cat.multi.1up1down.bal}).
	For $p = 2$, the MMCM statistic gives again very poor results, while the C2ST variants and Petrie's method work as intended but not very well.
	For $p > 2$, the C2ST methods clearly outperform MMCM and Petrie's method, which perform very poorly.
	Using a neural network performs better than or is comparable to using KNN for the C2ST in most cases. 
	The performances are again overall increasing in $N$ and $p$.
	Again, the C2ST methods suffer from unbalanced sample sizes and perform quite poorly for low $N$.
	C2ST~(KNN) is more heavily impacted than C2ST~(NN) (see Figure~\ref{fig:pow.cat.multi.multi.1u1d.unbal.3+1} to~\ref{fig:pow.cat.multi.multi.1u1d.unbal.1+1+1+1} in Appendix~\ref{app:add.figs.cat.multi.1up1down.unbal}).
	
	The clustering of the results (see Section~\ref{sec:clustering} in Appendix~\ref{app:add.figs}) again shows a distinction between the ``skewed'' and ``1 up, 1 down'' deviations, with the graph-based tests being better at detecting the former and the C2ST variants being better at detecting the latter.

	
	\section{Applicability in Practice}\label{sec:applicability}
	In the following, the runtime and memory consumption of the methods are compared, and the errors and numerical problems that occur are discussed. 
	
	\subsection{Runtime}
	The runtimes of the methods are compared, first for the two-sample setting and then for the multi-sample setting. 
	
	\subsubsection{Two-sample Setting}
\begin{figure}[!t]
		\centering
		\includegraphics[width=\linewidth]{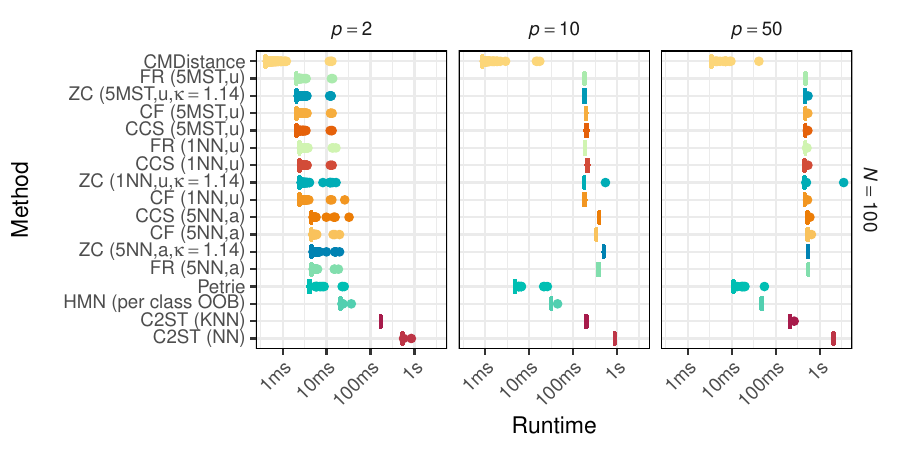}
		\caption{Runtime comparison for the scenario with two binary datasets with $N=100$ without target variables with balanced class probabilities and equal sample sizes. At least ten repetitions were performed for each method.} \label{fig:runtime.cat.no.y}
	\end{figure}

	Figure~\ref{fig:runtime.cat.no.y} shows boxplots of the runtimes for each of the pre-selected methods for the scenario with two binary datasets without a target variable with balanced class probabilities and equal sample sizes. 
	The full results for all methods can be found in Figure~\ref{fig:runtime.cat.no.y.full} in Appendix~\ref{app:add.figs.benchmarks}. 
	For $p = 2$, the runtimes of the CM distance are the lowest, followed by the edge count tests, Petrie's method, and HMN. 
	The C2ST variants take considerably longer, with C2ST (KNN) being faster than C2ST (NN). 
	The edge count tests' runtimes increase with increasing $p$ and $N$, while the other runtimes remain almost the same.
	Thus, for increasing $N$ and $p$, the edge count tests have higher runtimes than HMN, Petrie's method, and consequently also than the C2ST variants. 
	There are very slight differences in the runtimes of the edge count tests between the graph types. 
	The 1NN, ``u'' versions and 5MST, ``u'' have similar runtimes. 
	The 5NN, ``a'' takes the longest.
	There are no differences between the different edge count tests, which might be due to the implementation, in which the required quantities for all tests are calculated regardless of which test is actually performed.  
	
\begin{figure}[!t]
		\centering
		\includegraphics[width=\linewidth]{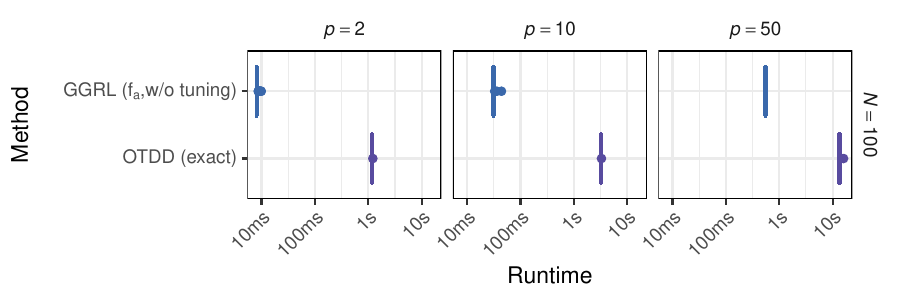}
		\caption{Runtime comparison for the scenario with $N=100$ and with two binary datasets that include target variables with balanced class probabilities and equal sample sizes. At least ten repetitions were performed for each method.} \label{fig:runtime.cat.y}
	\end{figure}

	Figure~\ref{fig:runtime.cat.y} shows boxplots of the runtimes for each of the pre-selected methods for the scenario with two binary datasets including a target variable with balanced class probabilities and equal sample sizes. 
	The full results for all methods can be found in Figure~\ref{fig:runtime.cat.y.full} in Appendix~\ref{app:add.figs.benchmarks}. 
	For both GGRL and the OTDD, the runtimes increase with increasing $N$ and $p$. 
	The runtime of the OTDD is considerably larger than that of GGRL, which would already be among the slower methods compared to those that do not take into account a target variable in the data. 
	For the highest $N$ and $p$ combination, a single calculation of the OTDD takes more than a quarter of an hour. 
	Note that the Sinkhorn variant of the OTDD, which is intended as a faster approximation, does not reduce the runtime notably in this case.
	
	\subsubsection{Multi-sample Setting}
\begin{figure}[!t]
		\centering
		\includegraphics[width=\linewidth]{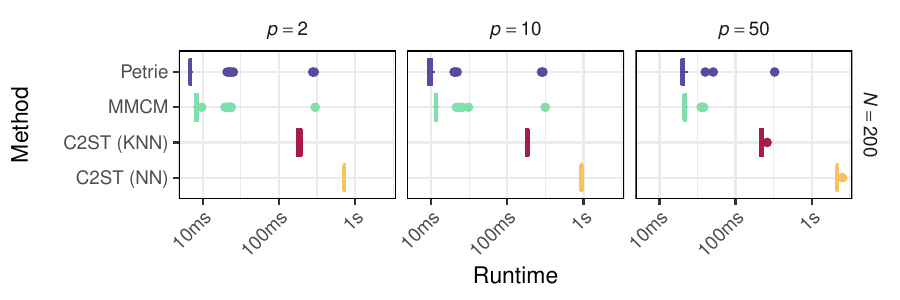}
		\caption{Runtime comparison for the scenario with four binary datasets with $N=200$ with balanced class probabilities and equal sample sizes. Ten repetitions were performed for each method.} \label{fig:runtime.cat.multi}
	\end{figure}

	Figure~\ref{fig:runtime.cat.multi} shows boxplots of the runtimes for each method for the scenario with four binary datasets with balanced class probabilities and equal sample sizes. 
	The results are similar to those for the two-sample setting.
	There is a clear method ranking with regard to runtime in the considered scenario. 
	The method of \textcite{petrie_graph-theoretic_2016} and the MMCM have the lowest runtimes. 
	The C2ST variants need considerably more time. 
	The C2ST (NN) again has considerably higher runtimes than the C2ST (KNN). 
	The runtimes of all methods do not seem to be impacted much by the dimension of the dataset in this setting.
	
	\subsection{Memory Consumption}
	The memory consumption is compared between the methods in addition to the runtime. 
	In the following, the results of that comparison are presented, again, first for the two-sample case and then for the multi-sample case. 
	
	\subsubsection{Two-sample Setting}
\begin{figure}[!t]
		\centering	\includegraphics[width=\linewidth]{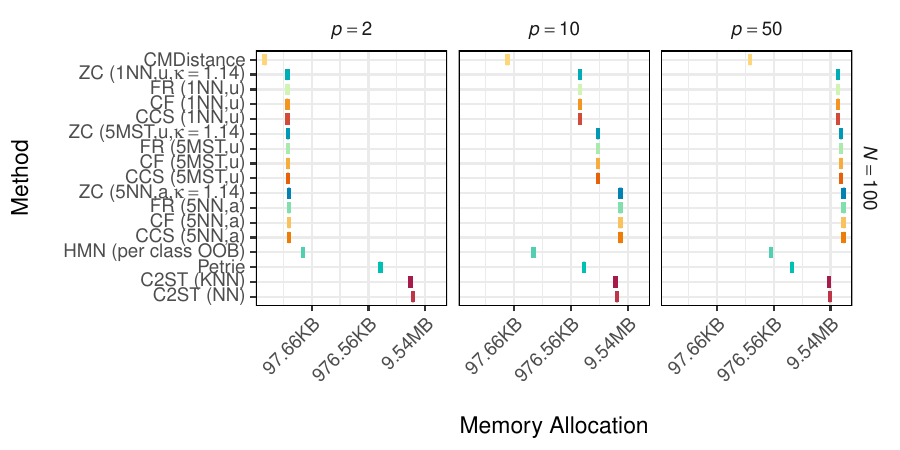}
		\caption{Memory consumption comparison for the scenario with two binary datasets with $N=100$ with balanced class probabilities and equal sample sizes. One repetition was performed for each method.} \label{fig:mem.cat.no.y}
	\end{figure}

	Figure~\ref{fig:mem.cat.no.y} shows boxplots of the memory allocation for each method for the scenario with two binary datasets with balanced class probabilities and equal sample sizes for the pre-selected methods that were compared before. 
	The full results for all methods can be found in Figure~\ref{fig:mem.cat.no.y.full} in Appendix~\ref{app:add.figs.benchmarks}.
	Except for the smallest $N$ and $p$ combination, the CM distance has the lowest memory allocation. 
	The remaining order is similar to that for the runtime: for low $p$, the edge count tests need the second least memory, followed by the HMN, and Petrie's method, and the C2ST needs the most. 
	For increasing $N$ and $p$, the memory allocation of the edge count tests increases while the other methods are not affected, so the edge count tests again need the most resources for the higher $p$ and $N$ combinations. 
	The memory requirements for Petrie's method increase with increasing $N$ such that for the highest $N$, this method requires more memory than both C2ST variants but still less than the edge count tests.
	
	With regard to memory, there are clearer differences between the graphs, at least for $p = 10$. 
	The 1NN, ``u'' versions use the least memory, followed by 5MST, ``u''. 
	The 5NN, ``a'' requires the most memory. 
	When looking at the full results, this is not a difference between ``a'' and ``u'', but the $K  = 5$ versions consistently require more memory than the $K = 1$ versions, and calculating the nearest neighbor graph allocates more memory than calculating the minimum spanning tree.
	
	The results for GGRL and the OTDD are shown in Figure~\ref{fig:mem.cat.y}. 
	The full results for all methods can be found in Figure~\ref{fig:mem.cat.y.full} in Appendix~\ref{app:add.figs.benchmarks}. 
	The memory consumption for both methods increases with increasing $N$ and $p$. 
	The calculation of the OTDD allocates more memory than that of GGRL.	
	Note that the Sinkhorn variant of the OTDD also does not reduce the memory consumption notably in this case.

	\begin{figure}[!t]
		\centering	\includegraphics[width=\linewidth]{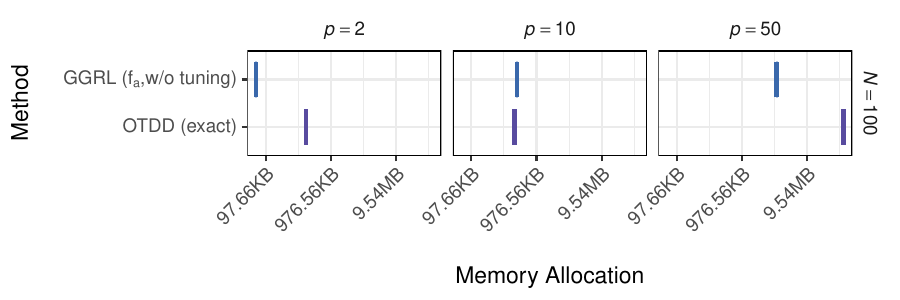}
		\caption{Memory consumption comparison for the scenario with two binary datasets including a target variable  with $N=100$ and with balanced class probabilities and equal sample sizes. One repetition was performed for each method.} \label{fig:mem.cat.y}
	\end{figure} 
	
	\subsubsection{Multi-sample setting}
\begin{figure}[!t]
		\centering
		\includegraphics[width=\linewidth]{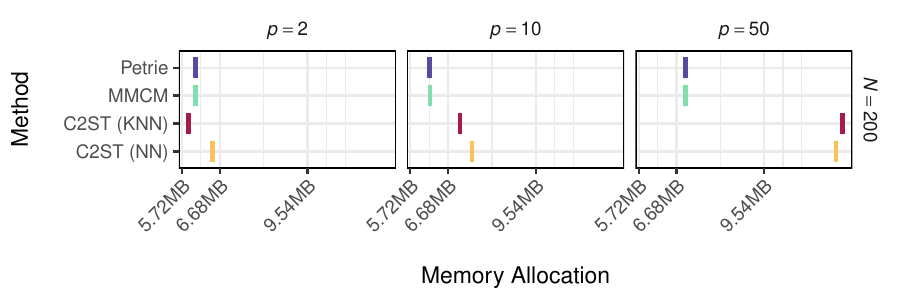}
		\caption{Memory consumption comparison for the scenario with four binary datasets with balanced class probabilities with $N=200$ and equal sample sizes. One repetition was performed for each method.} \label{fig:mem.cat.multi}
	\end{figure}

	The memory consumption of each method for the scenario with four binary datasets with balanced class probabilities and equal sample sizes is shown in Figure~\ref{fig:mem.cat.multi}.  
	The memory of the MMCM and the method of \textcite{petrie_graph-theoretic_2016} are lower than that for the C2ST variants for $N = 100, 200$ but higher for $N = 400$ as it again increases clearly with the number of observations. 
	For C2ST, the memory consumption also increases with the number of observations but additionally with the number of variables.

	\section{Summary of Best-performing Methods}\label{sec:best.meth}
	In the following, the sensitivity in detecting differences and the applicability in practice of the methods are summarized to guide the choice of the best method. 
	The results are again presented first for two samples and afterward for four samples.
	
	In each case, first, the rules for choosing the best method per scenario that can be derived from the results of Section~\ref{sec:sensitivity} are condensed. 
	These decision rules for determining the best method are displayed compactly in a decision tree. 
	The rules depend on the type of alternative (binary, multinomial ``skewed'', multinomial ``1 up, 1 down''), the balance of the sample sizes, the number of variables $p$, the number of observations $N$ in the pooled sample, and, for the multi-sample case, additionally on the grouping. 
	Depending on these parameters, the best-performing methods in terms of the highest PESR curve are given.
	Based on this, the best method for a given situation in which all the parameters are known can be easily determined.
	
	Since the true deviations between the datasets are often unknown in practice, a method for quantifying the similarity or distance of the datasets then has to be chosen independently of this true deviation. 
	Therefore, the question arises of how much worse a fixed method would perform across different scenarios than the ideal method choice for each scenario. 
	Therefore, for each alternative scenario, the difference of the PESR value of each method from the highest PESR value across all methods for that scenario is calculated. 
	The null scenarios are not considered for this comparison since, for these, the PESR should be at $0.05$  for all methods per definition.
	The median of the differences is calculated for each combination of the sample size balance, the number of categories, and $p$, i.e.\ the differences are aggregated over the different types of deviations, $N$, and in the multi-sample case, the grouping. 
	The aggregation over the deviations (and grouping) is performed since these would be unknown in real-world applications. 
	The aggregation over $N$ is performed for clarity since the best method rarely differs depending on $N$. 
	Missing PESR values are assigned the maximum difference of $1$ to penalize errors of the method.
	
	The results regarding runtime, memory, and errors from Section~\ref{sec:applicability} are summarized as follows. 
	The runtimes were measured for different $N$ and $p$ for the binary null scenario. 
	For the $i$-th $N$ and $p$ combination, the median runtimes are scaled to $[0, 1]$: 
	\[
	r_{i,j}^{\text{scaled}} = \frac{r_{i,j} - \min_{j}r_{i,j}}{\max_j r_{i,j} - \min_j r_{i,j}},
	\]
	where $r_{i,j}$ denotes the median runtime of the $j$-th method $j = 1,\dots,n_{\text{meth}}$ for the $i$-th $N$ and $p$ combination.
	Then, for each method, the median $\operatorname{med}_i r_{i,j}^{\text{scaled}}$ of these scaled values across all $N$ and $p$ combinations is taken. 
	Analogously, the measured memory consumption values are scaled to $[0,1]$ per $N$ and $p$ combination, and the median per method is taken. 
	To quantify the overall error susceptibility of the method, the maximum proportion of missing values due to computational errors or runtime issues is calculated across all scenarios of the simulation. 
	For the methods that can take a target variable into account, all aforementioned calculations are performed only for the scenarios where the OGM is not altered for comparability with the other methods.

	\subsection{Two-sample Setting}
\begin{figure}[!t]
		\centering
		\includegraphics[width=\linewidth]{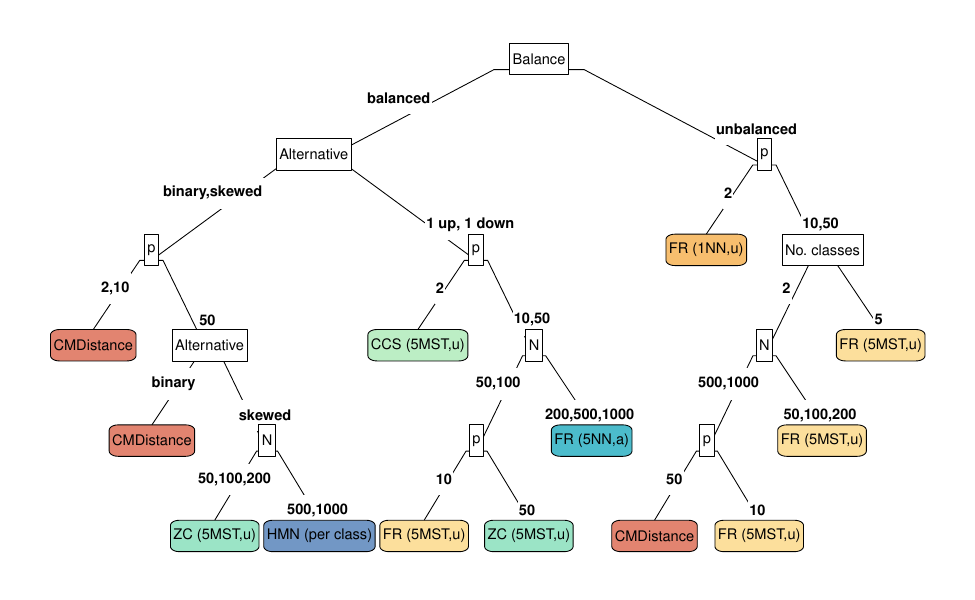}
		\caption{Summary of best-performing method per scenario for the two-sample case.}
		\label{fig:CART.cat.no.y}
	\end{figure}

	Figure~\ref{fig:CART.cat.no.y} shows a decision tree in which the decision rules for determining the best method are given.  
	For the binary and ``skewed'' case and balanced sample sizes, the CM distance is the best choice, except for $p = 50$ for which it cannot be calculated.
	In the latter case, the ZC (5MST, u) is best for low sample sizes and the HMN (per class OOB) for large sample sizes. 
	For the ``1 up, 1 down'' alternative and balanced sample sizes, different edge count tests using dense graphs are best depending on the dataset dimensions ($N$ and $p$). 
	For unbalanced sample sizes and $p = 2$, the FR (1NN, u) performs best. 
	For larger $p$, the FR (5MST, u) performs best, except for binary data, $p = 50$, and large $N$, where the CM distance is better. 
	Note that the OTDD and GGRL are never the best methods. 
	Therefore, it would lead to better results for detecting changes in the probability distributions to exclude the target variable and then use the best-performing method according to the rules presented above. 
	If the target variable must be included, OTDD can be recommended as it performed consistently better than GGRL.
	However, both performed very poorly for detecting the differences in the OGM here, or for the five categories. 

\begin{figure}[!t]
		\centering
		\includegraphics[width=\linewidth]{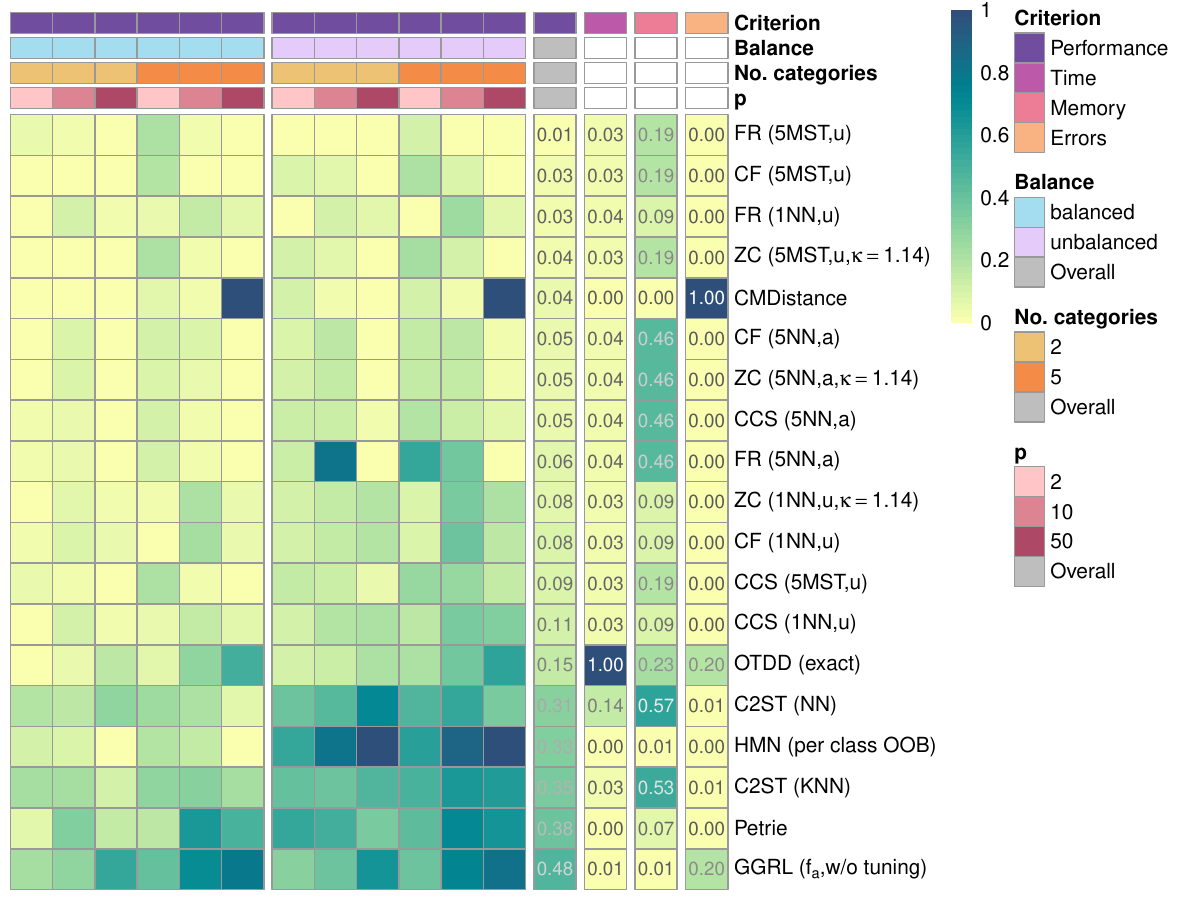}
		\caption{Performance: Median difference of the PESR values to that of the best-performing method per scenario for the two-sample case; Time: Median of per $N$-$p$-combination $[0,1]$-scaled median runtimes for binary null scenarios; Memory: Median of per $N$-$p$-combination $[0,1]$-scaled median memory consumption for binary null scenario; Errors: Maximum proportion of iterations with missing statistic values per scenario for the two-sample case. For all columns, lower values correspond to better performance.}
		\label{fig:diff.best.cat.no.y}
	\end{figure}

	Figure~\ref{fig:diff.best.cat.no.y} shows the median difference of the PESR values per method to the highest PESR value, for each combination of the sample size, balance, the number of categories, and $p$, and the overall median differences of the PESR values to the scenario-specific best method.
	It can be seen that overall, the FR (5MST, u) has the lowest median difference and is, therefore, typically not much worse than the best-performing method. 
	It is closely followed by the CF~(5MST, u) and the ZC~(5MST, u), and FR~(1NN, u). 
	Then comes the CM Distance, the other edge count tests, the OTDD, the classifier-based tests, GGRL, and last, Petrie's test.	
	With regard to runtime and errors, the best-performing methods are also among the best. 
	The edge-count tests show comparably higher values for the memory consumption, however. 
	The CM Distance would be the best choice considering performance and practical applicability for low enough $p$. 
	
	\subsection{Multi-Sample Setting}
\begin{figure}[!t]
		\centering
		\includegraphics[width=\linewidth]{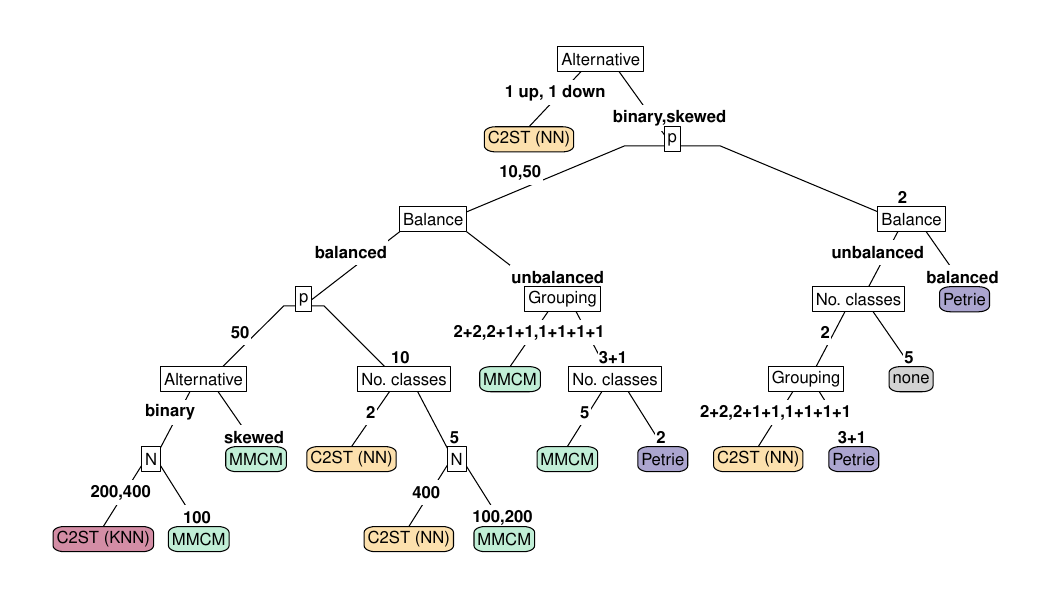}
		\caption{Summary of best-performing method per scenario for the multi-sample case.}
		\label{fig:CART.cat.multi}
	\end{figure}

	Figure~\ref{fig:CART.cat.multi} summarizes the findings for the multi-sample case. 
	For the ``1 up, 1 down'' alternative, the C2ST (NN) performs best. 
	For the binary and multinomial ``skewed'' case, $p > 2$, and balanced sample sizes, the MMCM performs best in high variable/low sample size settings; otherwise, the C2ST (NN/KNN) is better. 
	For the binary and multinomial ``skewed'' case, $p > 2$, and balanced sample sizes, the MMCM performs best except for the binary ``3+1'' case in which Petrie's method is better. 
	For the binary or the multinomial ``skewed'' case with equal sample sizes, and $p = 2$, Petrie's method performs best. 
	For binary, unbalanced data, $p = 2$, and the ``3+1'' grouping, Petrie's method is best. 
	For other groupings, the C2ST (NN) is best.
	For the multinomial ``skewed'' alternative, unbalanced data and $p = 2$, none of the methods performs well.
	
\begin{figure}[!t]
		\centering
		\includegraphics[width=\linewidth]{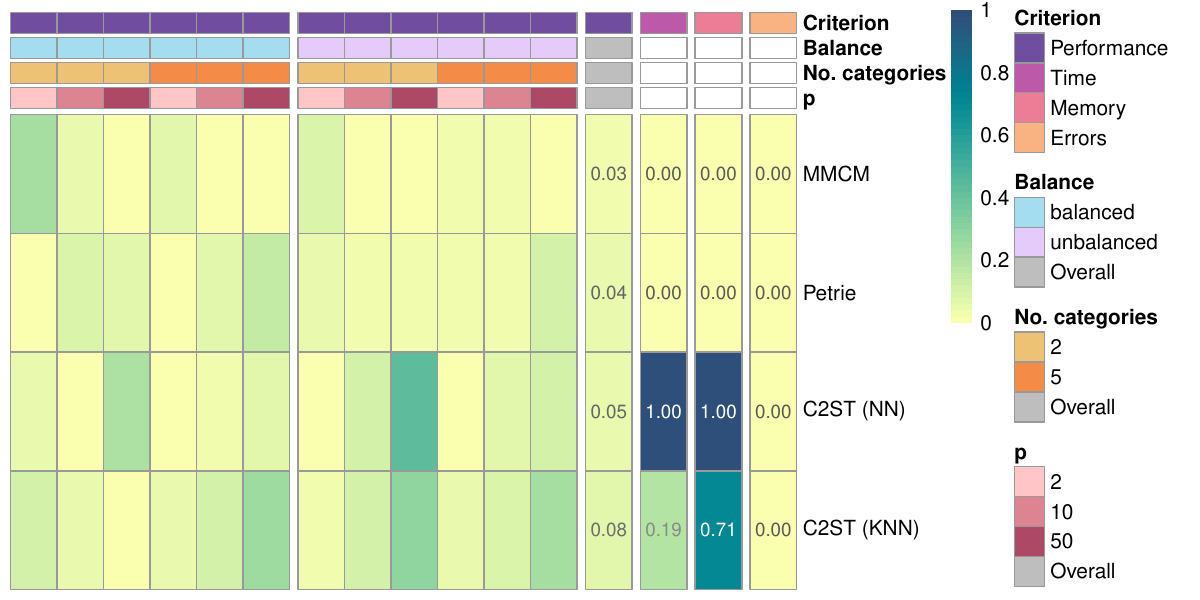}
		\caption{Performance: Median difference of the PESR values to that of the best-performing method per scenario for the multi-sample case; Time: Median of per $N$-$p$-combination $[0,1]$-scaled median runtimes for binary null scenarios; Memory: Median of per $N$-$p$-combination $[0,1]$-scaled median memory consumption for binary null scenario; Errors: Maximum proportion of iterations with missing statistic values per scenario for the two-sample case. For all columns, lower values correspond to better performance.}
		\label{fig:diff.best.cat.multi}
	\end{figure}

	As in the two-sample case, the median of the differences between the PESR of each method and that of the scenario-specific best-performing method is calculated for each combination of the sample size balance, the number of categories, and $p$. 
	Figure~\ref{fig:diff.best.cat.multi} shows these median differences and the overall median differences.
	It can be seen that overall, the MMCM has the lowest median difference and, therefore, performs best. 	
	It is followed by Petrie's method, the C2ST~(NN), and lastly the C2ST~(KNN). 
	With regard to memory and runtime, the C2ST variants are clearly worse than the graph-based methods. 
	Especially the C2ST~(NN) that shows the best performance has the longest runtimes and the highest memory consumption. 
	Therefore, the MMCM is overall best regarding performance and applicability in practice.
	However, Petrie's method might be a valuable alternative for low $p$ since the MMCM was unable to identify any deviations for $p = 2$.
	
	The ordering of the methods in the four-sample case is mostly consistent with that in the two-sample case. 
	However, it should be noted that all methods that are available in the four-sample case are among the worst-performing methods from the two-sample case.
	
	\section{Discussion and Conclusion}\label{sec:summary}
	Quantifying the similarity of two or more datasets is an important task in statistical and machine learning applications. 
	There are various methods for quantifying dataset similarity proposed in the literature, but there are no neutral comparisons of the empirical performance of such methods. 
	\textcite{stolte_methods_2024} presented a taxonomy and theoretical comparison of such methods. 
	That article, however, did not include method performance in practice. 
	Here, such a comparison is performed for selected methods from the aforementioned review that are applicable to categorical data and that performed well with regard to the theoretical criteria. 
	The aims of this comparison study are to:
	\begin{enumerate}
		\item Compare dataset similarity measures with respect to their performance in detecting differences of datasets drawn from distributions that differ in certain aspects, and identify groups of dataset similarity measures that act similarly across different alternatives.
		\item Compare dataset similarity measures with respect to their consumption of computational resources.
	\end{enumerate}
	For this, a simulation study was performed. 
	As no proper power comparison of the methods was feasible, the proportion of extreme simulation repetitions (PESR) is considered instead.
	In short, this is the proportion of repetitions in which the observed statistic value is more extreme than a threshold calculated from simulations under a null scenario in which the true distributions from which the datasets are generated do not differ. 
	Three main scenarios were considered for the two- and $k$-sample case with $k = 4$. 
	\begin{enumerate}
		\item Binary datasets with equal or differing class distributions. 
		\item Categorical dataset with equal class distributions or differing class distributions that gradually become more skewed. 
		\item Categorical dataset with equal class distributions or differing class distributions where the probability for one class increases while the probability of another class decreases accordingly (``1 up, 1 down''). 
	\end{enumerate}
	
	In the two-sample case, a total of 62 variants of 12 methods were included. 
	Out of these, 19 could be pre-selected for the overall comparisons by excluding variants that consistently performed worse than the selected variants. 
	The included methods were: 
	\begin{itemize}
		\item Graph-based methods
		\begin{itemize}
			\item Edge count tests of \textcite{friedman_multivariate_1979} (FR), \textcite{chen_new_2017} (CF), \textcite{chen_weighted_2018} (CCS), and \textcite{zhang_graph-based_2022} (ZC) each with a 5-minimum spanning tree (MST) and using the union of all optimal graphs due to ties, a 1-nearest neighbors graph (NN) and using the union, or a 5-NN and using averaging over all optimal graphs
			\item Cross-Match-based tests using the optimal non-bipartite matching by \textcite{petrie_graph-theoretic_2016} (Petrie) (only for multi-sample case) and \textcite{mukherjee_distribution-free_2022} (MMCM)
		\end{itemize}
		\item Classifier-based methods
		\begin{itemize}
			\item The Classifier Two-sample Test (C2ST) \textcite{lopez-paz_revisiting_2017} using a neural net (NN) or a $K$-nearest neighbors (KNN) classifier
			\item The random-forest based test by \textcite{hediger_use_2021} (HMN)
		\end{itemize}
		\item The Constrained Minimum (CM) Distance \textcite{tatti_distances_2007} that is based on summaries of the data
		\item The dataset distance based on decision trees by \textcite{ganti_framework_1999} (GGRL) that takes into account a target variable included in the dataset
		\item The Optimal Transport Dataset Distance (OTDD) by \textcite{alvarez-melis_geometric_2020} (GGRL) that takes into account a target variable included in the dataset
	\end{itemize}
	The results for these selected methods in the two-sample case can be summarized as follows. 
	In general, the PESR values for all methods increased with increasing sample sizes $N$. 
	For most methods, the PESR values also increased with increasing numbers of variables $p$. 
	Exceptions from this were the edge count tests using the 1NN or 1MST, for which the PESR decreased in most cases for increasing $p$, and the C2ST (NN) for which the PESR decreased in some cases for increasing $p$.
	The PESR values for each method and scenario are lower for unbalanced sample sizes than for balanced sample sizes.  
	For binary datasets and for multinomial datasets with the ``skewed'' alternative and small numbers of variables $p = 2, 10$ and with equal sample sizes, the constrained minimum (CM) distance \textcite{tatti_distances_2007} performed best. 
	It is, however, infeasible to calculate for categorical data with five categories and $p = 50$ variables, and it is affected by the imbalance of the sample sizes. 
	Moreover, it was unable to detect the ``1 up, 1 down'' alternatives in the multinomial case. 
	In the cases where the CM distance has its weaknesses, one of the edge count tests was best. 
	Otherwise, these were the next best alternatives to the CM distance.
	
	The tests using the 1NN, ``u'' were then best for $p=2$ and the tests using the 5MST, ``u'', or 5NN, ``a'' for $p = 10, 50$. 
	The differences between the edge count tests were small for balanced sample sizes. 
	For unbalanced sample sizes and binary data, the weighted edge count test \textcite{chen_weighted_2018} was best.
	This was expected as it was specifically intended for unbalanced sample sizes.  
	For unbalanced multinomial data, however, the original edge count test \textcite{friedman_multivariate_1979} was best.
	The HMN was competitive for balanced data and higher sample sizes $N$ and numbers of variables $p$, but its performance broke down in the case of unbalanced sample sizes. 
	The C2ST variants (using a KNN classifier or a multilayer perceptron) were typically outperformed by the CM distance, the edge count tests, and the HMN. 
	They were also more heavily affected by the imbalance of the sample sizes than the edge count tests. 
	Petrie's method was often the worst in the comparison. 
	
	For datasets that include a target variable, there were two methods considered that take the target variable into account, namely the OTDD and GGRL. 
	The OTDD consistently outperformed GGRL.
	For the case where the datasets only differ in the relationship between the covariates and the target variable, both methods were mostly unable to detect this kind of difference, except for the OTDD with very high $N$, low $p$ for cases where the coefficients for the outcome-generating model changed completely.
	In the case where the datasets only differ in the covariate distribution, for $p = 2$, the OTDD is among the better methods compared to the methods that do not consider the target variable, while GGRL is among the worst. 
	For higher $p$, OTDD is only in the middle field for binary data, and for multinomial data, sometimes also among the worst. 
	
	Overall, there was a tendency for the edge count tests to show high performance for all alternatives. 
	Especially the FR (5MST, u) can be recommended.
	The CM distance showed high performance for binary data or multinomial data with the ``skewed alternative'' but only for balanced sample sizes. 
	In those cases, it had the highest PESR values even for small deviations. 
	The HMN is somewhere in the middle field for balanced sample sizes and the worst for unbalanced sample sizes. 
	The C2ST had comparably low PESR values and was better for detecting the ``1 up, 1 down'' alternative than for the ``skewed'' alternative in the multinomial case. 
	The OTDD outperformed the GGRL, but neither showed advantages over methods ignoring the target variable, especially since they were both unable to detect the considered differences between the datasets in the outcome-generating model (OGM).
	Based on the results presented here, discarding the target variable and choosing one of the other well-performing methods would typically perform better. 
	
	The observation that denser graphs, such as the 5MST and 5NN, instead of the MST and 1NN, perform better empirically for higher-dimensional data is in line with earlier simulation studies. 
	For a detailed discussion of the use of denser graphs, refer to \textcite{zhu_limiting_2024}. 
	In particular, they derive less strict assumptions for the asymptotic distributions of the edge count statistics that allow for denser graphs than the assumptions made in the original articles. 
	
	For the multi-sample case, four datasets were considered, and all possible combinations of how many of these four datasets can differ from each other. 
	Only the C2ST, Petrie's method, and the MMCM were applicable to more than two samples. 
	The comparison of these can be summarized as follows. 
	Again, the PESR values increase with increasing $N$ and $p$. 
	They also increase with an increasing number of differing pairs of datasets.
	The PESR values for each method and scenario are again lower for unbalanced sample sizes than for balanced sample sizes. 
	For binary datasets, the MMCM or Petrie's method is often the best. 
	The MMCM does, however, not work for $p = 2$ due to the presence of many ties. 
	In the multinomial case, MMCM and Petrie's method are better at detecting the ``skewed'' alternatives while the C2ST is better at detecting ``1 up, 1 down''. 
	Often, the C2ST (NN) is better than the C2ST (KNN). 
	The C2ST is more heavily affected by unequal sample sizes than the MMCM and Petrie's method are. 
	
	With regard to runtime and memory, the CM distance, the MMCM, and Petrie's method performed best. 
	The edge count tests' performances depended on the number of samples and the number of variables. 
	For small $p$, their computational costs are lower than those of the HMN and the C2ST variants; for increasing $p$ (and $N$), the costs increase while those of the HMN and C2ST variants are almost constant, such that the resource consumption of the edge count tests exceeds the others. 
	For the classifier-based tests, the HMN had considerably lower resource consumption than the C2ST variants, and the C2ST (KNN) had lower runtime and memory allocation than the C2ST (NN). 
	The GGRL and especially the OTDD show very high resource consumption, especially with regard to the runtime. 
	These results should, however, be interpreted with caution as they might depend on the operating system, the configuration of the computer, and \texttt{R} \textcite{R} and the required packages. 
	Moreover, memory can only be measured for computations that are performed in \texttt{R} itself, while computations that are performed in other programming languages cannot be considered. 
	This might give an unfair advantage to the tests based on the optimal non-bipartite matching, where the matching itself is done in \texttt{Fortran} as well as for the HMN, where the random forest calculation is done in \texttt{C}. 
	This might be a reason that their memory did not depend on the dataset dimension. 
	The MST calculation for the edge count tests is also performed in \texttt{C}, while the NN calculation is performed in \texttt{R} itself. 
	The C2ST also uses internal \texttt{C} code for training the models. 
	
	In summary, with no knowledge about the true deviation of the datasets, the edge count tests, especially FR~(5MST,u), can be recommended overall in the two-sample case as a good compromise of high performance across many alternatives and acceptable resource consumption and computational error occurrences. 
	For low $p$, the CM Distance might be an even better-performing alternative with lower resource consumption. 
	
	For the multi-sample case, the MMCM can be recommended overall due to its comparably high performance and low runtimes, and memory consumption. 
	Petrie's method is an alternative with similarly low resource requirements for low $p$ where the MMCM was unable to detect any differences between datasets.
	
	One potential limitation of the current study is that the scenarios could not be chosen in a way that, with an increase in the number of variables, the true difference between the datasets increases as each variable follows the same distribution. 
	Therefore, the decreasing performance of methods for an increasing number of variables might be covered by the increasing true dataset difference. 
	Similarly, the maximum difference in the class probabilities increases with an increasing number of differing datasets, which might explain the increasing ability of the methods to detect differences for an increasing number of differing datasets. 
	
	Another aspect that could be improved is the differing implementations of the methods, which allowed for the choice of an appropriate distance measure for calculating the graph for the edge count tests, but not for the MMCM and Petrie's method, where only Euclidean distances could be used. 
	This could in particular influence the performance for the ``1 up, 1 down'' scenario since for the Euclidean distance, observing category 5 instead of 4 (which was the case that became more likely here in this scenario) is a smaller change than observing category 5 instead of 1, which was more likely in the ``skewed'' case. 
	For all other methods, these changes have equal distances, which would be appropriate for nominal data.
	This might, in part, explain the comparably bad performance of MMCM and Petrie's method for the former scenario and the comparably good performance for the latter. 
	This issue has been fixed in the new implementation used by the \texttt{DataSimilarity} package \textcite{DataSimilarity}, which allows choosing a distance function. 
	Moreover, for the CM distance, only the suggested feature functions were used. 
	Again, the coding in the ``1 up, 1 down'' scenario might influence the results here. 
	Finding a feature function that is more suitable for detecting the ``1 up, 1 down'' alternative might be considered in future research, as the CM distance performed so well in the other cases.
	This highlights the importance of choosing the distance function or feature function, respectively, appropriately for the coding and scale level of the data at hand.  
	
	Moreover, a more in-depth analysis of datasets including a target variable is needed to gain insight into the best way to compare them.
	Additional analysis on the influence of other outcome-generating models on the performance of the OTDD and GGRL could give more insights into how these methods behave in practice. 
	An analysis of the performance of the other methods when including the target variable in the same way as all covariates could provide additional guidance on how to best handle datasets with a target variable.
	
	Other aspects that could be considered in the future are the comparison of dataset similarity methods for numerical datasets. 
	Based on the comparisons, one could try to combine methods that worked well for different alternatives, like the MMCM / Petrie's method and the C2ST in the multi-sample case, to create a new method or a group of methods that can detect various types of differences between datasets. 

	\section*{Acknowledgments}
	This work has been supported (in part) by the Research Training Group ``Biostatistical Methods for High-Dimensional Data in Toxicology'' (RTG 2624, Project P1) funded by the Deutsche Forschungsgemeinschaft (DFG, German Research Foundation -- Project Number 427806116).\\
	The authors gratefully acknowledge the computing time provided on the Linux HPC cluster at TU Dortmund University (LiDO3), partially funded in the course of the Large-Scale Equipment Initiative by the Deutsche Forschungsgemeinschaft (DFG, German Research Foundation) as project 271512359.
	
	\section*{Declaration of Interests}
	The authors report there are no competing interests to declare. 
	
	\section*{Author CRediT Roles}
	Conceptualization: M.S., J.R., A.B.; Formal analysis: M.S.; Methodology: M.S.; Project administration: J.R., A.B.; Software: M.S.; Supervision: J.R., A.B.; Visualization: M.S.; Writing -- original draft: M.S.; Writing -- review \& editing: J.R., A.B.

	\begin{sloppypar}
		\begin{footnotesize}
			\printbibliography[title = References]
		\end{footnotesize}
	\end{sloppypar}

	\clearpage
	\appendix
	\section{Scenario Parameter Settings}\label{app:scen.tabs}
	\begin{table}[H]
		\centering
		\begin{tabular}{lp{4.5cm}p{6.5cm}}
			\toprule
			Distribution & Alternative & Parameters\\
			\midrule
			Bernoulli & Null: $F_j = \text{Bin}(0.5)$, $ j = 1, 2$ & -- \\
			& Unbalanced: $F_j = \text{Bin}(\pi^{(j)}),$ $ j = 1, 2$& $\pi^{(1)} = 0.5$, $\pi^{(2)} = \frac{1 + \delta}{2 + \delta}$, $\delta = 0.1, $$0.2,$$ 0.3, $$ 0.4, $$0.5, $$0.6, 0.7, $$ 0.8,$$ 0.9,$$ 1, $$1.5,$$ 2$ \\
			Multinomial & Null: $F_j = \text{Mult}(1/5\cdot \mathbf{1}_5)$, $ j = 1, 2$ & -- \\
			& Skewed: $F_j = Mult(\pi^{(j)})$ & $\pi^{(1)} = 1/5\cdot \mathbf{1}_5$, $\pi^{(2)} = w / \sum_{l=c}^5 w_c$, $w_c = 1 + (c-1)\delta, c = 1,\dots,5,$  $\delta = 0.1, 0.2, 0.3, $$ 0.4, 0.5, 1$ \\
			& 1 up, 1 down: $F_j = \text{Mult}(\pi^{(j)}),$ $ j = 1, 2$& $\pi^{(1)} = 1/5 \cdot \mathbf{1}_5$, $\pi^{(2)} = w / 5$, $w_1 = w_2 = w_3 = 1,$ $w_4 = 1-\delta$, $w_5 = 1 + \delta$, $\delta = 0.1, 0.2, 0.3, $$ 0.4, 0.5, 0.6, 0.7, 0.8, 0.9$ \\
			\bottomrule
		\end{tabular}
		\caption{Scenarios for categorical data generation for $k = 2$. $X_i^{(j)}\stackrel{\text{iid}}{\sim}F_j$, $i = 1, \dots, n_j,$ $j = 1, 2$.}
		\label{tab:scen.cat.no.y}
	\end{table}
	
	\begin{table}[H]
		\centering
		\begin{tabular}{lp{6cm}p{5.4cm}}
			\toprule
			Distribution & Alternative & Parameters\\
			\midrule
			Each & Null: $\Prob\left(Y^{(j)}_i = 1\right) = \exp(\eta_i^{(j)}) / (1+\exp(\eta_i^{(j)}))$, $\eta_i^{(j)} = -1/2 + \tilde{X}^{(j)}_i\beta$  & $\beta = 1/2\cdot(\mathbf{1}_{(C-1)p/2}^T, -\mathbf{1}_{(C-1)p/2}^T)^T$ \\
			& Wrong sign: $\Prob\left(Y^{(j)}_i = 1\right) = \exp(\eta_i^{(j)}) / (1+\exp(\eta_i^{(j)}))$, $\eta_i^{(j)} = -1/2 + \tilde{X}^{(j)}_i\beta^{(j)}$ & $\beta^{(1)} = 1/2\cdot(\mathbf{1}_{(C-1)p/2}^T, -\mathbf{1}_{(C-1)p/2}^T)^T$, $\beta^{(2)} = 1/2\cdot(-\mathbf{1}_{p/2}^T, \mathbf{1}_{p/2}^T)^T$ \\
			& Wrong size: $\Prob\left(Y^{(j)}_i = 1\right) = \exp(\eta_i^{(j)}) / (1+\exp(\eta_i^{(j)}))$, $\eta_i^{(j)} = -1/2 + \tilde{X}^{(j)}_i\beta^{(j)}$ & $\beta^{(1)} = 1/2\cdot(\mathbf{1}_{(C-1)p/2}^T, -\mathbf{1}_{(C-1)p/2}^T)^T$, $\beta^{(2)} = 1/4\cdot(\mathbf{1}_{(C-1)p/2}^T, -\mathbf{1}_{(C-1)p/2}^T)^T$ \\
			& Wrong coef: $\Prob\left(Y^{(j)}_i = 1\right) = \exp(\eta_i^{(j)}) / (1+\exp(\eta_i^{(j)}))$, $\eta_i^{(j)} = \beta_0^{(j)} + \tilde{X}^{(j)}_i\beta^{(j)}$ & $\beta_0^{(1)} = - 1/2$, $\beta^{(1)} = 1/2\cdot(\mathbf{1}_{(C-1)p/2}^T, -\mathbf{1}_{(C-1)p/2}^T)^T$, $\beta_0^{(1)} = 1$, $\beta^{(2)} = \cdot(\mathbf{0}_{(C-1)p/2}^T, -\mathbf{2}_{(C-1)p/2}^T)^T$ \\
			\bottomrule
		\end{tabular}
		\caption{Additional scenarios for categorical data generation for $k = 2$ with target variable. $X_i^{(j)}\stackrel{\text{iid}}{\sim}F_j$, $i = 1, \dots, n_j,$ $j = 1, 2$. $\tilde{X}^{(j)} \in \mathbb{R}^{(C-1)p}$ is the vector in which all observations of $X_i^{(j)}\in \mathbb{R}^{p}$ are dummy coded. $C$ denotes the number of categories.}
		\label{tab:scen.cat.y}
	\end{table}
	
	\begin{figure}[H]
		\centering
		\includegraphics[width=\linewidth]{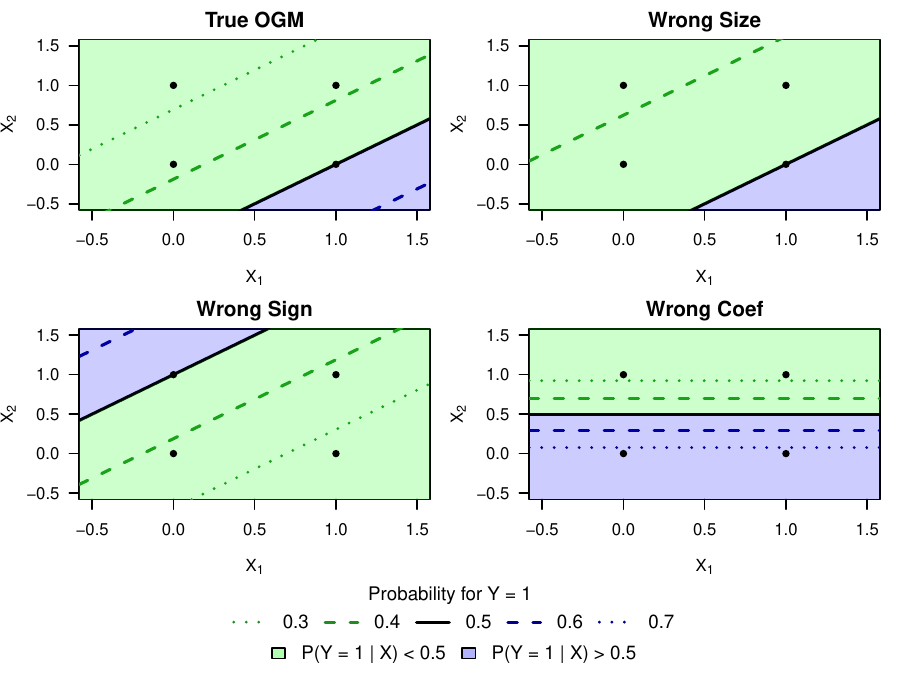}
		\caption{Visualization of the decision boundaries for the different OGMs for generating a binary target variable $Y$ for datasets with $p = 2$ binary covariates $X_1$ and $X_2$.}
		\label{fig:dec.bound}
	\end{figure}
	
	\FloatBarrier
	\begin{longtable}{lp{4.5cm}p{6.5cm}}
		\toprule
		Distribution & Alternative & Parameters\\
		\midrule
		\endfirsthead
		\toprule
		Distribution & Alternative & Parameters\\
		\midrule
		\endhead
		\bottomrule
		\endfoot
		\bottomrule
		\caption{Scenarios for categorical data generation for $k = 4$. $X_i^{(j)}\stackrel{\text{iid}}{\sim}F_j$, $i = 1, \dots, n_j,$ $j = 1,\dots, 4$.}
		\label{tab:scen.cat.multi}\\
		\endlastfoot
		Bernoulli & Null: $F_j = \text{Bin}(0.5)$, $ j = 1,\dots, 4$ & -- \\
		& Unbalanced: $F_j = \text{Bin}(\pi^{(j)}), $ $ j = 1,\dots, 4$& $\pi^{(1)} = \pi^{(2)} = \pi^{(3)} = 0.5$, $\pi^{(4)} = \frac{1 + \delta}{2 + \delta}$, $\delta = 0.1,$$ 0.2, $$0.3, $$ 0.4, $$0.5, $$0.6, $$0.7,  $$0.8,$$ 0.9,$$ 1,$ $ 1.5,$$ 2$ \\
		& & $\pi^{(1)} = \pi^{(2)} = 0.5$, $\pi^{(3)} = \pi^{(4)} = \frac{1 + \delta}{2 + \delta}$, $\delta = 0.1, 0.2, 0.3, $$ 0.4, 0.5, 0.6, 0.7, $$ 0.8,$$ 0.9, $ $1, $$1.5, 2$ \\
		& & $\pi^{(1)} = \pi^{(2)} = 0.5$, $\pi^{(3)} = \frac{1 + \delta}{2 + \delta},$ $\pi^{(4)} = \frac{1 + 2\delta}{2 + 2\delta}$, $\delta = 0.1, 0.2, 0.3, $$ 0.4, 0.5,$$ 0.6, $ $0.7,  0.8, 0.9, 1, 1.5, 2$ \\
		& & $\pi^{(j)} = \frac{1 + (j-1)\delta}{2 + (j-1)\delta}$, $\delta = 0.1, 0.2, 0.3, $$ 0.4, 0.5,$ $ 0.6, 0.7,  0.8, 0.9, 1, 1.5, 2$ \\
		Multinomial & Null: $F_j = \text{Mult}(1/5\cdot \mathbf{1}_5)$, $ j = 1,\dots, 4$ & -- \\
		& Skewed: $F_j = \text{Mult}(\pi^{(j)})$ & $\pi^{(1)} = \pi^{(2)} = \pi^{(3)} = 1/5\cdot \mathbf{1}_5$, $\pi^{(4)} = w / \sum_{l=c}^5 w_c$, $w_c = 1 + (c-1)\delta,$ $ c = 1,\dots,5,$  $\delta = 0.1, 0.2, 0.3, $$ 0.4, $$0.5,$ $ 0.6, 0.7, 0.8, 0.9, 1$ \\
		& & $\pi^{(1)} = \pi^{(2)} = 1/5\cdot \mathbf{1}_5$, $\pi^{(3)} = \pi^{(4)} = w / \sum_{l=c}^5 w_c$, $w_c = 1 + (c-1)\delta,$ $ c = 1,\dots,5,$  $\delta = 0.1, 0.2, 0.3, $$ 0.4, 0.5, $ $0.6, 0.7, 0.8, 0.9, 1$ \\
		& & $\pi^{(1)} = \pi^{(2)} = 1/5\cdot \mathbf{1}_5$, $\pi^{(3)} = w / \sum_{l=c}^5 w_c$, $w_c = 1 + (c-1)\delta$, $\pi^{(4)} = w / \sum_{l=c}^5 w_c$, $w_c = 1 + (c-1)(\delta+0.1),$ $ c = 1,\dots,5,$  $\delta = 0.1, 0.2, 0.3, $$ 0.4, 0.5,$$ 0.6,$ $ 0.7, $$0.8, 0.9, 1$ \\
		& & $\pi^{(1)} = 1/5\cdot \mathbf{1}_5$,$\pi^{(j)} = w / \sum_{l=c}^5 w_c$, $w_c = 1 + (c-1)(\delta + (j-1)0.1),$ $c = 1,\dots,5,$ $j = 2, 3, 4,$  $\delta = 0.1, 0.2, 0.3, $$ 0.4, 0.5,$ $ 0.6, 0.7, 0.8, 0.9, 1$ \\
		& 1 up, 1 down: $F_j = \text{Mult}(\pi^{(j)}),$ $ j = 1,\dots, 4$& $\pi^{(1)} = \pi^{(2)} =  \pi^{(3)} = 1/5 \cdot \mathbf{1}_5$, $\pi^{(4)} = w / 5$, $w_1 = w_2 = w_3 = 1,$ $w_4 = 1-\delta$, $w_5 = 1 + \delta,$ $\delta = 0.1, 0.2, 0.3, $$ 0.4, 0.5,$$ 0.6, 0.7,$ $ 0.8, 0.9$ \\
		& & $\pi^{(1)} = \pi^{(2)} =  1/5 \cdot \mathbf{1}_5$, $\pi^{(3)} = \pi^{(4)} = w / 5$, $w_1 = w_2 = w_3 = 1,$ $w_4 = 1-\delta$, $w_5 = 1 + \delta,$ $\delta = 0.1, 0.2, 0.3, $$ 0.4, 0.5, 0.6, 0.7,$ $ 0.8, 0.9$ \\
		& & $\pi^{(1)} = \pi^{(2)} =  1/5 \cdot \mathbf{1}_5$, $\pi^{(3)} = w / 5$, $w_1 = w_2 = w_3 = 1,$ $w_4 = 1-\delta$, $w_5 = 1 + \delta,$ $\pi^{(4)} = w / 5$, $w_1 = w_2 = w_3 = 1,$ $w_4 = 1-\delta-0.1$, $w_5 = 1 + \delta + 0.1,$ $\delta = 0.1, 0.2, 0.3, $$ 0.4, 0.5, 0.6, 0.7,$ $ 0.8, 0.9$ \\
		& & $\pi^{(1)} = 1/5 \cdot \mathbf{1}_5,$ $\pi^{(j)} = w / 5$, $w_1 = w_2 = w_3 = 1,$ $w_4 = 1-\delta - (j-1) 0.1$, $w_5 = 1 + \delta + (j-1) 0.1,$$\delta = 0.1, 0.2, 0.3,$$ 0.4, 0.5, 0.6, 0.7,$ $ 0.8, 0.9$ \\
	\end{longtable}
	\FloatBarrier
	
	\section{Complete Method list}\label{app:meth.tabs}
	\begin{landscape}
		\begin{longtable}{p{3cm}p{3cm}p{1.3cm}p{4.8cm}p{2.8cm}p{0.4cm}p{0.9cm}p{0.8cm}p{1.2cm}}
			\toprule
			Method &  (Sub)class & No. fulfilled & Implementation & Inclusion & $y$? & Num.? & Cat.? & $K > 2$?\\
			\midrule
			\endhead
			\bottomrule
			\endfoot
			\bottomrule
			\caption{Methods from theoretical comparison chosen for the empirical comparison. Methods applicable to two or more categorical datasets without target variables are highlighted. These are compared in the current study. No. fulfilled: Number of fulfilled criteria in theoretical comparison. $y$?: Can the method deal with a target variable in the dataset? Num.?: Is the method as implemented applicable to numeric data? Cat.?: Is the method as implemented applicable to categorical data? $K > 2$?: Is the method as implemented applicable to more than two datasets at a time? \faTimes$^*$: Method is, in theory, applicable, but implementation is not. \faCheck$^*$: Implementation is applicable, although this case is not described in the literature.}\label{tab:methods}\\
			\endlastfoot
			KMD \textcite{huang_kernel_2022} & Kernel-based & 15 & R package KMD \textcite{KMD} & Implemented \& $\ge 11$ criteria & \faTimes & \faCheck & \faTimes$^*$ & \faCheck\\
			\rowcolor{lightgray} \textcite{mukherjee_distribution-free_2022} & Graph-based & 13 & R package multicross \textcite{multicross} & Implemented \& $\ge 11$ criteria & \faTimes & \faCheck & \faCheck & \faCheck \\
			\textcite{biswas_distribution-free_2014} & Graph-based & 12 & Own implementation & $\ge 11$ criteria & \faTimes & \faCheck & \faTimes & \faCheck\\
			\rowcolor{lightgray} \textcite{friedman_multivariate_1979} & Graph-based & 13 & R package gTests \textcite{gTests} & Implemented \& $\ge 11$ criteria & \faTimes & \faCheck & \faCheck & \faTimes \\
			Cross-match test \textcite{rosenbaum_exact_2005} & Graph-based & 13 & R package crossmatch \textcite{crossmatch} & Implemented \& $\ge 11$ criteria & \faTimes & \faCheck & \faTimes & \faTimes \\
			Cramér test \textcite{baringhaus_new_2004} & Inter-point distances & 11 & R package cramer \textcite{cramer} & Implemented \& $\ge 11$ criteria & \faTimes & \faCheck & \faTimes & \faTimes \\
			Energy statistic \textcite{szekely_energy_2017} & Inter-point distances & 13 & R package energy \textcite{energy} & Implemented \& $\ge 11$ criteria & \faTimes & \faCheck & \faTimes & \faCheck \\
			\textcite{deb_multivariate_2021} & Inter-point distances / Rank-based & 12 & Implementation based on R code for paper (\url{https://github.com/NabarunD/MultiDistFree}) & Implemented \& $\ge 11$ criteria & \faTimes & \faCheck & \faTimes & \faTimes \\
			\textcite{ntoutsi_general_2008} & Comparison of density functions & 11 & Own implementation & $\ge 11$ criteria & \faCheck & \faCheck & \faTimes & \faTimes \\
			\rowcolor{lightgray} \textcite{ganti_framework_1999} & Comparison of density functions & 11 & Own implementation & $\ge 11$ criteria & \faCheck & \faCheck & \faTimes$^*$ & \faTimes \\
			\rowcolor{lightgray} \textcite{hediger_use_2021} & Binary classification & 11 & R package hypoRF \textcite{hypoRF} & Implemented \& $\ge 11$ criteria & \faTimes & \faCheck & \faCheck & \faTimes \\
			\rowcolor{lightgray} \textcite{petrie_graph-theoretic_2016} & Graph-based & 13 & R package multicross \textcite{multicross} & Implemented \& $\ge 11$ criteria & \faTimes & \faCheck & \faCheck$^*$ & \faCheck \\
			\rowcolor{lightgray} OTDD \textcite{alvarez-melis_geometric_2020} & Distance/ similarity measure for datasets & 11 & own implementation based on python implementation (\url{https://github.com/microsoft/otdd}) & $\ge 11$ criteria & \faCheck & \faCheck & \faCheck & \faTimes \\
			Jeffreys divergence & Divergence & 11 & Own implementation &  $\ge 11$ criteria & \faTimes & \faCheck & \faTimes & \faTimes \\
			\textcite{baringhaus_rigid_2010} & Inter-point distances & 11 & R package cramer \textcite{cramer} & Implemented \& $\ge 11$ criteria & \faTimes & \faCheck & \faTimes & \faTimes \\
			\textcite{bahr_ein_1996} & Inter-point distances & 11 & R package cramer \textcite{cramer} & Implemented \& $\ge 11$ criteria & \faTimes & \faCheck & \faTimes & \faTimes \\
			\textcite{biswas_nonparametric_2014} & Inter-point distances & 11 & Own implementation &  $\ge 11$ criteria & \faTimes & \faCheck & \faTimes & \faTimes \\
			\textcite{schilling_multivariate_1986} and \textcite{henze_multivariate_1988} & Graph-based & 11 & Own implementation &  $\ge 11$ criteria & \faTimes & \faCheck & \faTimes & \faTimes \\
			\rowcolor{lightgray} \textcite{yu_two-sample_2007} & Binary classification & 11 & Own implementation using R package Ecume \textcite{Ecume} & (Almost) implemented \& $\ge 11$ criteria & \faTimes & \faCheck & \faCheck & \faTimes \\
			Wasserstein distance & Probability metric & 9 & R package Ecume \textcite{Ecume} & Implemented & \faTimes & \faCheck & \faTimes & \faTimes \\   
			\rowcolor{lightgray} \textcite{chen_new_2017} & Graph-based & 11 & R package gTests \textcite{gTests} & Implemented \& $\ge 11$ & \faTimes & \faCheck & \faCheck & \faTimes \\
			\rowcolor{lightgray} \textcite{chen_weighted_2018} & Graph-based & 12 & R package gTests \textcite{gTests} & Implemented \& $\ge 11$ & \faTimes & \faCheck & \faCheck & \faTimes\\
			Ball divergence \textcite{pan_ball_2018} & Testing & 9 & R package ball \textcite{ball} & Implemented & \faTimes & \faCheck & \faTimes & \faCheck \\
			\textcite{song_new_2022} & Graph-based & 11 & R package gTestsMulti \textcite{gTestsMulti} & Implemented \& $\ge 11$ & \faTimes & \faCheck & \faTimes & \faCheck\\
			DISCO \textcite{rizzo_disco_2010} & Inter-point distances & 10 & R package energy \textcite{energy} & Implemented & \faTimes & \faCheck & \faTimes & \faCheck\\
			\textcite{li_measuring_2022} & Comparison of cha\-rac\-teris\-tic func\-tions & 9 & Own implementation & Best in (sub)class & \faTimes & \faCheck & \faTimes & \faTimes \\
			Maximum Mean Discrepancy (MMD) \textcite{gretton_kernel_2006} & Kernel-based (MMD) & 9 & R packages kernlab \textcite{kernlab} and Ecume \textcite{Ecume} & Implemented & \faTimes & \faCheck & \faTimes$^*$ & \faTimes\\
			\textcite{mukhopadhyay_nonparametric_2020} & Graph-based & 9 & R package LPKsample \textcite{LPKsample} & Implemented & \faTimes & \faCheck & \faTimes$^*$ & \faCheck \\
			\rowcolor{lightgray} \textcite{chen_ensemble_2013} & Graph-based & 9 & R packages gTests \textcite{gTests}, gCat \textcite{gCat} & Implemented & \faTimes & \faCheck & \faCheck & \faTimes \\
			Block MMD \textcite{zaremba_b-test_2013} & Kernel-based (MMD) & 8 & R implementation based on matlab code for paper (\url{https://github.com/wojzaremba/btest}) & Implemented & \faTimes & \faCheck & \faTimes$^*$ & \faTimes \\
			\textcite{song_generalized_2021} & Kernel-based (MMD) & 8 & R package kerTests \textcite{kerTests} & Implemented & \faTimes & \faCheck & \faTimes$^*$ & \faTimes \\
			\rowcolor{lightgray} Constrained Minimum Distance \textcite{tatti_distances_2007} & Comparison based on summary statistics & 8 & Own implementation & Best in (sub)class & \faTimes & \faTimes & \faCheck & \faTimes \\
			\textcite{biau_asymptotic_2005} & Comparison of CDFs & 8 & Own implementation & Best in (sub)class & \faTimes & \faCheck & \faTimes & \faCheck \\
			\rowcolor{lightgray} Classifier Two-Sample Test \textcite{lopez-paz_revisiting_2017} & Binary classification & 7 & R package Ecume \textcite{Ecume} & Implemented & \faTimes & \faCheck & \faCheck & \faCheck \\
			&  &  &  &  &  & &  \\
			DiProPerm test \textcite{wei_direction-projection-permutation_2016} & Binary classification & 5 & Own implementation & Implemented & \faTimes & \faCheck & \faTimes & \faTimes\\
		\end{longtable}
	\end{landscape}
	\FloatBarrier
	
	\section{Method Parameter Settings}\label{app:meth.pars}
	The classifier for the classifier two-sample test (C2ST) of \textcite{lopez-paz_revisiting_2017} is chosen as a $K$-nearest neighbor classifier (KNN) or as a multilayer perceptron network (neural net, NN) optimized by stochastic gradient descent. 
	A $K$-nearest neighbor classifier, as well as a one-layer neural network, were compared in the simulations in the original publication. 
	The neural network performed superior there. 
	The $K$NN classifier is the default in the R implementation of the test. 
	Note that \textcite{lopez-paz_revisiting_2017} did not tune the hyperparameters of the classifiers, which might affect the performance. 
	The implementation that is used here tunes the hyperparameters.
	The categorical data is dummy-coded in the present implementation. 
	The $K$NN implementation uses the Euclidean distances of these dummy-coded data, which is equivalent to using the Hamming distance on the original categorical data. 
	
	For the random forest-based test of \textcite{hediger_use_2021}, both methods of estimating the prediction accuracy, namely taking the overall OOB accuracy or averaging the per-class OOB accuracies, are compared.
	
	For the edge count tests, a similarity graph has to be chosen. 
	Typical choices are the $K$-minimum spanning tree (MST) and the $K$-nearest neighbor (NN) graph. 
	The choice of $K$ is difficult, and there are no guidelines on how to choose this parameter. 
	Many of the tests were originally proposed for $K = 1$, therefore, this value is used. 
	Previous simulation studies show better empirical performance for higher values of $K$ \textcite{zhu_limiting_2024}. 
	Often, $K = 5$ is used, so this is included as a second choice here.
	
	For the max-type test of \textcite{zhang_graph-based_2022}, all recommended values for the parameter $\kappa$ are compared, i.e.\ $\kappa \in\{1,$$ 1.14,$$1.31\}$.
	
	For the tree-based methods of \textcite{ganti_framework_1999}, using trees with default parameters and tuning the parameters of the trees are compared. 
	For \textcite{ganti_framework_1999}, both $f_a$ and $f_s$ are used as difference functions.
	As the specification of $f_a$ and $f_s$ still leaves some freedom to define which proportions are compared using the absolute or scaled difference, respectively, two different versions are compared. 
	Initially, only the proportion of data points in each section of the greatest common refinement (GCR) is used. 
	Since this is expected to be unable to identify differences that only affect the outcome-generating model, a second version was tested later in addition. 
	This second version compares the proportions for each but one class in the target variable separately, i.e.\ for a binary target, only the proportions of ones per section in the GCR are compared, like in the computation example in \textcite{ganti_framework_1999}. 
	This did however not lead to substantial improvements in the detection of the changes in the OGM but notably reduced the performance for detecting differences in the distribution of the covariates. 
	Therefore, only results for the initial variant are shown, where the proportion of data points is used without distinguishing between the target values. 
	Only the sum is used as the aggregate function since it is the only known aggregate function with guaranteed optimality of the greatest common refinement (GCR). 
	
	For the graph-based tests for categorical data \textcite{zhang_graph-based_2022}, both strategies of averaging the results of all optimal graphs and using the union of all optimal graphs are compared.
	
	For the CM distance, a feature function has to be chosen. 
	There are no clear re\-com\-men\-dat\-ions for this choice in the literature. 
	The original paper uses the conjunction function for binary data. 
	In a comparison of different feature functions on real data, the CM distance values for the independent means only and the independent means along with pairwise correlations were used, and were highly correlated. 
	The other feature functions considered focused on frequent itemsets, which are not part of this study. 
	Therefore, here, the independent means are used since they are computationally less demanding than calculating pairwise correlations in addition. 
	
	For the optimal transport dataset distance (OTDD) \textcite{alvarez-melis_geometric_2020}, the exact version is used along with different speed-ups. 
	One version where both the inner and outer optimal transport problems are approximated using the Sinkhorn divergence is compared as well as the data augmentation version, the Gaussian approximation for the inner OT problem, the Gaussian approximation assuming equal covariances for the inner OT problem, and the naive upper bound for the inner OT problem. 
	All of these are approximations of the problem that can potentially speed up the computation, but some come with restrictive assumptions. 
	Therefore, it is of interest how much computation speed can be gained and how much performance of the dataset distance in recognizing differences between datasets is lost. 
	However, for categorical data, any Gaussian approximation is inappropriate, and therefore, only the exact and the Sinkhorn versions are compared. 
	The Hamming distance is used as the distance function for the covariates, as it is a commonly used distance function for categorical data. 
	
	\section{Comparison of PESR and Asymptotic Power}\label{app:comp.pesr.power}
\begin{figure}[H]
		\centering
		\includegraphics[width=\linewidth]{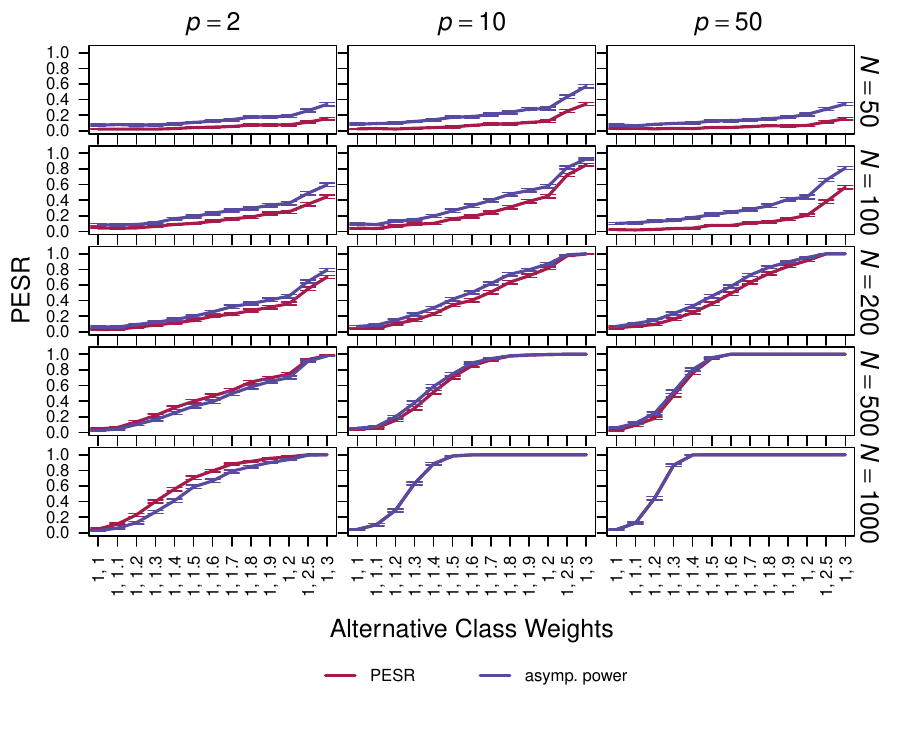}
		\caption{Comparison of the PESR to the asymptotic power for two datasets of the same sample sizes with binary variables for \textbf{C2ST~(NN)}. The class weights give the unnormalized probabilities $(1, 1+\delta)$ for the values 0 and 1 for each variable in the second dataset. The weights in the first dataset are always set to $(1, 1)$. Error bars indicate Monte Carlo standard errors.}
		\label{fig:comp.pesr.power.bin.bal.C2ST.NN}
	\end{figure}

\begin{figure}[H]
		\centering
		\includegraphics[width=\linewidth]{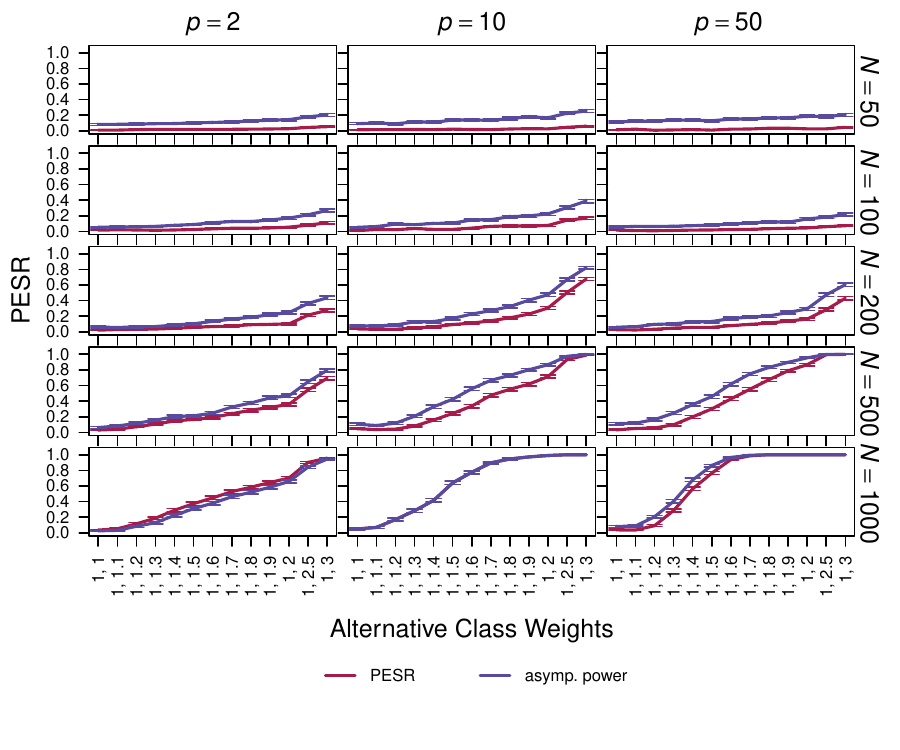}
		\caption{Comparison of the PESR to the asymptotic power for two datasets of unequal sample sizes with binary variables for \textbf{C2ST~(NN)}. The class weights give the unnormalized probabilities $(1, 1+\delta)$ for the values 0 and 1 for each variable in the second dataset. The weights in the first dataset are always set to $(1, 1)$. Error bars indicate Monte Carlo standard errors.}
		\label{fig:comp.pesr.power.bin.unbal.C2ST.NN}
	\end{figure}

\begin{figure}[H]
		\centering
		\includegraphics[width=\linewidth]{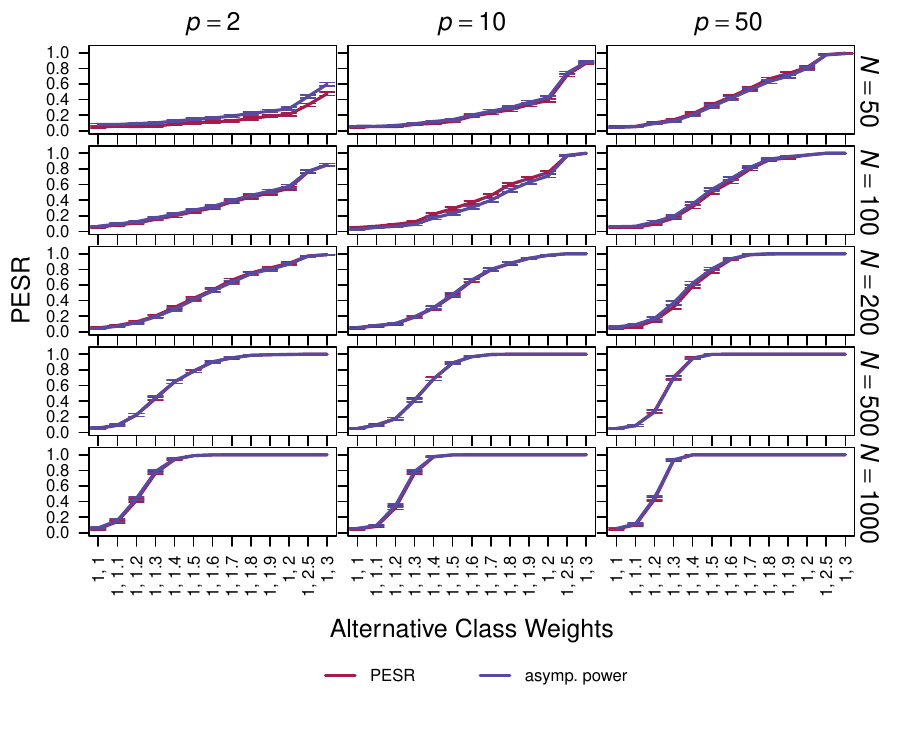}
		\caption{Comparison of the PESR to the asymptotic power for two datasets of the same sample sizes with binary variables for \textbf{FR (1NN, u)}. The class weights give the unnormalized probabilities $(1, 1+\delta)$ for the values 0 and 1 for each variable in the second dataset. The weights in the first dataset are always set to $(1, 1)$. Error bars indicate Monte Carlo standard errors.}
		\label{fig:comp.pesr.power.bin.bal.FR.1NN.u}
	\end{figure}

\begin{figure}[H]
		\centering
		\includegraphics[width=\linewidth]{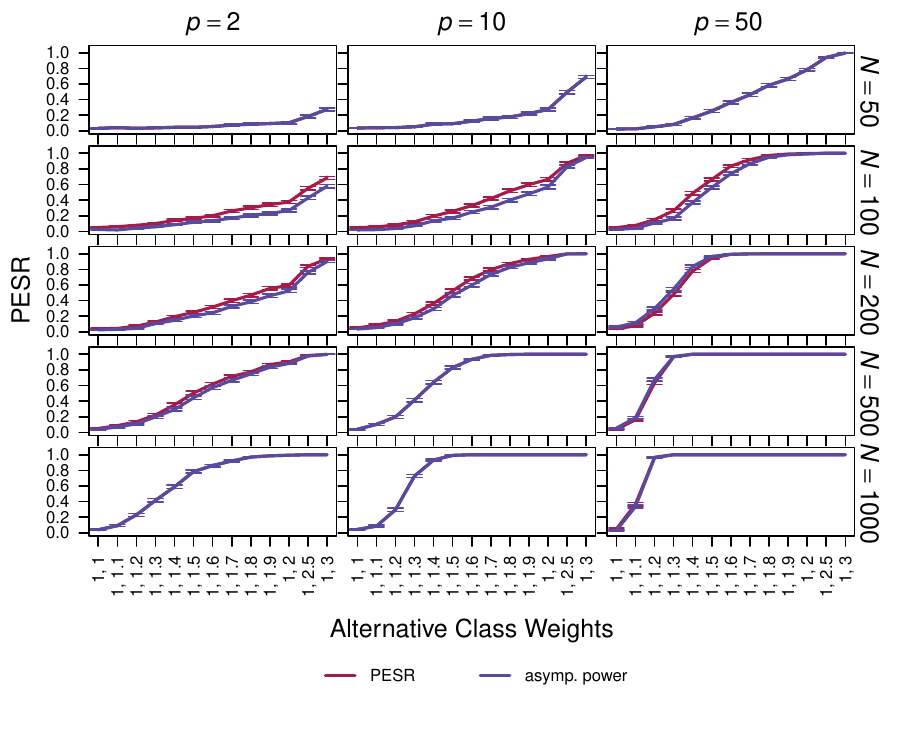}
		\caption{Comparison of the PESR to the asymptotic power for two datasets of the same sample sizes with binary variables for \textbf{HMN (overall OOB)}. The class weights give the unnormalized probabilities $(1, 1+\delta)$ for the values 0 and 1 for each variable in the second dataset. The weights in the first dataset are always set to $(1, 1)$. Error bars indicate Monte Carlo standard errors.}
		\label{fig:comp.pesr.power.bin.bal.HMN.OverallOOB}
	\end{figure}

\begin{figure}[H]
		\centering
		\includegraphics[width=\linewidth]{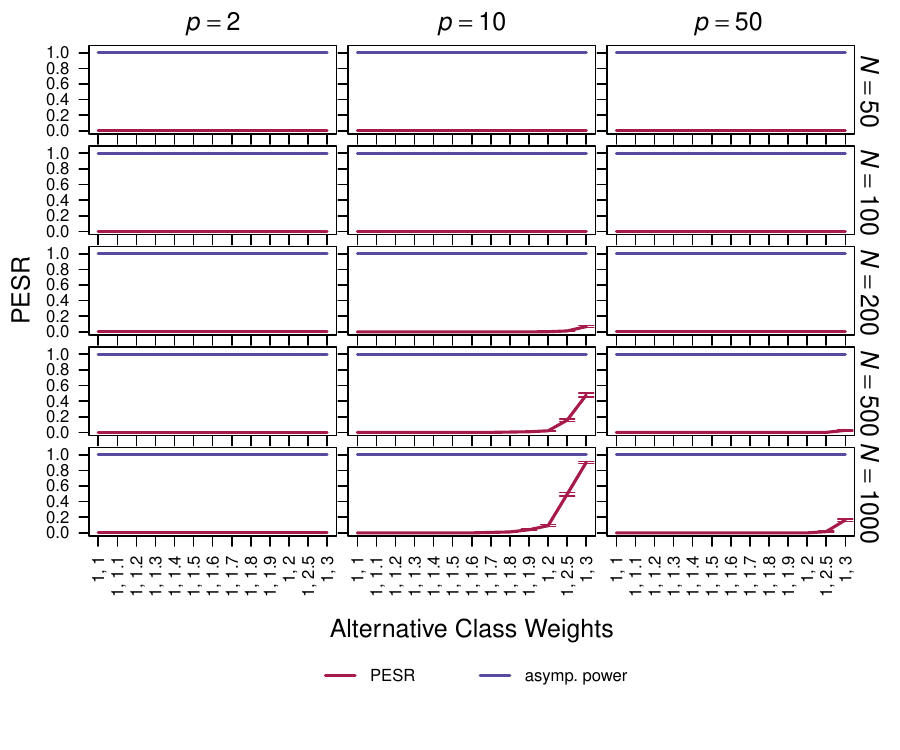}
		\caption{Comparison of the PESR to the asymptotic power for two datasets of unequal sample sizes with binary variables for \textbf{HMN (overall OOB)}. The class weights give the unnormalized probabilities $(1, 1+\delta)$ for the values 0 and 1 for each variable in the second dataset. The weights in the first dataset are always set to $(1, 1)$. Error bars indicate Monte Carlo standard errors.}
		\label{fig:comp.pesr.power.bin.unbal.HMN.OverallOOB}
	\end{figure}

\begin{figure}[H]
		\centering
		\includegraphics[width=\linewidth]{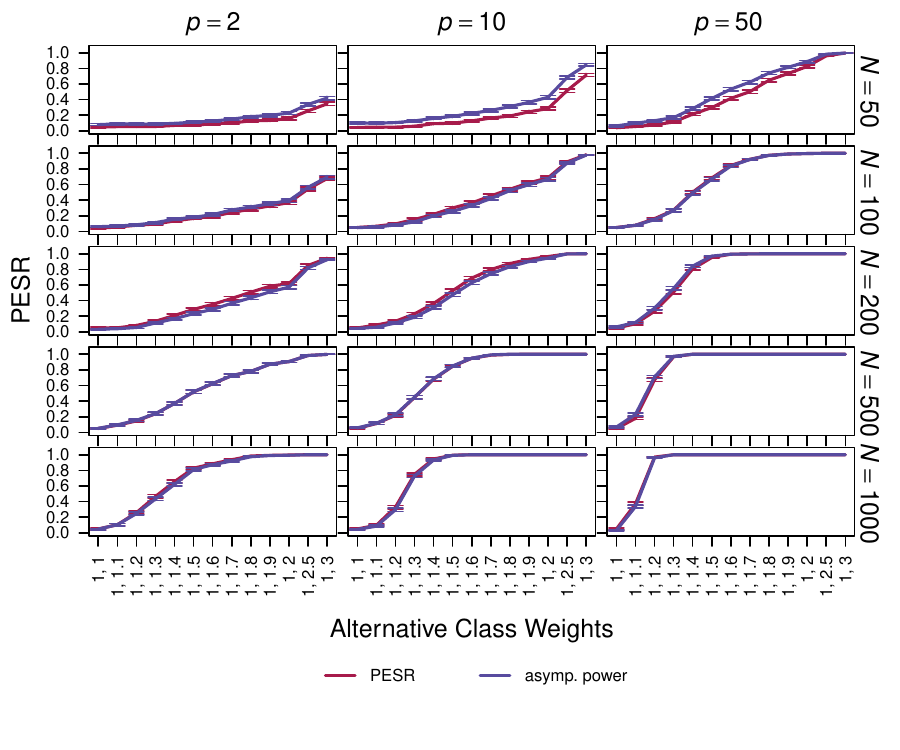}
		\caption{Comparison of the PESR to the asymptotic power for two datasets of the same sample sizes with binary variables for \textbf{HMN (per class OOB)}. The class weights give the unnormalized probabilities $(1, 1+\delta)$ for the values 0 and 1 for each variable in the second dataset. The weights in the first dataset are always set to $(1, 1)$. Error bars indicate Monte Carlo standard errors.}
		\label{fig:comp.pesr.power.bin.bal.HMN.PerClassOOB}
	\end{figure}

\begin{figure}[H]
		\centering
		\includegraphics[width=\linewidth]{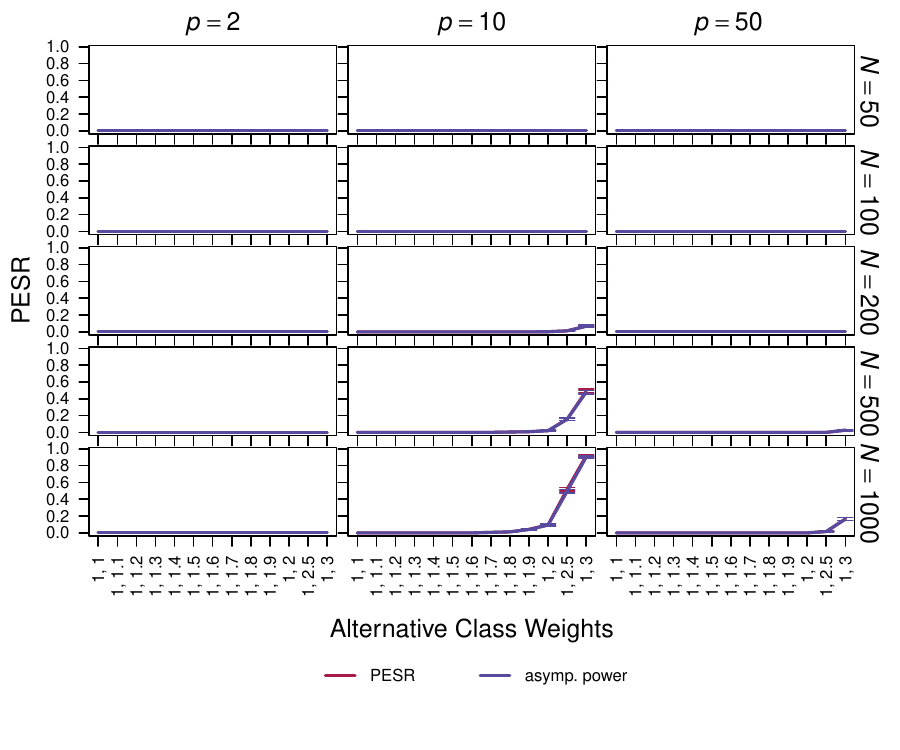}
		\caption{Comparison of the PESR to the asymptotic power for two datasets of unequal sample sizes with binary variables for \textbf{HMN (per class OOB)}. The class weights give the unnormalized probabilities $(1, 1+\delta)$ for the values 0 and 1 for each variable in the second dataset. The weights in the first dataset are always set to $(1, 1)$. Error bars indicate Monte Carlo standard errors.}
		\label{fig:comp.pesr.power.bin.unbal.HMN.PerClassOOB}
	\end{figure}

\begin{figure}[H]
		\centering
		\includegraphics[width=\linewidth]{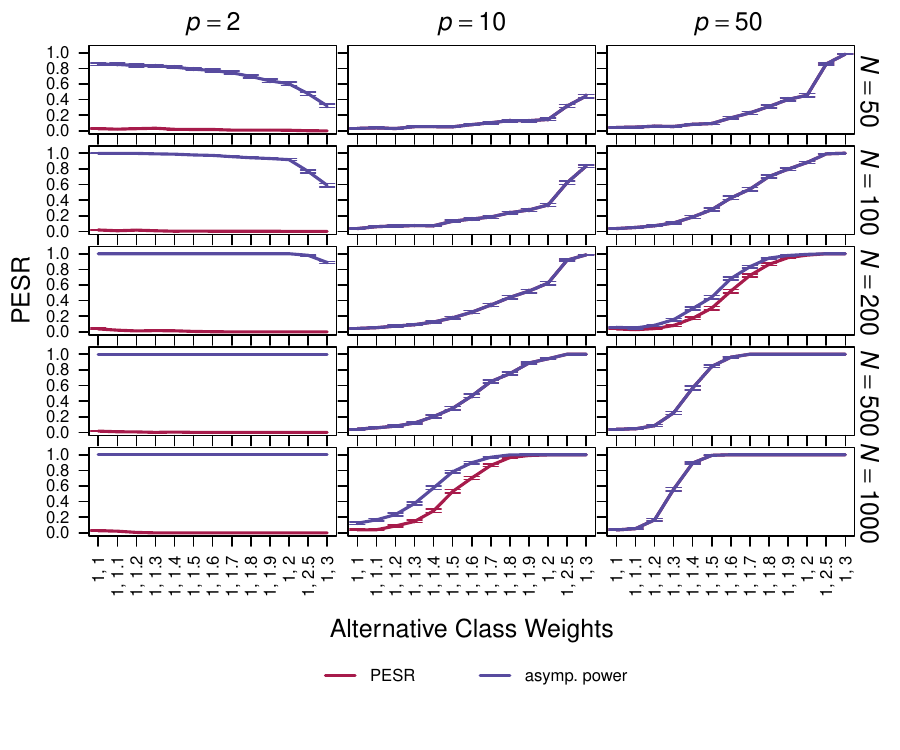}
		\caption{Comparison of the PESR to the asymptotic power for two datasets of the same sample sizes with binary variables for \textbf{MMCM}. The class weights give the unnormalized probabilities $(1, 1+\delta)$ for the values 0 and 1 for each variable in the second dataset. The weights in the first dataset are always set to $(1, 1)$. Error bars indicate Monte Carlo standard errors.}
		\label{fig:comp.pesr.power.bin.bal.MMCM}
	\end{figure}

\begin{figure}[H]
		\centering
		\includegraphics[width=\linewidth]{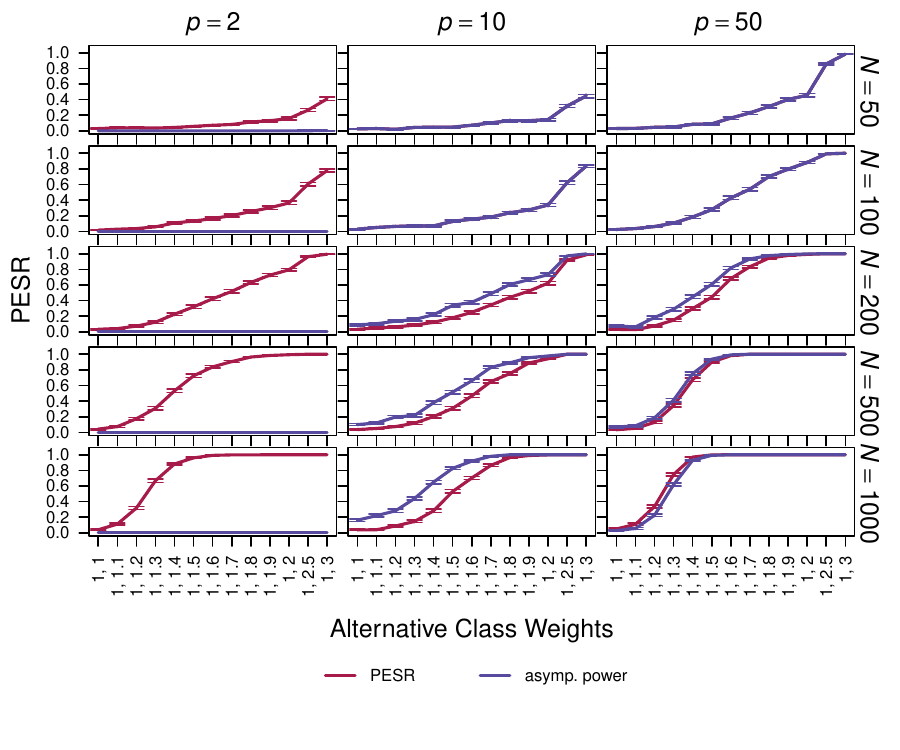}
		\caption{Comparison of the PESR to the asymptotic power for two datasets of the same sample sizes with binary variables for \textbf{Petrie}. The class weights give the unnormalized probabilities $(1, 1+\delta)$ for the values 0 and 1 for each variable in the second dataset. The weights in the first dataset are always set to $(1, 1)$. Error bars indicate Monte Carlo standard errors.}
		\label{fig:comp.pesr.power.bin.bal.Petrie}
	\end{figure}

\begin{figure}[H]
		\centering
		\includegraphics[width=\linewidth]{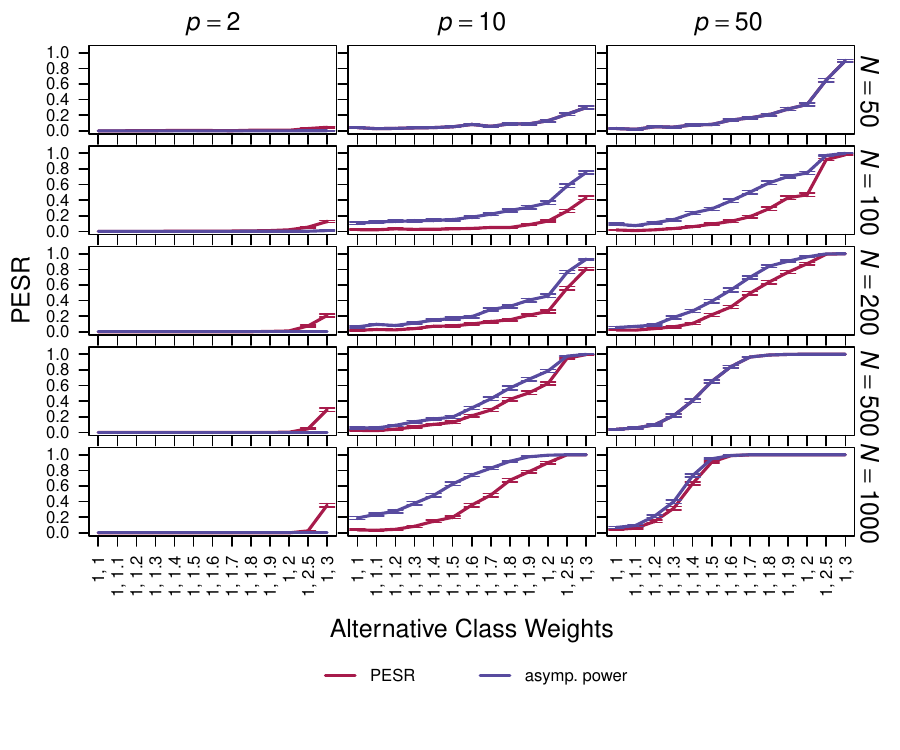}
		\caption{Comparison of the PESR to the asymptotic power for two datasets of unequal sample sizes with binary variables for \textbf{Petrie}. The class weights give the unnormalized probabilities $(1, 1+\delta)$ for the values 0 and 1 for each variable in the second dataset. The weights in the first dataset are always set to $(1, 1)$. Error bars indicate Monte Carlo standard errors.}
		\label{fig:comp.pesr.power.bin.unbal.Petrie}
	\end{figure}

	Figure~\ref{fig:comp.pesr.power.bin.bal.C2ST.NN} to~\ref{fig:comp.pesr.power.bin.unbal.Petrie} show the simulated power of the asymptotic test and PESR curves for methods where an asymptotic test is defined. 
	The power of the asymptotic test is estimated by the proportion of simulation repetitions in which the asymptotic $p$~value is smaller than $0.05$. 
	As an example, the two-sample case with binary data and balanced sample sizes is selected. 
	The unbalanced sample size case is shown additionally for methods for which it makes a difference in the comparison of power and PESR.
	The C2ST (NN) is selected as a representative for C2ST (KNN) and YMRZL, and the FR (1NN, u) was selected as a representative for the edge count tests since the methods within these groups acted similarly.
	
	For the C2ST (Figures~\ref{fig:comp.pesr.power.bin.bal.C2ST.NN} and~\ref{fig:comp.pesr.power.bin.unbal.C2ST.NN}) and YMRZL, there are some differences between the asymptotic power and the PESR. 
	In most cases, the PESR is lower than the simulated asymptotic power. 
	These differences vanish for increasing $N$, where the asymptotic test is expected to be most accurate.
	For unbalanced sample sizes, however, there are still some differences for the largest $N$.

	For the edge count tests, the power and PESR curves are very similar with some small differences for low $N$ (see e.g.\ Figure~\ref{fig:comp.pesr.power.bin.bal.FR.1NN.u}).

	For the HMN and balanced sample sizes, the PESR and power curves are also typically close with some differences for small to moderate $N$ (Figure~\ref{fig:comp.pesr.power.bin.bal.HMN.OverallOOB} and ~\ref{fig:comp.pesr.power.bin.bal.HMN.PerClassOOB}). 
	For unbalanced sample sizes and the HMN (overall OOB), the PESR is almost constantly zero, while the simulated power is constantly equal to one (Figure~\ref{fig:comp.pesr.power.bin.unbal.HMN.OverallOOB}). 
	The reason for this is that the test statistics in the simulation are high (even under the null) but almost constant. 
	Therefore, the test rejects in all cases, but the PESR does not find a difference between the distributions under the alternatives and the null distribution. 
	For the HMN (per class OOB), on the other hand, the PESR and power are close in the unbalanced sample size, and both are mostly zero (Figure~\ref{fig:comp.pesr.power.bin.unbal.HMN.PerClassOOB}). 
	The test statistic here is almost constant but close to zero as expected under the null, so the test does not reject, and the PESR also does not find differences between the test statistics under the null and the alternatives. 
	
	Similar observations can be made for the MMCM and Petrie's method for $p=2$, where very large test statistic values are observed regardless of whether there are differences between the distributions or not (Figure~\ref{fig:comp.pesr.power.bin.bal.MMCM} to~\ref{fig:comp.pesr.power.bin.unbal.Petrie}). 
	For Petrie's method in the binary case, the PESR is increasing while the power is constantly zero (Figure~\ref{fig:comp.pesr.power.bin.bal.Petrie}). 
	The reason for this is that the observed test statistic values are small with low variance but slightly decreasing with the alternatives.
	Thus, the test does not find any differences, but the PESR picks up a slight decrease. 
	For $p > 2$, there are some differences between the PESR and the power, even for higher $N$. 
	Typically, the power is higher than the PESR. 
	For $p = 10$, the tests are even liberal. 
	
	\FloatBarrier
	\newpage
	
		\section{Pre-Selection of Methods}\label{app:presel}
	Counting all variants, a total of 62 methods are applied in the two-sample setting. 
	In the following, the best variants of similar methods are pre-selected to facilitate the comparison. 
	For that, the variants are compared, and variants that are inferior to others in all considered scenarios are eliminated from the following analyses. 
	
	The first group is the classifier two-sample tests.
	The C2ST itself is used with two classifiers, the multilayer perceptron (NN) and the $K$-nearest neighbor classifier (KNN).
	Moreover, the YMRZL method of \textcite{yu_two-sample_2007} can also be seen as a variant of the C2ST that uses a decision tree as the classifier. 
	When comparing these three methods, the YMRZL is worse than the C2ST~(NN) and the C2ST~(KNN) in almost all cases. 
	Which of the C2ST~(NN) and the C2ST~(KNN) performs better depends on the scenario. 
	The comparison of the three methods for binary data and balanced sample sizes is shown as an example in Figure~\ref{fig:pow.cat.no.y.bin.bal.C2ST} in Appendix~\ref{app:add.figs.cat.no.y.sel.meths}. 
	In the following analyses, the results for the YMRZL method will not be shown. 
	
	For each of the edge-count tests by \textcite{friedman_multivariate_1979}, \textcite{chen_new_2017}, \textcite{chen_weighted_2018}, and \textcite{zhang_graph-based_2022}, the graph-type is varied as a $K$-minimum spanning tree ($K$MST), or a $K$-nearest neighbor-graph ($K$NN) with $K = 1$ or $5$. 
	Moreover, for each graph, averaging (``a'') or the union (``u'') is applied to handle ties. 
	Therefore, there are in total six variants for each of the methods: FR, CF, and CCS. 
	The ZC method has an additional parameter $\kappa$ that is varied over the three recommended values $\kappa = 1, 1.14, 1.31$. 
	Thus, there are in total 18 variants for ZC.
	The PESR comparisons between the different graphs and averaging and union for each method are shown for binary datasets with balanced sample sizes as an example in Figures~\ref{fig:pow.cat.no.y.bin.bal.FR} to~\ref{fig:pow.cat.no.y.bin.bal.ZC.kappa.1.31} in Appendix~\ref{app:add.figs.cat.no.y.sel.meths}. 
	The comparisons can be summarized over all scenarios as follows.
	For $p = 2$, the 5NN, ``u'' versions fail, resulting in either only \texttt{NaN} (not a number) or only 0 test statistic values due to standardization with variances that are analytically equal to zero. 
	The same holds for the CF and ZC 1NN, ``a''.
	These issues are further discussed in Appendix~\ref{app:fail.p.2}.
	For binary data, the differences between the remaining methods are very small.
	For multinomial data, the differences are clearer, and typically, the 1NN, ``u'' version performs best.
	For $p = 10$, the $K = 5$ graphs typically perform better than the $K = 1$ graphs. 
	Typically, either the 5MST, ``u'' is best, or it is only best for small $N$, and the 5NN, ``u'' is better (or very similar to the 5MST) for large $N$.
	For $p = 50$, there are no differences between the ``a'' and ``u'' versions. 
	One reason for this might be that with so many variables, there are fewer tied observations and therefore fewer or even only one optimal graph, such that the differences between averaging over the optimal graphs or using their union diminish. 
	The superiority of the $K = 5$ versions over the $K = 1$ versions is even clearer for $p = 50$ than for $p = 10$. 
	Often, the 5NN performs best. 
	Sometimes, the 5MST performs better for small $N$. 
	Generally speaking, the 5MST performs best for CF and ZC in general and FR in the case of unbalanced data. 
	So overall, the $K = 1$ methods lose PESR with increasing $p$, while the $K = 5$ methods gain PESR with increasing $p$. 
	``u'' tends to work better than ``a'' except for the $p = 1$ and 5NN case, but in many cases the differences between ``a'' and ``u'' are small or even negligible. 
	The differences are clearer for unbalanced sample sizes where the ``a'' variants are slightly more affected than the ``u'' variants. 
	Regarding the choice of $\kappa$ for ZC, the differences between the PESR values for a given graph and ``a''/``u'' combination are negligible (see e.g.\ Figure~\ref{fig:pow.cat.no.y.bin.bal.ZC.comp.kappas} in Appendix~\ref{app:add.figs.cat.no.y.sel.meths}). 
	Therefore, for the remaining analysis, the methods are restricted to three versions: the 1NN, ``u'' performs best for $p = 2$, the 5MST, ``u'' performs best in many cases for $p > 2$, and the 5NN, ``a'' is considered since 5NN sometimes outperforms 5MST for $p > 2$ and 5NN, ``u'' fails for $p = 2$. 
	For ZC, only the default $\kappa=1.14$ is used. 
	
	For the random forest-based test HMN by \textcite{hediger_use_2021}, the classification error can be calculated per class or overall. 
	There are often no differences with regard to PESR between these variants (see e.g.\ Figure~\ref{fig:pow.cat.no.y.bin.bal.HMN} in Appendix~\ref{app:add.figs.cat.no.y.sel.meths}), but if there are differences, the per-class OOB version shows a higher PESR. 
	Therefore, only that version is used in the following. 
	
	The method of \textcite{petrie_graph-theoretic_2016} and the MMCM \textcite{mukherjee_distribution-free_2022} are based on the same graph-based quantities and can, therefore, be seen as variants of each other. 
	In many cases, there are no differences between the two, but if there are, Petrie's method typically performs better (see e.g.\ Figure~\ref{fig:pow.cat.no.y.bin.bal.MMCM} in Appendix~\ref{app:add.figs.cat.no.y.sel.meths}). 
	Both methods mostly fail for $p = 2$, which is discussed in Appendix~\ref{app:fail.p.2}.
	Thus, the MMCM is not shown in the following.

	For the methods that use a target variable present in the data, the following options are pre-selected. 
	For the method GGRL \textcite{ganti_framework_1999}, the difference function $f_a$ consistently outperforms the difference function $f_s$. 
	Not tuning the parameters of the decision trees gives better results than tuning the decision trees. 
	Moreover, not tuning the trees is also considerably faster (see Figure~\ref{fig:runtime.cat.y.full} in Appendix~\ref{app:add.figs.benchmarks}) and results in fewer computational errors (see Section~\ref{sec:err}). 
	So, overall GGRL ($f_a$, w/o tuning) is the best version that is used in the following.
	
	For the optimal transport dataset distance OTDD \textcite{alvarez-melis_geometric_2020}, the exact and the Sinkhorn version achieve almost identical PESR values. 
	However, for the Sinkhorn version, the inner optimization frequently fails, while for the exact version, no computational errors occur (see Section~\ref{sec:err}). 
	Therefore, the exact version is used in the following.
	
	So, in total, 19 methods and variants are considered: FR, CF, CCS, and ZC~($\kappa = 1.14$) each with the three versions 1NN, ``u'', 5MST, ``u'', and 5NN, ``a'', C2ST~(KNN) and C2ST~(NN), HMN (per class OOB), Petrie, GGRL~($f_a$, w/o tuning), OTDD~(Sinkhorn), and the CM distance, for which no variants were tested. 
	For balanced sample sizes, CCS and FR are equivalent, such that the methods can be reduced further to 16 methods, and then only FR is shown.
	
	\FloatBarrier
	
	\section{Computational Problems}\label{sec:err}
	In the following, the computational problems that occurred in the simulation study are discussed. 
	
	\subsection{Two-sample Setting}
	Table~\ref{tab.errors} gives an overview of the number of errors per scenario and method that occurred during the simulations for $k = 2$, considering only methods that do not consider a target variable. 
	For most method and scenario combinations, no errors occurred. 
	For some combinations, one to eight errors occurred, which is negligible considering the total number of $500$ iterations. 
	For $160$ method and scenario combinations, errors occurred in all repetitions such that no results could be retrieved. 
	These are exactly the scenarios with $p = 50$ and $5$ categories for the CM distance where the enumeration of the sample space is infeasible. 
	
	\begin{table}[!b]
		\centering 
		\begin{tabular}{lrrrrrrrr}
			\toprule
			No.\ Errors & 0 & 1 & 2 & 5 & 7 & 7 & 8 & 500\\
			No.\ Scenarios &  376026 &  94 &  7 &  13 &  13 &  6 &  1 & 160 \\
			\bottomrule
		\end{tabular}
		\caption{Frequency of numbers of occurring errors for scenarios without the target variable. No.\ Errors: Number of repetitions of the simulation in which an error occurred. No.\ Scenarios: Number of methods and scenarios (combination of $N$, $p$, balance, deviation) in which a certain number or errors occurred.}\label{tab.errors}
	\end{table}
	
	The remaining errors are related to technical problems while running the simulation study. 
	There were sporadic problems with the file system or the parallel loading of packages. 
	These are not related to the methods themselves and are unlikely to occur again when repeating the simulation study. 
	
	In addition to the occurring errors, there were also infinite, missing, and \texttt{NaN} (not a number) values for the test statistics that were not the result of an error. 
	For HMN (per class OOB), the test statistic is not defined for perfect classification since the variance of the classification error used to standardize the statistic is zero in that case. 
	The test statistic is then set to infinity in the implementation. 
	This occurred $50$ times within the simulation study, mostly for the highest deviations but spread across scenarios such that for each individual scenario, only a few repetitions were affected.
	For $p = 2$, the test statistics of all edge count tests using the 5NN, ``u'' are undefined due to standardization with a variance of zero. 
	For CF and ZC, this also holds for the 1NN, ``a''. 
	For more details on this issue, see Appendix~\ref{app:fail.p.2}.
	
	Table~\ref{tab.errors.y} gives an overview of the number of errors per scenario and method that occurred during the simulations for $k = 2$, for the methods that consider a target variable. 
	In most scenarios, no errors occurred. 
	However, in some scenarios, errors occurred in almost all iterations. 
	These are cases with the OGM with completely wrong coefficients in which only zeroes are generated, and the GGRL cannot be calculated.
	For the scenarios with the other OGMs, at most, errors occurred in $57$ iterations for one scenario. 
	
	\begin{table}[!b]
		\centering
		\begin{tabular}{l|rrrrrrrrrrrrrr}
			\toprule
			No.\ Errors & 0 & 1 & 2 & 3 & 4 & 5 & 6 & 7 & 8 & 9 & 10 & 11 & 13 & 15 \\
			No.\ Scenarios & 41179 & 100 &   3 &  40 &  39 &  73 &   5 &   1 &   2 &  37 &   7 &   4 &   2 &   1  \\ 
			\midrule
			No.\ Errors & 17 & 19 & 20 & 21 & 24 & 30 & 31 & 32 & 33 & 34 & 35 & 36 & 37 & 38   \\ 
			No.\ Scenarios  &  30 &   2 &  36 &  36 &   2 &  33 &   1 &   1 &   1 &   5 &   2 &   1 &   2 &  36 \\
			\midrule
			No.\ Errors & 39 & 42 & 44 & 57 & 62 & 81 & 140 & 497 & 498 & 499 & 500 &&&\\
			No.\ Scenarios &   1 &  30 &   1 &   1 &   2 &   2 &   2 &   2 &   6 &   6 &  26&&&\\
			\bottomrule
		\end{tabular}
		\caption{Frequency of numbers of occurring errors for scenarios with a target variable. No.\ Errors: Number of repetitions of the simulation in which an error occurred. No.\ Scenarios: Number of combinations of methods and data-generating mechanisms (combination of $N$, $p$, balance, deviation) in which a certain number or errors occurred.}\label{tab.errors.y}
	\end{table}
	
	For the OTDD with the Sinkhorn approximation, in 895 iterations across all scenarios, the inner optimal transport problem could not be solved, resulting in an error. 
	For the exact calculation of the OTDD, no errors occurred.
	In 2033 iterations, either no ones or no zeroes were generated. 
	This happened only for the OGM with completely different coefficients in almost all iterations for $p = 50$. 
	In these cases, none of the GGRL variants can be calculated. 
	Additionally, for the GGRL variants with tuning, too few zeroes or ones were generated for the target variable in 6074 iterations across all scenarios. 
	In these cases, the tree cannot be fitted, and an error is thrown. 
	This concerns especially datasets with low numbers of observations. 
	For the untuned versions, this happened in only 3513 iterations across all scenarios. 
	The higher error numbers for the tuned version can be attributed to the cross-validation for tuning, in which the data is further split into smaller datasets, such that even if the whole dataset includes enough zero and one observations in the target, the dataset after splitting might not. 
	Most of these errors also occurred for the OGM with completely wrong coefficients. 
	
	Very high runtimes were another issue that occurred in the simulation for the methods that consider a target variable. 
	Due to this, it was unfortunately not possible to obtain results for all iterations in this case.
	Simulations were performed in parallel in batches of 50--500 iterations. 
	In the most extreme cases, batches of 50 iterations did not finish within 14 days of computing time. 
	These were then not repeated with even longer runtimes but counted as missing. 
	Table~\ref{tab.miss.iter.y} summarizes the scenario specifications, where the iteration number consequently had to be reduced to 450. 
	The scenarios all include $p = 2$ covariates and $N = 100$ or $N = 200$ observations. 
	Most scenarios have unbalanced sample sizes and multinomial data. 
	
	\begin{table}[!t]
		\centering
		\begin{tabular}{lllll}
			\toprule
			Balance & $p$ &$ N$ & $c$ & Deviation \\ 
			\midrule
			unbalanced & 2 & 100 & 2 & none \\ 
			unbalanced & 2 & 100 & 2 & 1, 1.6 \\ 
			unbalanced & 2 & 100 & 2 & 1, 2.5 \\ 
			unbalanced & 2 & 100 & 5 & 1, 1, 1, 0.2, 1.8 \\ 
			unbalanced & 2 & 100 & 5 & 1, 1, 1, 0.3, 1.7 \\ 
			unbalanced & 2 & 100 & 5 & 1, 1.3, 1.6, 1.9, 2.2 \\ 
			balanced & 2 & 200 & 5 & 1, 1, 1, 0.2, 1.8 \\ 
			unbalanced & 2 & 200 & 5 & 1, 1, 1, 0.3, 1.7 \\ 
			unbalanced & 2 & 200 & 5 & 1, 1, 1, 0.5, 1.5 \\ 
			unbalanced & 2 & 200 & 5 & 1, 1.4, 1.8, 2.2, 2.6 \\ 
			unbalanced & 2 & 200 & 5 & 1, 1.5, 2, 2.5, 3 \\ 
			unbalanced & 2 & 100 & 2 & ogm \\ 
			\bottomrule
		\end{tabular}
		\caption{Scenario specifications for scenarios with reduced numbers of iterations. $c$ refers to the number of categories for each covariate.}\label{tab.miss.iter.y}
	\end{table}
	
	The missing results could not be replaced in a meaningful way, as there is no reasonable fallback strategy for the methods.
	Therefore, the reported performances are based on the results that could be obtained. 
	The reduced number of statistic values in the calculation is, however, reflected by the reported Monte Carlo Standard Errors that increase for a decreasing number of iterations. 
	For more than 100 iterations with missing values due to runtime or computational errors, no performance was reported.

	\subsection{Multi-sample Setting}
	No errors occurred in any repetition of any scenario.
	All methods appear to be numerically stable in the chosen scenarios for the four-sample case.	
	
	\FloatBarrier
	\section{Graph-Based Methods for Low-Dimensional Categorical Data}\label{app:fail.p.2}
\begin{figure}[H]
		\centering
		\includegraphics[width=\linewidth]{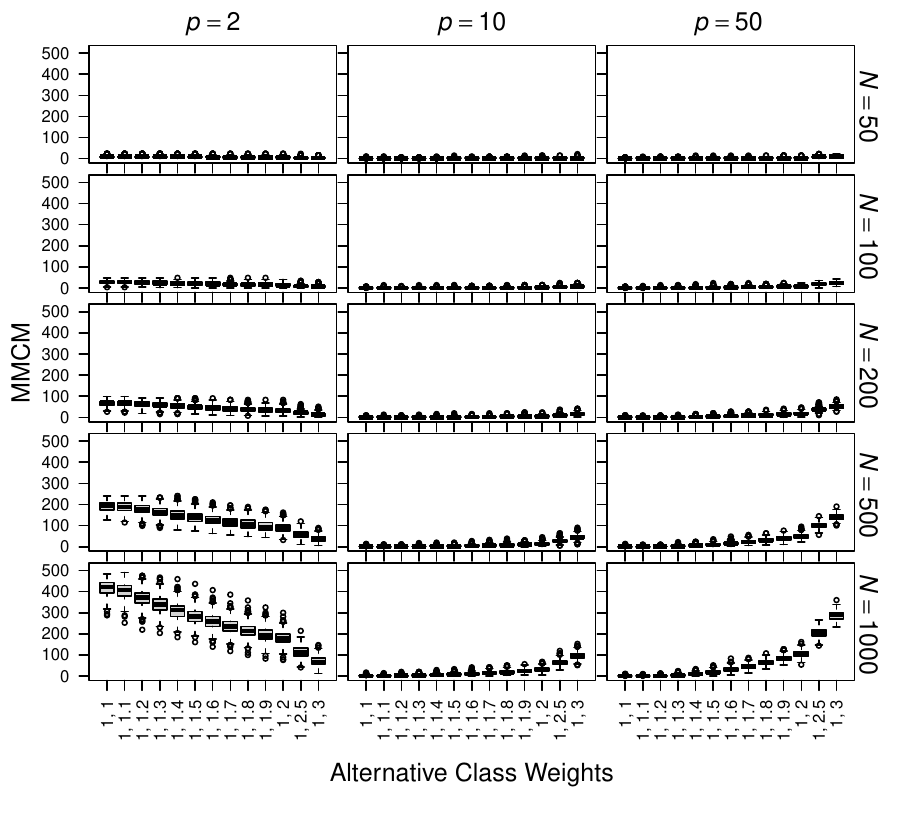}
		\caption{MMCM test statistic values for $k = 2$ datasets of the same sample sizes with binary variables. The class weights give the unnormalized probabilities $(1, 1+\delta)$ for the values 0 and 1 for each variable in the second dataset. The weights in the first dataset are always set to $(1, 1)$. Error bars indicate Monte Carlo standard errors.}
		\label{fig:boxplot.MMCM.cat.no.y.bin.bal}
	\end{figure}

\begin{figure}[!t]
		\centering
		\includegraphics[width=\linewidth]{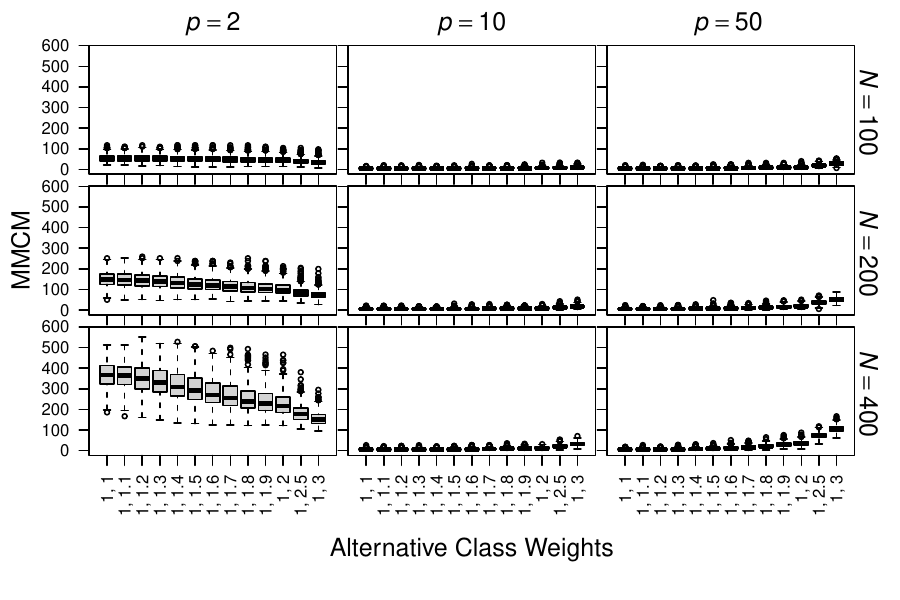}
		\caption{MMCM test statistic values for $k = 4$ datasets of the same sample sizes with binary variables. The class weights give the unnormalized probabilities $(1, 1+\delta)$ for the values 0 and 1 for each variable in the fourth dataset. The weights in the first to third datasets are always set to $(1, 1)$. Error bars indicate Monte Carlo standard errors.}
		\label{fig:boxplot.MMCM.cat.multi.bin.bal.3+1}
	\end{figure}

	In all settings, clear problems with the performance of MMCM and sometimes also for the method of \textcite{petrie_graph-theoretic_2016} were visible for $p = 2$. 
	In the following, the reasons for this issue are investigated.  
	Both methods use the optimal non-bipartite matching as a basis where, based on the Euclidean distances, pairs of observations are matched such that the sum of the edge lengths, i.e.\ the sum of the Euclidean distances of matched observations, is minimized.
	With categorical data and few variables, there are only a few possible observations, which leads to ties in the distance matrix. 
	For $p = 2$ variables, only $c^2$ combinations are possible, where $c$ denotes the number of categories, which is either $2$ or $5$ here. 
	When calculating the Euclidean distances as is done for the two above-mentioned methods, there are even fewer possible distance values. 
	For $c = 2$, the possible distances are only $0, 1, \sqrt{2}$. 
	Therefore, when calculating the optimal non-bipartite matching that is used in both methods, there are many optimal solutions. 
	The implemented matching algorithm goes through the observations in the order of the samples and starts looking for a match in the reverse order. 
	Therefore, for the two-sample case with ties, the observations from the first dataset mostly get matched with observations from the second sample. 
	In the four-sample case, they mostly get matched with observations from the fourth sample, then from the third, then from the second, and only lastly from the first. 
	The observations from the second sample mostly get matched with the third sample, as most observations from the fourth sample already had a perfect match after going through the first sample. 
	Therefore, in the two-sample case, the cross-counts of the first and second datasets are very high, while the number of counts within the samples is very low. 
	In the multi-sample case, the cross counts of the first and last datasets and the second and third datasets are very high; the cross counts of the first and third datasets and the second and fourth datasets are already lower than expected under the null, and the within-sample counts are typically very low. 
	Thus, even under the null, the observed cross counts deviate largely from the expected values, resulting in very large test statistic values. 
	When changing the class weights of the last sample, fewer observations from the first sample get matched with observations of the last sample, which leads to a redistribution of the cross matches that is more evenly than under the null, so counterintuitively, the test statistic values often decrease for increasing differences in the class weights, as can be seen for the MMCM test statistic in Figure~\ref{fig:boxplot.MMCM.cat.no.y.bin.bal} for $k = 2$ and in Figure~\ref{fig:boxplot.MMCM.cat.multi.bin.bal.3+1} for $k = 4$.
	For Petrie's test, this decrease in the binary and balanced setting might result in good PESR values, as decreasing values are what would be expected under the alternative, and the PESR does not detect the too-high values in the null situation, which is a shortcoming of this approach.
	However, for $p = 2$, both asymptotic tests would reject the null for all considered class weights in almost all simulation repetitions.
	For more than two categories, the problem is less severe but still present.
	
	Randomly permuting the order of the observations before calculating the matching could prevent the peculiar matching due to the ties, but it introduces randomness in the calculation of the test statistic value.
	
\begin{figure}[!t]
		\centering
		\begin{tikzpicture}[scale=1.5]
			\draw[thin, gray, ->] (-0.2, 0) -- (1.5, 0); 
			\draw[thin, gray, ->] (0, -0.2) -- (0, 1.5); 
			
			\node at (1, -0.25) {1};
			\node at (0.01, -0.29) {0};
			\node at (-0.2, 1) {1};
			\node at (-0.26, 0) {0};
			
			\node at (1.5, -0.2) [right] {$X_1$};  
			\node at (-0.25, 1.8) [below] {$X_2$};  
			
			\coordinate (A) at (0, 0);
			\coordinate (B) at (0, 1);
			\coordinate (C) at (1, 0);
			\coordinate (D) at (1, 1);
			
			\filldraw (A) circle (2pt) node[below left] {};
			\filldraw (B) circle (2pt) node[above left] {};
			\filldraw (C) circle (2pt) node[below right] {};
			\filldraw (D) circle (2pt) node[above right] {};
			
			\draw (A) -- (C);
			\draw (B) -- (D);
			\draw (A) -- (B);
			\draw (C) -- (D);	
		\end{tikzpicture}
		\caption{Undirected 1NN graph on the distinct values for two binary variables.}\label{fig:1NN}
	\end{figure}
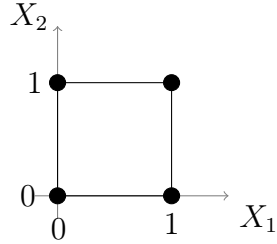

	In the case with two categories and two variables, the CF and ZC using 1NN, ``a'' break down with test statistic values mostly simulated as \texttt{NaN} (not a number) or zeros. 
	The (undirected) 1NN graph on the distinct values that results in this case is displayed in Figure~\ref{fig:1NN}. 
	The CF and ZC both use the standardized difference between the edge counts within the first and second samples. 
	The variance under the null hypothesis used for standardization can be calculated as
	\[
	\Var_{H_0}(R_{d,a}) = \frac{4n_1n_2}{N(N-1)}\left[\sum_{u = 1}^c \frac{\left(\left|\mathcal{E}_u^{C_0}\right| - 2\right)^2}{4 m_u} - \frac{\left(\left|C_0\right|-4\right)^2}{N}\right], 
	\]
	where $u$ takes on all distinct values.
	$\left|\mathcal{E}_u^{C_0}\right|$ denotes the number of edges of node $u$ in the graph on the distinct values, which is $2$ here, see Figure~\ref{fig:1NN}.
	$\left|C_0\right|$ denotes the number of edges in the graph on the distinct values, which is $4$ for the 1NN graph. 
	Thus, $\Var(R_{d,a}) = 0$. 
	However, the variance is calculated by a different but equivalent expression in the implementation and is sometimes numerically not exactly equal to zero but some very small positive number in which case the resulting test statistic value is equal to zero as the observed and expected $R_{d,a}$, whose difference is the numerator of the test statistic, are also equal to zero here with similar arguments. 
	
	For the 5NN, ``u'', a similar problem is given with $p =2$. 
	In the extreme case for $c = 2$, the 5NN is always the full graph, as for each distinct value, there are only 3 other distinct values such that it is connected with all three of those. 
	In the $c = 5$ case, there are $5^2 = 25$ distinct values, so it is also very likely that the union of all optimal 5NN graphs is the full graph. 
	In the case where the 5NN, ``u'' is equal to the full graph, the variance of each test statistic is equal to zero.
	Since the parts of the test statistics of FR, CCS, CF, and ZC that are standardized can be expressed in terms of linear combinations of the edge counts within the first and second sample, $R_{1, u}$ and $R_{2, u}$, respectively, in the union graph $\bar{G}$ \textcite{zhang_graph-based_2022}, it suffices to show $\Var_{H_0}(R_{1, u}) = \Var_{H_0}(R_{2, u}) = \Cov_{H_0}(R_{1, u}, R_{2, u}) = 0$ to show that the null variance of these statistics is equal to zero. 
	For the (undirected) full graph, it holds that the number of edges is given by $|\bar{G}| = \frac{N (N-1)}{2}$ and the degree of each node $i$ is given by $|\mathcal{E}_i^{\bar{G}}| = N - 1, i = 1,\dots,N$.
	Inserting these quantities in the expressions for the variances and covariances given in the supplemental material of \textcite{zhang_graph-based_2022} yields 
	{\allowdisplaybreaks
		\begin{align*}
			\Var_{H_0}(R_{1, u}) =& \left[\frac{n_1(n_1-1)}{N(N-1)} - \frac{n_1(n_1-1)(n_1-2)(n_1-3)}{N(N-1)(N-2)(N-3)}\right] \frac{N(N-1)}{2}\\
			& + \left[\frac{n_1(n_1-1)(n_1-2)}{N(N-1)(N-2)} - \frac{n_1(n_1-1)(n_1-2)(n_1-3)}{N(N-1)(N-2)(N-3)}\right]N(N-1)(N-2)\\
			& + \left[\frac{n_1(n_1-1)(n_1-2)(n_1-3)}{N(N-1)(N-2)(N-3)} - \left(\frac{n_1(n_1-1)}{N(N-1)}\right)^2\right] \left(\frac{N(N-1)}{2}\right)^2\\
			=&  \frac{n_1(n_1-1)}{2} - \frac{n_1(n_1-1)(n_1-2)(n_1-3)}{2(N-2)(N-3)} \\
			& + n_1(n_1-1)(n_1-2) - \frac{n_1(n_1-1)(n_1-2)(n_1-3)}{N-3}\\
			& + \frac{n_1(n_1-1)(n_1-2)(n_1-3)}{4(N-2)(N-3)}N(N-1)-\left(\frac{n_1(n_1-1)}{2}\right)^2\\
			=& \frac{n_1(n_1-1)}{2} - \left(\frac{n_1(n_1-1)}{2}\right)^2 + n_1(n_1-1)(n_1-2)\\
			& - \frac{2+4(N-2)-N(N-1)}{4(N-2)(N-3)}\left[n_1(n_1-1)(n_1-2)(n_1-3)\right]\\
			=& n_1(n_1-1) \left[\frac{1}{2}-\frac{1}{4}n_1(n_1-1)+n_1-2\right]\\
			& - \frac{2+4N-8-N^2+N}{4(N^2-5N+6)}\left[n_1(n_1-1)(n_1-2)(n_1-3)\right]\\
			=& n_1(n_1-1) \left[-\frac{1}{4}n_1^2 + \frac{5}{4}n_1-\frac{3}{2}\right] + \frac{1}{4}\left[n_1(n_1-1)(n_1-2)(n_1-3)\right]\\
			=& \frac{n_1(n_1-1)}{4} \left[-n_1^2 + 5n_1-6\right] + \frac{n_1(n_1-1)}{4}\left[n_1^2-5n_1+6\right]\\
			=& 0.
		\end{align*} 
	}
	Analogously, $\Var_{H_0}(R_{1, u}) = 0$ can be shown by replacing the sample size $n_1$ of the first sample with the sample size $n_2$ of the second sample in the above calculation.\\
	For the covariance, it holds
	\begin{align*}
		\Cov_{H_0}(R_{1, u}, R_{2, u}) =& \frac{n_1(n_1-1)n_2(n_2-1)}{N(N-1)(N-2)(N-3)}\\
		&\,\, \cdot \left[\left(\frac{N(N-1)}{2}\right)^2 - \frac{N(N-1)}{2} - N(N-1)(N-2)\right]\\
		&-\frac{n_1(n_1-1)}{N(N-1)}\frac{n_2(n_2-1)}{N(N-1)}\left(\frac{N(N-1)}{2}\right)^2\\
		=& \frac{n_1(n_1-1)n_2(n_2-1)}{4(N-2)(N-3)}N(N-1) - \frac{n_1(n_1-1)n_2(n_2-1)}{2(N-2)(N-3)}\\
		&-\frac{n_1(n_1-1)n_2(n_2-1)}{(N-3)} -\frac{n_1(n_1-1)n_2(n_2-1)}{4}\\
		=& n_1(n_1-1)n_2(n_2-1)\left[\frac{N(N-1) - 2 - 4(N-2)}{4(N-2)(N-3)} - \frac{1}{4}\right]\\
		=& n_1(n_1-1)n_2(n_2-1)\left[\frac{N^2-N-2-4N+8}{4(N^2-5N+6)} - \frac{1}{4}\right]\\
		=& n_1(n_1-1)n_2(n_2-1)\left[\frac{1}{4} - \frac{1}{4}\right]\\
		=& 0.
	\end{align*}
	Therefore, all variances used for standardization in the calculation of the edge count test statistics are analytically equal to zero, which results in the \texttt{NaN} results. 
	Numerically, the calculated variances are sometimes not exactly equal to zero but always have a very small absolute value. 
	If this very small value is positive, the resulting test statistics are equal to zero. 
	This happens especially for unequal sample sizes and small $N$. 
	Sometimes, the numerically calculated variance estimate is even slightly negative, which also results in \texttt{NaN} values of the test statistic since the square root of this value is taken in the standardization.

	\FloatBarrier
	\section{Additional Figures}\label{app:add.figs}
	\subsection[Pre-Selection of Methods for k = 2]{Pre-Selection of Methods for $k = 2$}\label{app:add.figs.cat.no.y.sel.meths}
	\begin{figure}[H]
		\centering
		\includegraphics[width=\linewidth]{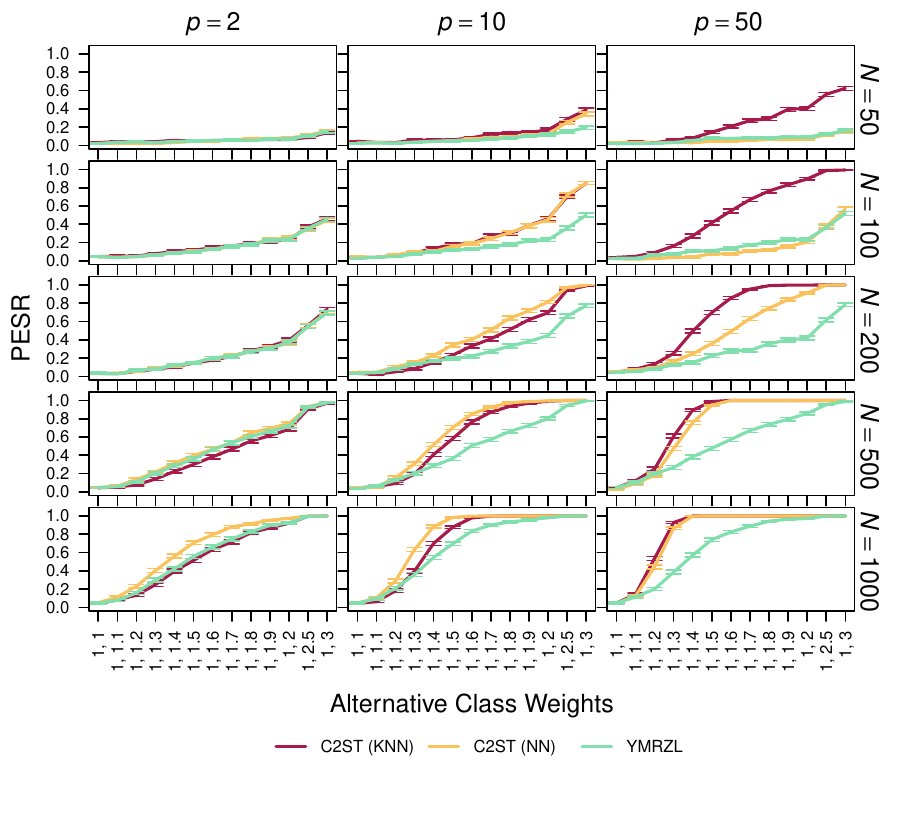}
		\caption{Proportion of extreme simulation repetitions (PESR) for two datasets of the same sample sizes with binary variables. The class weights give the unnormalized probabilities $(1, 1+\delta)$ for the values 0 and 1 for each variable in the second dataset. This means the weights in the first dataset are set to $(1, 1)$, and in the second dataset to $(1, 1+\delta)$. Error bars indicate Monte Carlo standard errors.}
		\label{fig:pow.cat.no.y.bin.bal.C2ST}
	\end{figure}
	
	\begin{figure}[H]
		\centering
		\includegraphics[width=\linewidth]{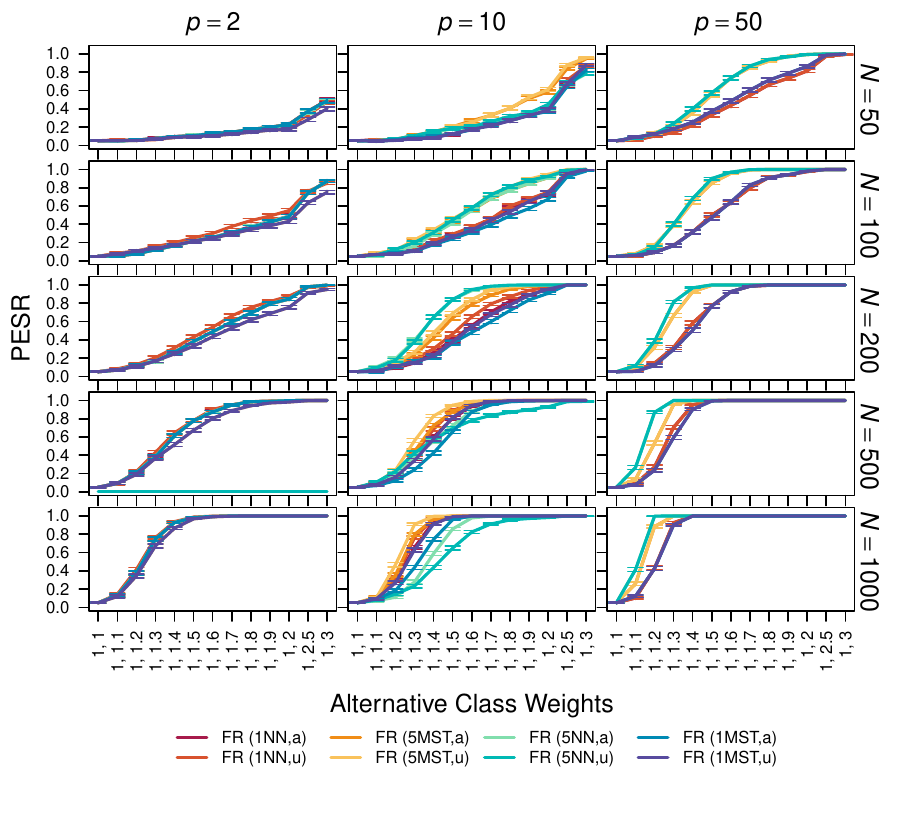}
		\caption{Proportion of extreme simulation repetitions (PESR) for two datasets of the same sample sizes with binary variables. The class weights give the unnormalized probabilities $(1, 1+\delta)$ for the values 0 and 1 for each variable in the second dataset. This means the weights in the first dataset are set to $(1, 1)$, and in the second dataset to $(1, 1+\delta)$. Error bars indicate Monte Carlo standard errors.}
		\label{fig:pow.cat.no.y.bin.bal.FR}
	\end{figure}
	
	\begin{figure}[H]
		\centering
		\includegraphics[width=\linewidth]{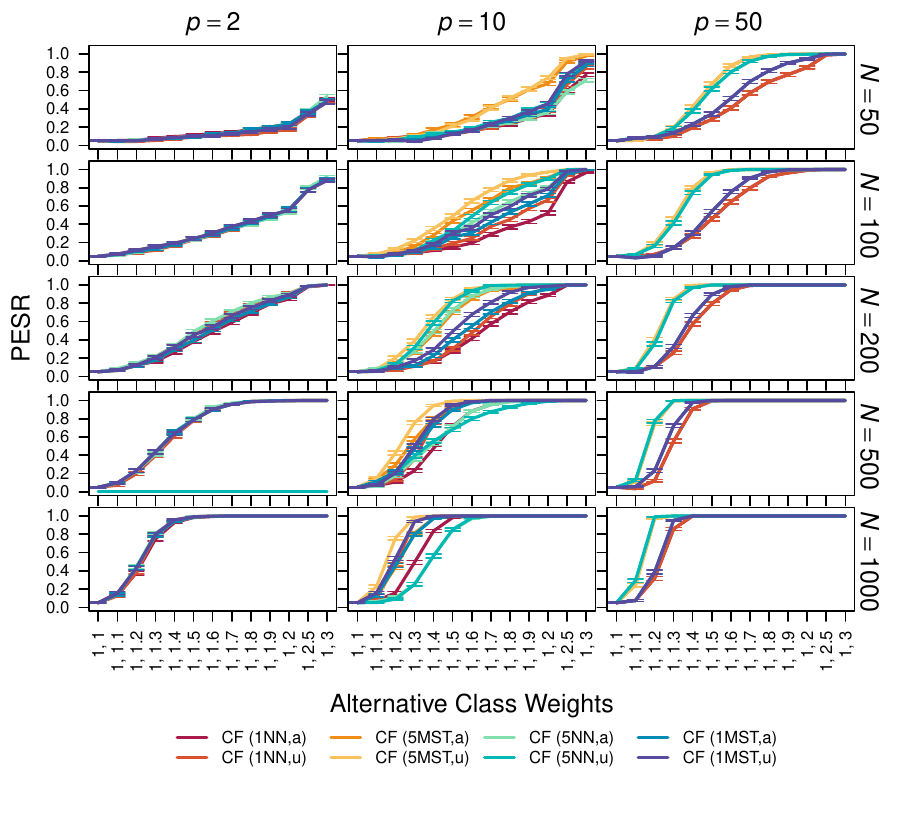}
		\caption{Proportion of extreme simulation repetitions (PESR) for two datasets of the same sample sizes with binary variables. The class weights give the unnormalized probabilities $(1, 1+\delta)$ for the values 0 and 1 for each variable in the second dataset. This means the weights in the first dataset are set to $(1, 1)$, and in the second dataset to $(1, 1+\delta)$. Error bars indicate Monte Carlo standard errors.}
		\label{fig:pow.cat.no.y.bin.bal.CF}
	\end{figure}
	
	\begin{figure}[H]
		\centering
		\includegraphics[width=\linewidth]{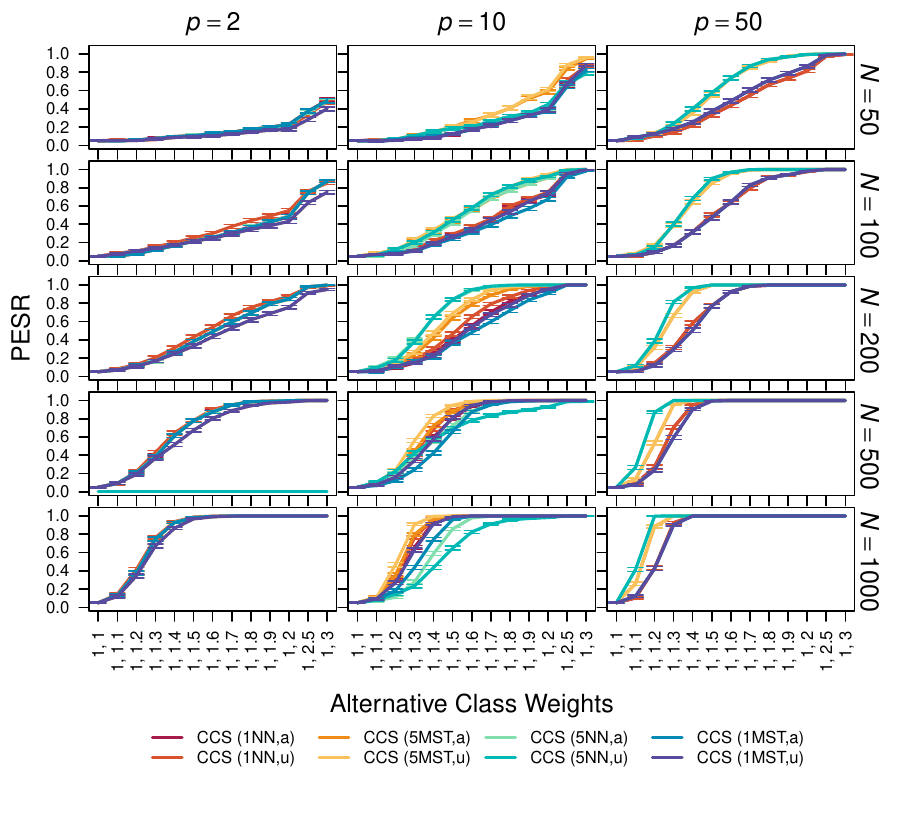}
		\caption{Proportion of extreme simulation repetitions (PESR) for two datasets of the same sample sizes with binary variables. The class weights give the unnormalized probabilities $(1, 1+\delta)$ for the values 0 and 1 for each variable in the second dataset. This means the weights in the first dataset are set to $(1, 1)$, and in the second dataset to $(1, 1+\delta)$. Error bars indicate Monte Carlo standard errors.}
		\label{fig:pow.cat.no.y.bin.bal.CCS}
	\end{figure}
	
	\begin{figure}[H]
		\centering
		\includegraphics[width=\linewidth]{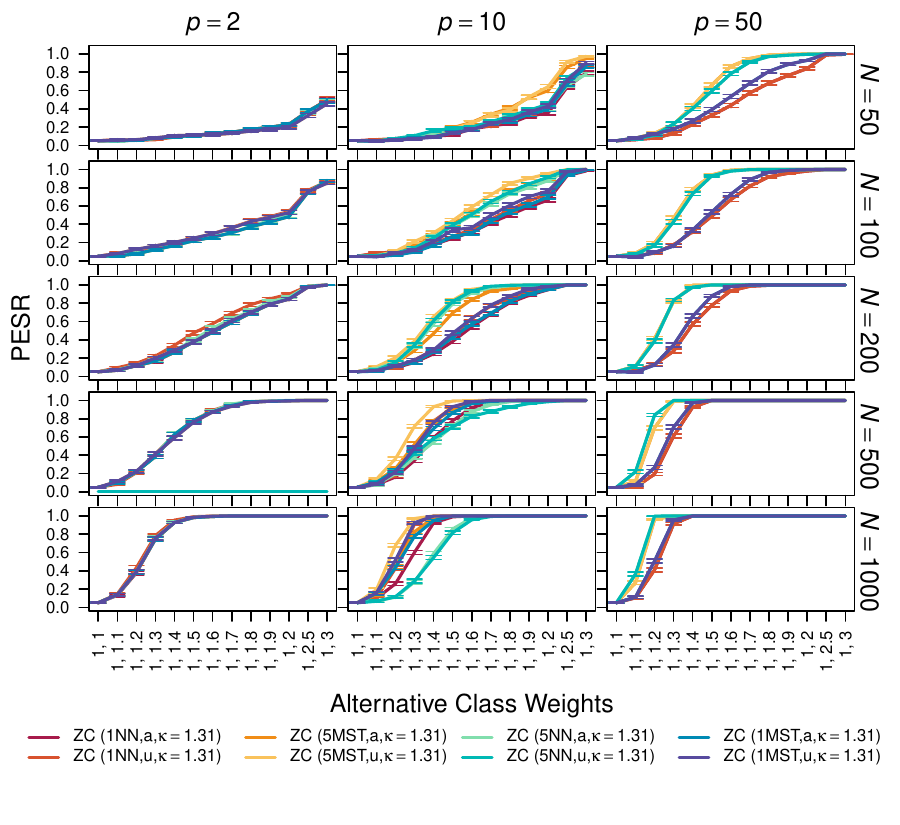}
		\caption{Proportion of extreme simulation repetitions (PESR) for two datasets of the same sample sizes with binary variables. The class weights give the unnormalized probabilities $(1, 1+\delta)$ for the values 0 and 1 for each variable in the second dataset. This means the weights in the first dataset are set to $(1, 1)$, and in the second dataset to $(1, 1+\delta)$. Error bars indicate Monte Carlo standard errors.}
		\label{fig:pow.cat.no.y.bin.bal.ZC.kappa.1.31}
	\end{figure}
	
	\begin{figure}[H]
		\centering
		\includegraphics[width=\linewidth]{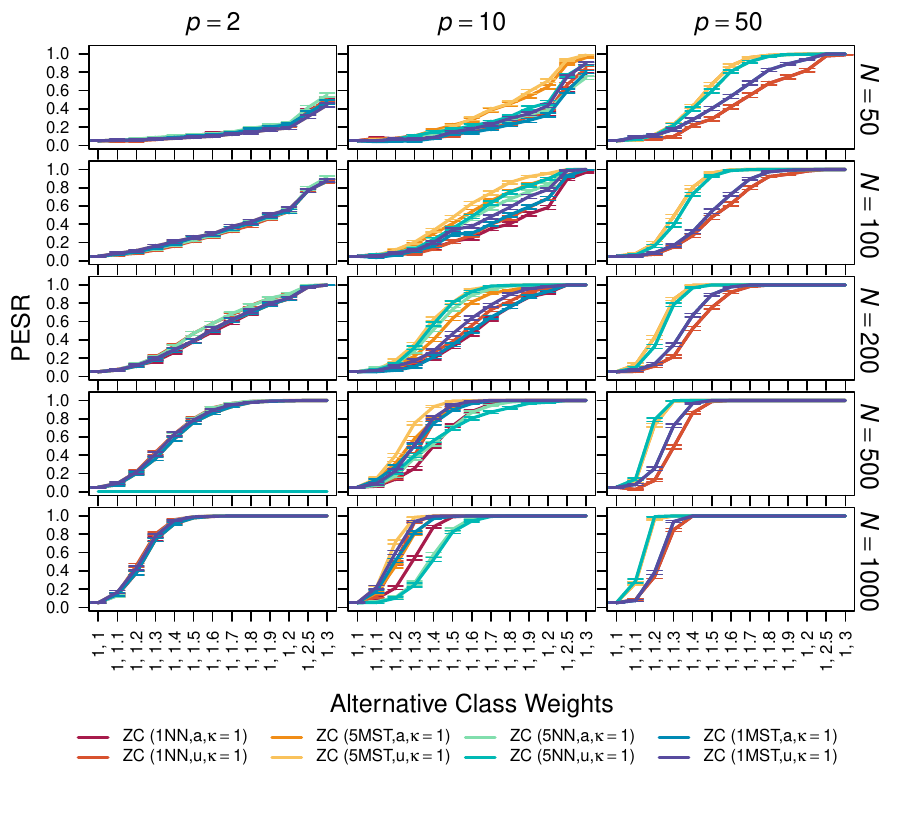}
		\caption{Proportion of extreme simulation repetitions (PESR) for two datasets of the same sample sizes with binary variables. The class weights give the unnormalized probabilities $(1, 1+\delta)$ for the values 0 and 1 for each variable in the second dataset. This means the weights in the first dataset are set to $(1, 1)$, and in the second dataset to $(1, 1+\delta)$. Error bars indicate Monte Carlo standard errors.}
		\label{fig:pow.cat.no.y.bin.bal.ZC.kappa.1}
	\end{figure}
	
	\begin{figure}[H]
		\centering
		\includegraphics[width=\linewidth]{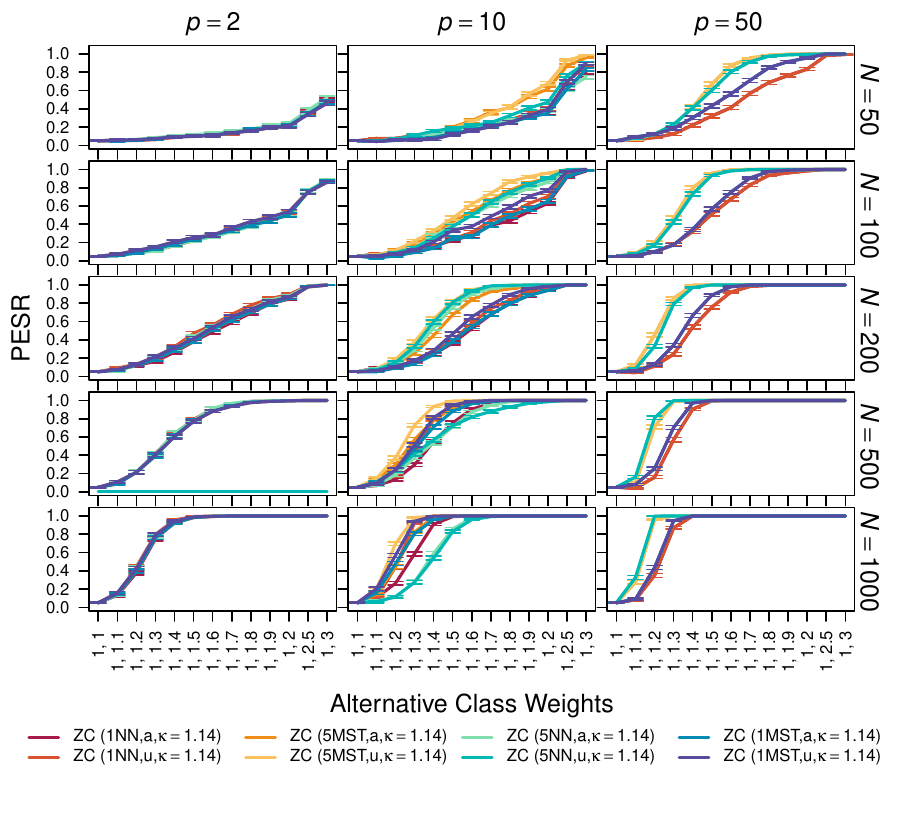}
		\caption{Proportion of extreme simulation repetitions (PESR) for two datasets of the same sample sizes with binary variables. The class weights give the unnormalized probabilities $(1, 1+\delta)$ for the values 0 and 1 for each variable in the second dataset. This means the weights in the first dataset are set to $(1, 1)$, and in the second dataset to $(1, 1+\delta)$. Error bars indicate Monte Carlo standard errors.}
		\label{fig:pow.cat.no.y.bin.bal.ZC.kappa.1.14}
	\end{figure}
	
	\begin{figure}[H]
		\centering
		\includegraphics[width=\linewidth]{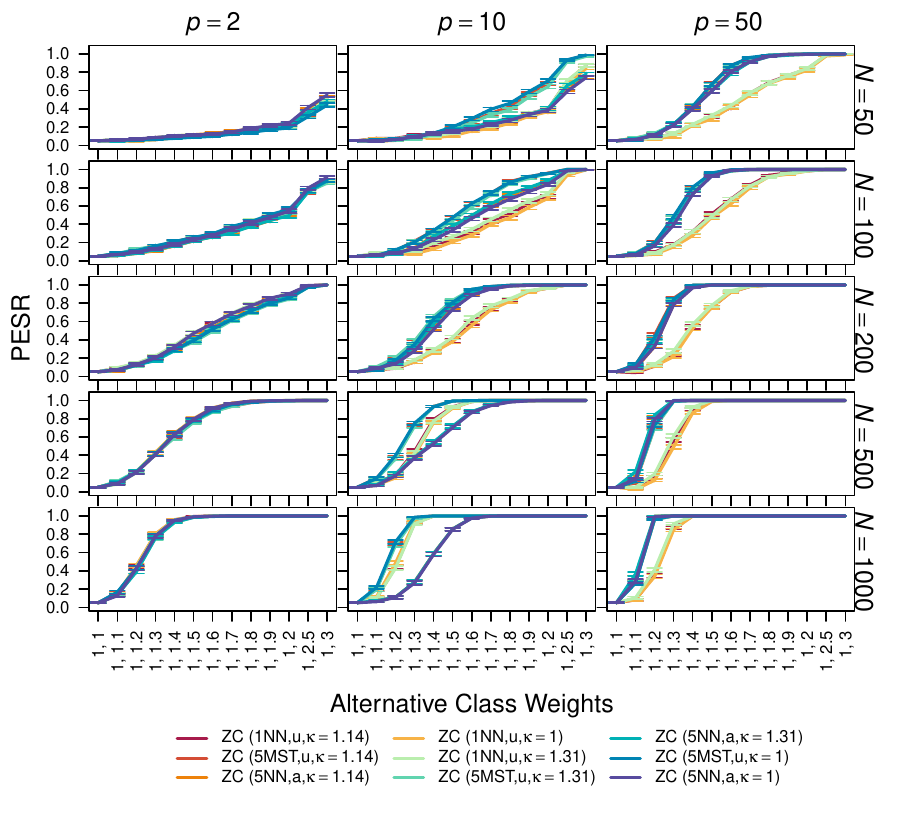}
		\caption{Proportion of extreme simulation repetitions (PESR) for two datasets of the same sample sizes with binary variables. The class weights give the unnormalized probabilities $(1, 1+\delta)$ for the values 0 and 1 for each variable in the second dataset. This means the weights in the first dataset are set to $(1, 1)$, and in the second dataset to $(1, 1+\delta)$. Error bars indicate Monte Carlo standard errors.}
		\label{fig:pow.cat.no.y.bin.bal.ZC.comp.kappas}
	\end{figure}
	
	\begin{figure}[H]
		\centering
		\includegraphics[width=\linewidth]{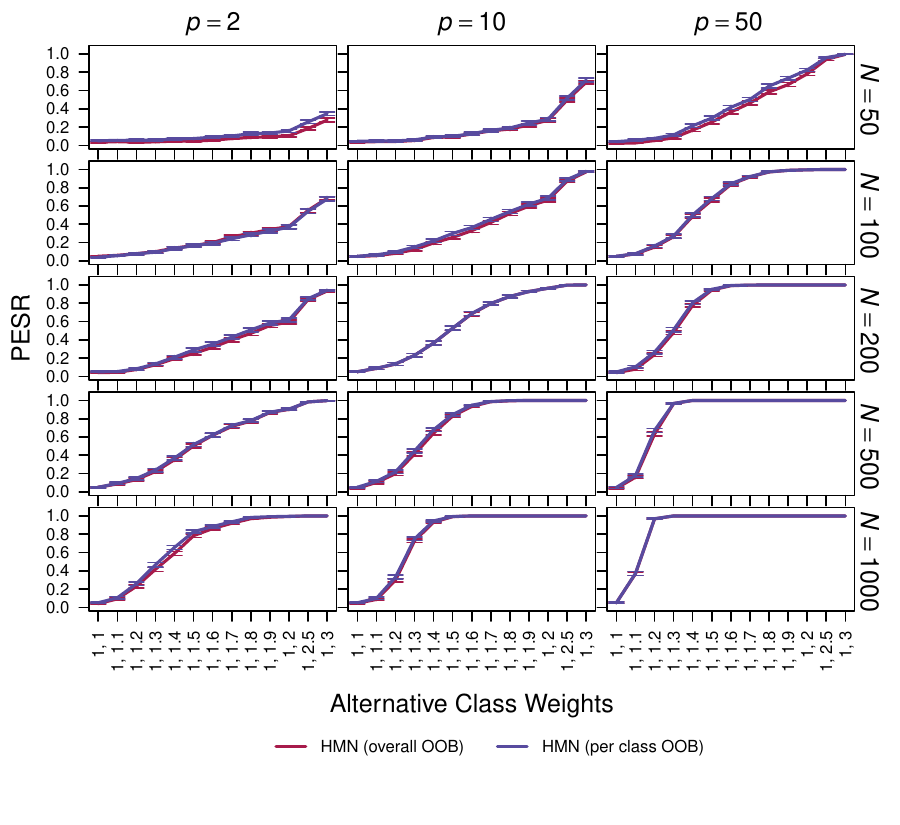}
		\caption{Proportion of extreme simulation repetitions (PESR) for two datasets of the same sample sizes with binary variables. The class weights give the unnormalized probabilities $(1, 1+\delta)$ for the values 0 and 1 for each variable in the second dataset. This means the weights in the first dataset are set to $(1, 1)$, and in the second dataset to $(1, 1+\delta)$. Error bars indicate Monte Carlo standard errors.}
		\label{fig:pow.cat.no.y.bin.bal.HMN}
	\end{figure}
	
	\begin{figure}[H]
		\centering
		\includegraphics[width=\linewidth]{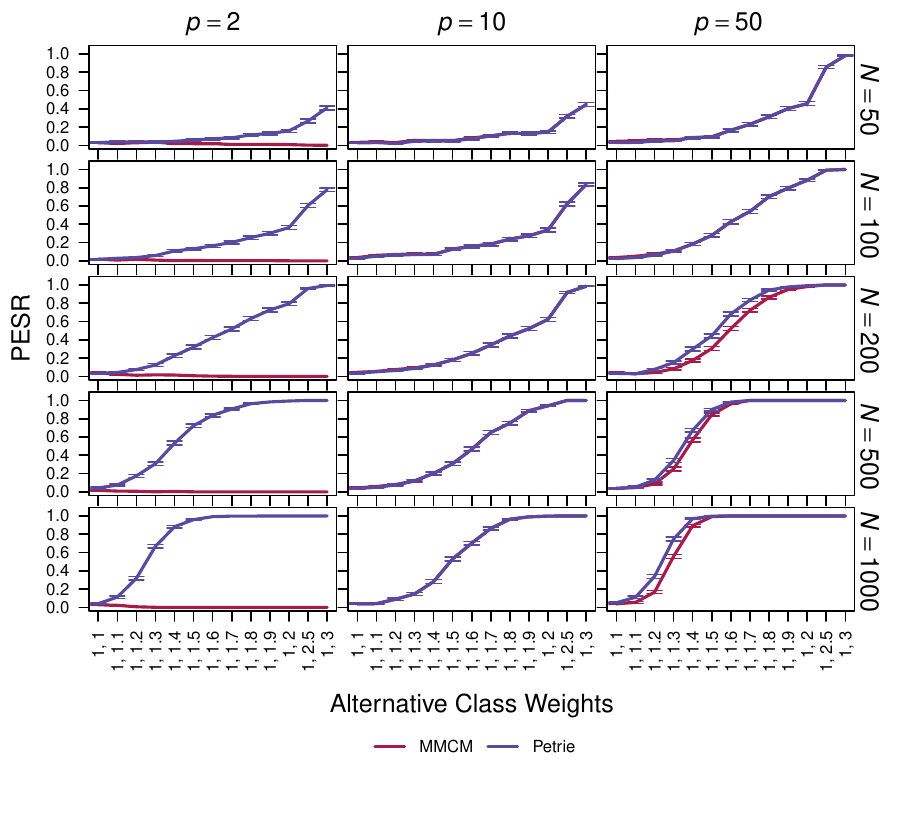}
		\caption{Proportion of extreme simulation repetitions (PESR) for two datasets of the same sample sizes with binary variables. The class weights give the unnormalized probabilities $(1, 1+\delta)$ for the values 0 and 1 for each variable in the second dataset. This means the weights in the first dataset are set to $(1, 1)$, and in the second dataset to $(1, 1+\delta)$. Error bars indicate Monte Carlo standard errors.}
		\label{fig:pow.cat.no.y.bin.bal.MMCM}
	\end{figure}
	
	\subsection[k = 2, Unbalanced Sample Sizes]{$k = 2$, Unbalanced Sample Sizes}\label{app:add.figs.cat.no.y.unbal}
	\begin{figure}[H]
		\centering
		\includegraphics[width=\linewidth]{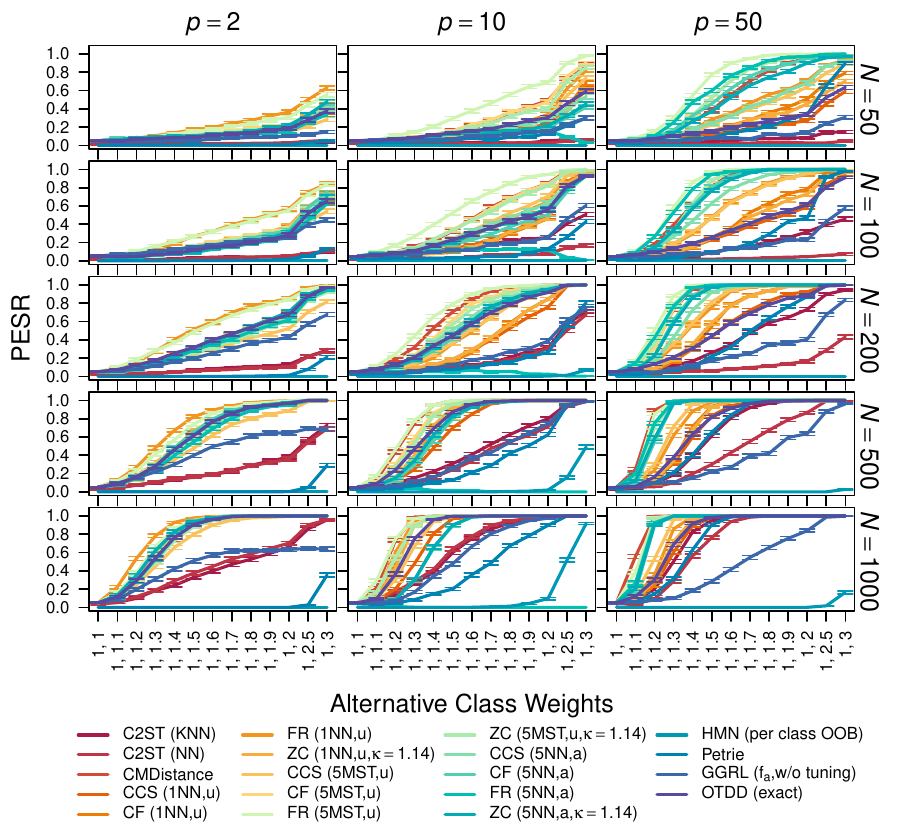}
		\caption{Proportion of extreme simulation repetitions (PESR) for two datasets of unequal sample sizes with binary variables. The class weights give the unnormalized probabilities $(1, 1+\delta)$ for the values 0 and 1 for each variable in the second dataset. This means the weights in the first dataset are set to $(1, 1)$, and in the second dataset to $(1, 1+\delta)$. Error bars indicate Monte Carlo standard errors.}
		\label{fig:pow.cat.no.y.bin.unbal}
	\end{figure}
	
	\begin{figure}[H]
		\centering
		\includegraphics[width=\linewidth]{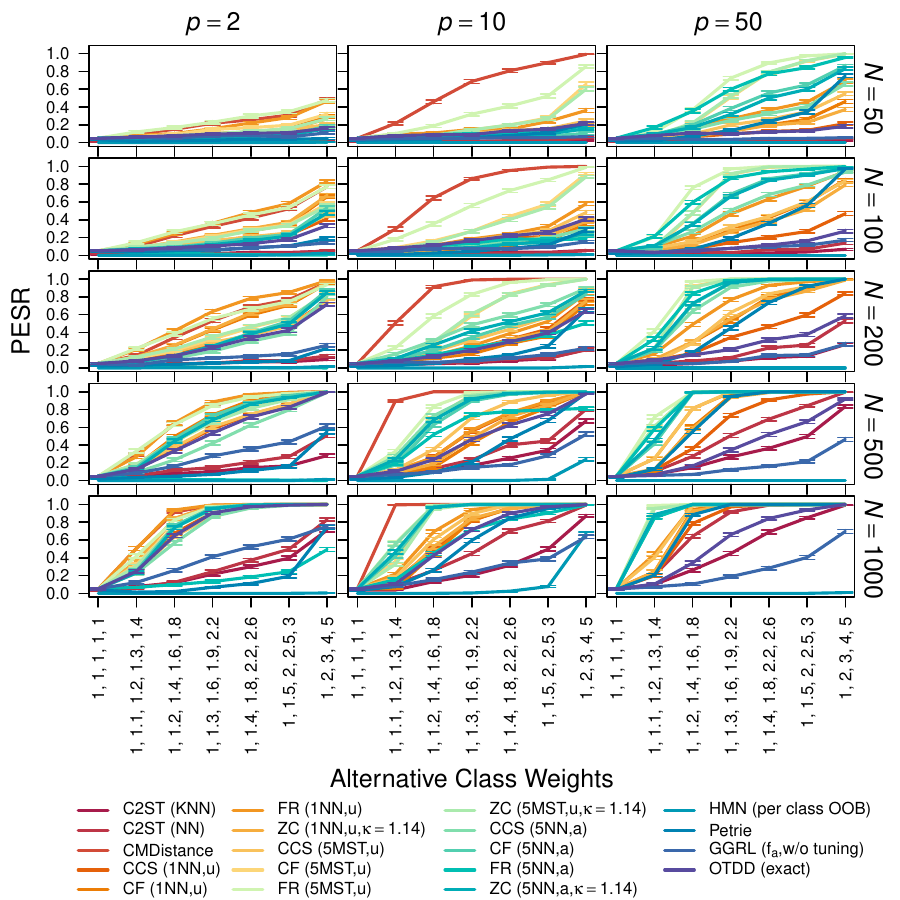}
		\caption{Proportion of extreme simulation repetitions (PESR) for two datasets of unequal sample sizes. The class weights give the unnormalized probabilities $(1, 1+\delta, 1+2\delta, 1+3\delta, 1+4\delta)$ for the values $1$ to $5$ for each variable in the second dataset. The weights in the first dataset are always set to $(1, 1, 1, 1, 1)$. Error bars indicate Monte Carlo standard errors.}
		\label{fig:pow.cat.no.y.multinom.skewed.unbal}
	\end{figure}
	
	\begin{figure}[H]
		\centering
		\includegraphics[width=\linewidth]{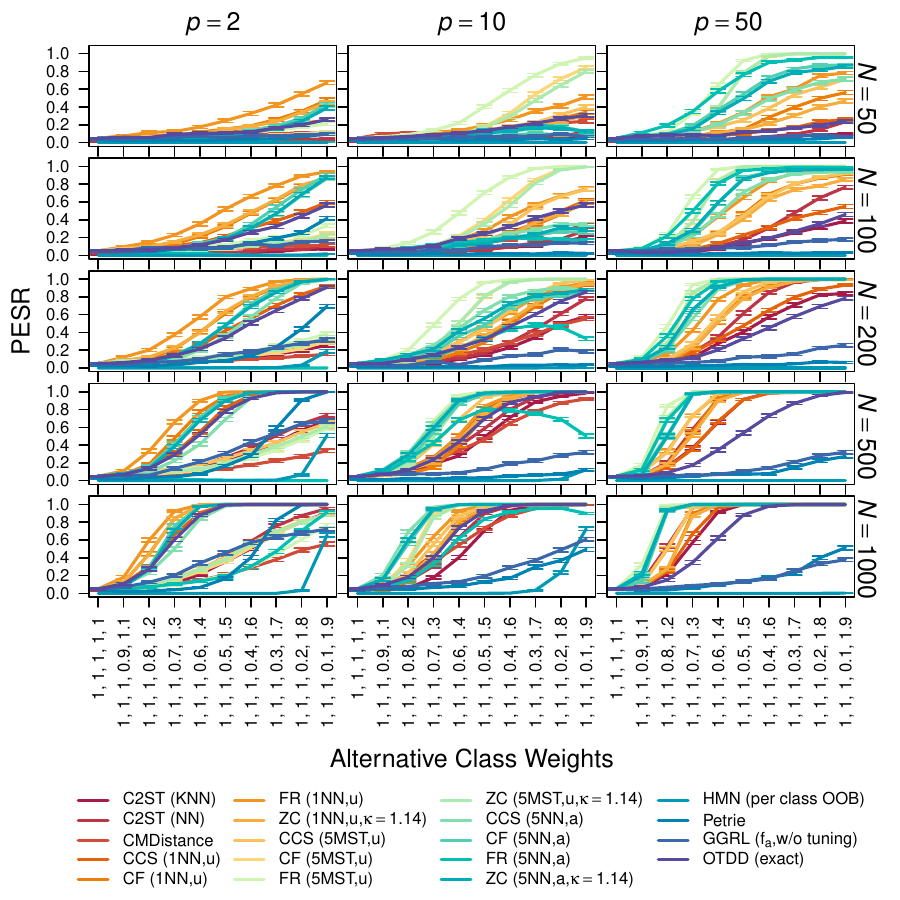}
		\caption{Proportion of extreme simulation repetitions (PESR) for two datasets of unequal sizes. The class weights give the unnormalized probabilities $(1, 1, 1, 1+\delta, 1-\delta)$ for the values $1$ to $5$ for each variable in the second dataset. The weights in the first dataset are always set to $(1, 1, 1, 1, 1)$. Error bars indicate Monte Carlo standard errors.}
		\label{fig:pow.cat.no.y.multinom.1u1d.unbal}
	\end{figure}
	
	\subsection[k = 2, Target Variable]{$k = 2$, Target Variable}\label{app:add.figs.cat.y}
	\begin{figure}[H]
		\centering
		\includegraphics[width=\linewidth]{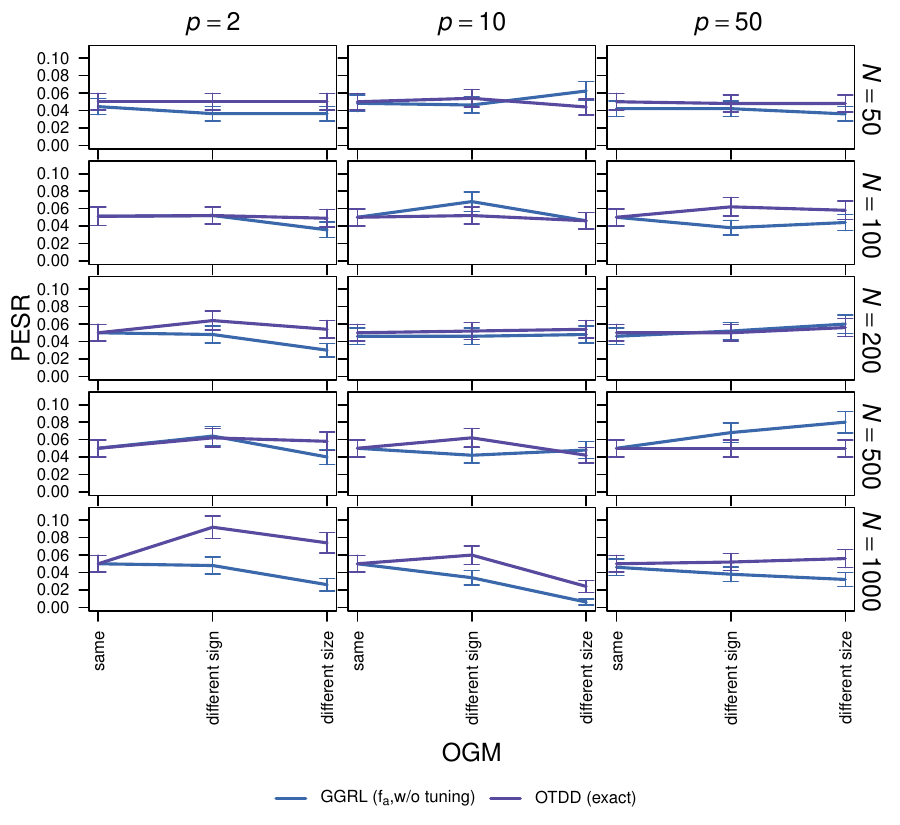}
		\caption{PESR (proportion of extreme simulation repetitions) for $k = 2$ binary datasets of unequal sample sizes. A target variable is generated in both datasets, either using the same outcome-generating model (OGM) or the sign or size of the coefficients in the logistic model in the second dataset are changed. The OGM in the first dataset is a logistic model with the first half of the coefficients equal to $0.5$ and the second half and the intercept equal to $-0.5$. For ``different sign'' the signs of all coefficients except for the intercept are changed in the second dataset. For ``different size'' all coefficients except for the intercept are halved in the second dataset. Error bars indicate Monte Carlo standard errors.}
		\label{fig:pow.cat.y.bin.unbal}
	\end{figure}
	
	\begin{figure}[H]
		\centering
		\includegraphics[width=\linewidth]{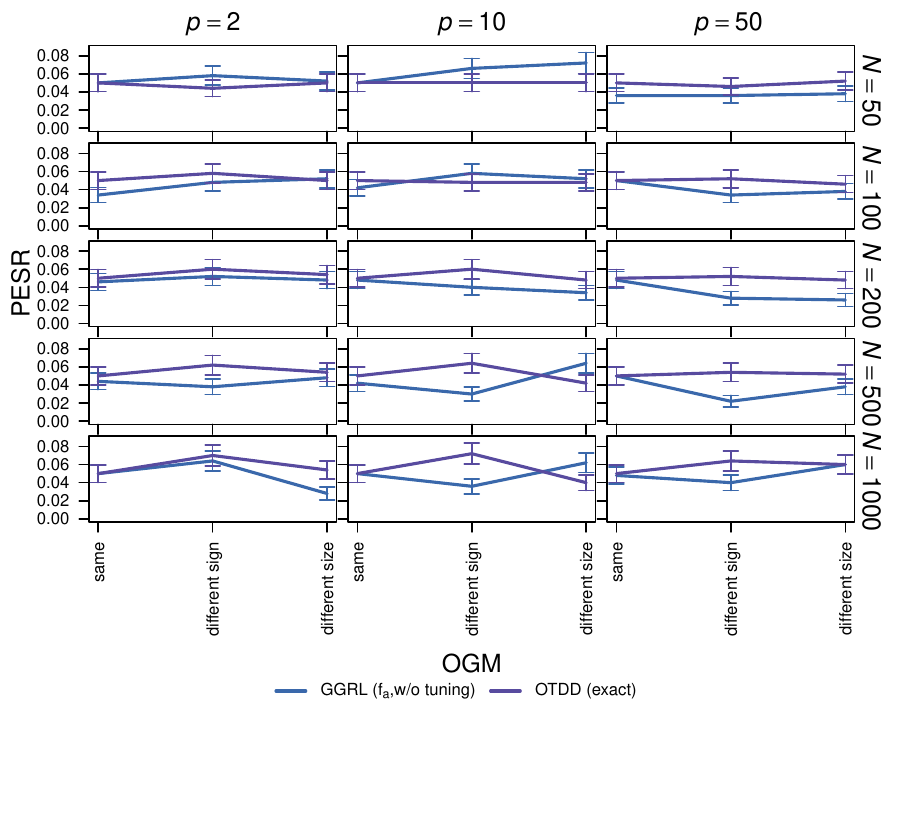}
		\caption{PESR (proportion of extreme simulation repetitions) for $k = 2$ multinomial datasets of equal sample sizes. A target variable is generated in both datasets, either using the same outcome-generating model (OGM) or the sign or size of the coefficients in the logistic model in the second dataset are changed. The OGM in the first dataset is a logistic model with the first half of the coefficients equal to $0.5$ and the second half and the intercept equal to $-0.5$. For ``different sign'' the signs of all coefficients except for the intercept are changed in the second dataset. For ``different size'' all coefficients except for the intercept are halved in the second dataset. Error bars indicate Monte Carlo standard errors.}
		\label{fig:pow.y.multi.bal}
	\end{figure}
	
	\begin{figure}[H]
		\centering
		\includegraphics[width=\linewidth]{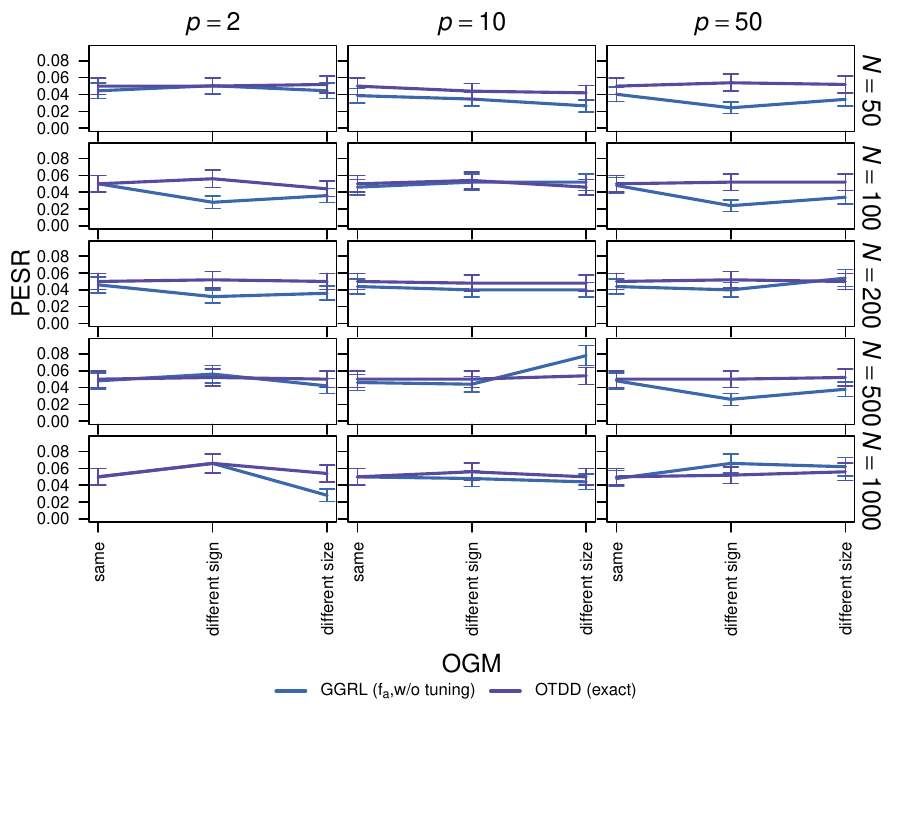}
		\caption{PESR (proportion of extreme simulation repetitions) for $k = 2$ multinomial datasets of unequal sample sizes. A target variable is generated in both datasets, either using the same outcome-generating model (OGM) or the sign or size of the coefficients in the logistic model in the second dataset are changed. The OGM in the first dataset is a logistic model with the first half of the coefficients equal to $0.5$ and the second half and the intercept equal to $-0.5$. For ``different sign'' the signs of all coefficients except for the intercept are changed in the second dataset. For ``different size'' all coefficients except for the intercept are halved in the second dataset. Error bars indicate Monte Carlo standard errors.}
		\label{fig:pow.y.multi.unbal}
	\end{figure}

	\subsection[k = 4, Binary Data, Balanced Sample Sizes]{$k = 4$, Binary Data, Balanced Sample Sizes}\label{app:add.figs.cat.multi.bin.bal}
	\begin{figure}[H]
		\centering
		\includegraphics[width=\linewidth]{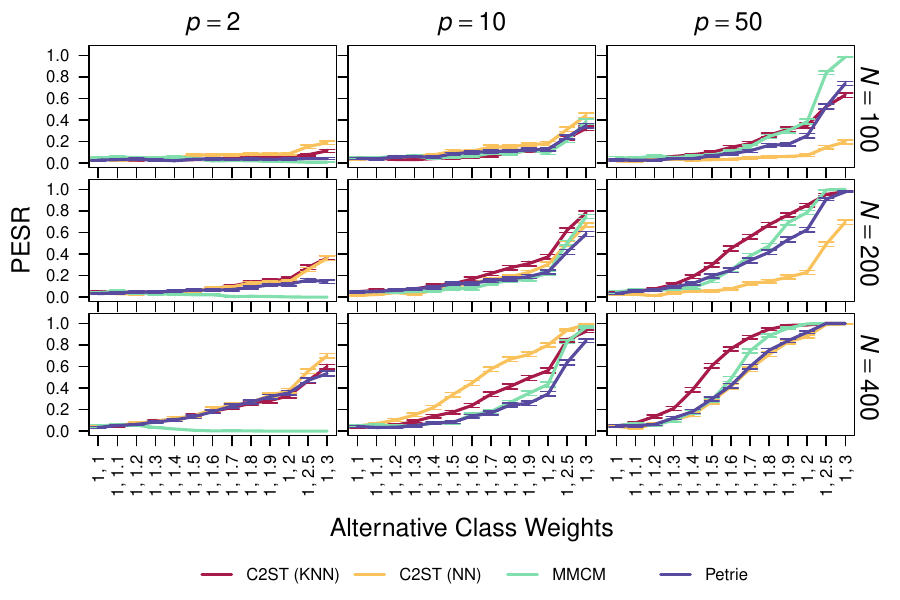}
		\caption{Proportion of extreme simulation repetitions (PESR) for four datasets of the same sample sizes with binary variables. The class weights, i.e.\ the unnormalized probabilities for the values 0 and 1, in the first, second, and third datasets are always set to $(1, 1)$. The class weights on the $x$-axis give the unnormalized probabilities $(1, 1+\delta)$ for each variable in the fourth dataset. Error bars indicate Monte Carlo standard errors.}
		\label{fig:pow.cat.multi.bin.bal.3+1}
	\end{figure}
	
	\begin{figure}[H]
		\centering
		\includegraphics[width=\linewidth]{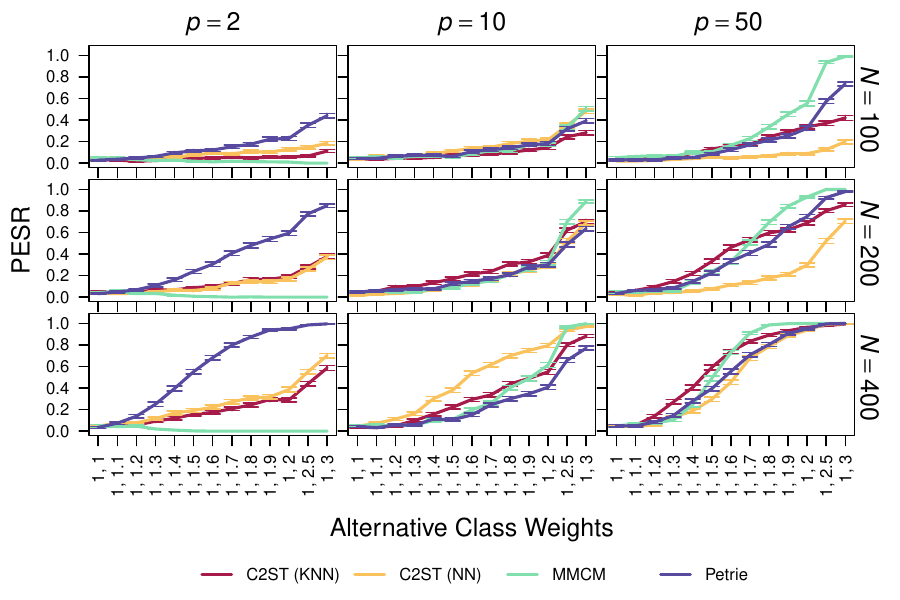}
		\caption{Proportion of extreme simulation repetitions (PESR) for four datasets of the same sample sizes with binary variables. The class weights, i.e.\ the unnormalized probabilities for the values 0 and 1, in the first and second datasets, are always set to $(1, 1)$. The class weights on the $x$-axis give the unnormalized probabilities $(1, 1+\delta)$for each variable in the third and fourth datasets. Error bars indicate Monte Carlo standard errors.}
		\label{fig:pow.cat.multi.bin.bal.2+2}
	\end{figure}
	
	\begin{figure}[H]
		\centering
		\includegraphics[width=\linewidth]{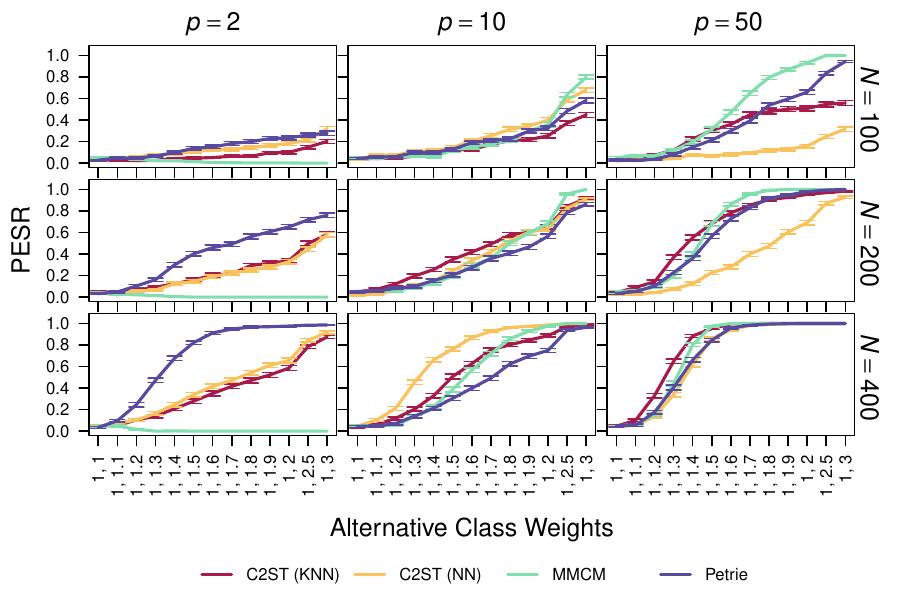}
		\caption{Proportion of extreme simulation repetitions (PESR) for four datasets of the same sample sizes with binary variables. The class weights, i.e.\ the unnormalized probabilities for the values 0 and 1, in the first and second datasets, are always set to $(1, 1)$. The class weights on the $x$-axis give the unnormalized probabilities $(1, 1+\delta)$ for each variable in the third dataset. The weights in the fourth dataset are set to $(1, 1+2\delta)$. Error bars indicate Monte Carlo standard errors.}
		\label{fig:pow.cat.multi.bin.bal.2+1+1}
	\end{figure}

	\subsection[k = 4, Binary Data, Unbalanced Sample Sizes]{$k = 4$, Binary Data, Unbalanced Sample Sizes}\label{app:add.figs.cat.multi.bin.unbal}
	\begin{figure}[H]
		\centering
		\includegraphics[width=\linewidth]{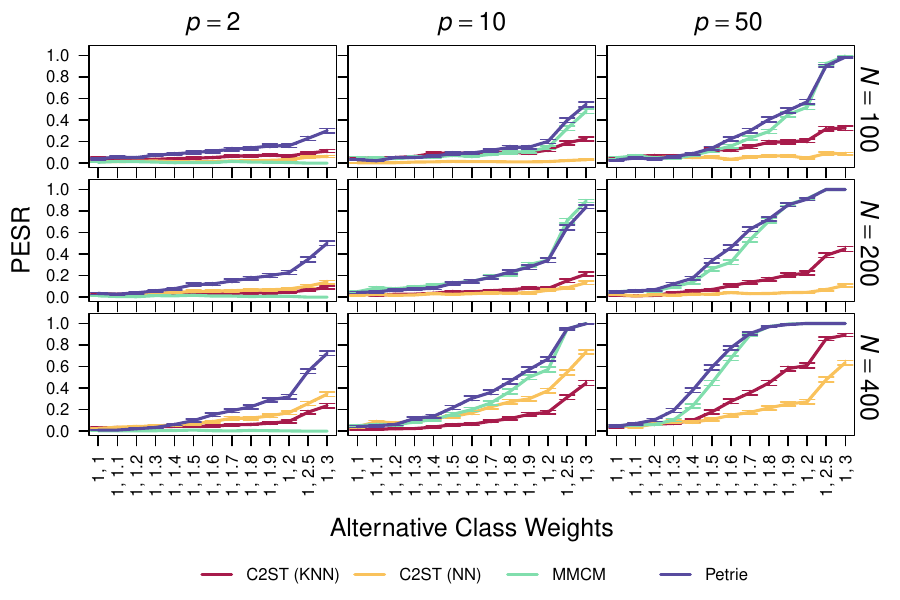}
		\caption{Proportion of extreme simulation repetitions (PESR) for four datasets of unequal sample sizes with binary variables. The class weights, i.e.\ the unnormalized probabilities for the values 0 and 1, in the first, second, and third datasets are always set to $(1, 1)$. The class weights on the $x$-axis give the unnormalized probabilities $(1, 1+\delta)$ for each variable in the fourth dataset. Error bars indicate Monte Carlo standard errors.}
		\label{fig:pow.cat.multi.bin.unbal.3+1}
	\end{figure}
	
	\begin{figure}[H]
		\centering
		\includegraphics[width=\linewidth]{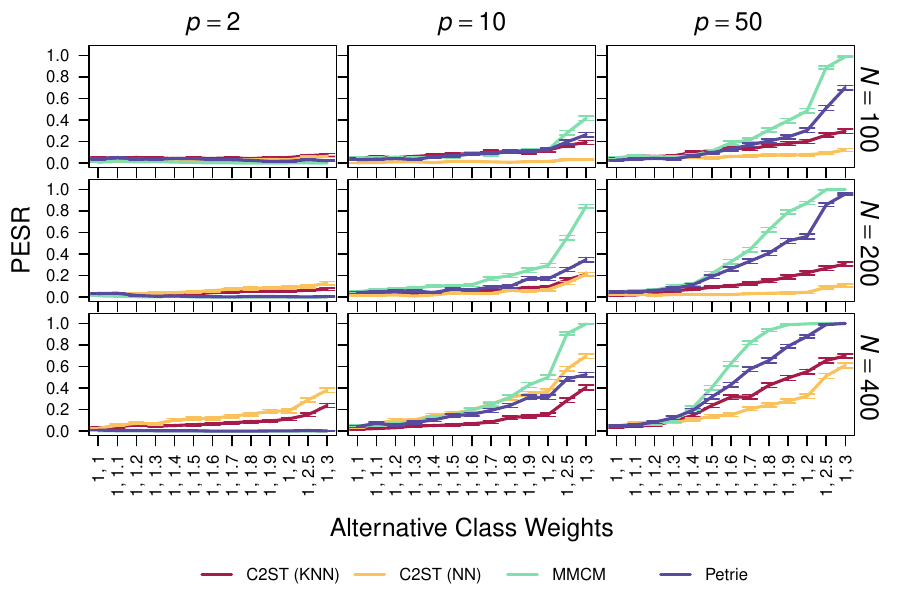}
		\caption{Proportion of extreme simulation repetitions (PESR) for four datasets of unequal sample sizes with binary variables. The class weights, i.e.\ the unnormalized probabilities for the values 0 and 1, in the first and second datasets, are always set to $(1, 1)$. The class weights on the $x$-axis give the unnormalized probabilities $(1, 1+\delta)$for each variable in the third and fourth datasets. Error bars indicate Monte Carlo standard errors.}
		\label{fig:pow.cat.multi.bin.unbal.2+2}
	\end{figure}
	
	\begin{figure}[H]
		\centering
		\includegraphics[width=\linewidth]{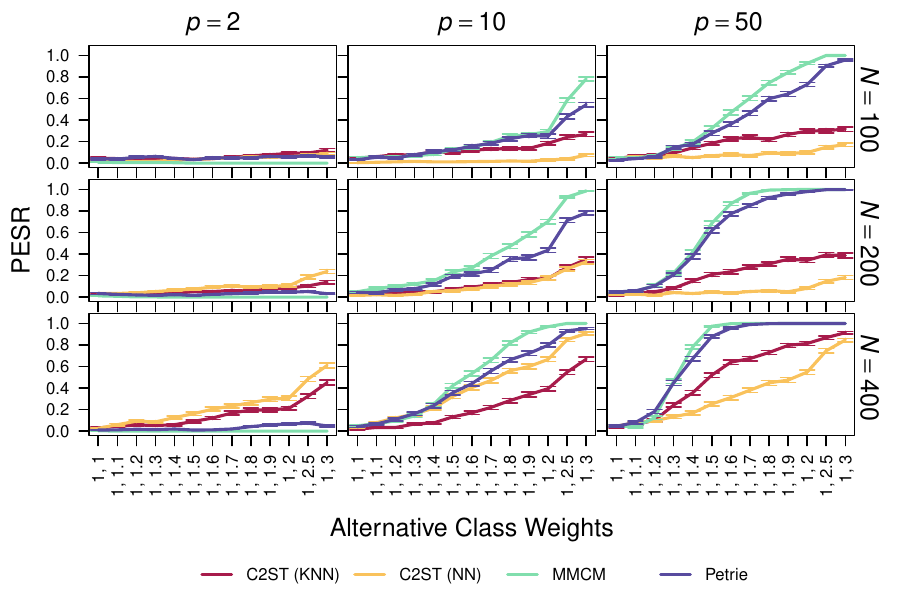}
		\caption{Proportion of extreme simulation repetitions (PESR) for four datasets of unequal sample sizes with binary variables. The class weights give the unnormalized probabilities $(1, 1+\delta)$ for the values 0 and 1 for each variable in the third dataset. The weights in the first and second datasets are always set to $(1, 1)$. The weights in the fourth dataset are set to $(1, 1+2\delta)$. Error bars indicate Monte Carlo standard errors.}
		\label{fig:pow.cat.multi.bin.unbal.2+1+1}
	\end{figure}
	
	\begin{figure}[H]
		\centering
		\includegraphics[width=\linewidth]{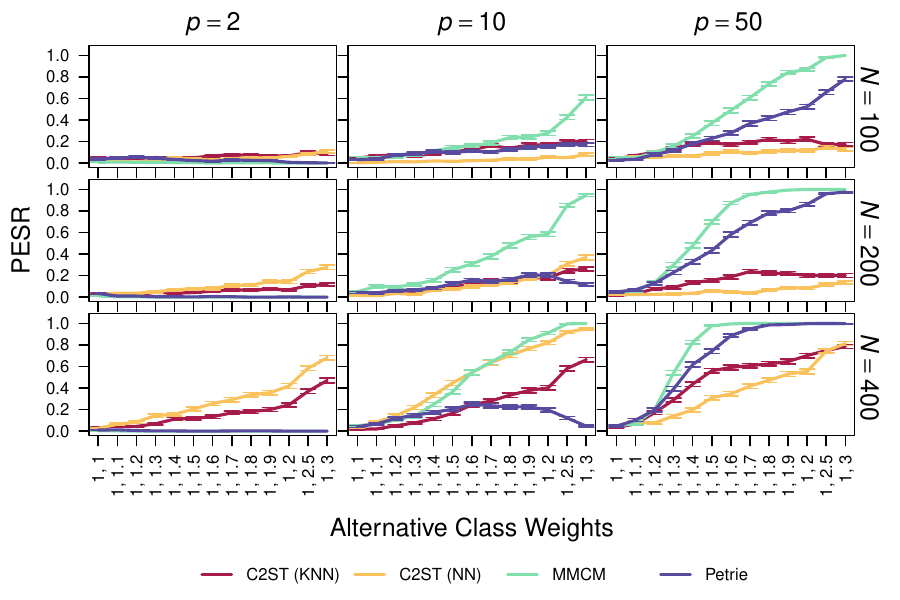}
		\caption{Proportion of extreme simulation repetitions (PESR) for four datasets of unequal sample sizes with binary variables. The class weights, i.e.\ the unnormalized probabilities for the values 0 and 1, in the first dataset, are always set to $(1, 1)$. The class weights on the $x$-axis give the unnormalized probabilities $(1, 1+\delta)$ for each variable in the second dataset. The weights in the third dataset are set to $(1, 1+2\delta)$.  The weights in the fourth dataset are set to $(1, 1+3\delta)$. Error bars indicate Monte Carlo standard errors.}
		\label{fig:pow.cat.multi.bin.unbal.1+1+1+1}
	\end{figure}
	
	\subsection[k = 4, Multinomial Data, Skewed Probability Distribution, Balanced Sample Sizes]{$k = 4$, Multinomial Data, Skewed Probability Distribution, Balanced Sample Sizes}\label{app:add.figs.cat.multi.skew.bal}
	
	\begin{figure}[H]
		\centering
		\includegraphics[width=\linewidth]{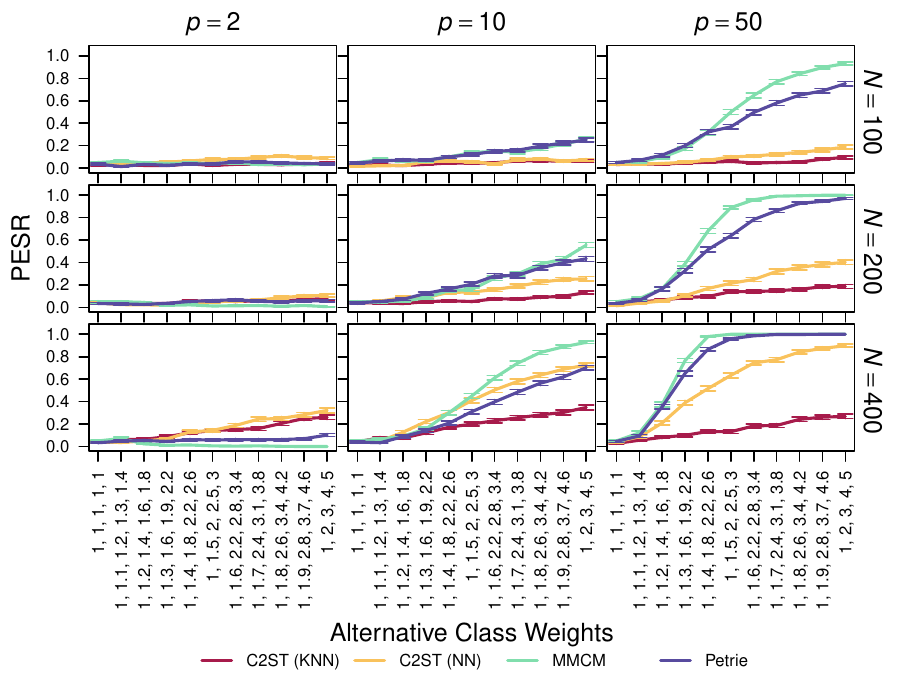}
		\caption{Proportion of extreme simulation repetitions (PESR) for four multinomial datasets of the same sample sizes. The class weights, i.e.\ the unnormalized probabilities for the values $1$ to $5$, in the first to the third dataset, are always set to $(1, 1, 1, 1, 1)$. The class weights on the $x$-axis give the unnormalized probabilities $(1, 1+\delta, 1+2\delta, 1+3\delta, 1+4\delta)$ for each variable in the fourth dataset. Error bars indicate Monte Carlo standard errors.}
		\label{fig:pow.cat.multi.multi.skew.bal.3+1}
	\end{figure}
	
	\begin{figure}[H]
		\centering
		\includegraphics[width=\linewidth]{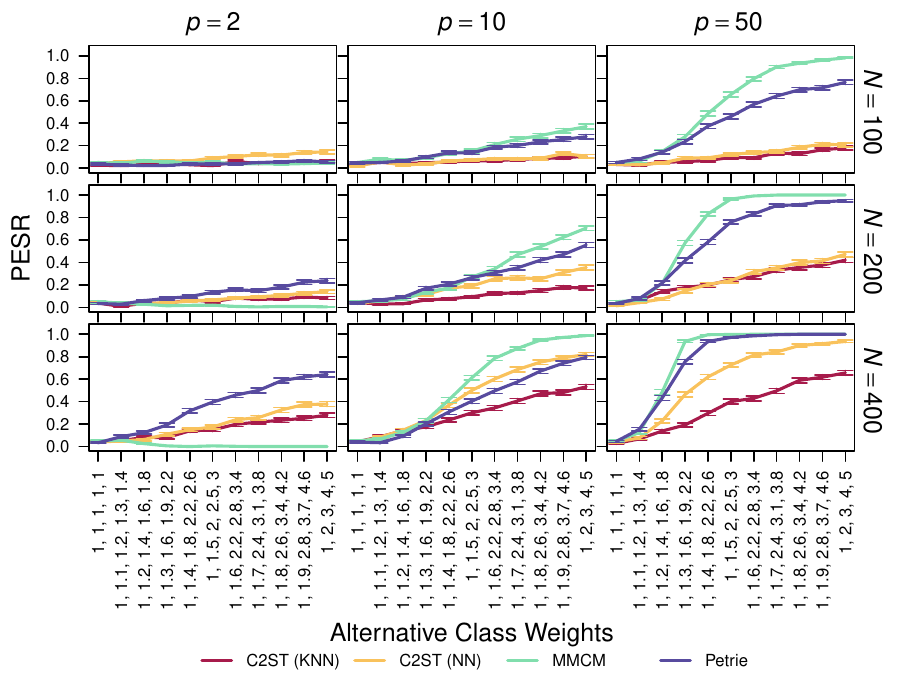}
		\caption{Proportion of extreme simulation repetitions (PESR) for four multinomial datasets of the same sample sizes. The class weights, i.e.\ the unnormalized probabilities for the values $1$ to $5$, in the first and second datasets, are always set to $(1, 1, 1, 1, 1)$. The class weights on the $x$-axis give the unnormalized probabilities $(1, 1+\delta, 1+2\delta, 1+3\delta, 1+4\delta)$ for each variable in the third and fourth datasets. Error bars indicate Monte Carlo standard errors.}
		\label{fig:pow.cat.multi.multi.skew.bal.2+2}
	\end{figure}

	\begin{figure}[H]
		\centering
		\includegraphics[width=\linewidth]{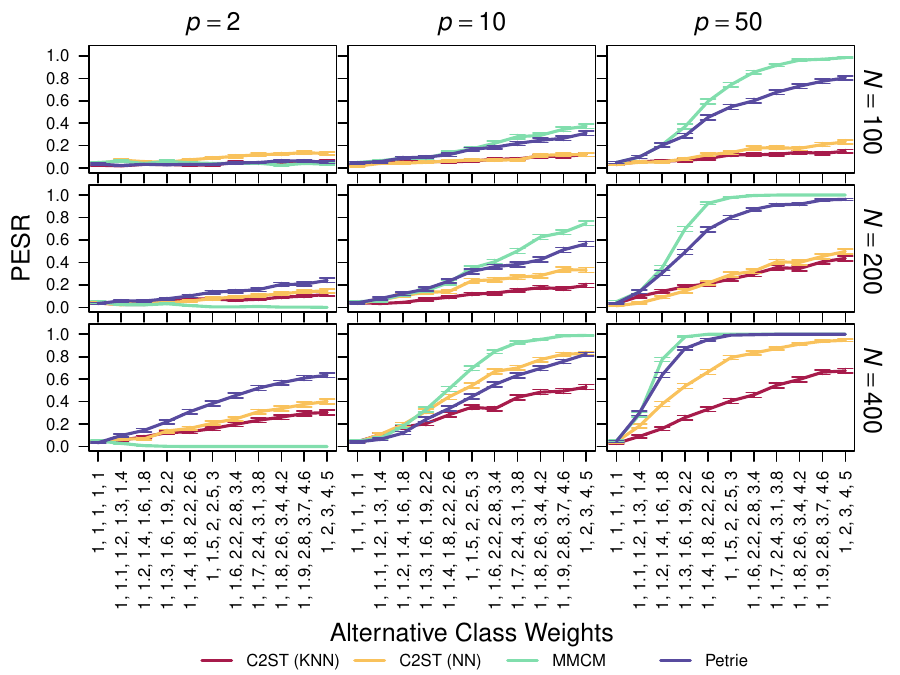}
		\caption{Proportion of extreme simulation repetitions (PESR) for four multinomial datasets of the same sample sizes. The class weights, i.e.\ the unnormalized probabilities for the values $1$ to $5$, in the first and second datasets, are always set to $(1, 1, 1, 1, 1)$. The class weights on the $x$-axis give the unnormalized probabilities $(1, 1+\delta, 1+2\delta, 1+3\delta, 1+4\delta)$ for each variable in the third dataset. The weights in the fourth dataset are given by $(1, 1+(\delta+0.1), 1+2(\delta+0.1), 1+3(\delta+0.1), 1+4(\delta+0.1))$. Error bars indicate Monte Carlo standard errors.}
		\label{fig:pow.cat.multi.multi.skew.bal.2+1+1}
	\end{figure}
	
	\subsection[k = 4, Multinomial Data, Skewed Probability Distribution, Unbalanced Sample Sizes]{$k = 4$, Multinomial Data, Skewed Probability Distribution, Unbalanced Sample Sizes}\label{app:add.figs.cat.multi.skew.unbal}
	
	\begin{figure}[H]
		\centering
		\includegraphics[width=\linewidth]{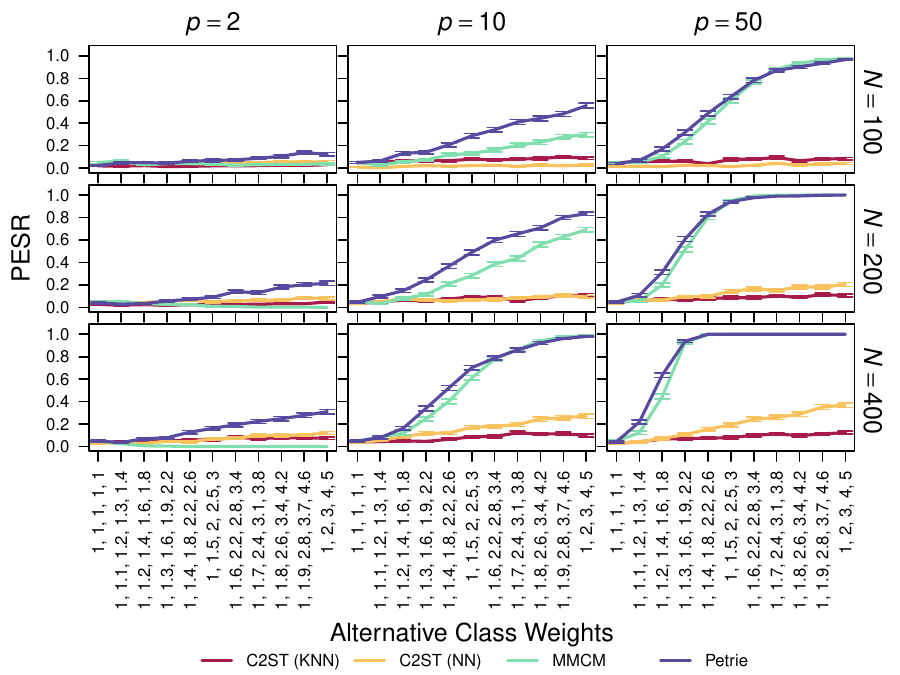}
		\caption{Proportion of extreme simulation repetitions (PESR) for four multinomial datasets of unequal sample sizes. The class weights, i.e.\ the unnormalized probabilities for the values $1$ to $5$, in the first to the third dataset, are always set to $(1, 1, 1, 1, 1)$. The class weights on the $x$-axis give the unnormalized probabilities $(1, 1+\delta, 1+2\delta, 1+3\delta, 1+4\delta)$ for each variable in the fourth dataset. Error bars indicate Monte Carlo standard errors.}
		\label{fig:pow.cat.multi.multi.skew.unbal.3+1}
	\end{figure}
	
	\begin{figure}[H]
		\centering
		\includegraphics[width=\linewidth]{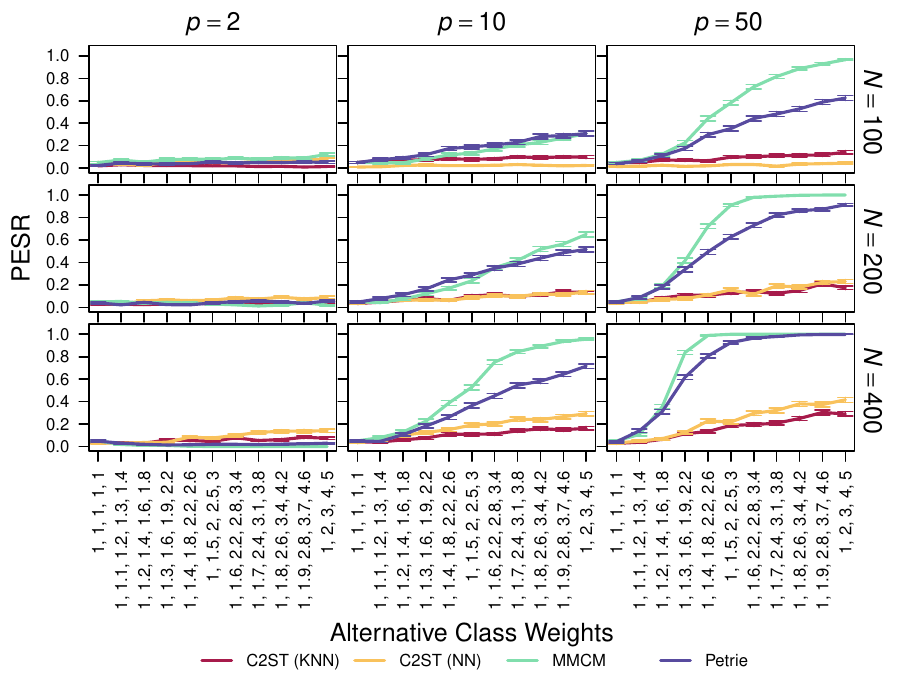}
		\caption{Proportion of extreme simulation repetitions (PESR) for four multinomial datasets of unequal sample sizes. The class weights, i.e.\ the unnormalized probabilities for the values $1$ to $5$, in the first and second datasets, are always set to $(1, 1, 1, 1, 1)$. The class weights on the $x$-axis give the unnormalized probabilities $(1, 1+\delta, 1+2\delta, 1+3\delta, 1+4\delta)$ for each variable in the third and fourth datasets. Error bars indicate Monte Carlo standard errors.}
		\label{fig:pow.cat.multi.multi.skew.unbal.2+2}
	\end{figure}

	\begin{figure}[H]
		\centering
		\includegraphics[width=\linewidth]{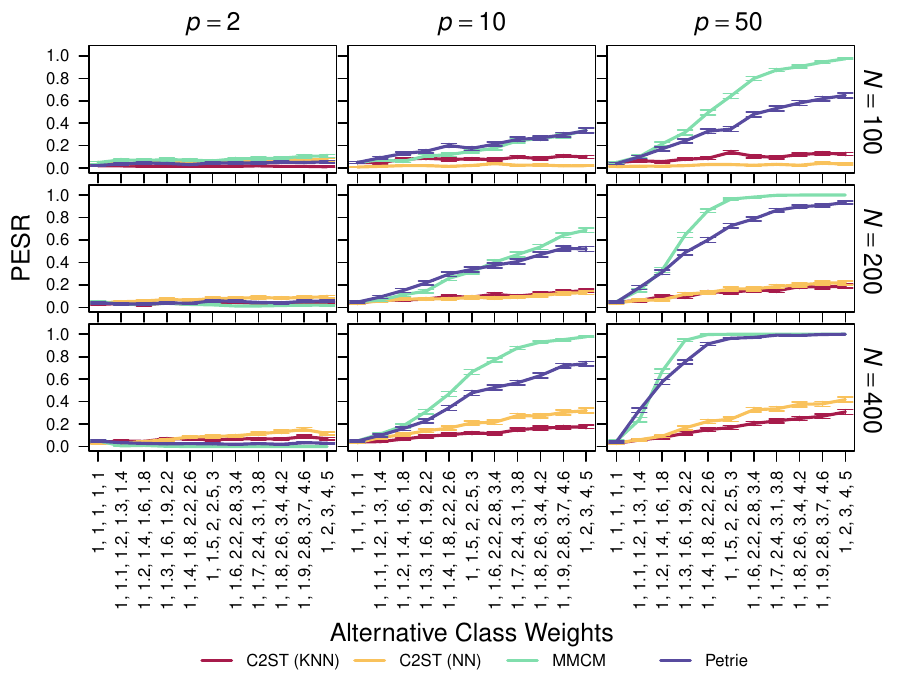}
		\caption{Proportion of extreme simulation repetitions (PESR) for four multinomial datasets of unequal sample sizes. The class weights, i.e.\ the unnormalized probabilities for the values $1$ to $5$, in the first and second datasets, are always set to $(1, 1, 1, 1, 1)$. The class weights on the $x$-axis give the unnormalized probabilities $(1, 1+\delta, 1+2\delta, 1+3\delta, 1+4\delta)$ for each variable in the third dataset. The weights in the fourth dataset are given by $(1, 1+(\delta+0.1), 1+2(\delta+0.1), 1+3(\delta+0.1), 1+4(\delta+0.1))$. Error bars indicate Monte Carlo standard errors.}
		\label{fig:pow.cat.multi.multi.skew.unbal.2+1+1}
	\end{figure}

	\begin{figure}[H]
		\centering
		\includegraphics[width=\linewidth]{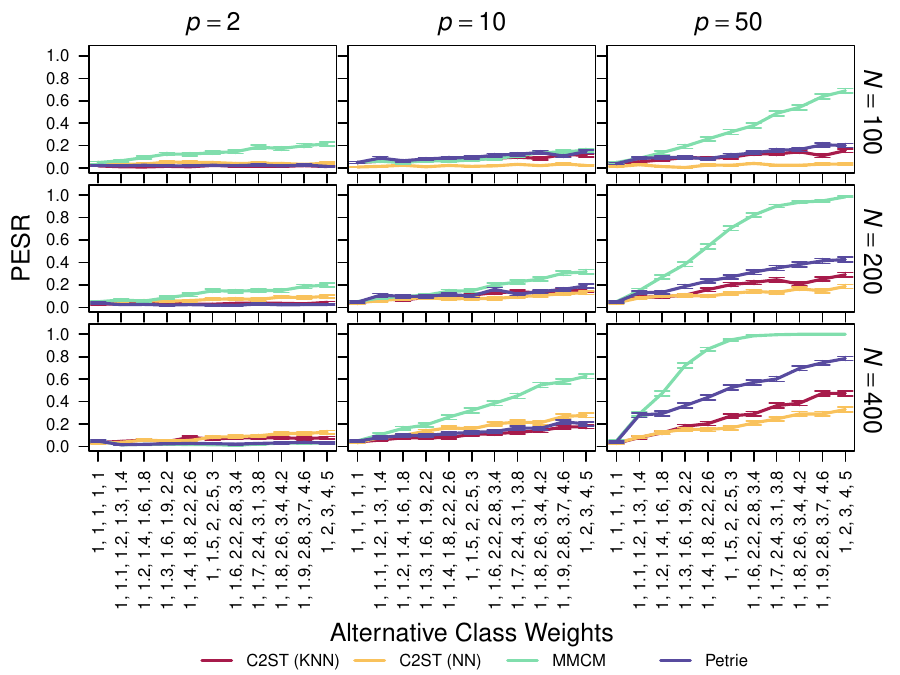}
		\caption{Proportion of extreme simulation repetitions (PESR) for four multinomial datasets of unequal sample sizes. The class weights, i.e.\ the unnormalized probabilities for the values $1$ to $5$, in the first dataset are always set to $(1, 1, 1, 1, 1)$. The class weights on the $x$-axis give the unnormalized probabilities $(1, 1+\delta, 1+2\delta, 1+3\delta, 1+4\delta)$ for each variable in the second dataset. The weights in the third dataset are given by $(1, 1+(\delta+0.1), 1+2(\delta+0.1), 1+3(\delta+0.1), 1+4(\delta+0.1))$. The weights in the fourth dataset are given by $(1, 1+(\delta+0.2), 1+2(\delta+0.2), 1+3(\delta+0.2), 1+4(\delta+0.2))$. Error bars indicate Monte Carlo standard errors.}
		\label{fig:pow.cat.multi.multi.skew.unbal.1+1+1+1}
	\end{figure}
	
	\subsection[k = 4, Multinomial Data, One Class Up, One Class Down, Balanced Sample Sizes]{$k = 4$, Multinomial Data, One Class Up, One Class Down, Balanced Sample Sizes}\label{app:add.figs.cat.multi.1up1down.bal}
	
	\begin{figure}[H]
		\centering
		\includegraphics[width=\linewidth]{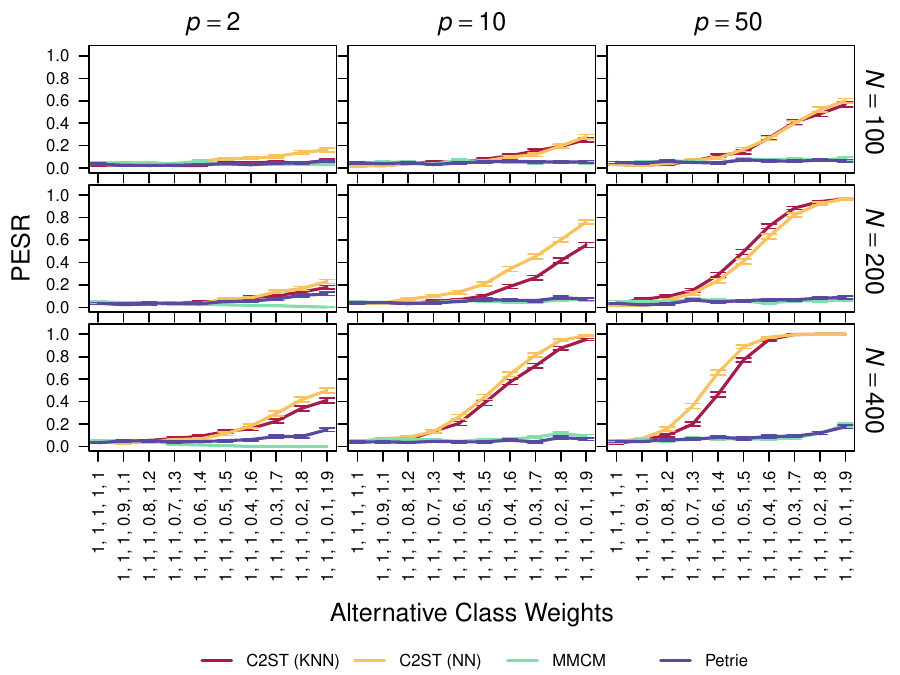}
		\caption{Proportion of extreme simulation repetitions (PESR) for four multinomial datasets of the same sample sizes. The class weights, i.e.\ the unnormalized probabilities for the values $1$ to $5$, in the first to the third dataset, are always set to $(1, 1, 1, 1, 1)$. The class weights on the $x$-axis give the unnormalized probabilities $(1, 1, 1, 1+\delta, 1-\delta)$ in the fourth dataset. Error bars indicate Monte Carlo standard errors.}
		\label{fig:pow.cat.multi.multi.1u1d.bal.3+1}
	\end{figure}
	
	\begin{figure}[H]
		\centering
		\includegraphics[width=\linewidth]{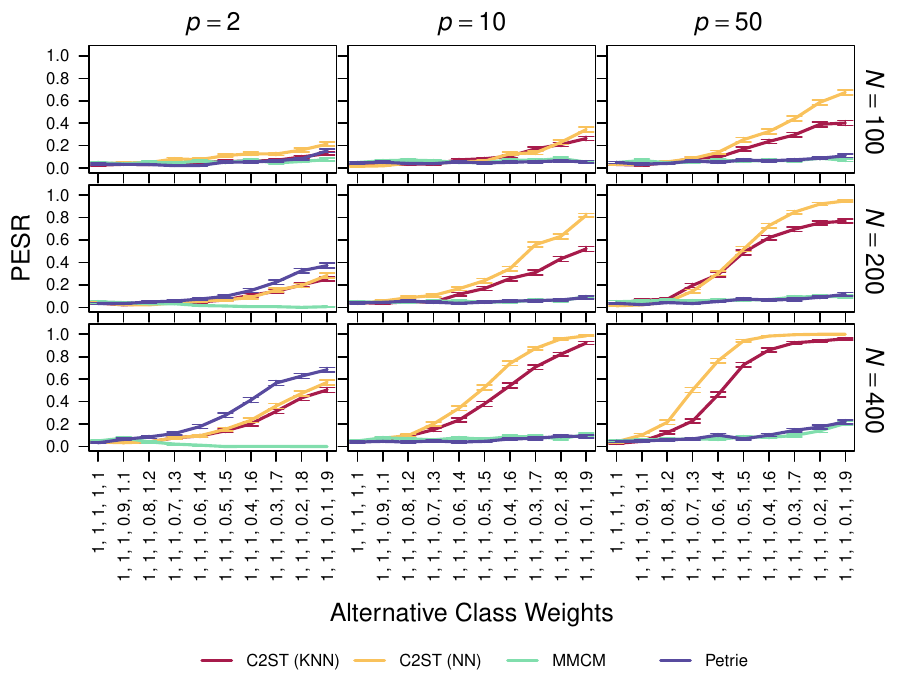}
		\caption{Proportion of extreme simulation repetitions (PESR) for four multinomial datasets of the same sample sizes. The class weights, i.e.\ the unnormalized probabilities for the values $1$ to $5$, in the first and second datasets, are always set to $(1, 1, 1, 1, 1)$. The class weights on the $x$-axis give the unnormalized probabilities $(1, 1, 1, 1+\delta, 1-\delta)$ for each variable in the third and fourth datasets. Error bars indicate Monte Carlo standard errors.}
		\label{fig:pow.cat.multi.multi.1u1d.bal.2+2}
	\end{figure}

	\begin{figure}[H]
		\centering
		\includegraphics[width=\linewidth]{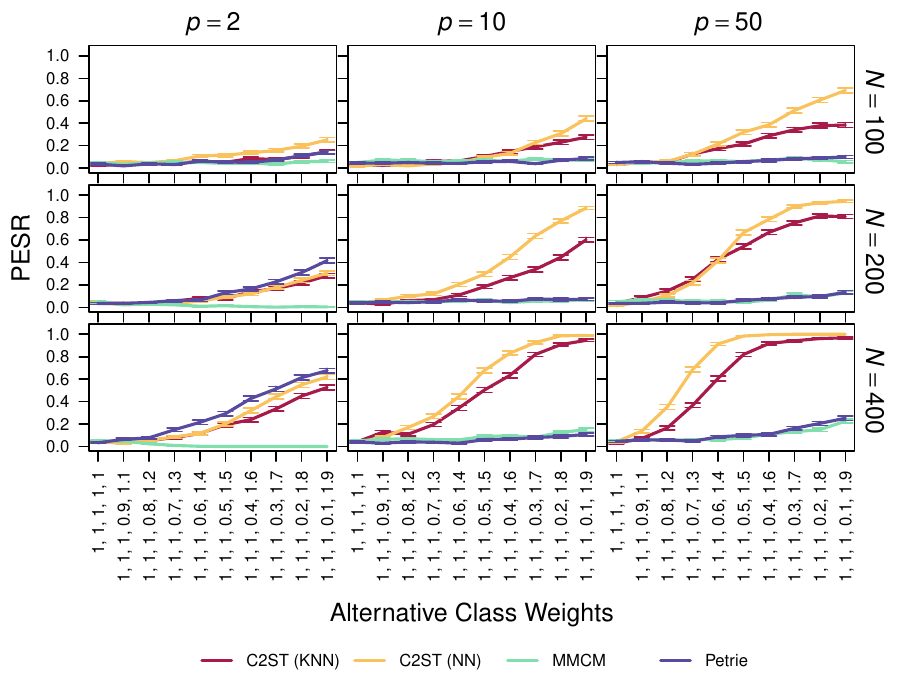}
		\caption{Proportion of extreme simulation repetitions (PESR) for four multinomial datasets of the same sample sizes. The class weights, i.e.\ the unnormalized probabilities for the values $1$ to $5$, in the first and second datasets, are always set to $(1, 1, 1, 1, 1)$. The class weights on the $x$-axis give the unnormalized probabilities  $(1, 1, 1, 1+\delta, 1-\delta)$  for each variable in the third dataset. The weights in the fourth dataset are given by $(1, 1, 1, 1+\delta+0.1, 1-\delta-0.1)$. Error bars indicate Monte Carlo standard errors.}
		\label{fig:pow.cat.multi.multi.1u1d.bal.2+1+1}
	\end{figure}
	
	\subsection[k = 4, Multinomial Data, One Class Up, One Class Down, Unbalanced Sample Sizes]{$k = 4$, Multinomial Data, One Class Up, One Class Down, Unbalanced Sample Sizes}\label{app:add.figs.cat.multi.1up1down.unbal}
	
	\begin{figure}[H]
		\centering
		\includegraphics[width=\linewidth]{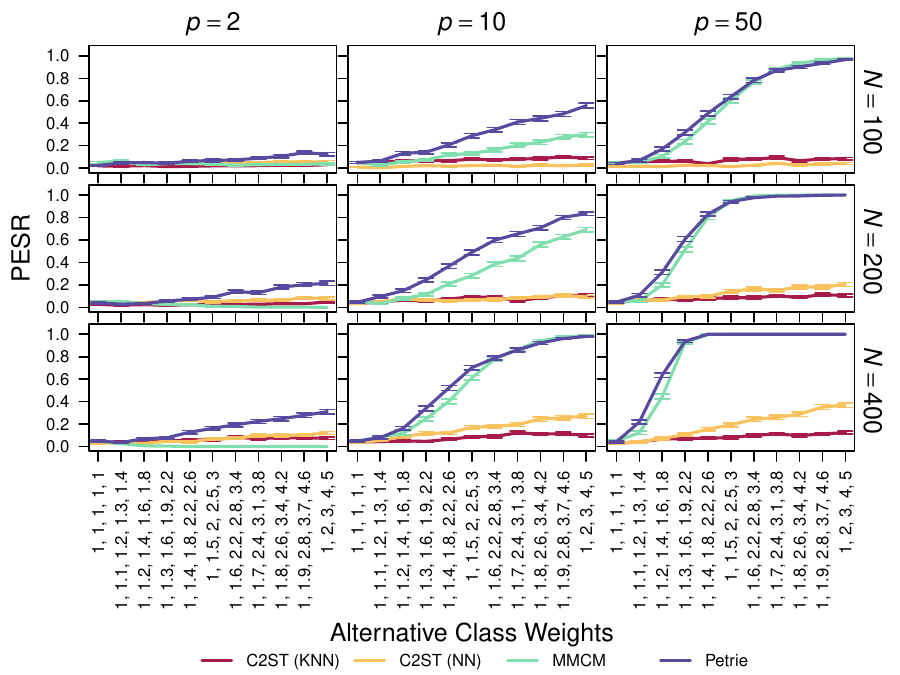}
		\caption{Proportion of extreme simulation repetitions (PESR) for four multinomial datasets of unequal sample sizes. The class weights, i.e.\ the unnormalized probabilities for the values $1$ to $5$, in the first to the third dataset, are always set to $(1, 1, 1, 1, 1)$. The class weights on the $x$-axis give the unnormalized probabilities $(1, 1, 1, 1+\delta, 1-\delta)$ in the fourth dataset. Error bars indicate Monte Carlo standard errors.}
		\label{fig:pow.cat.multi.multi.1u1d.unbal.3+1}
	\end{figure}
	
	\begin{figure}[H]
		\centering
		\includegraphics[width=\linewidth]{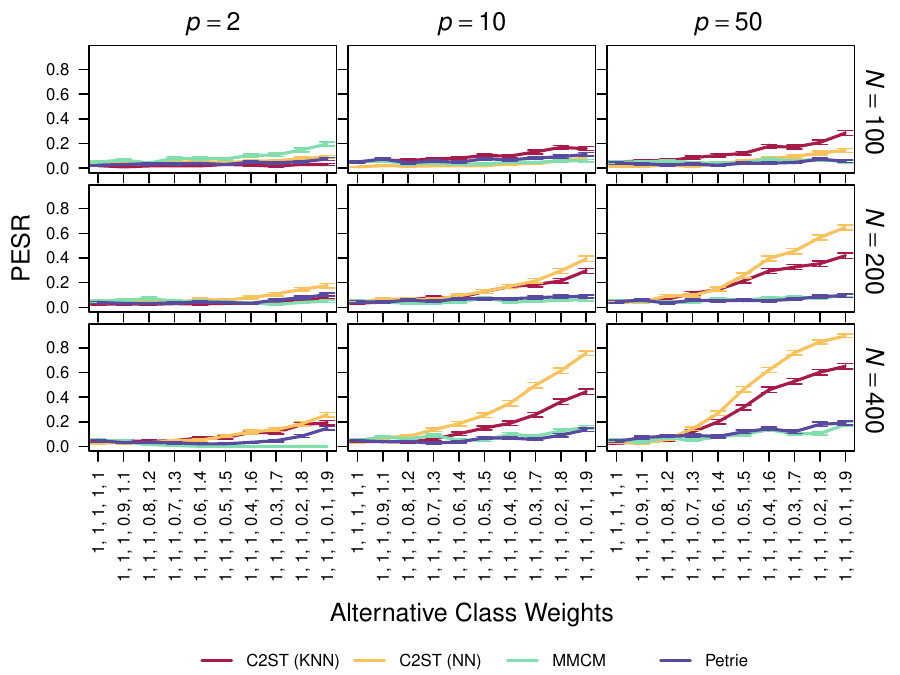}
		\caption{Proportion of extreme simulation repetitions (PESR) for four multinomial datasets of unequal sample sizes. The class weights, i.e.\ the unnormalized probabilities for the values $1$ to $5$, in the first and second datasets, are always set to $(1, 1, 1, 1, 1)$. The class weights on the $x$-axis give the unnormalized probabilities $(1, 1, 1, 1+\delta, 1-\delta)$ for each variable in the third and fourth datasets. Error bars indicate Monte Carlo standard errors.}
		\label{fig:pow.cat.multi.multi.1u1d.unbal.2+2}
	\end{figure}

	\begin{figure}[H]
		\centering
		\includegraphics[width=\linewidth]{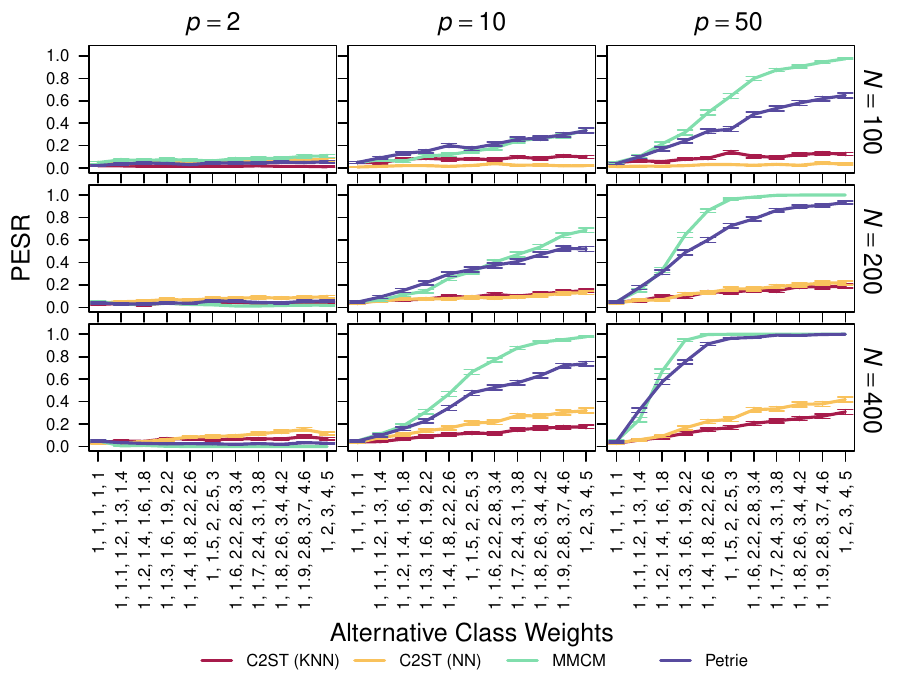}
		\caption{Proportion of extreme simulation repetitions (PESR) for four multinomial datasets of unequal sample sizes. The class weights, i.e.\ the unnormalized probabilities for the values $1$ to $5$, in the first and second datasets, are always set to $(1, 1, 1, 1, 1)$. The class weights on the $x$-axis give the unnormalized probabilities  $(1, 1, 1, 1+\delta, 1-\delta)$  for each variable in the third dataset. The weights in the fourth dataset are given by $(1, 1, 1, 1+\delta+0.1, 1-\delta-0.1)$. Error bars indicate Monte Carlo standard errors.}
		\label{fig:pow.cat.multi.multi.1u1d.unbal.2+1+1}
	\end{figure}

	\begin{figure}[H]
		\centering
		\includegraphics[width=\linewidth]{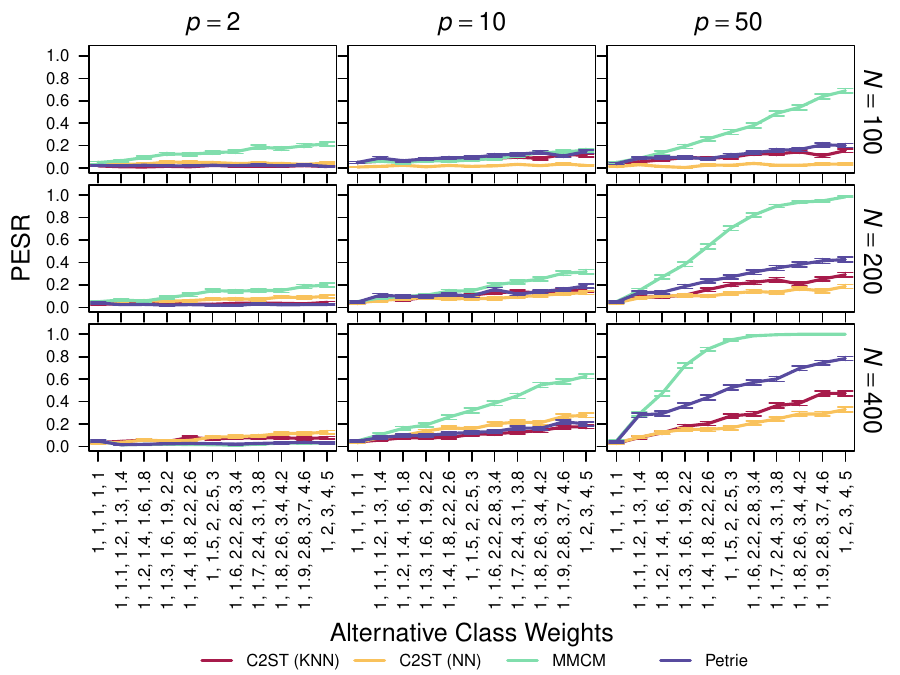}
		\caption{Proportion of extreme simulation repetitions (PESR) for four multinomial datasets of unequal sample sizes. The class weights, i.e.\ the unnormalized probabilities for the values $1$ to $5$, in the first and second datasets, are always set to $(1, 1, 1, 1, 1)$. The class weights on the $x$-axis give the unnormalized probabilities  $(1, 1, 1, 1+\delta, 1-\delta)$  for each variable in the third dataset. The weights in the third dataset are given by $(1, 1, 1, 1+\delta+0.1, 1-\delta-0.1)$, and the weights for the fourth dataset are given by $(1, 1, 1, 1+\delta+0.2, 1-\delta-0.2)$. Error bars indicate Monte Carlo standard errors.}
		\label{fig:pow.cat.multi.multi.1u1d.unbal.1+1+1+1}
	\end{figure}
	
	\subsection{Clustering of Methods}\label{sec:clustering}
	In the following, the aim is to identify groups of methods that perform similarly across deviations. 
	To find such groups, the proportions of extreme simulation repetitions (PESR) values are clustered using hierarchical clustering with complete linkage and Euclidean distances. 
	For clustering, the PESR values per combination of a deviation and grouping are considered as variables. 
	The combinations of the method, $N$, $p$, and sample size balance are considered as observations. 
	The idea behind this division is that the deviation and grouping would be unknown in a real-world dataset comparison, while the dimensions of the datasets and the chosen method are known. 
	Both observations (rows) and variables (columns) are clustered to see which methods act similarly for different $N$, $p$, and balance combinations across different alternatives. 
	The clustering of the deviations thereby facilitates identifying groups of deviations for which the identified groups of methods perform particularly well or poorly. 
	
	The PESR value clustering is visualized using heatmaps. 
	The clustering is shown in dendrograms. 
	The dendrograms are ordered using an optimal leaf ordering that minimizes the sum of the distances along the path connecting the leaves in the given order \textcite{bar-joseph_fast_2001}. 
	This optimal ordering is used since it is expected to give a sensible order of the deviations in increasing strength. 
	
	\subsubsection{Two-sample Setting}
	For $k = 2$, the clustering is performed for each $N$ and $p$ combination separately since the large number of methods makes it impossible to show all methods for all $N$ and $p$ simultaneously, and the results from the PESR curve comparisons suggest that using the dataset dimensions represents a sensible stratification.
	The resulting clusterings are discussed in the following, first for binary and afterward for multinomial data. 
	All methods are considered; no pre-selection of methods is used here. 
	
\begin{figure}[!tb]
		\centering
		\includegraphics[width=\linewidth]{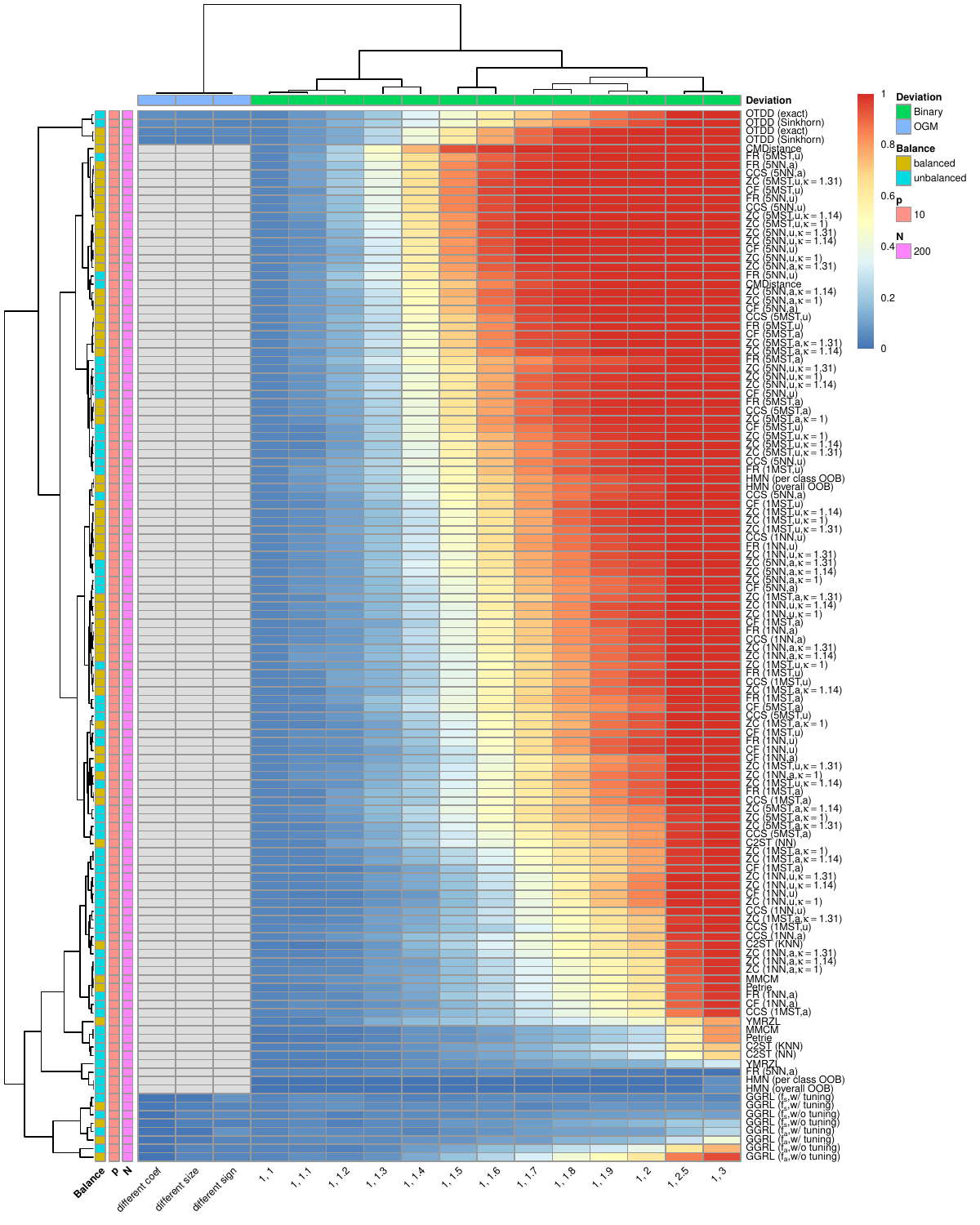}
		\caption{Clustering of PESR values per deviation ($x$-axis) and per method and sample size balance ($y$-axis) for two binary datasets with $N = 200$ and $p = 10$. The labels on the $x$-axis give the weight vector (unnormalized class probabilities) of the first deviating dataset. High PESR values (red) correspond to good performance.}
		\label{fig:heatmap.cat.no.y.bin.N200.p10}
	\end{figure}

\begin{figure}[H]
		\centering
		\includegraphics[width=\linewidth]{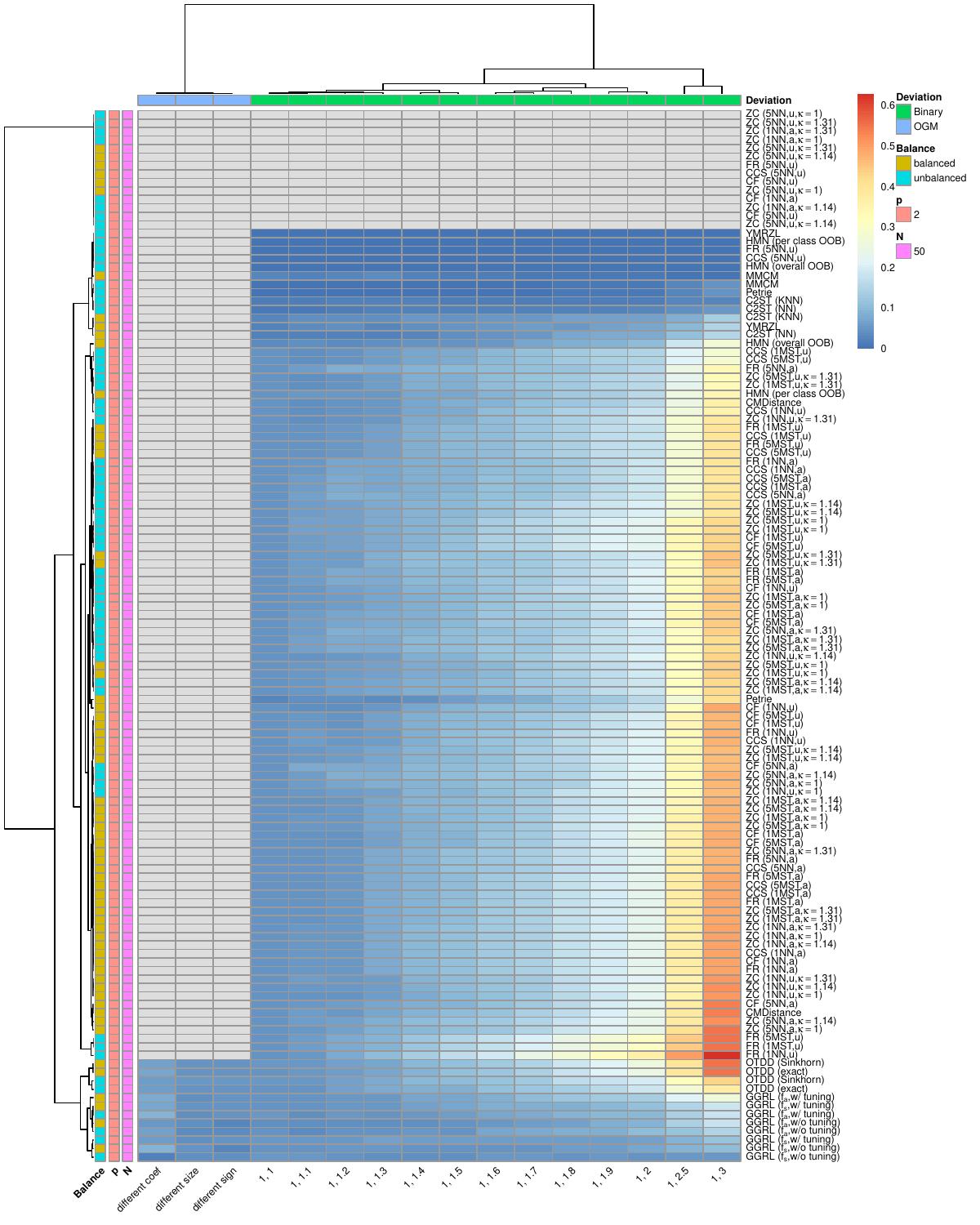}
		\caption{Clustering of PESR values per deviation ($x$-axis) and per method and sample size balance ($y$-axis) for two binary datasets with $N =50$ and $p = 2$. The values on the $x$-axis give the weight vector (unnormalized class probabilities) of the first deviating dataset. High PESR values (red) correspond to good performance.}
		\label{fig:heatmap.cat.no.y.bin.N50.p2}
	\end{figure}

\begin{figure}[H]
		\centering
		\includegraphics[width=\linewidth]{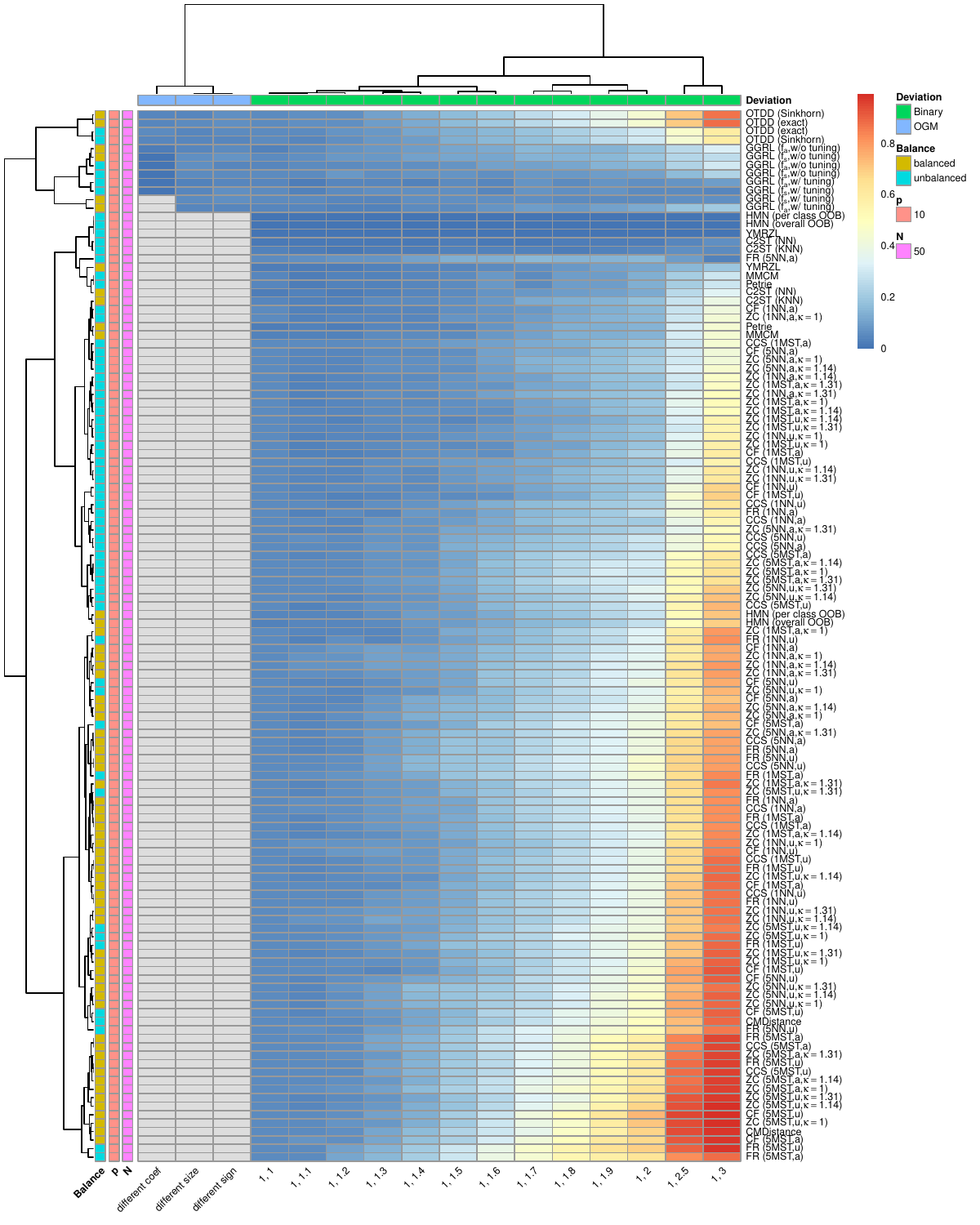}
		\caption{Clustering of PESR values per deviation ($x$-axis) and per method and sample size balance ($y$-axis) for two binary datasets with $N =50$ and $p = 10$. The values on the $x$-axis give the weight vector (unnormalized class probabilities) of the first deviating dataset. High PESR values (red) correspond to good performance.}
		\label{fig:heatmap.cat.no.y.bin.N50.p10}
	\end{figure}

\begin{figure}[H]
		\centering
		\includegraphics[width=\linewidth]{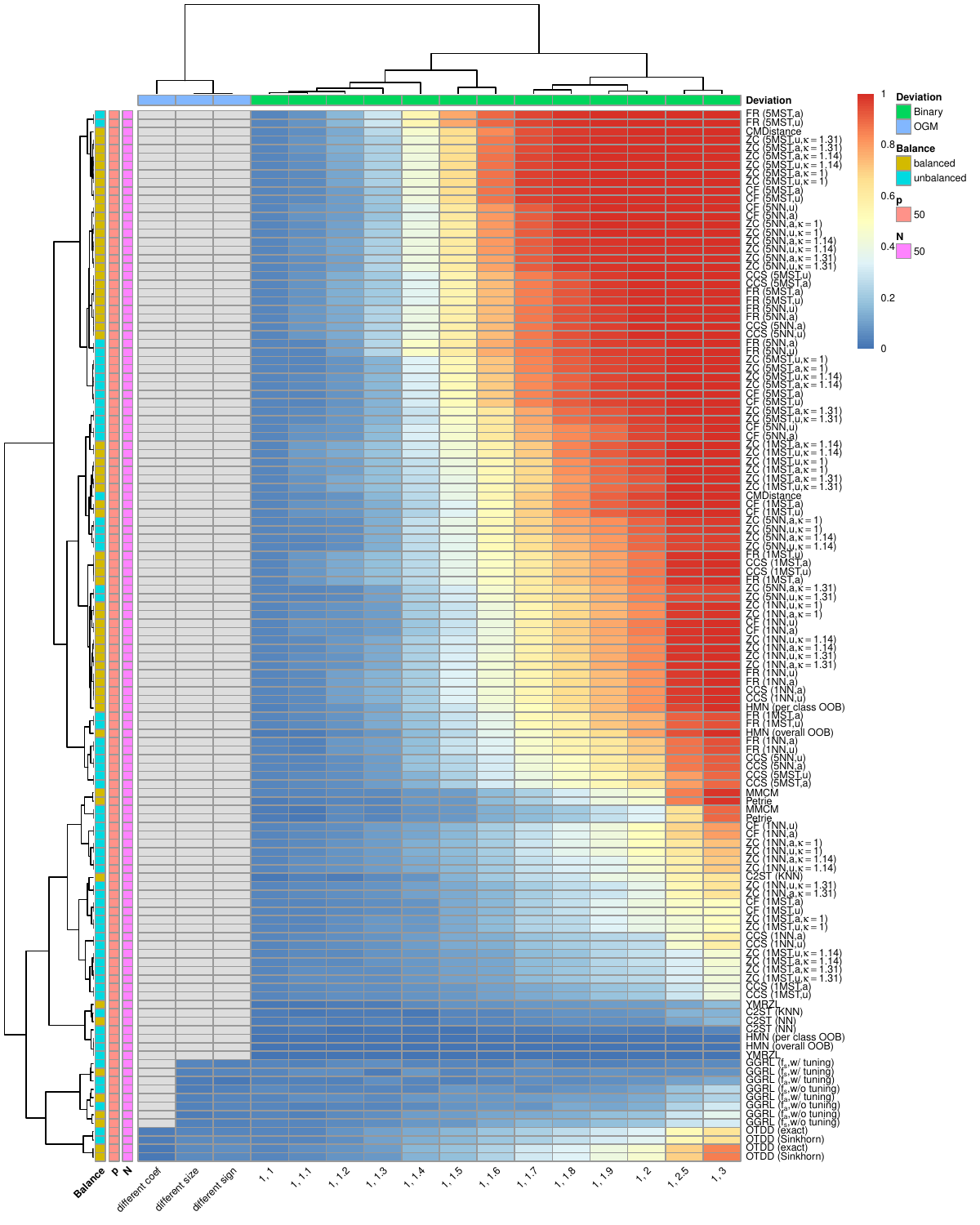}
		\caption{Clustering of PESR values per deviation ($x$-axis) and per method and sample size balance ($y$-axis) for two binary datasets with $N =50$ and $p = 50$. The values on the $x$-axis give the weight vector (unnormalized class probabilities) of the first deviating dataset. High PESR values (red) correspond to good performance.}
		\label{fig:heatmap.cat.no.y.bin.N50.p50}
	\end{figure}

\begin{figure}[H]
		\centering
		\includegraphics[width=\linewidth]{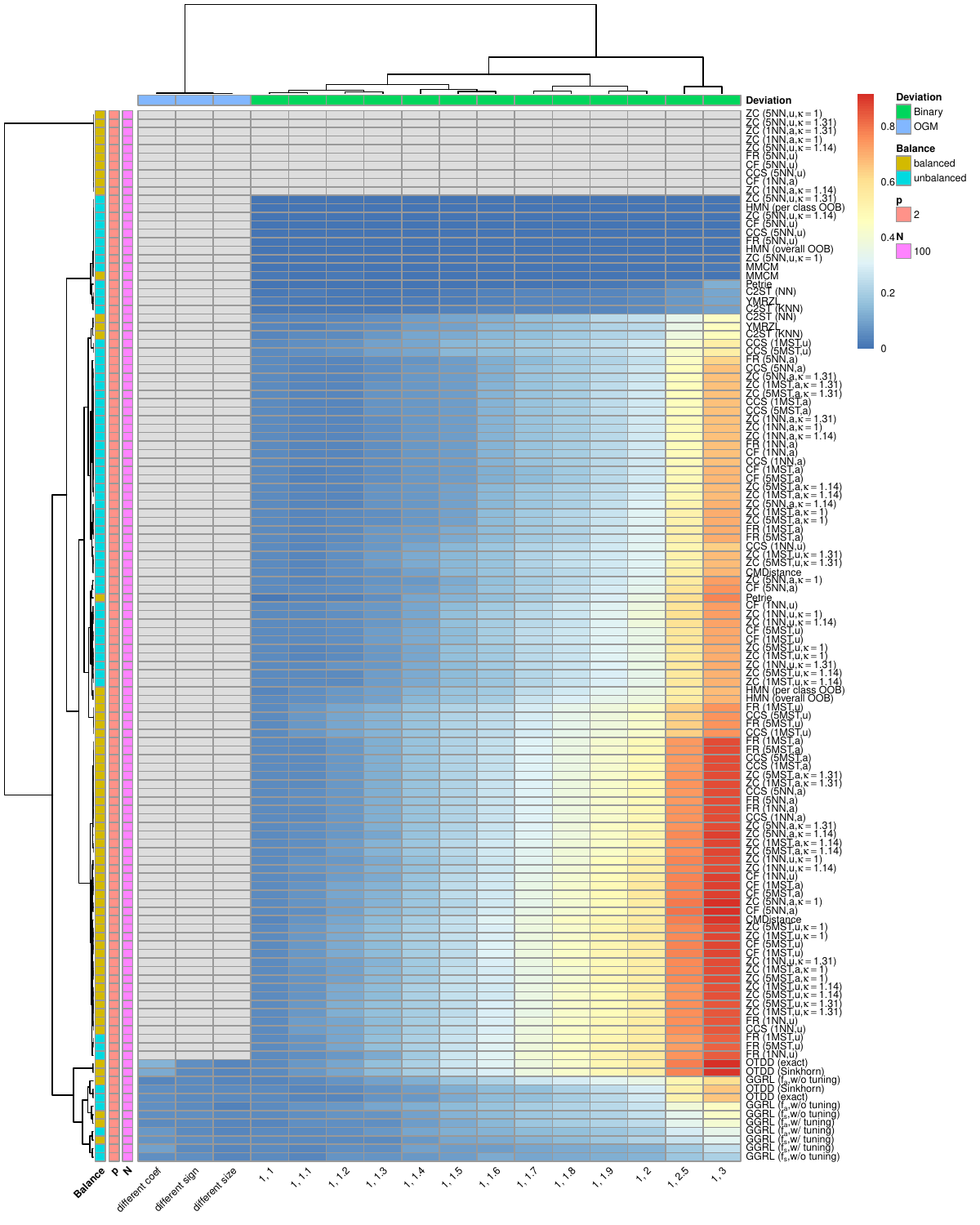}
		\caption{Clustering of PESR values per deviation ($x$-axis) and per method and sample size balance ($y$-axis) for two binary datasets with $N =100$ and $p = 2$. The values on the $x$-axis give the weight vector (unnormalized class probabilities) of the first deviating dataset. High PESR values (red) correspond to good performance.}
		\label{fig:heatmap.cat.no.y.bin.N100.p2}
	\end{figure}

\begin{figure}[H]
		\centering
		\includegraphics[width=\linewidth]{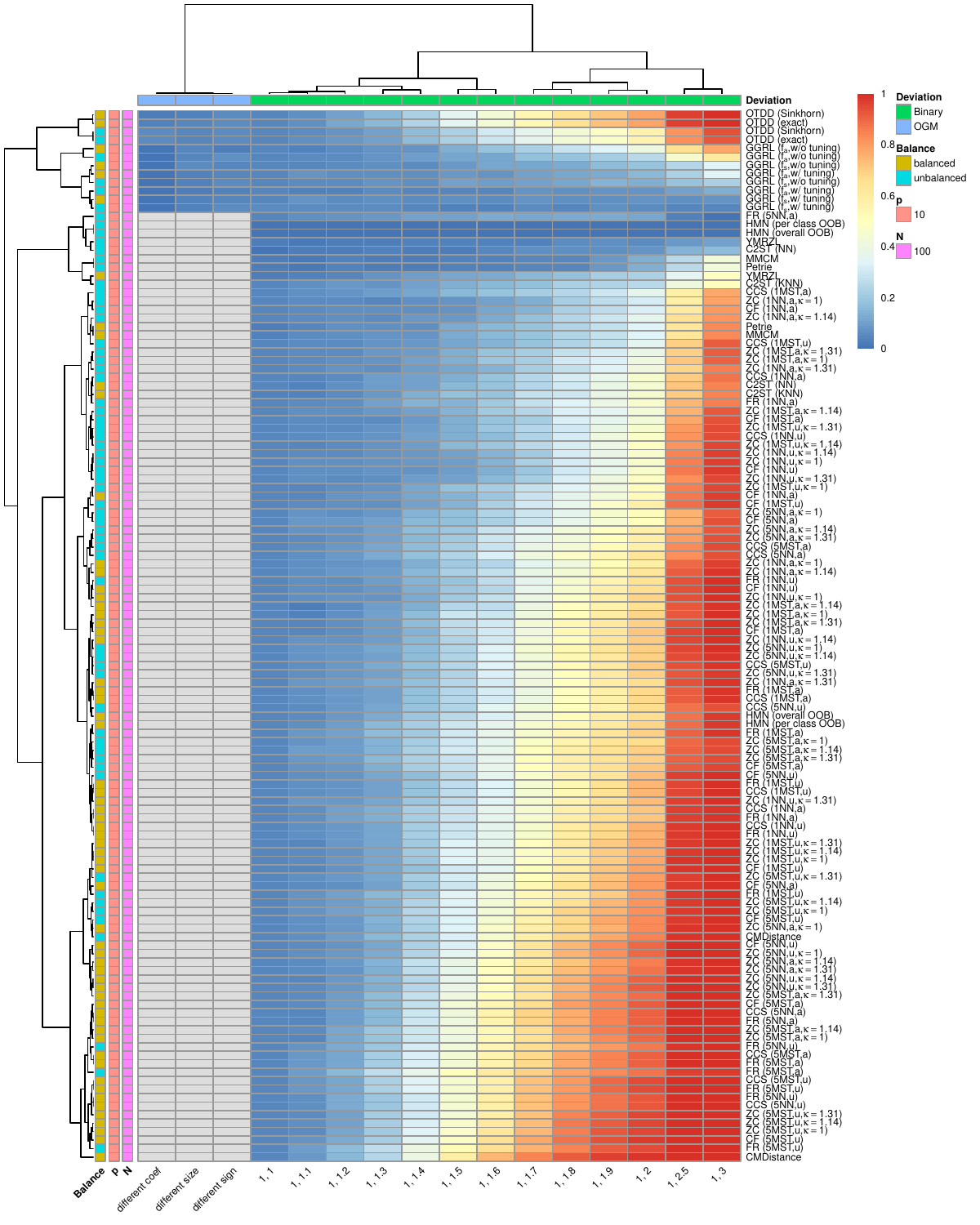}
		\caption{Clustering of PESR values per deviation ($x$-axis) and per method and sample size balance ($y$-axis) for two binary datasets with $N =100$ and $p = 10$. The values on the $x$-axis give the weight vector (unnormalized class probabilities) of the first deviating dataset. High PESR values (red) correspond to good performance.}
		\label{fig:heatmap.cat.no.y.bin.N100.p10}
	\end{figure}

\begin{figure}[H]
		\centering
		\includegraphics[width=\linewidth]{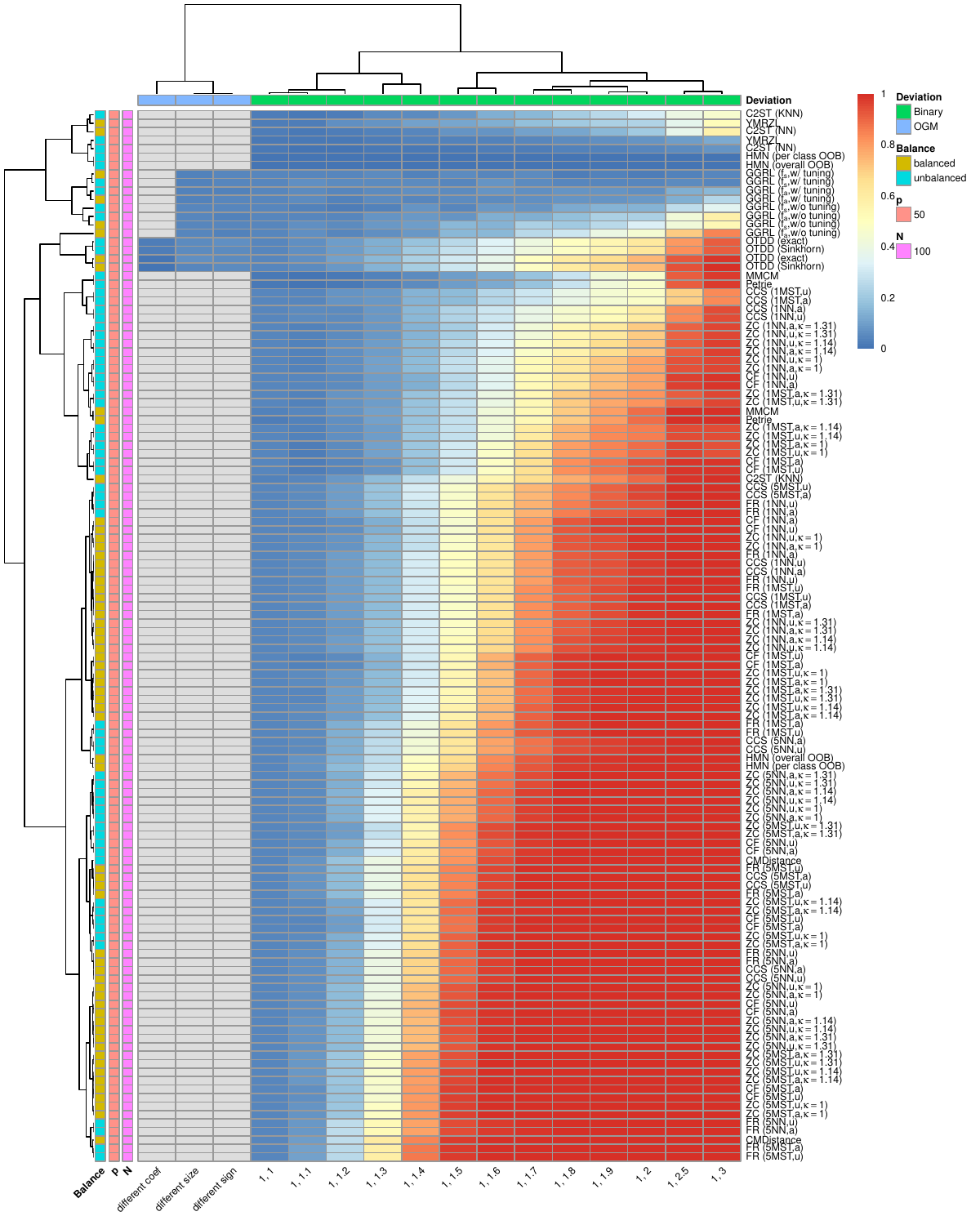}
		\caption{Clustering of PESR values per deviation ($x$-axis) and per method and sample size balance ($y$-axis) for two binary datasets with $N =100$ and $p = 50$. The values on the $x$-axis give the weight vector (unnormalized class probabilities) of the first deviating dataset. High PESR values (red) correspond to good performance.}
		\label{fig:heatmap.cat.no.y.bin.N100.p50}
	\end{figure}

\begin{figure}[H]
		\centering
		\includegraphics[width=\linewidth]{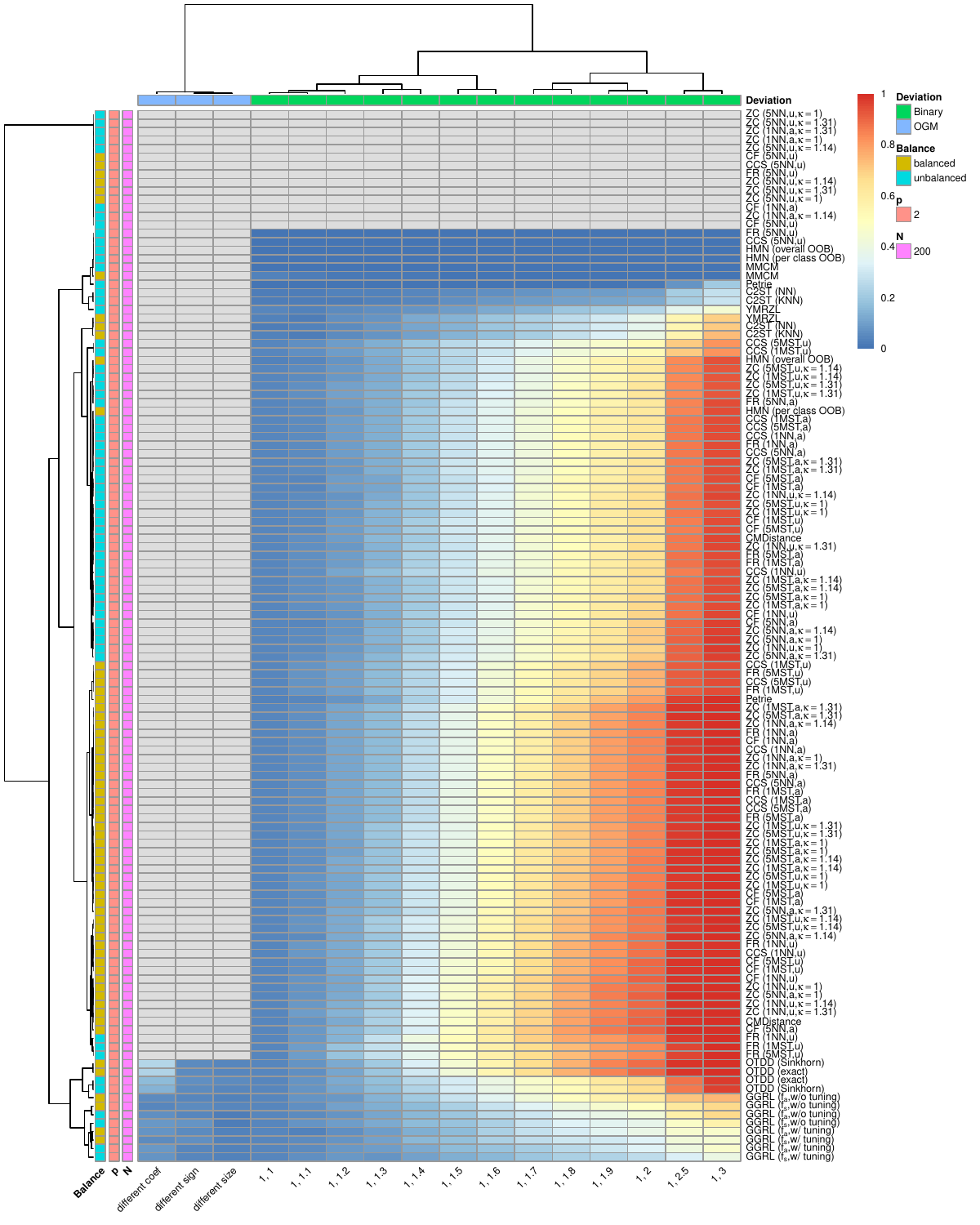}
		\caption{Clustering of PESR values per deviation ($x$-axis) and per method and sample size balance ($y$-axis) for two binary datasets with $N =200$ and $p = 2$. The values on the $x$-axis give the weight vector (unnormalized class probabilities) of the first deviating dataset. High PESR values (red) correspond to good performance.}
		\label{fig:heatmap.cat.no.y.bin.N200.p2}
	\end{figure}

\begin{figure}[H]
		\centering
		\includegraphics[width=\linewidth]{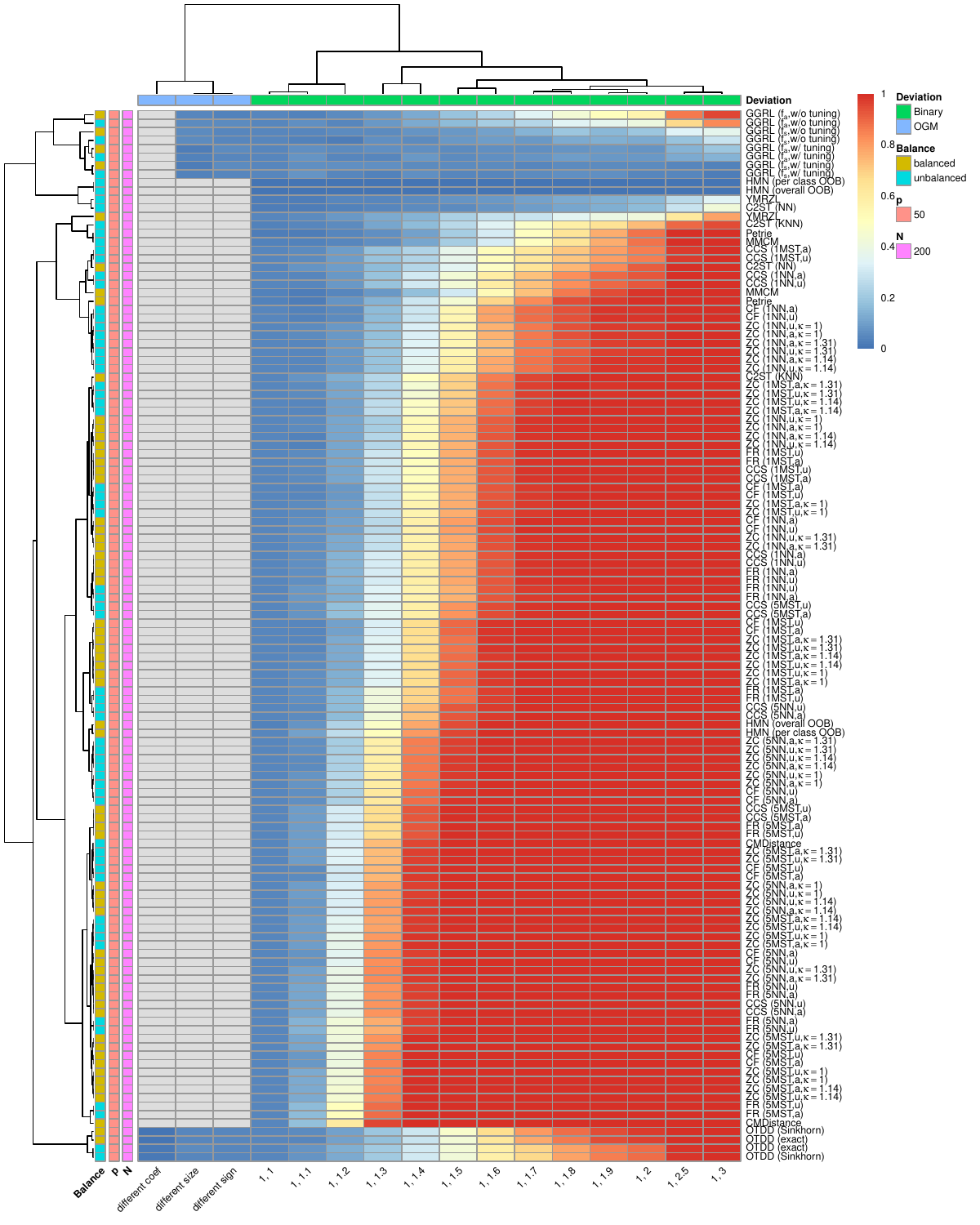}
		\caption{Clustering of PESR values per deviation ($x$-axis) and per method and sample size balance ($y$-axis) for two binary datasets with $N =200$ and $p = 50$. The values on the $x$-axis give the weight vector (unnormalized class probabilities) of the first deviating dataset. High PESR values (red) correspond to good performance.}
		\label{fig:heatmap.cat.no.y.bin.N200.p50}
	\end{figure}

\begin{figure}[H]
		\centering
		\includegraphics[width=\linewidth]{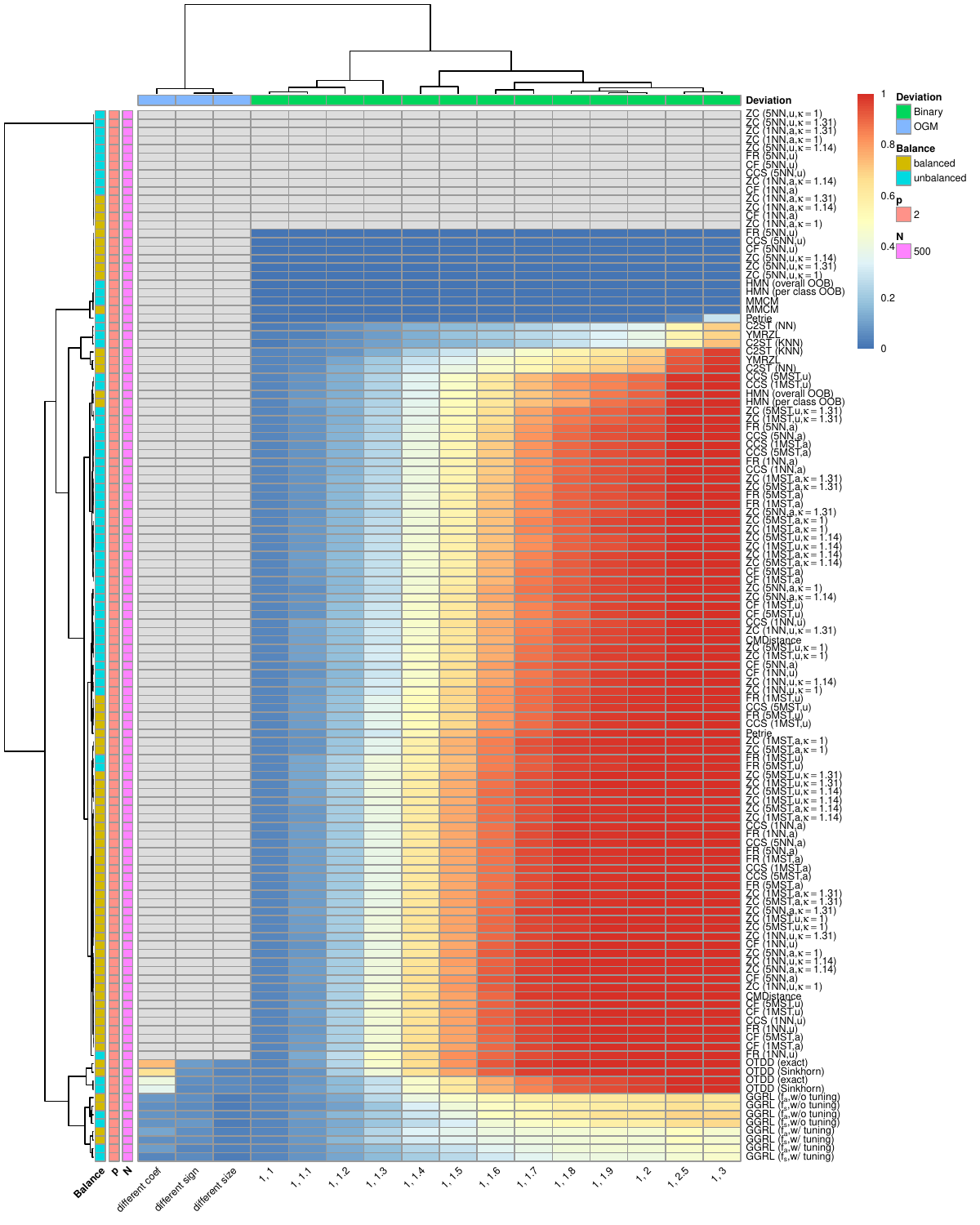}
		\caption{Clustering of PESR values per deviation ($x$-axis) and per method and sample size balance ($y$-axis) for two binary datasets with $N =500$ and $p = 2$. The values on the $x$-axis give the weight vector (unnormalized class probabilities) of the first deviating dataset. High PESR values (red) correspond to good performance.}
		\label{fig:heatmap.cat.no.y.bin.N500.p2}
	\end{figure}

\begin{figure}[H]
		\centering
		\includegraphics[width=\linewidth]{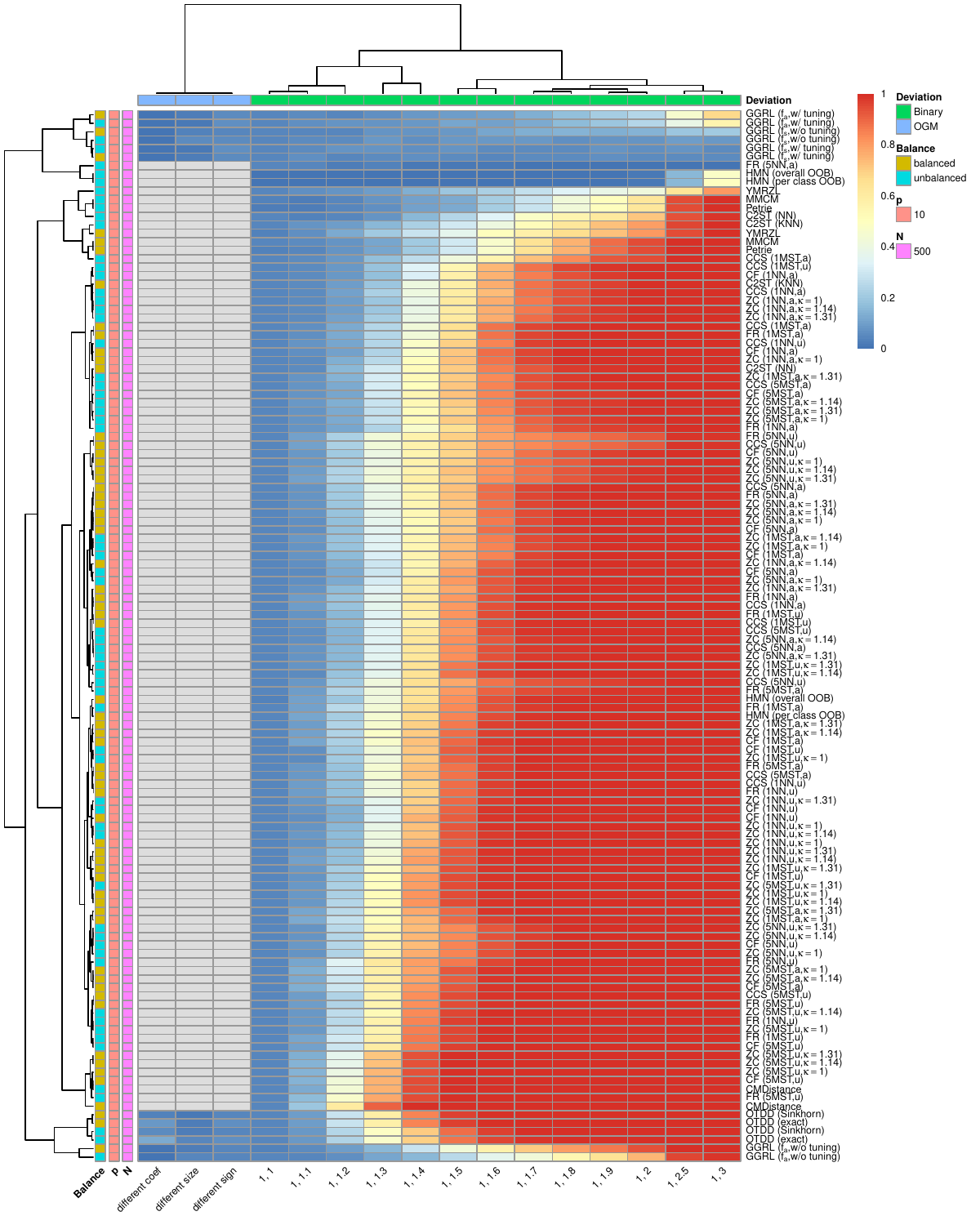}
		\caption{Clustering of PESR values per deviation ($x$-axis) and per method and sample size balance ($y$-axis) for two binary datasets with $N =500$ and $p = 10$. The values on the $x$-axis give the weight vector (unnormalized class probabilities) of the first deviating dataset. High PESR values (red) correspond to good performance.}
		\label{fig:heatmap.cat.no.y.bin.N500.p10}
	\end{figure}

\begin{figure}[H]
		\centering
		\includegraphics[width=\linewidth]{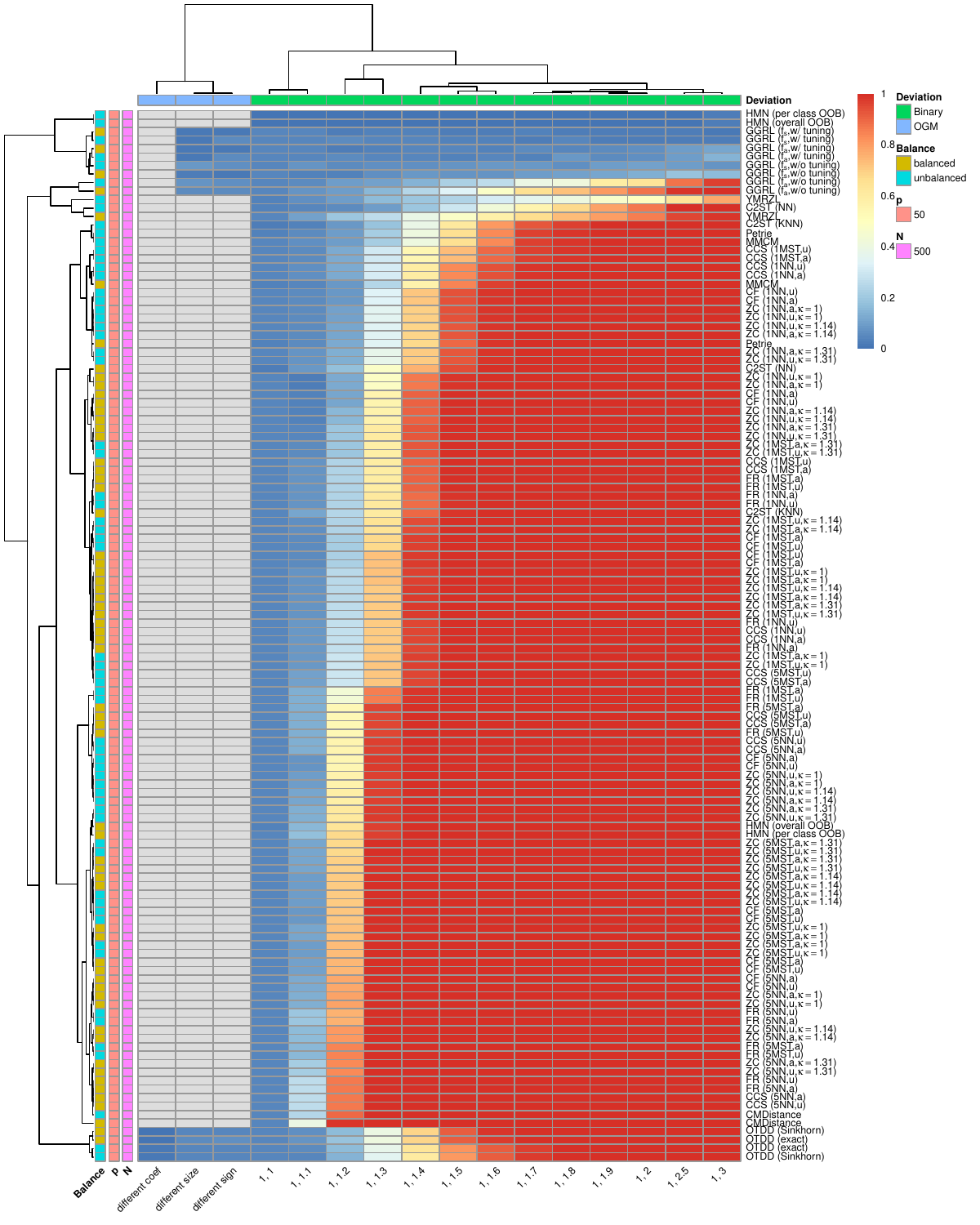}
		\caption{Clustering of PESR values per deviation ($x$-axis) and per method and sample size balance ($y$-axis) for two binary datasets with $N =500$ and $p = 50$. The values on the $x$-axis give the weight vector (unnormalized class probabilities) of the first deviating dataset. High PESR values (red) correspond to good performance.}
		\label{fig:heatmap.cat.no.y.bin.N500.p50}
	\end{figure}

\begin{figure}[H]
		\centering
		\includegraphics[width=\linewidth]{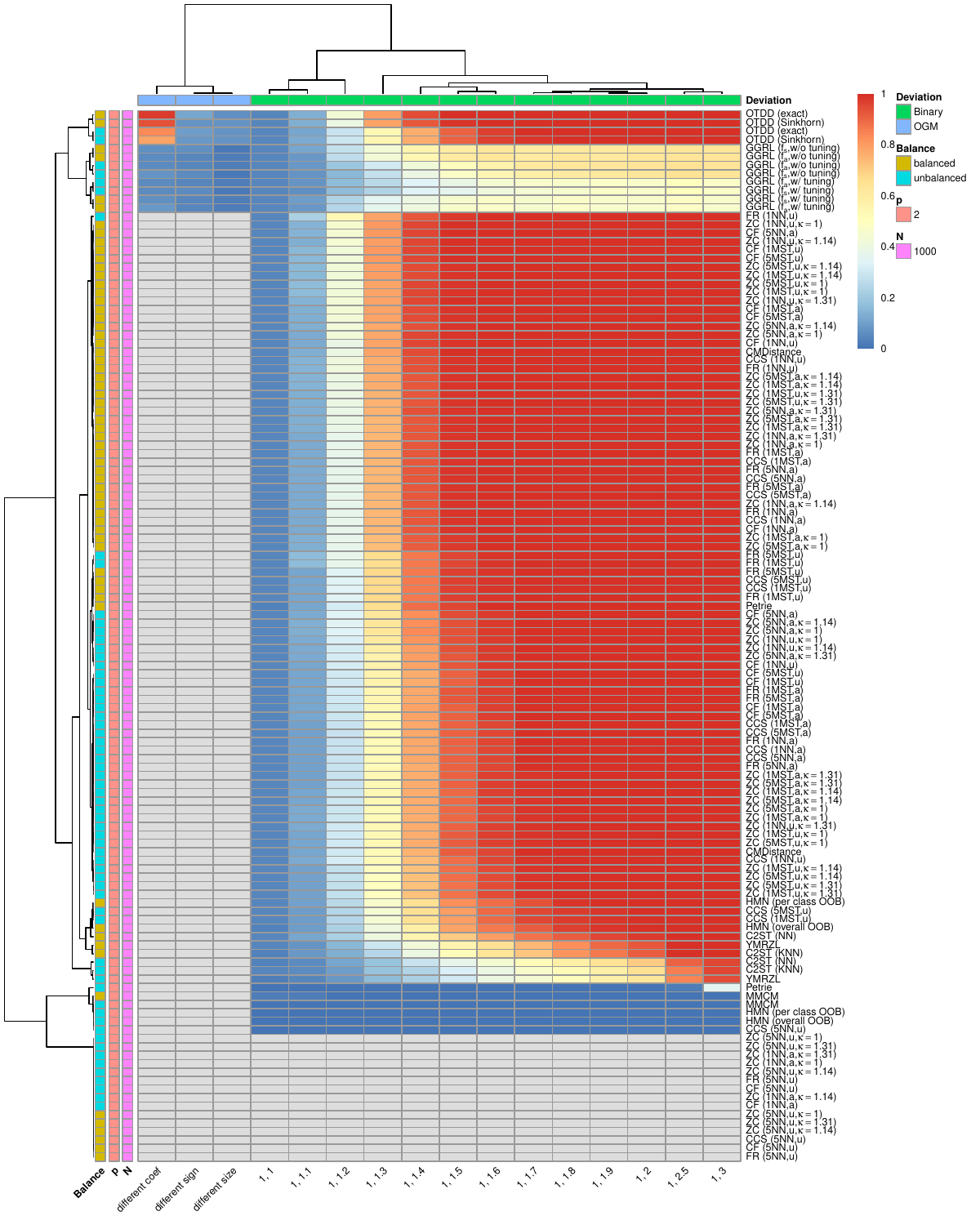}
		\caption{Clustering of PESR values per deviation ($x$-axis) and per method and sample size balance ($y$-axis) for two binary datasets with $N =1000$ and $p = 2$. The values on the $x$-axis give the weight vector (unnormalized class probabilities) of the first deviating dataset. High PESR values (red) correspond to good performance.}
		\label{fig:heatmap.cat.no.y.bin.N1000.p2}
	\end{figure}

\begin{figure}[H]
		\centering
		\includegraphics[width=\linewidth]{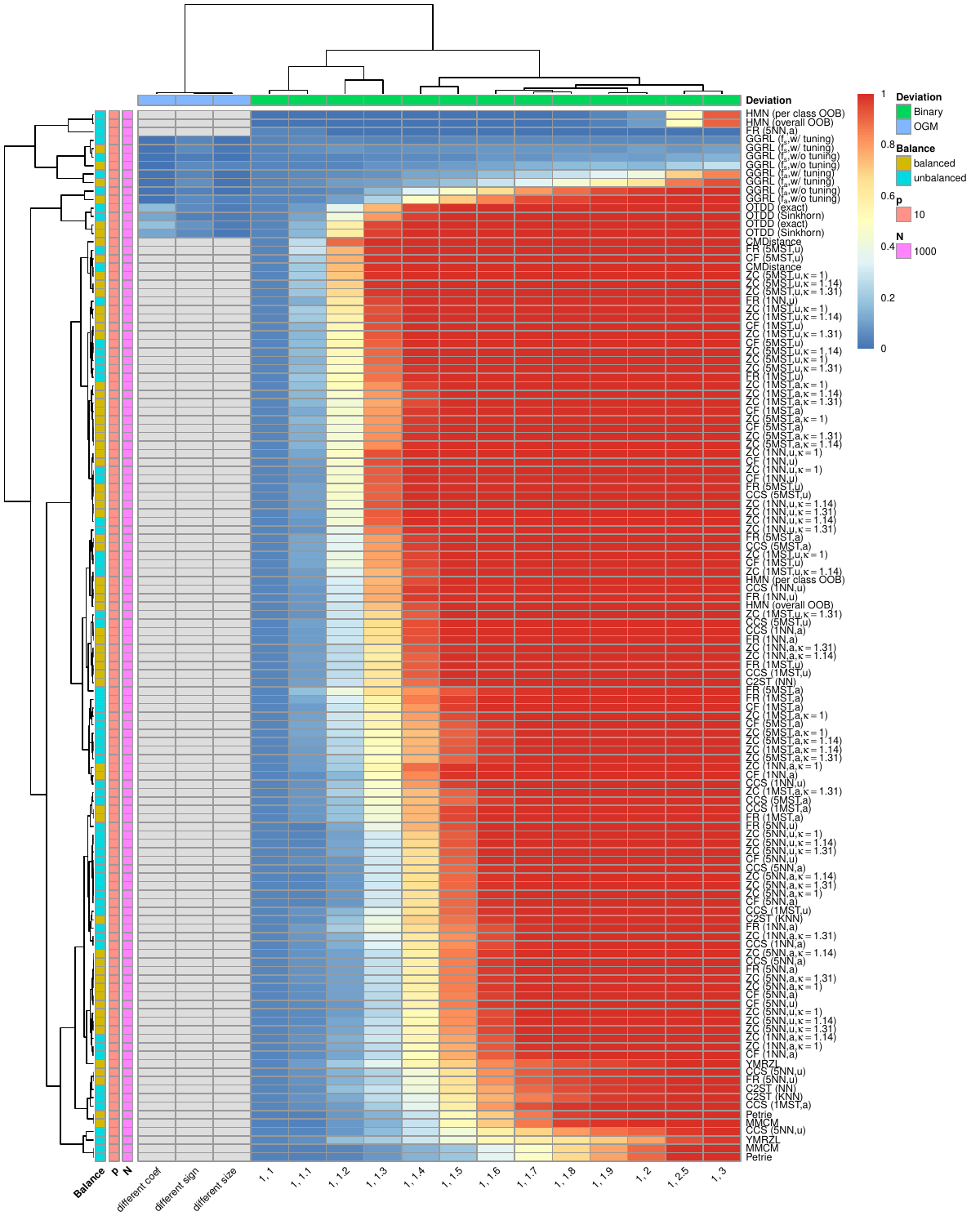}
		\caption{Clustering of PESR values per deviation ($x$-axis) and per method and sample size balance ($y$-axis) for two binary datasets with $N =1000$ and $p = 10$. The values on the $x$-axis give the weight vector (unnormalized class probabilities) of the first deviating dataset. High PESR values (red) correspond to good performance.}
		\label{fig:heatmap.cat.no.y.bin.N1000.p10}
	\end{figure}

\begin{figure}[H]
		\centering
		\includegraphics[width=\linewidth]{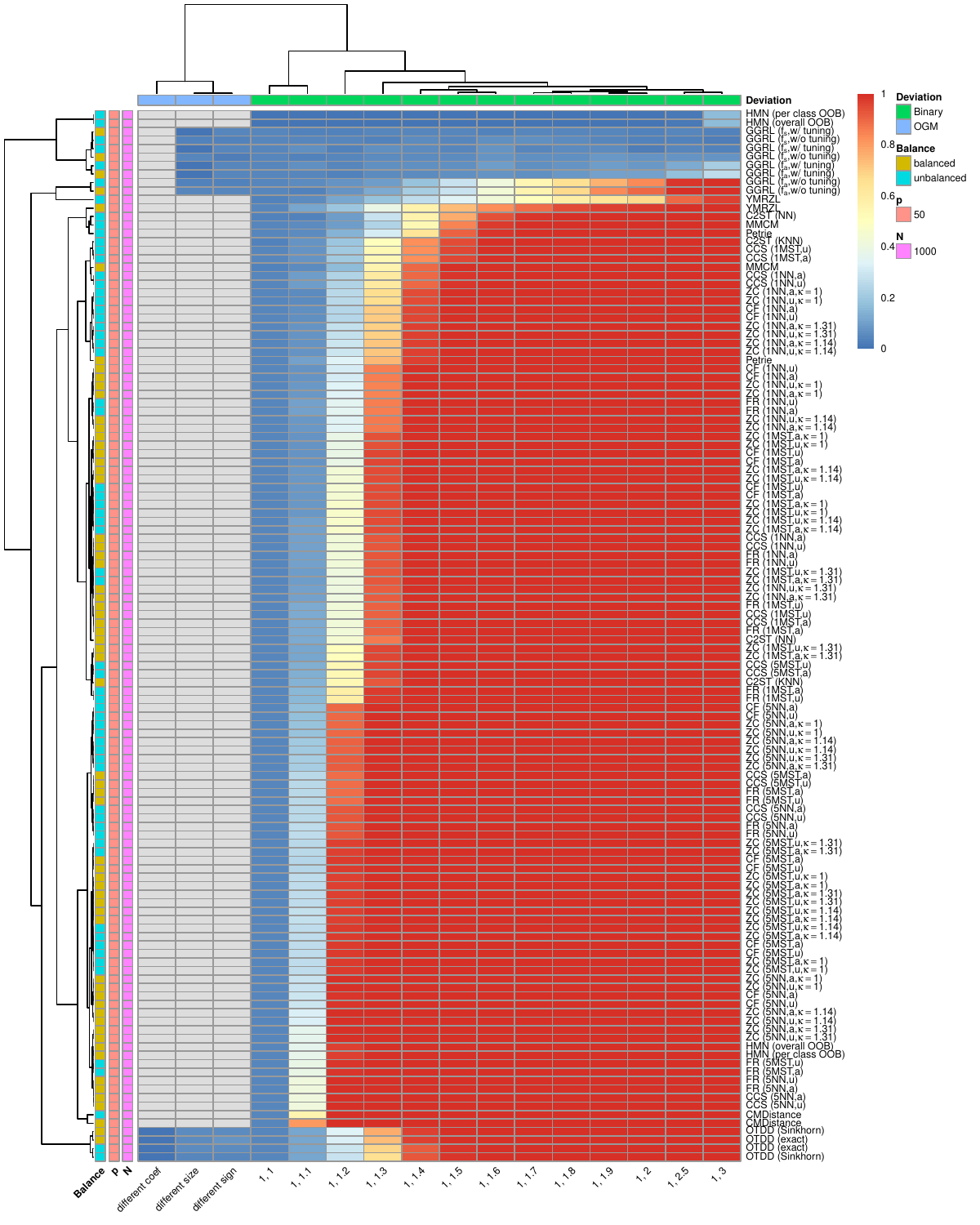}
		\caption{Clustering of PESR values per deviation ($x$-axis) and per method and sample size balance ($y$-axis) for two binary datasets with $N =1000$ and $p = 50$. The values on the $x$-axis give the weight vector (unnormalized class probabilities) of the first deviating dataset. High PESR values (red) correspond to good performance.}
		\label{fig:heatmap.cat.no.y.bin.N1000.p50}
	\end{figure}

	\paragraph{Binary Data}
	Figure~\ref{fig:heatmap.cat.no.y.bin.N200.p10} shows the clustering results for two binary datasets with $N = 200$ and $p = 10$ as a heatmap with dendrograms. 
	The results for other $N$ and $p$ are similar, with the overall trend of increasing PESR values for increasing $N$ and $p$ (see Figures~\ref{fig:heatmap.cat.no.y.bin.N50.p2} to~\ref{fig:heatmap.cat.no.y.bin.N1000.p50}).
	There are no distinct clusters visible. 
	The optimal sorting of the deviations corresponds to an increasing (or in some cases decreasing) ordering of the deviations on the $x$-axis. 
	The methods are sorted from top to bottom by decreasing PESR values.
	The only exception is the methods that can handle a target variable that are in a separate group that does not align with the overall sorting, but are again sorted within the group.  
	Typically, more balanced scenarios are among the higher-performing part of the ordering, and more unbalanced scenarios are among the lower-performing part. 
	The ordering of the methods matches the comparison of PESR curves discussed before. 
	The edge count methods are clearly sorted by the used graph type, with the $K = 5$ versions in the well-performing methods for $p>2$ and the $K=1$ versions in the well-performing methods for $p=2$.
	Overall, the CM distance and the edge count tests perform better than the MMCM, Petrie, C2ST, and YMRZL. 
	HMN is competitive in the balanced sample size case but breaks down completely in the unbalanced sample size case. 
	Out of the methods that take a target variable into account, the OTDD performs considerably better than the GGRL.
	
\begin{figure}[!tb]
		\centering
		\includegraphics[width=\linewidth]{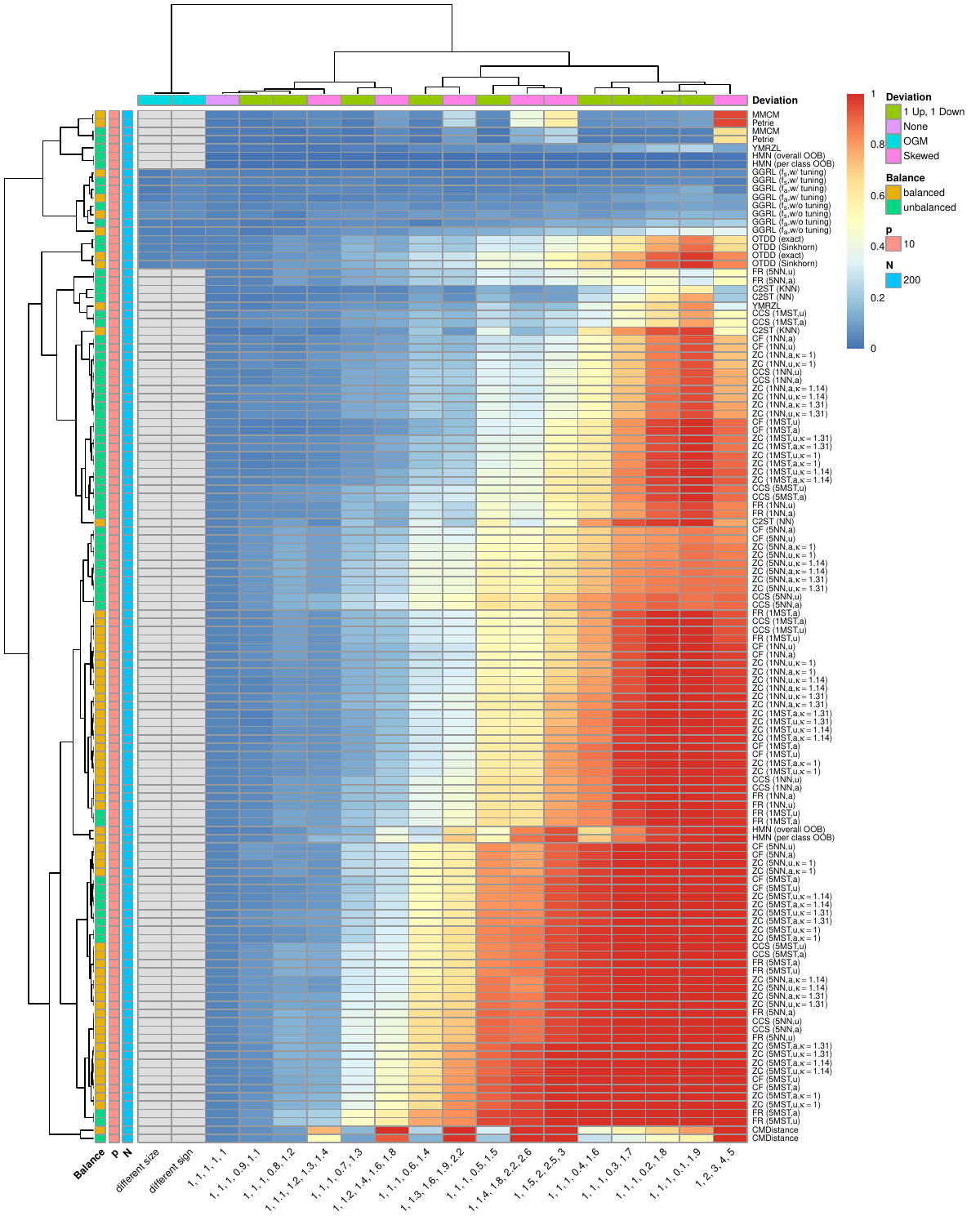}
		\caption{Clustering of PESR values per deviation ($x$-axis) and per method and sample size balance ($y$-axis) for two multinomial datasets with $N = 200$ and $p = 10$. The values on the $x$-axis give the weight vector (unnormalized class probabilities) of the first deviating dataset. High PESR values (red) correspond to good performance.}
		\label{fig:heatmap.cat.no.y.multi.N200.p10}
	\end{figure}

\begin{figure}[H]
		\centering
		\includegraphics[width=\linewidth]{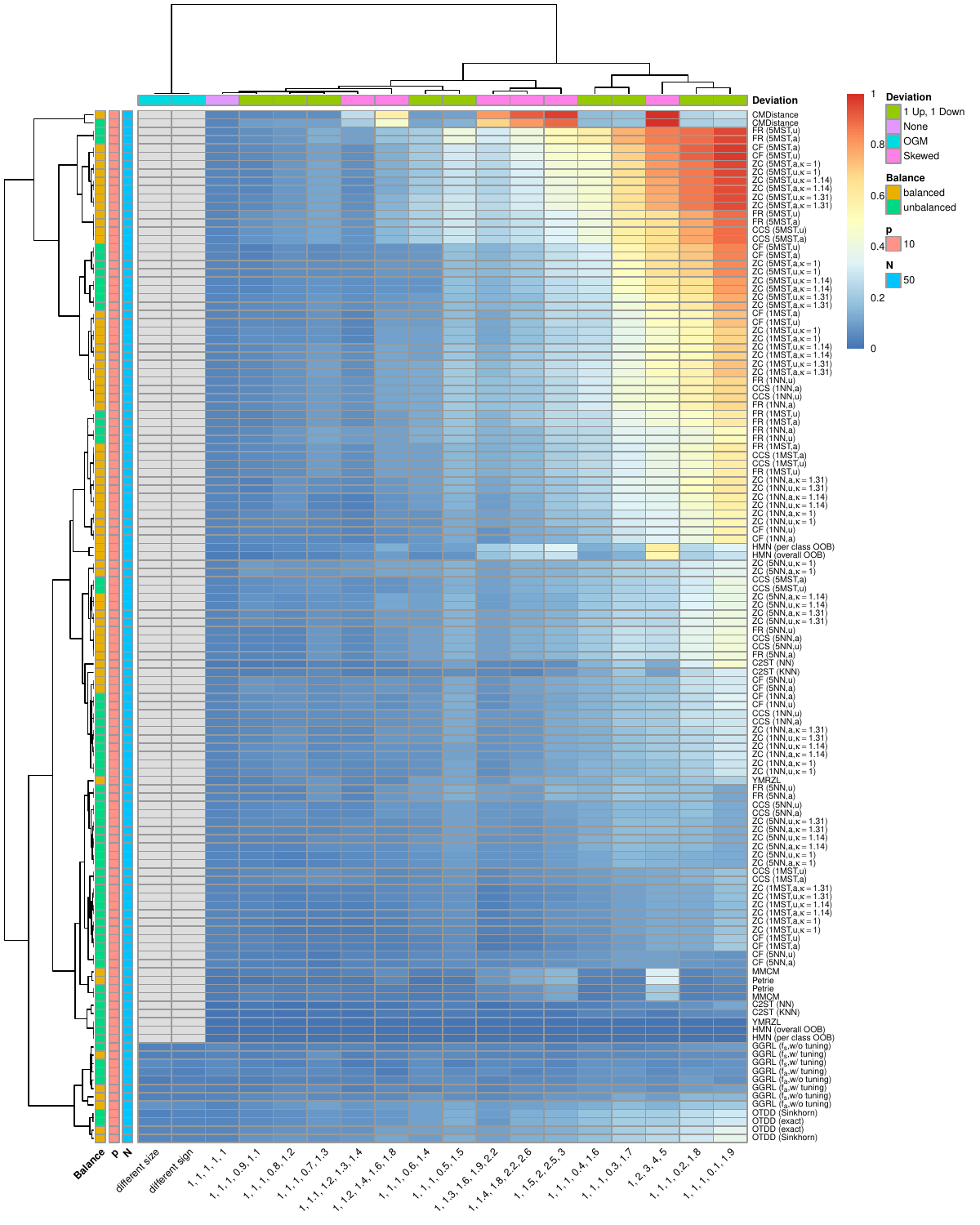}
		\caption{Clustering of PESR values per deviation ($x$-axis) and per method and sample size balance ($y$-axis) for two multinomial datasets with $N =50$ and $p = 10$. The values on the $x$-axis give the weight vector (unnormalized class probabilities) of the first deviating dataset. High PESR values (red) correspond to good performance.}
		\label{fig:heatmap.cat.no.y.multinom.N50.p10}
	\end{figure}

\begin{figure}[H]
		\centering
		\includegraphics[width=\linewidth]{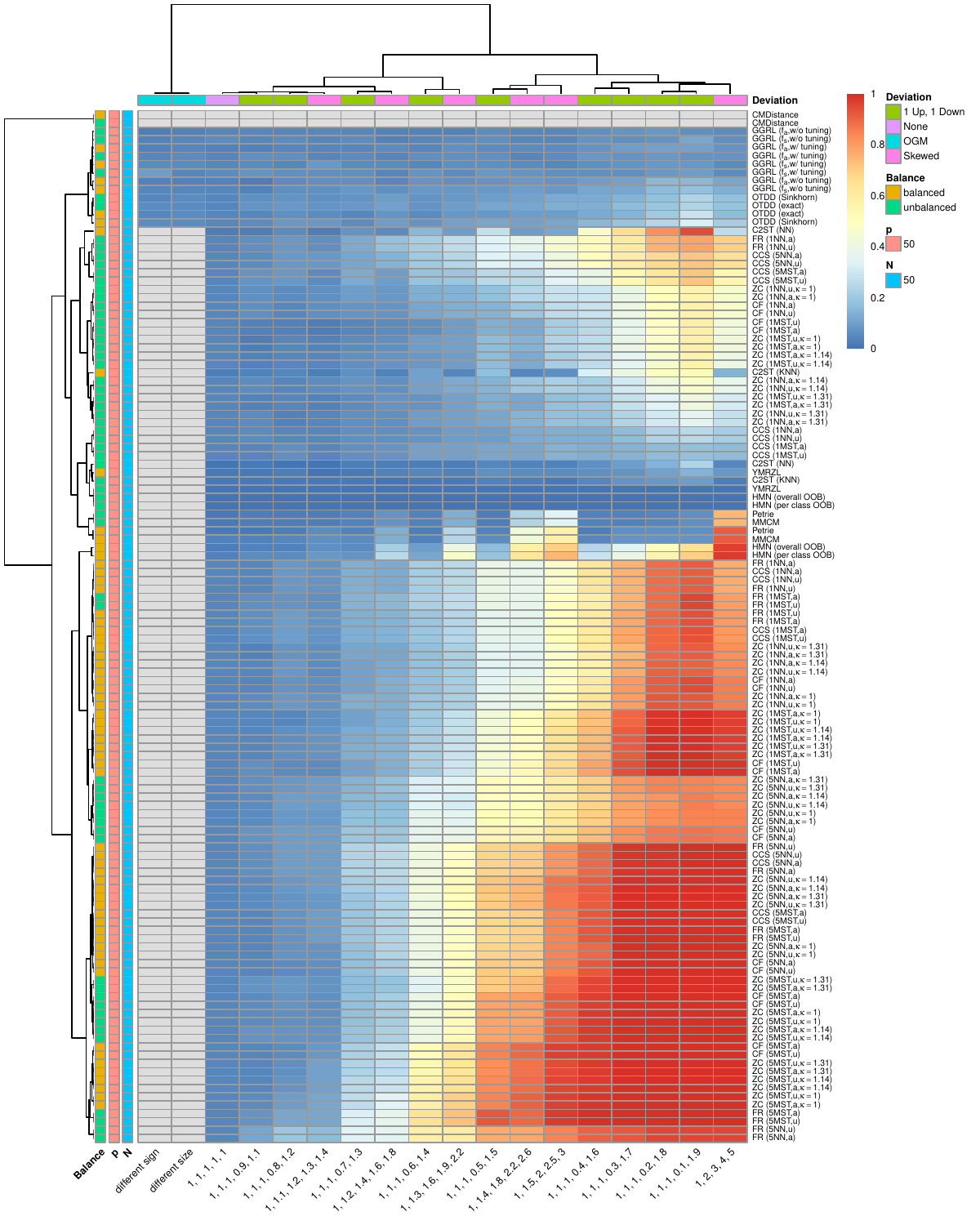}
		\caption{Clustering of PESR values per deviation ($x$-axis) and per method and sample size balance ($y$-axis) for two multinomial datasets with $N =50$ and $p = 50$. The values on the $x$-axis give the weight vector (unnormalized class probabilities) of the first deviating dataset. High PESR values (red) correspond to good performance.}
		\label{fig:heatmap.cat.no.y.multinom.N50.p50}
	\end{figure}

\begin{figure}[H]
		\centering
		\includegraphics[width=\linewidth]{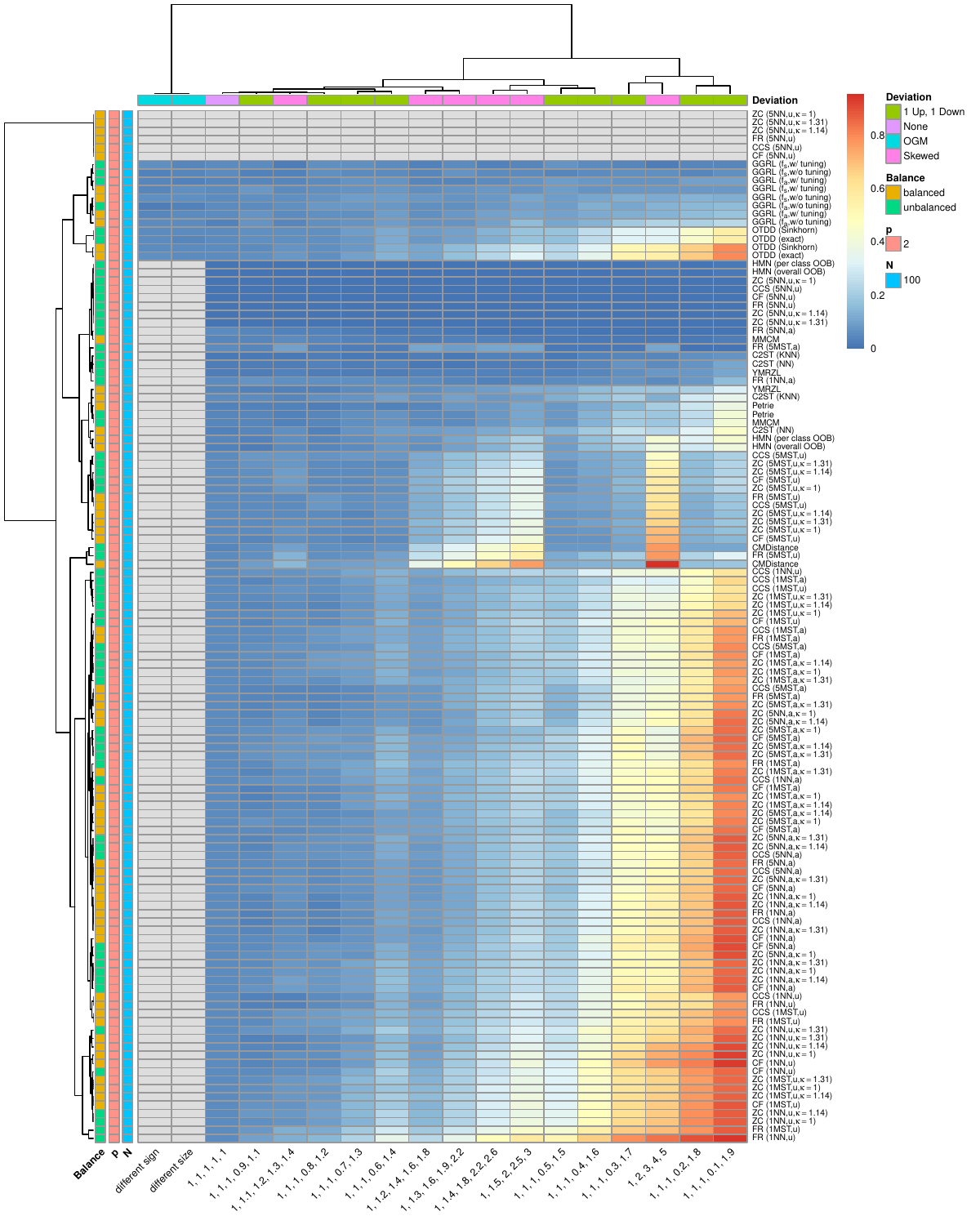}
		\caption{Clustering of PESR values per deviation ($x$-axis) and per method and sample size balance ($y$-axis) for two multinomial datasets with $N =100$ and $p = 2$. The values on the $x$-axis give the weight vector (unnormalized class probabilities) of the first deviating dataset. High PESR values (red) correspond to good performance.}
		\label{fig:heatmap.cat.no.y.multinom.N100.p2}
	\end{figure}

\begin{figure}[H]
		\centering
		\includegraphics[width=\linewidth]{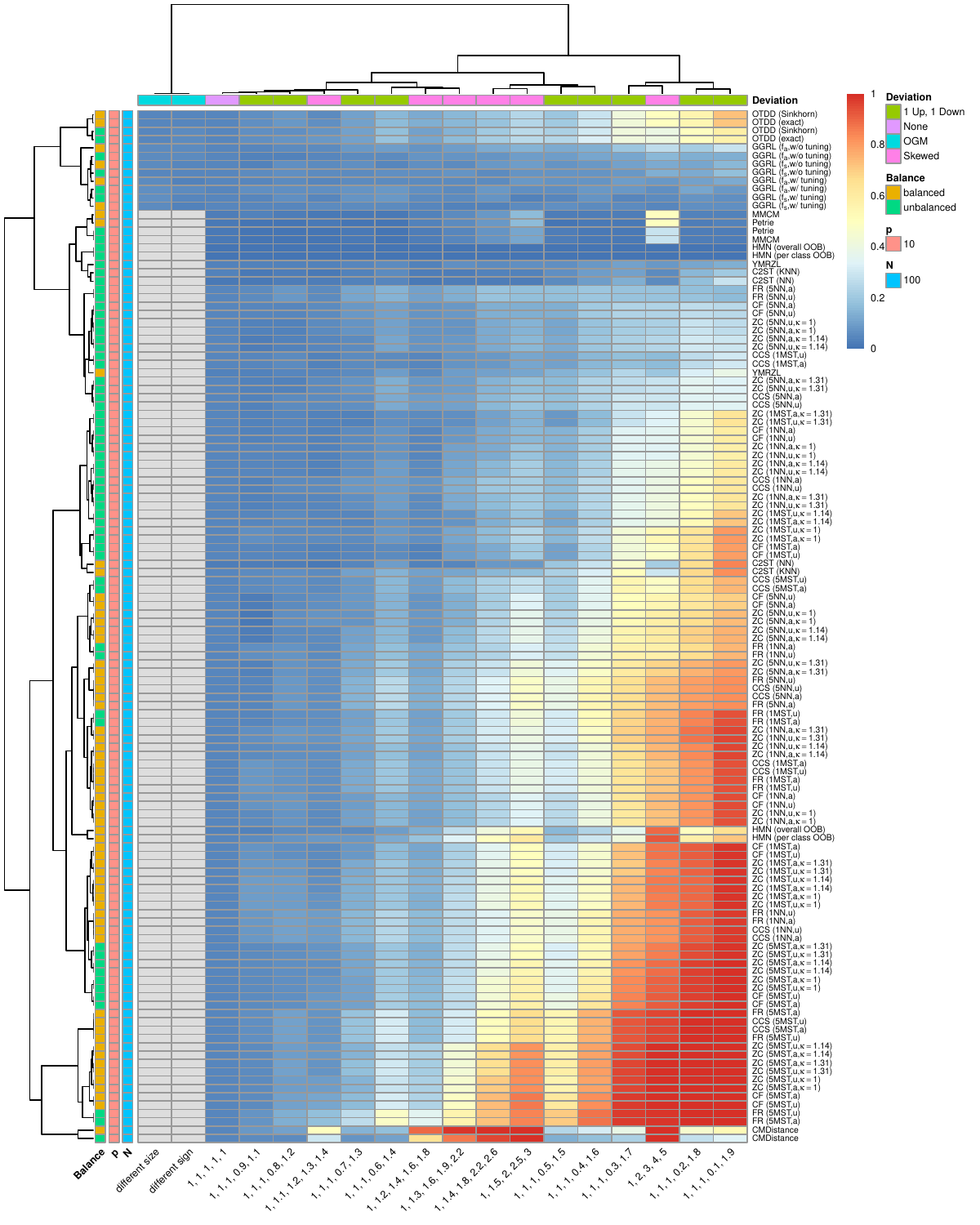}
		\caption{Clustering of PESR values per deviation ($x$-axis) and per method and sample size balance ($y$-axis) for two multinomial datasets with $N =100$ and $p = 10$. The values on the $x$-axis give the weight vector (unnormalized class probabilities) of the first deviating dataset. High PESR values (red) correspond to good performance.}
		\label{fig:heatmap.cat.no.y.multinom.N100.p10}
	\end{figure}

\begin{figure}[H]
		\centering
		\includegraphics[width=\linewidth]{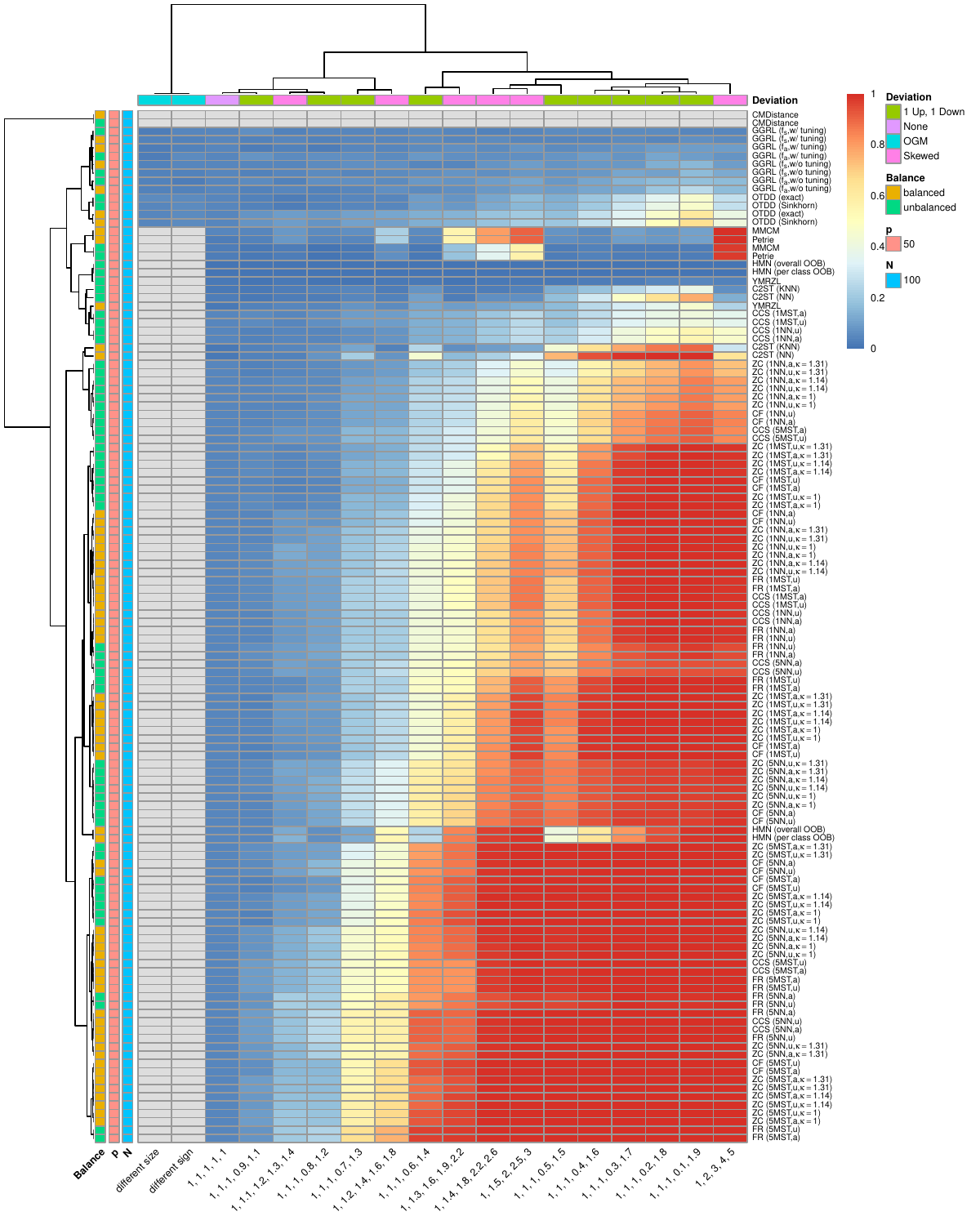}
		\caption{Clustering of PESR values per deviation ($x$-axis) and per method and sample size balance ($y$-axis) for two multinomial datasets with $N =100$ and $p = 50$. The values on the $x$-axis give the weight vector (unnormalized class probabilities) of the first deviating dataset. High PESR values (red) correspond to good performance.}
		\label{fig:heatmap.cat.no.y.multinom.N100.p50}
	\end{figure}

\begin{figure}[H]
		\centering
		\includegraphics[width=\linewidth]{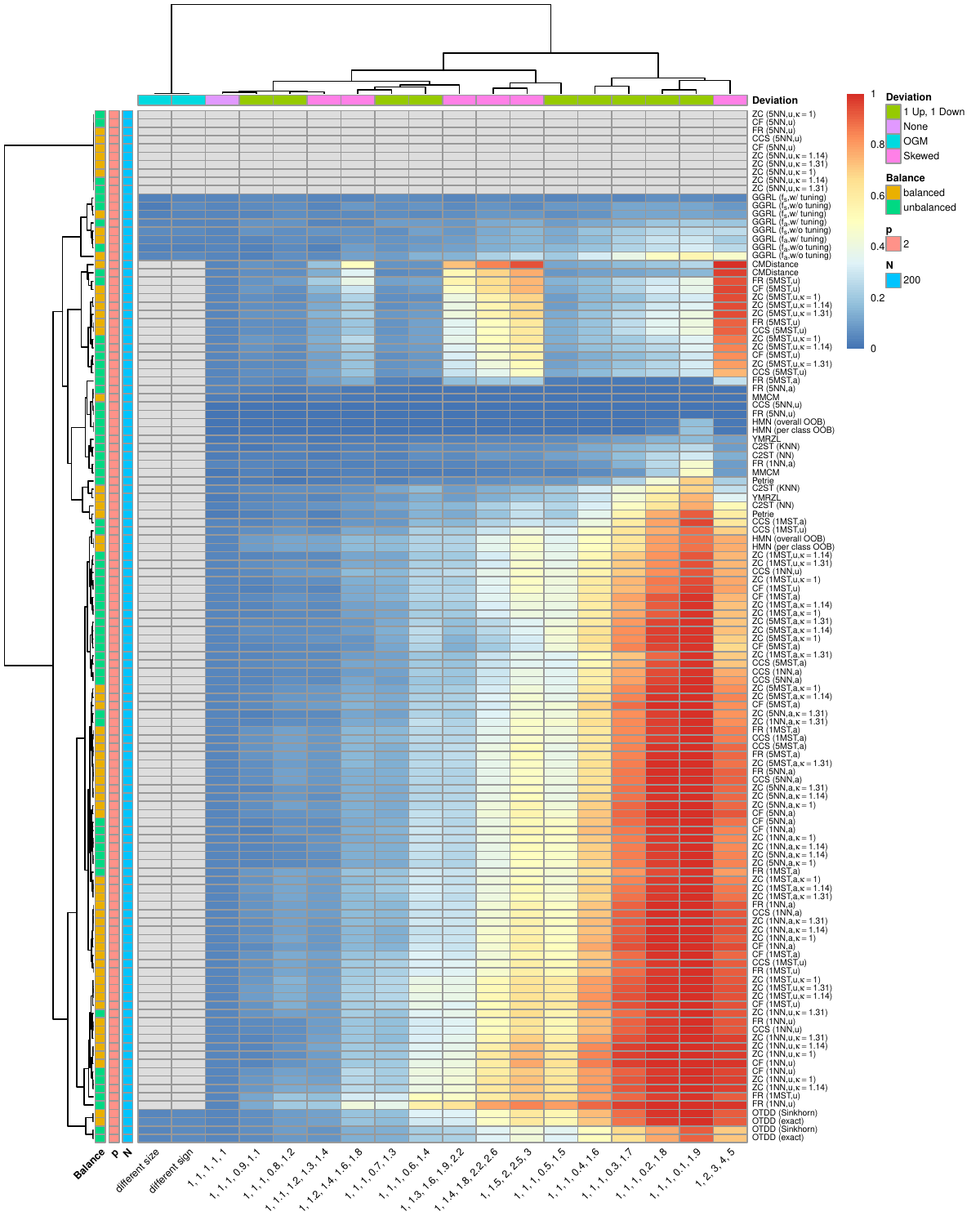}
		\caption{Clustering of PESR values per deviation ($x$-axis) and per method and sample size balance ($y$-axis) for two multinomial datasets with $N =200$ and $p = 2$. The values on the $x$-axis give the weight vector (unnormalized class probabilities) of the first deviating dataset. High PESR values (red) correspond to good performance.}
		\label{fig:heatmap.cat.no.y.multinom.N200.p2}
	\end{figure}

\begin{figure}[H]
		\centering
		\includegraphics[width=\linewidth]{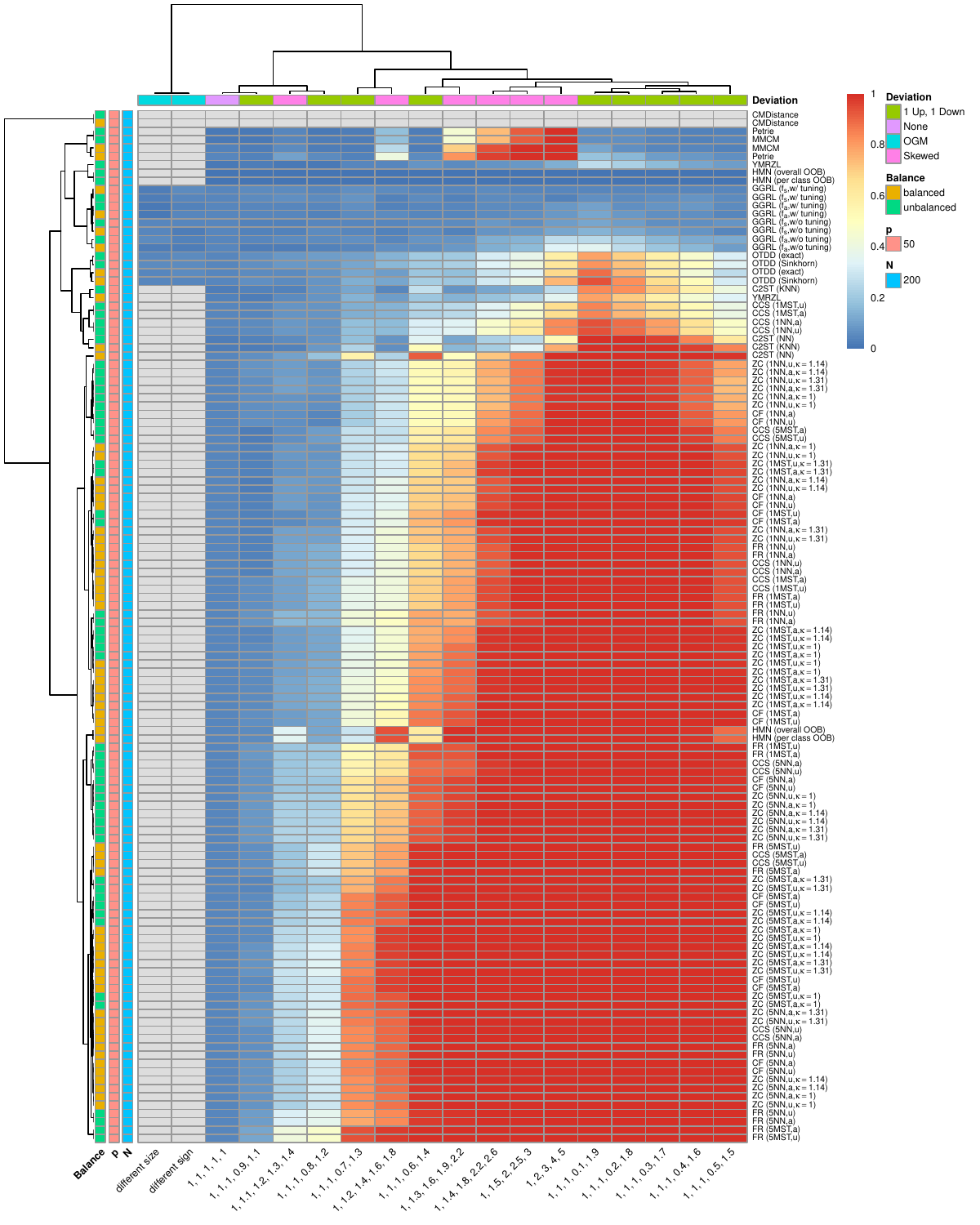}
		\caption{Clustering of PESR values per deviation ($x$-axis) and per method and sample size balance ($y$-axis) for two multinomial datasets with $N =200$ and $p = 50$. The values on the $x$-axis give the weight vector (unnormalized class probabilities) of the first deviating dataset. High PESR values (red) correspond to good performance.}
		\label{fig:heatmap.cat.no.y.multinom.N200.p50}
	\end{figure}

\begin{figure}[H]
		\centering
		\includegraphics[width=\linewidth]{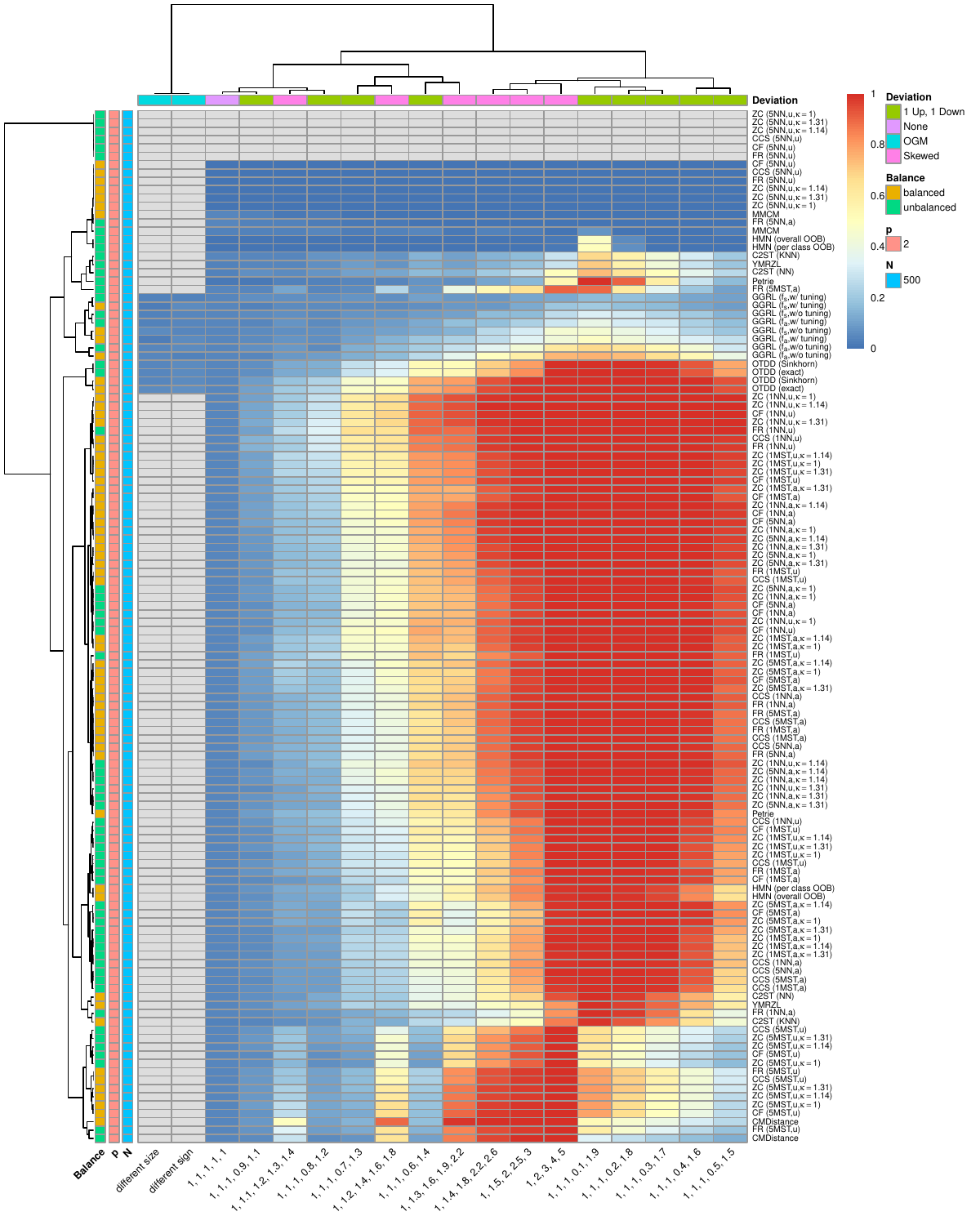}
		\caption{Clustering of PESR values per deviation ($x$-axis) and per method and sample size balance ($y$-axis) for two multinomial datasets with $N =500$ and $p = 2$. The values on the $x$-axis give the weight vector (unnormalized class probabilities) of the first deviating dataset. High PESR values (red) correspond to good performance.}
		\label{fig:heatmap.cat.no.y.multinom.N500.p2}
	\end{figure}

\begin{figure}[H]
		\centering
		\includegraphics[width=\linewidth]{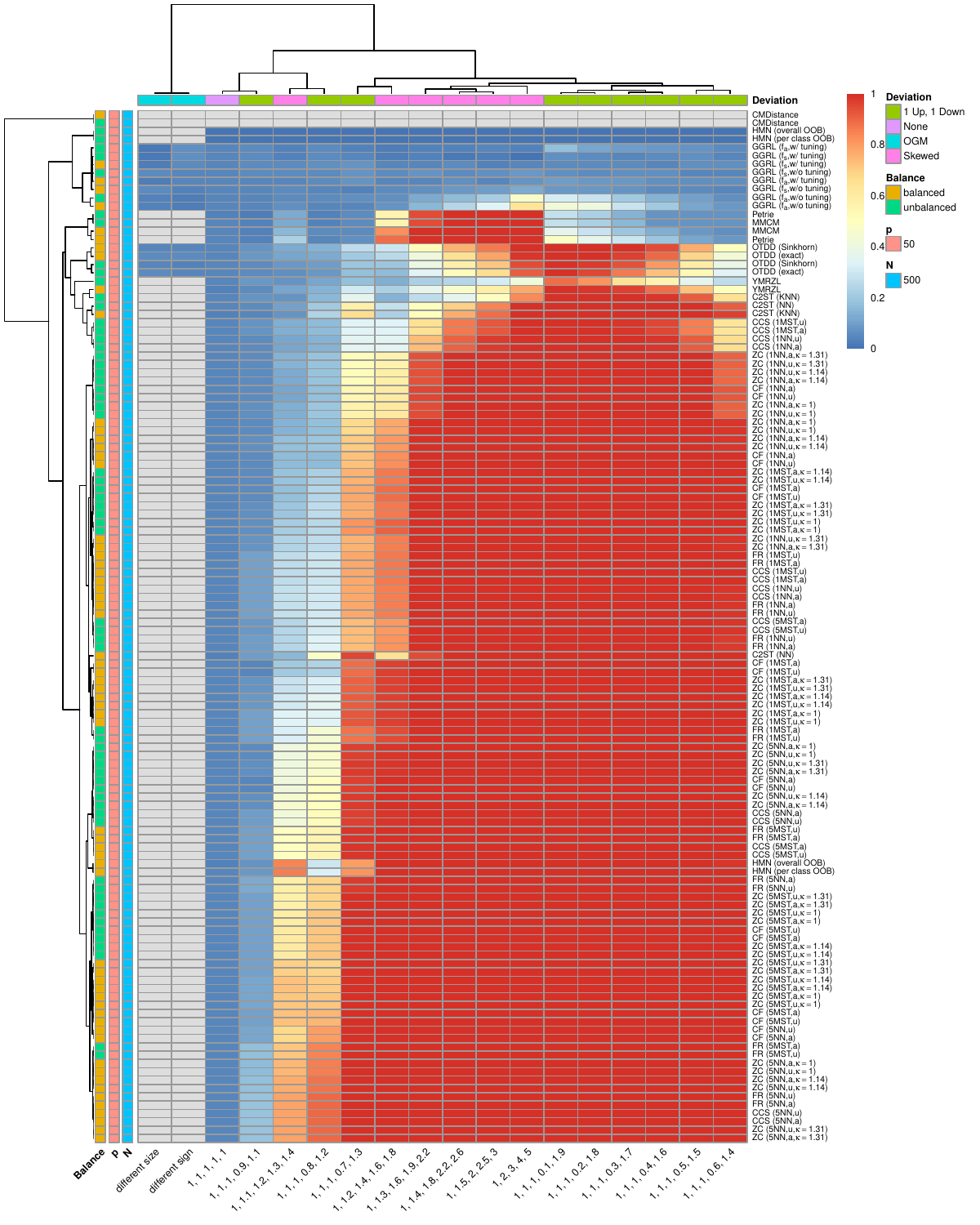}
		\caption{Clustering of PESR values per deviation ($x$-axis) and per method and sample size balance ($y$-axis) for two multinomial datasets with $N =500$ and $p = 50$. The values on the $x$-axis give the weight vector (unnormalized class probabilities) of the first deviating dataset. High PESR values (red) correspond to good performance.}
		\label{fig:heatmap.cat.no.y.multinom.N500.p50}
	\end{figure}

\begin{figure}[H]
		\centering
		\includegraphics[width=\linewidth]{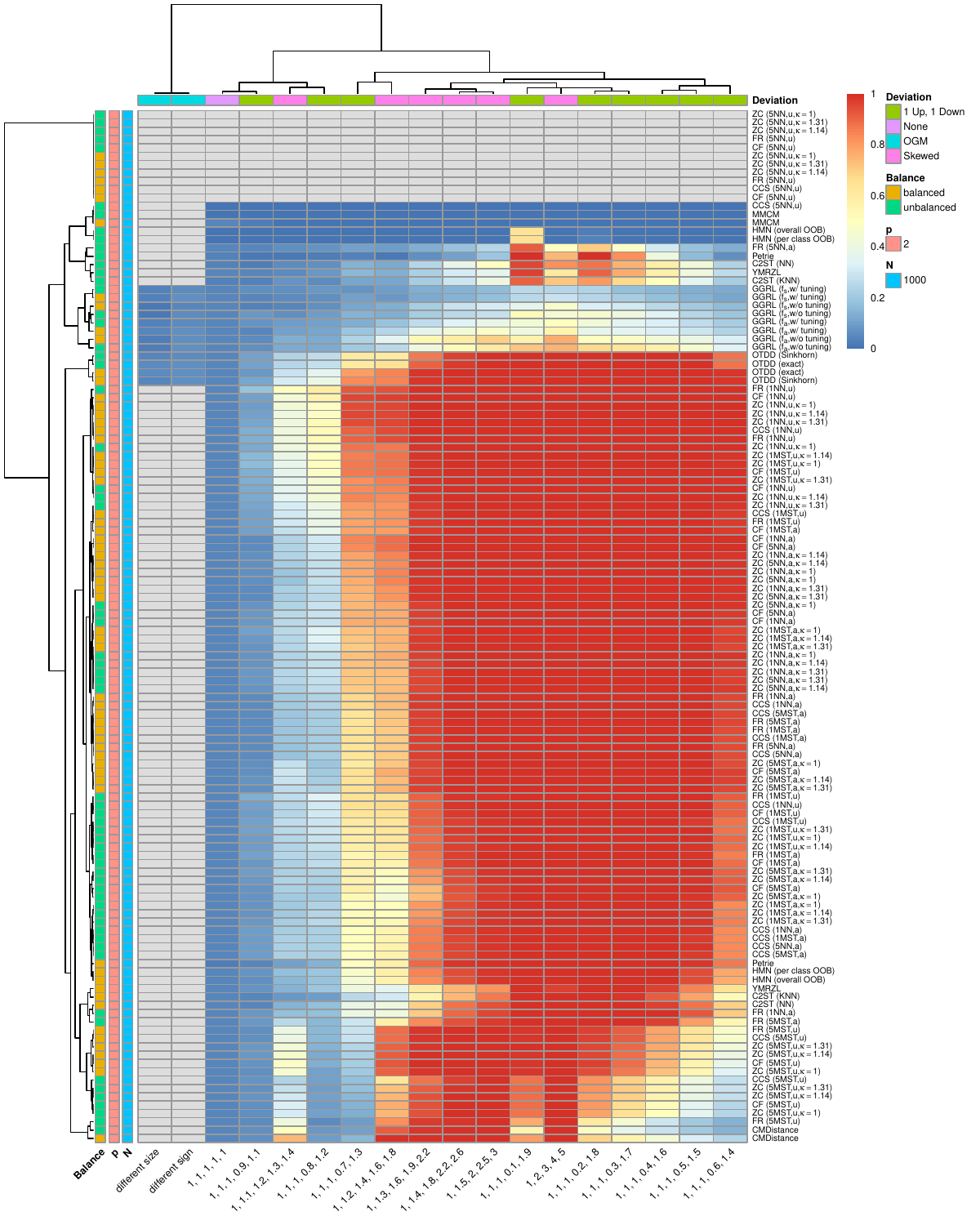}
		\caption{Clustering of PESR values per deviation ($x$-axis) and per method and sample size balance ($y$-axis) for two multinomial datasets with $N =1000$ and $p = 2$. The values on the $x$-axis give the weight vector (unnormalized class probabilities) of the first deviating dataset. High PESR values (red) correspond to good performance.}
		\label{fig:heatmap.cat.no.y.multinom.N1000.p2}
	\end{figure}

\begin{figure}[H]
		\centering
		\includegraphics[width=\linewidth]{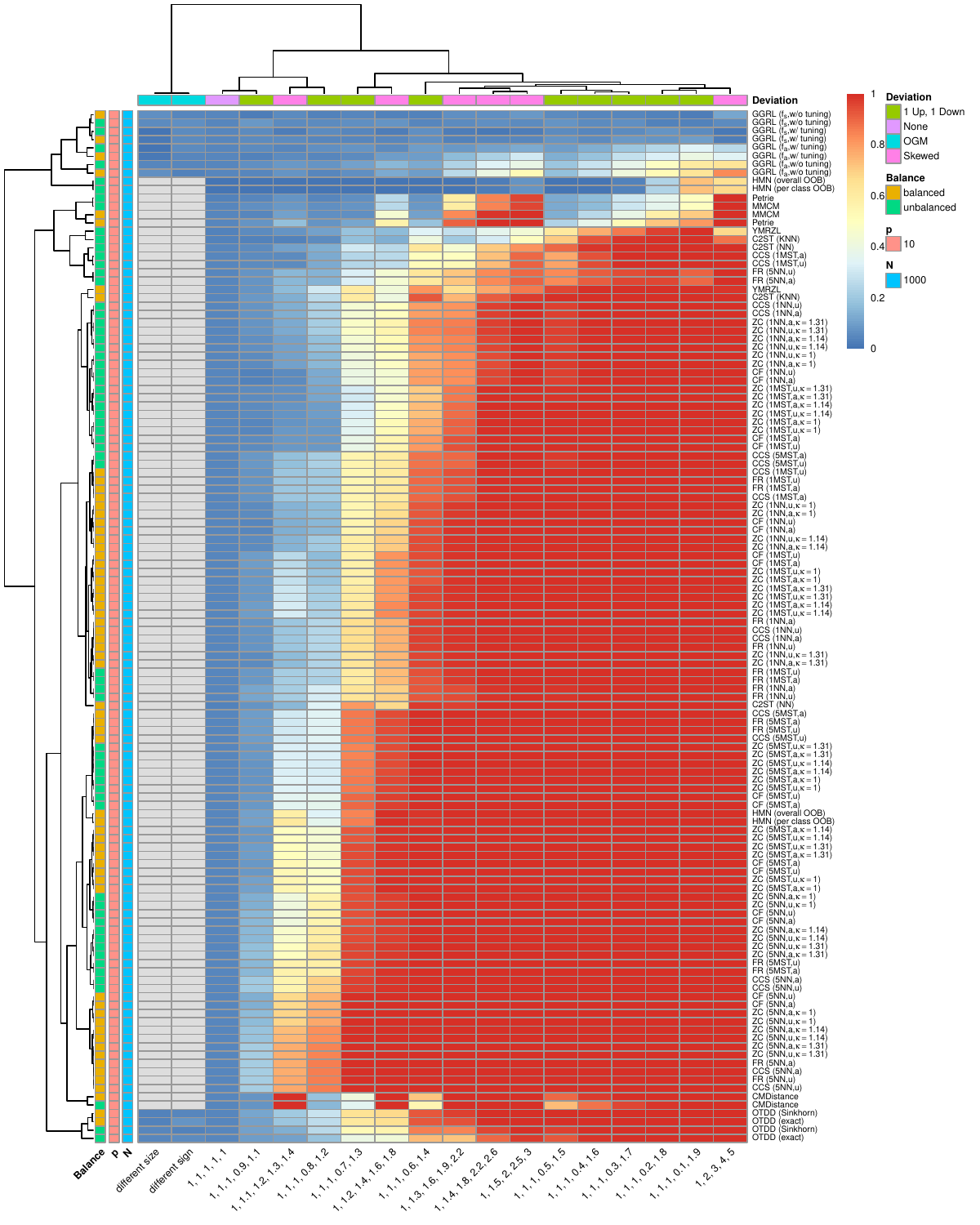}
		\caption{Clustering of PESR values per deviation ($x$-axis) and per method and sample size balance ($y$-axis) for two multinomial datasets with $N =1000$ and $p = 10$. The values on the $x$-axis give the weight vector (unnormalized class probabilities) of the first deviating dataset. High PESR values (red) correspond to good performance.}
		\label{fig:heatmap.cat.no.y.multinom.N1000.p10}
	\end{figure}

\begin{figure}[H]
		\centering
		\includegraphics[width=\linewidth]{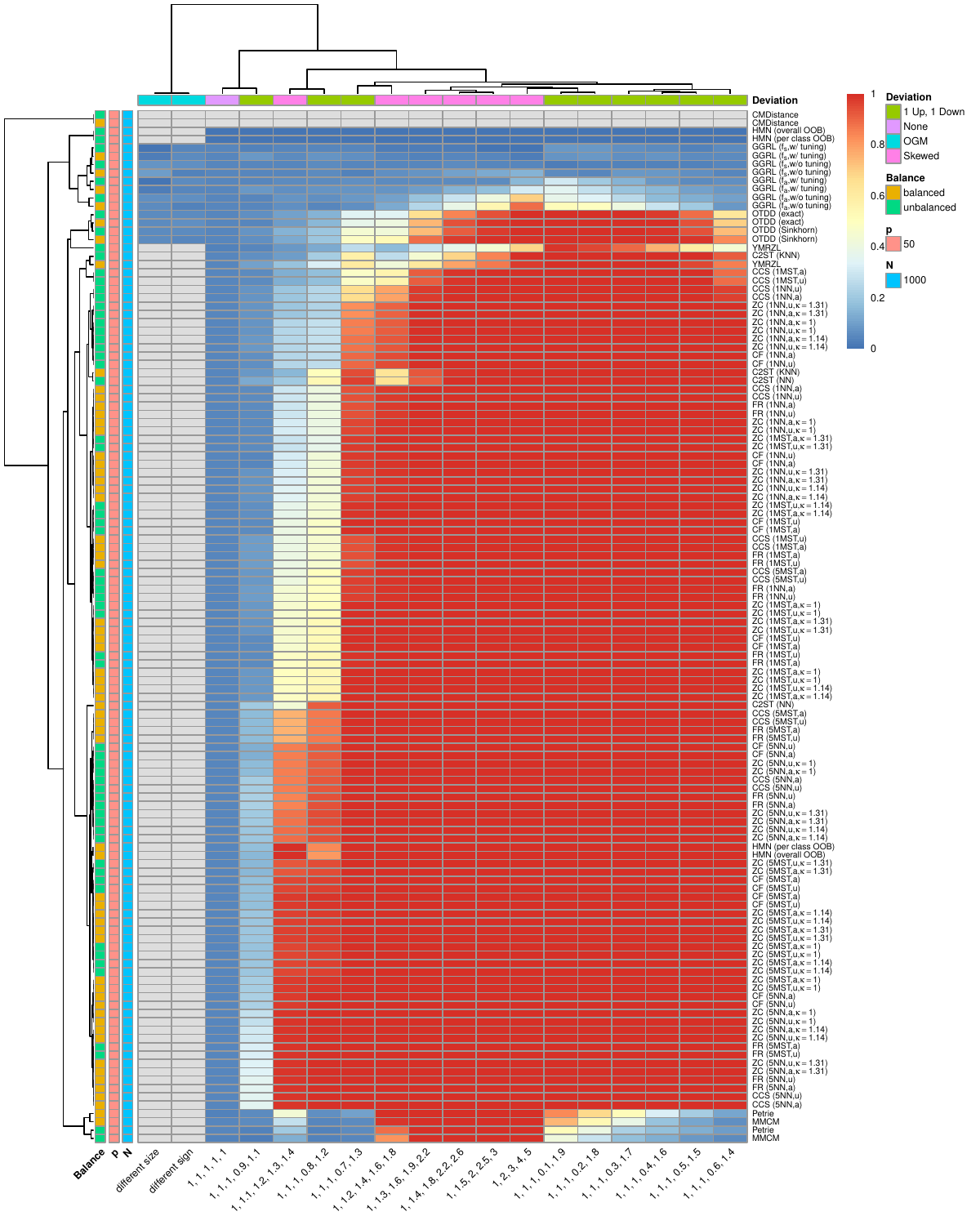}
		\caption{Clustering of PESR values per deviation ($x$-axis) and per method and sample size balance ($y$-axis) for two multinomial datasets with $N =1000$ and $p = 50$. The values on the $x$-axis give the weight vector (unnormalized class probabilities) of the first deviating dataset. High PESR values (red) correspond to good performance.}
		\label{fig:heatmap.cat.no.y.multinom.N1000.p50}
	\end{figure}

	\paragraph{Multinomial Data}
	For categorical data with five classes, the resulting clusterings differ depending on $N$ and $p$. 
	For lower $N$ and $p$, the results are similar to the binary case in the sense that there is no clear clustering of the methods, but they are rather ordered by gradually decreasing PESR. 
	This is, for example, shown in Figure~\ref{fig:heatmap.cat.no.y.multi.N200.p10} as a heatmap with dendrograms for  $N = 200$ and $p = 10$.
	The other heatmaps can be found in Appendix~\ref{app:add.figs.heatmaps.multinom}, Figures~\ref{fig:heatmap.cat.no.y.multinom.N50.p2}--\ref{fig:heatmap.cat.no.y.multinom.N1000.p50}.
	Again, the scenarios with higher PESR are mostly ones with balanced sample sizes, while most of the unbalanced scenarios are among the lower-performing cases. 
	In contrast to the binary case, for five categories, two types of deviations are considered: the class probability distribution becoming more and more skewed, or the probability of one class going up and the probability of another class going down. 
	For lower $N$ and $p$, these are not well separated in the clustering but rather mixed between these two cases, and instead sorted by the strength of deviation. 
	One noteworthy exception for the different deviations is the CM distance, which only shows a high PESR for the ``skewed'' alternatives but a low PESR for the ``1 up, 1 down'' alternatives. 
	As before, the OTDD and GGRL, which consider a target variable in the data, are clustered separately from the remaining methods. 
	
\begin{figure}[!tb]
		\centering
		\includegraphics[width=\linewidth]{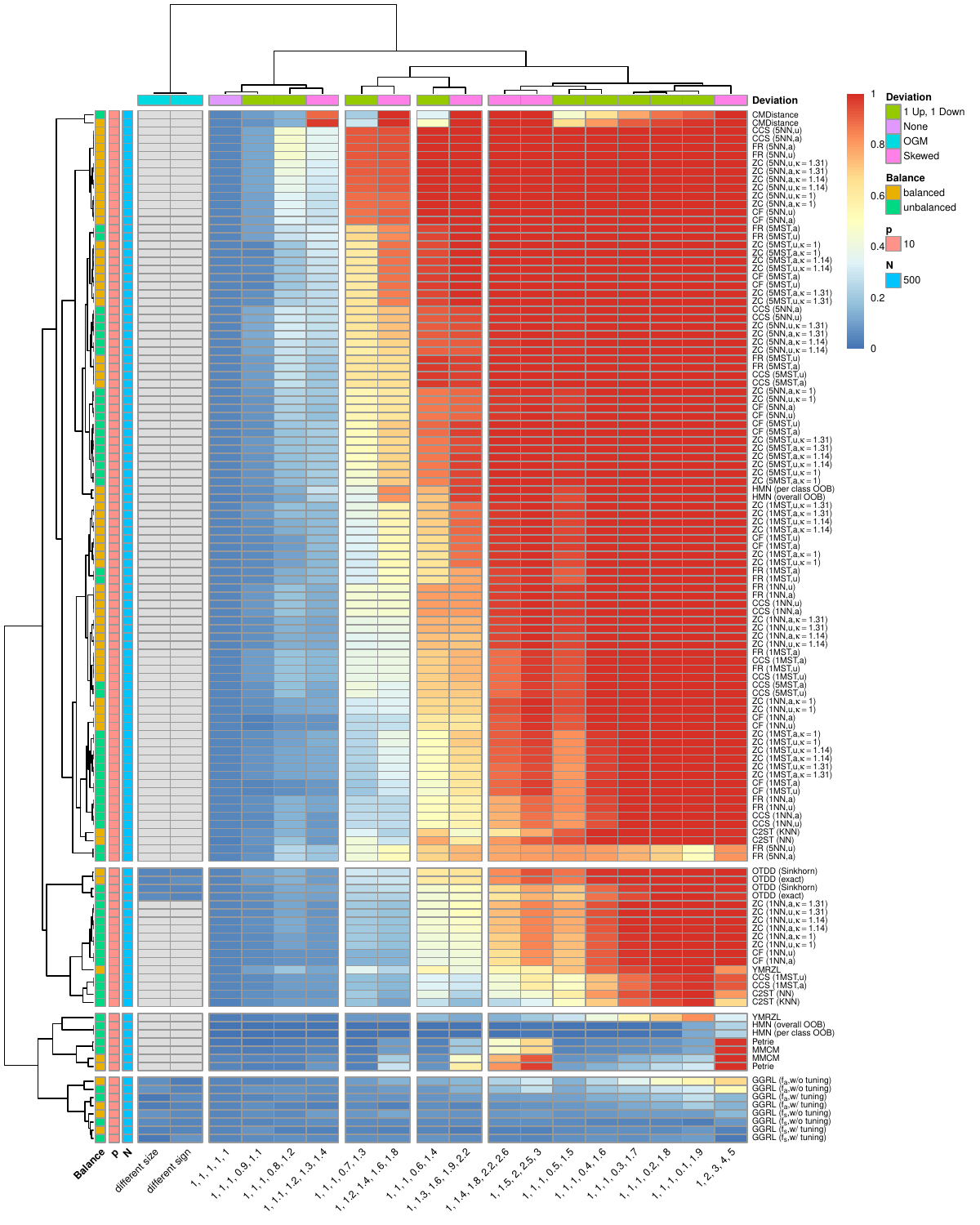}
		\caption{Clustering of PESR values per deviation ($x$-axis) and per method and sample size balance ($y$-axis) for two multinomial datasets with $N = 500$ and $p = 10$. The values on the $x$-axis give the weight vector (unnormalized class probabilities) of the first deviating dataset. High PESR values (red) correspond to good performance.}
		\label{fig:heatmap.cat.no.y.multi.N500.p10}
	\end{figure}

	For higher $N$ and $p$, there is a clearer distinction between the deviations. 
	This is, for example, shown in Figure~\ref{fig:heatmap.cat.no.y.multi.N500.p10} as a heatmap with dendrograms for $N = 500$ and $p = 10$. 
	With respect to the deviations, there are five groups: deviations of the OGM for which only the GGRL and OTDD can be applied, low, medium, and high deviations (each a mix of ``skewed'' and ``1 up, 1 down''). 
	
	The edge count tests and the HMN for balanced data show a high PESR for all medium to high deviations. 
	These are again mostly sorted by graph type and balance of the sample sizes, with $K = 5$ performing better than $K = 1$ for $p>2$ (vice versa for $p = 2$) and higher PESR for balanced than for unbalanced settings. 
	The C2ST variants and the YMRZL show mostly higher PESR for ``1 up, 1 down'' and lower PESR for ``skewed'' deviations. 
	In contrast, Petrie's method, the MMCM, and the CM distance show high PESR values for ``skewed'' but very low PESR values for ``1 up, 1 down'' deviations. 
	The HMN and YRZL for unbalanced sample sizes show (very) low PESR values across all deviations. 
	The GGRL shows very low PESR for almost all deviations. 
	
	Overall, the edge count tests perform best across all deviations for categorical data with five categories. 
	HMN is also competitive in the case of balanced sample sizes, but useless for unbalanced sample sizes.
	For only detecting ``skewed'' deviations, the CM distance also performs very well. 
	For only detecting ``1 up, 1 down'' deviations, the C2ST variants are a competitive alternative.
	The OTDD performs okay for detecting high deviations, but does not detect the lower deviations.
	The GGRL always performs poorly. 
	\FloatBarrier
	
	\subsubsection{Multi-sample Setting}
	In the following, the clusterings for $k = 4$ are presented, first for binary and afterward for multinomial data. 
	Here, all $N$ and $p$ are considered together, which is possible due to the considerably lower number of methods. 
	
\begin{figure}[!tb]
		\centering
		\includegraphics[width=1\linewidth]{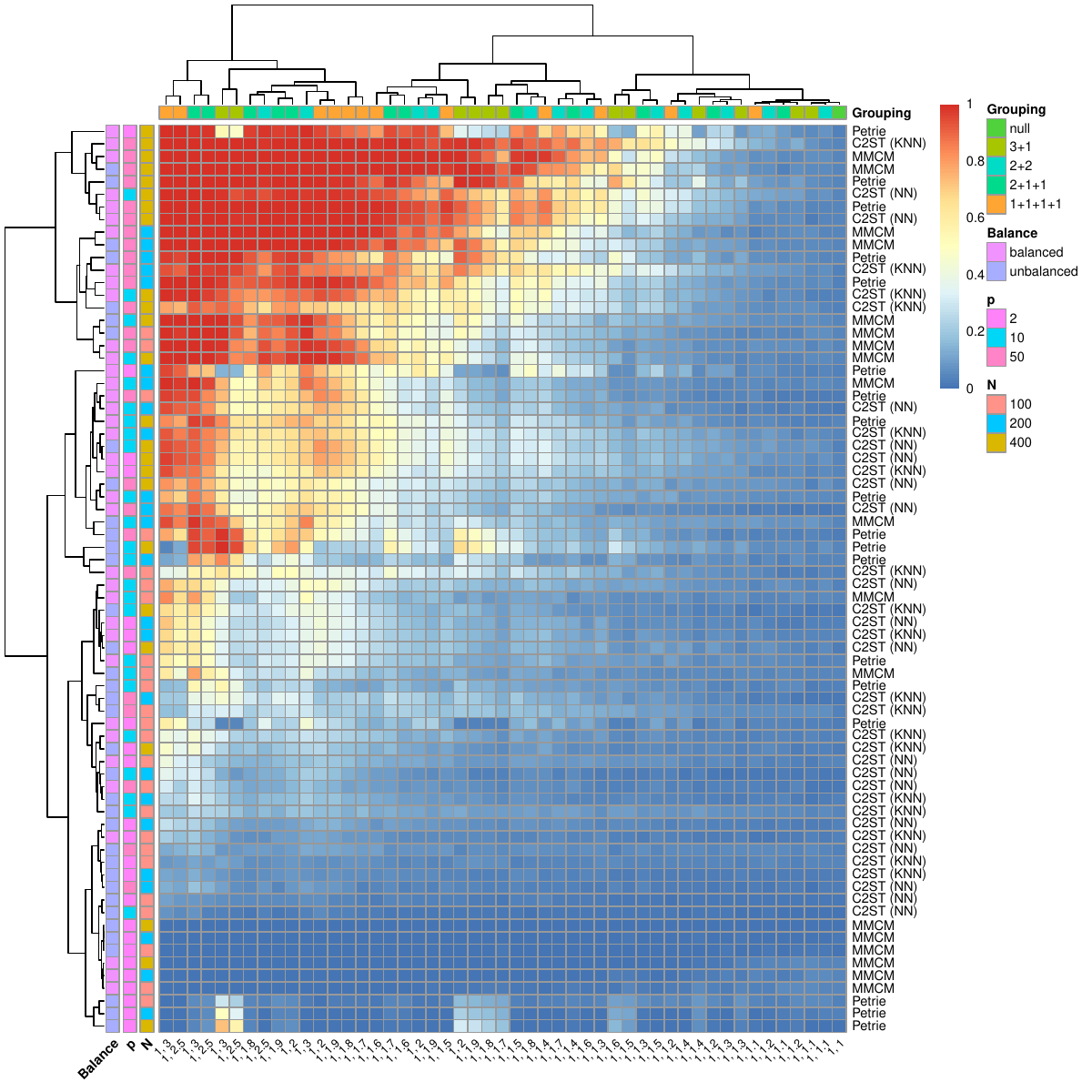}
		\caption{Clustering of PESR values per deviation ($x$-axis) and per method and dataset dimension ($y$-axis) for four binary datasets. The values on the $x$-axis give the weight vector (unnormalized class probabilities) of the first deviating dataset. High PESR values (red) correspond to good performance. }
		\label{fig:heatmap.cat.multi.bin}
	\end{figure}

	\paragraph{Binary Data}
	The resulting clustering for binary data is shown in the heatmap with dendrograms in Figure~\ref{fig:heatmap.cat.multi.bin}. 
	As for the two-sample case, there is no clear clustering visible. 
	The scenarios are again ordered by strength.
	Scenarios with large deviations are on the left, scenarios with medium deviations in the middle, and scenarios with high deviations on the right. 
	The strength of a deviation here seems to depend both on the alternative class weights and on the grouping. 
	Scenarios for groupings with a lower number of differing datasets and high class weights for ones are comparable to scenarios for groupings with a higher number of differing datasets and lower class weights. 
	
	The methods are ordered from top to bottom according to decreasing PESR. 
	The MMCM and Petrie's method for medium and high-dimensional datasets, as well as the C2ST (NN or KNN) for the highest dimensional datasets with balanced sample sizes, are among the best-performing methods with high PESR for medium and high deviations
	
	Next in the ordering are MMCM and Petrie's method for medium-dimensional data (mostly unbalanced sample sizes) and the C2ST for high-dimensional data with balanced sample sizes, which have high PESR for high deviations but low PESR for medium and low deviations.
	
	Last in the ordering are mostly the MMCM and Petrie for $p = 2$ and C2ST for low-dimensional data, and especially unbalanced sample sizes. 
	
\begin{figure}[!tb]
		\centering
		\includegraphics[width=1\linewidth]{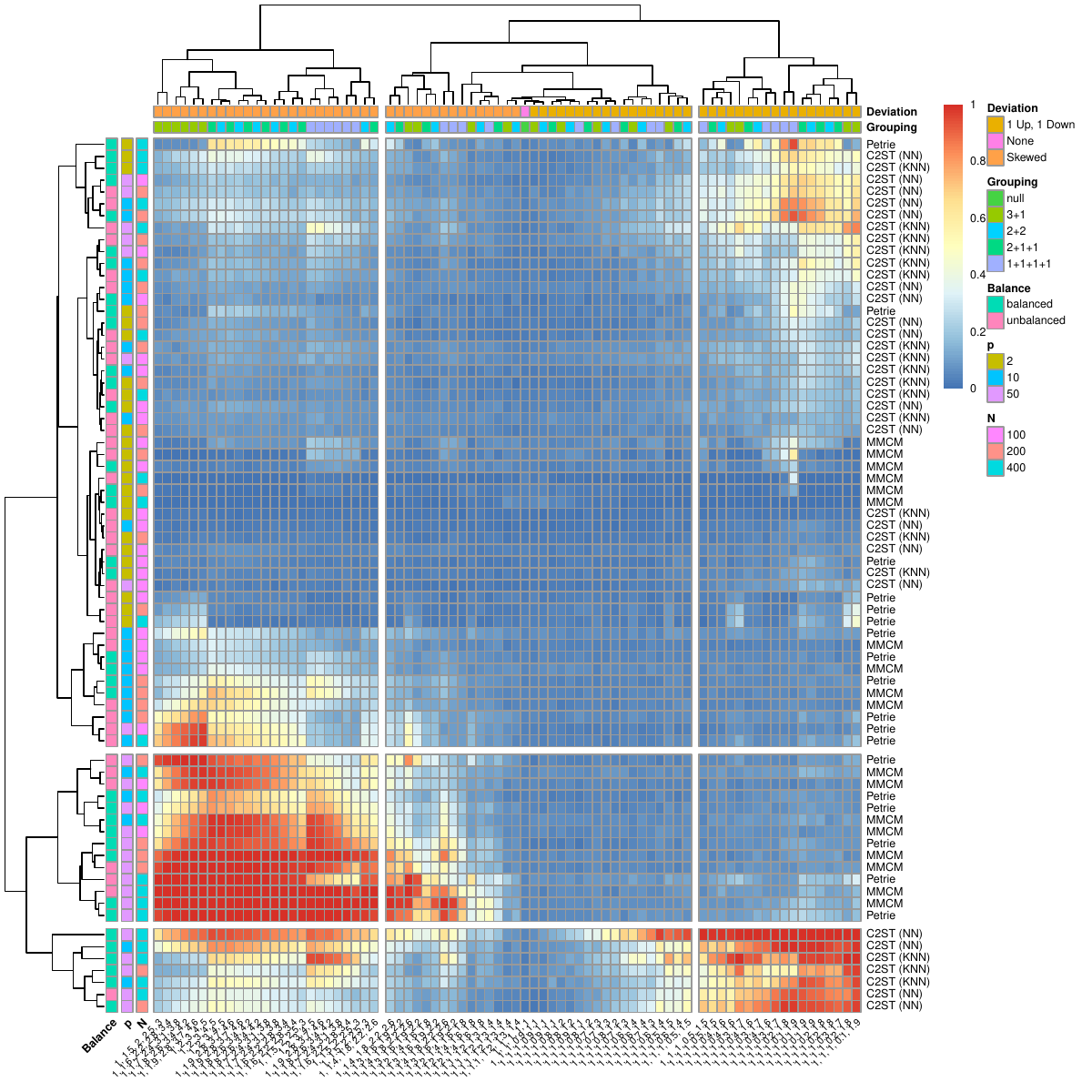}
		\caption{Clustering of PESR values per deviation ($x$-axis) and per method and dataset dimension ($y$-axis) for four categorical datasets where each variable is drawn from a multinomial distribution. The values on the $x$-axis give the weight vector (unnormalized class probabilities) of the first deviating dataset. High PESR values (red) correspond to good performance.}
		\label{fig:heatmap.cat.multi.multinom}
	\end{figure}

	\paragraph{Multinomial data}
	Figure~\ref{fig:heatmap.cat.multi.multinom} shows the resulting clustering for the case of five classes as a heatmap with dendrograms. 
	The scenarios are clustered into three overall blocks based on the severity of the deviation. 
	The left block includes large deviations with skewed class distribution, the block in the middle includes small deviations or groupings ``2+2'' and ``3+1'', and the block on the right includes large deviations with one class probability up and one class probability down. 
	The deviations are very clearly split into ``skewed'' (on the left) and ``1 up, 1 down'' (on the right). 
	
	There are three clusters of methods and $N$, $p$, balance combinations.
	The first block includes methods with mostly low PESR for both kinds of deviations. 
	These are the C2ST variants for unbalanced sample sizes, low $N$ or high $p$, as well as Petrie's method and the MMCM for $p = 2$. 
	
	The second block includes methods with high PESR for skewed class distribution deviations but low PESR for one up, one down deviations.
	These include the MMCM and Petrie's method for high $p$ or $N$ settings. 
	
	The last block includes methods with high PESR for detecting one up, one down deviations, but (mostly) low PESR for skewed class distributions. 
	These consist of the C2ST variants for high $N$ or $p$ and mostly balanced sample sizes.  
	
	Across all blocks of methods, the PESR values for the cluster of small deviations are low. 
	Overall, the clustering shows that the graph-based methods MMCM and Petrie are better at detecting skewed class distribution, while the C2ST variants are better at detecting one up, one down alternatives.
	Moreover, the graph-based methods suffer more heavily from low $p$, while the classifier-based methods suffer more heavily from unbalanced dataset sizes and low $N$ and $p$. 
	
	\newpage
	\vspace*{5cm}
	\subsection[Heatmaps for k = 2, Binary Data]{Heatmaps for $k = 2$, Binary Data} \label{app:add.figs.heatmaps.bin}
	
	\vspace*{-0.5em}
	\subsection[Heatmaps for k = 2, Multinomial Data]{Heatmaps for $k = 2$, Multinomial Data} \label{app:add.figs.heatmaps.multinom}
	\begin{figure}[H]
		\centering
		\includegraphics[width=\linewidth]{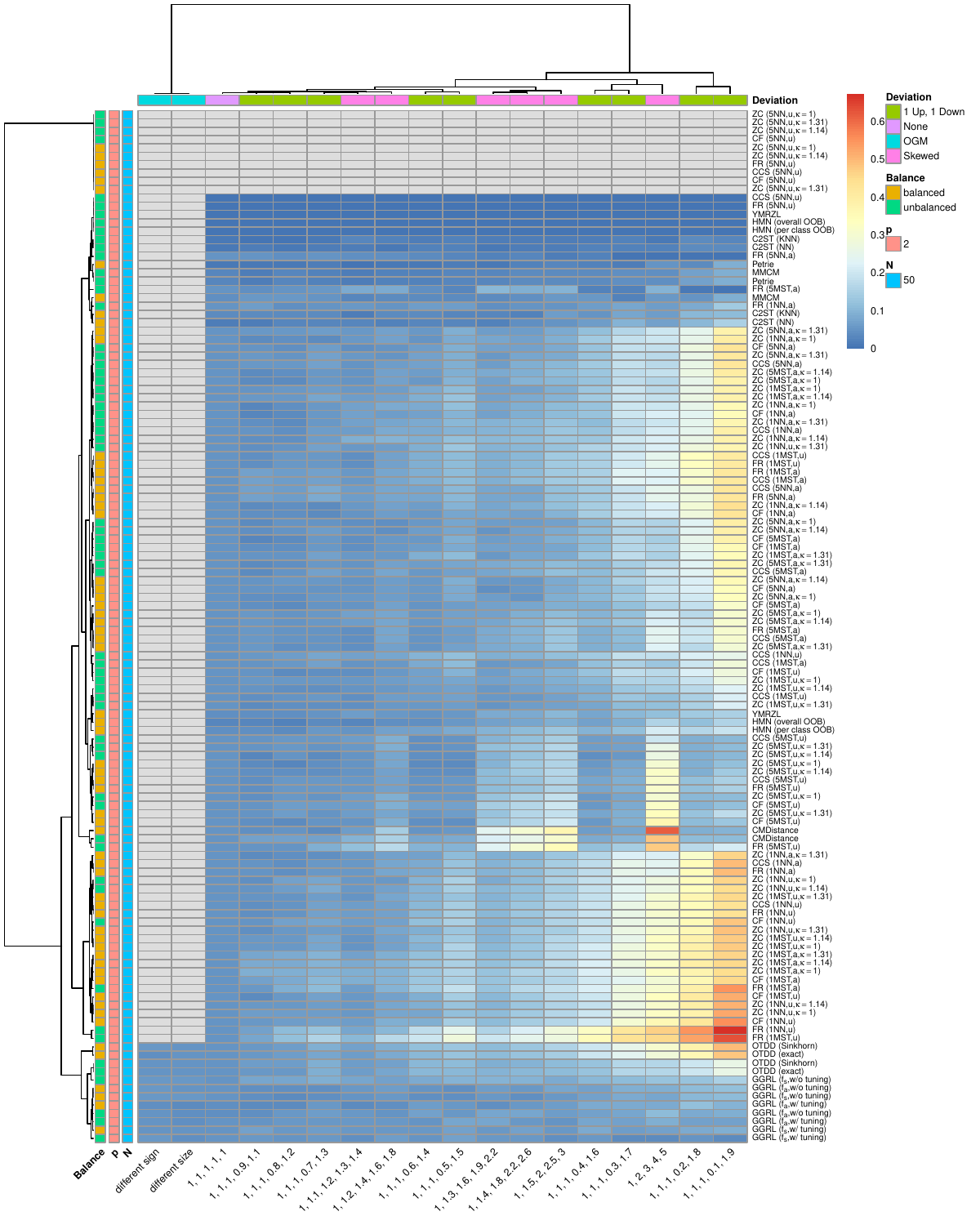}
		\caption{Clustering of PESR values per deviation ($x$-axis) and per method and sample size balance ($y$-axis) for two multinomial datasets with $N =50$ and $p = 2$. The values on the $x$-axis give the weight vector (unnormalized class probabilities) of the first deviating dataset. High PESR values (red) correspond to good performance.}
		\label{fig:heatmap.cat.no.y.multinom.N50.p2}
	\end{figure}
	
	\vspace*{-1cm}
	\subsection[Memory and Runtime for k=2]{Memory and Runtime for $k=2$} \label{app:add.figs.benchmarks}
	\begin{figure}[H]
		\centering
		\resizebox{!}{0.78\paperheight}{%
			\includegraphics[width=\linewidth]{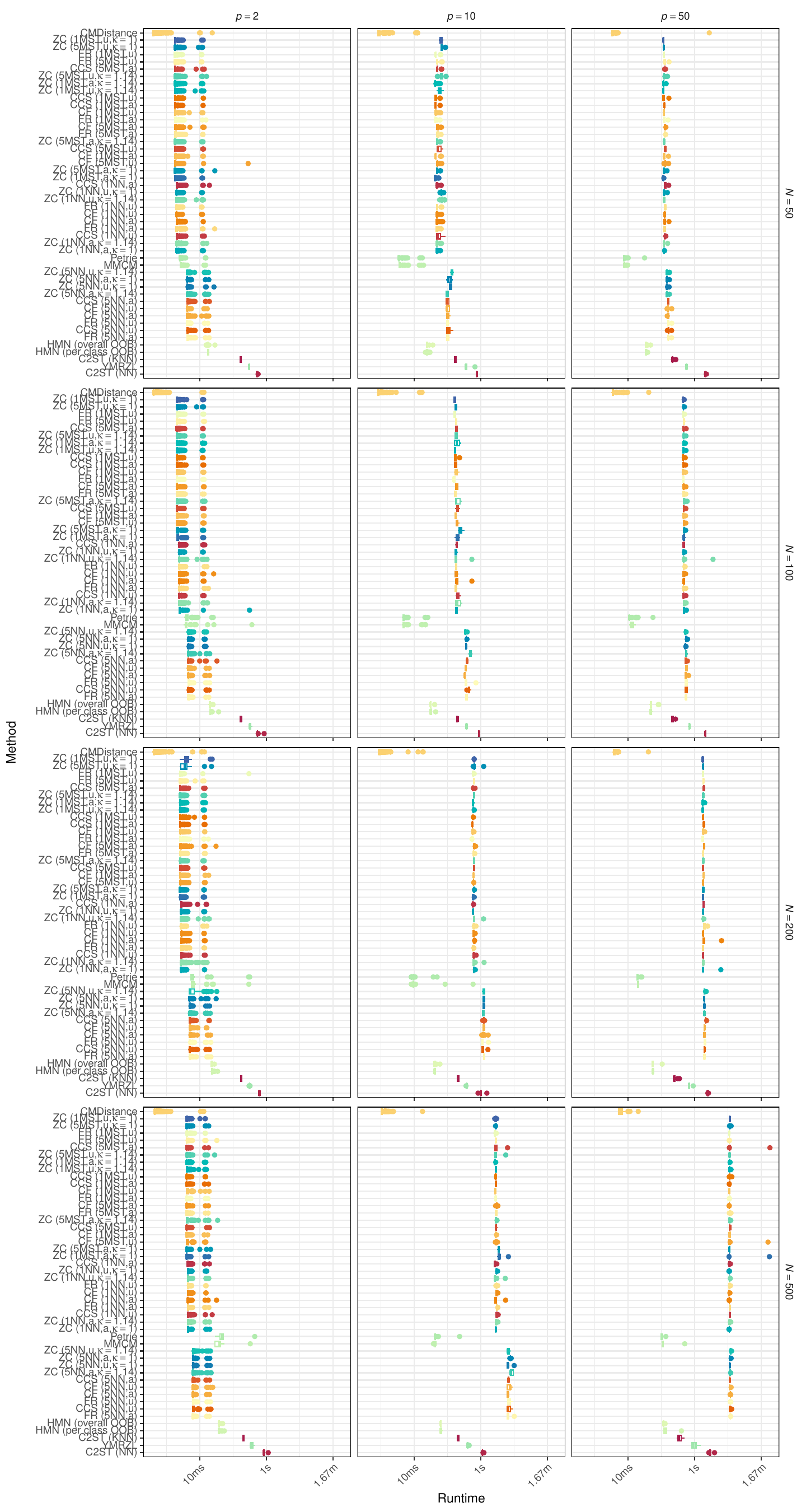}
		}
		\caption{Runtime comparison for the scenario with two binary datasets with balanced class probabilities and equal sample sizes. Ten repetitions were performed for each method.} \label{fig:runtime.cat.no.y.full}
	\end{figure}
	
	\begin{figure}[H]
		\centering
		\resizebox{!}{0.78\paperheight}{%
			\includegraphics[width=\linewidth]{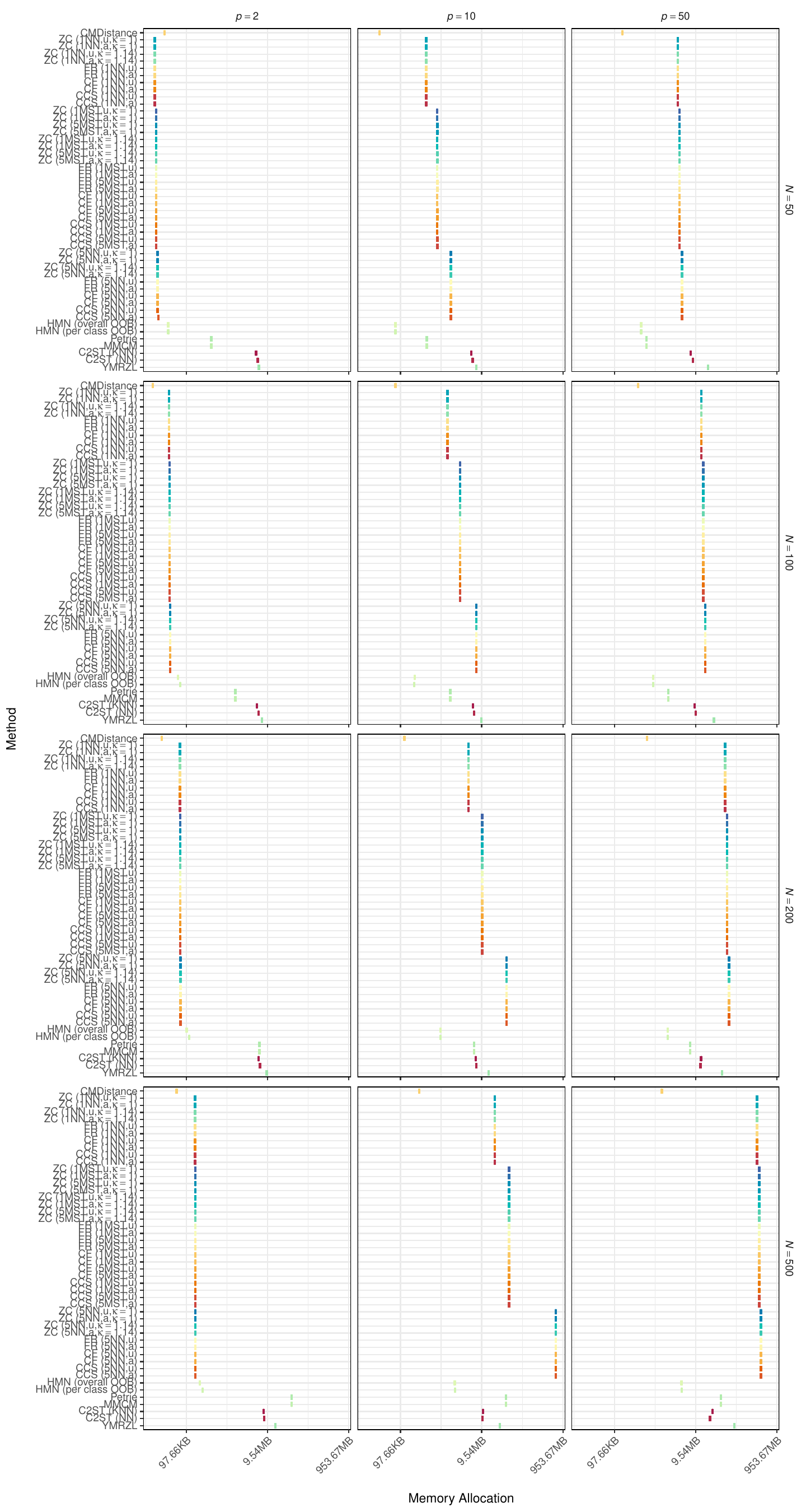}
		}
		\caption{Memory consumption comparison for the scenario with two binary datasets with balanced class probabilities and equal sample sizes. One repetition was performed for each method.} \label{fig:mem.cat.no.y.full}
	\end{figure}
	
	\begin{figure}[H]
		\centering
		\includegraphics[width=\linewidth]{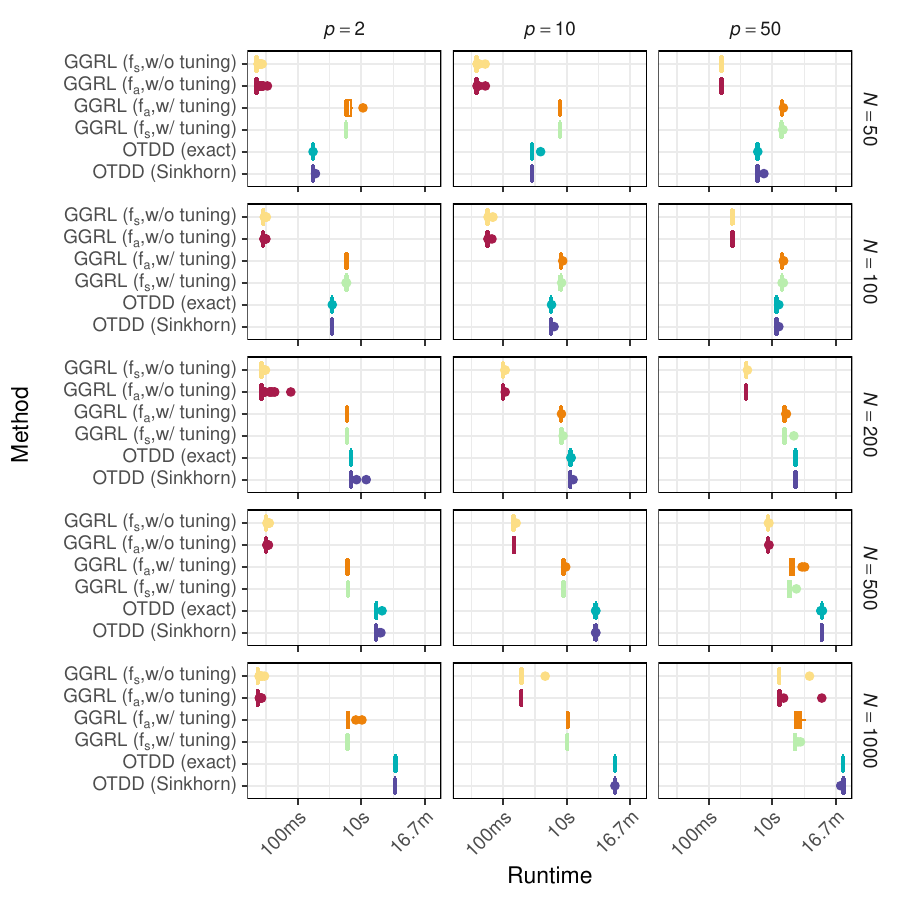}
		\caption{Runtime comparison for the scenario with two binary datasets with balanced class probabilities and equal sample sizes and with a target variable. Ten repetitions were performed for each method.} \label{fig:runtime.cat.y.full}
	\end{figure}
	
	\begin{figure}[H]
		\centering
		\includegraphics[width=\linewidth]{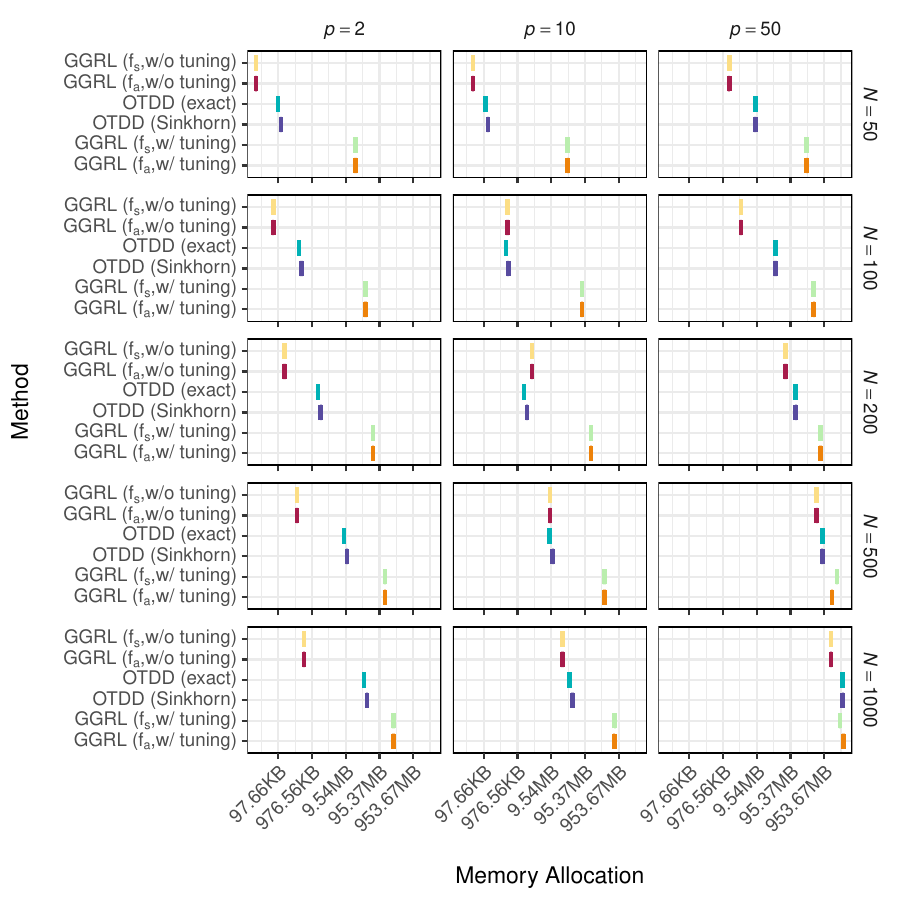}
		\caption{Memory consumption comparison for the scenario with two binary datasets with balanced class probabilities and equal sample sizes and with a target variable. One repetition was performed for each method.} \label{fig:mem.cat.y.full}
	\end{figure}

\end{document}
\typeout{get arXiv to do 4 passes: Label(s) may have changed. Rerun}